\documentclass[12pt]{book}

\usepackage{graphicx,color}
\usepackage{amssymb}
\usepackage[dvipdfm]{hyperref}

\newcommand{\D}{{\cal D}({\bf R})}
\newcommand{\E}{{\cal E}}
\newcommand{\F}{{\bf F}}
\newcommand{\FF}{{{\cal F}_2}}
\newcommand{\EE}{{\E_2}}
\newcommand{\dR}{d{\bf R}}
\newcommand{\intint}{\int\!\!\!\!\int}
\newcommand{\K}{{\cal K}}
\renewcommand{\k}{{\vec k}}
\newcommand{\dk}{d{\vec k}}

\newcommand{\R}{{\bf R}}
\newcommand{\Rc}{{\vec{\cal R}}}
\newcommand{\Rm}{R_{m}}
\renewcommand{\r}{{\vec r}}
\newcommand{\dr}{{d\vec r}}
\newcommand{\ri}{\r_i}
\newcommand{\rj}{\r_j}
\newcommand{\rk}{\r_k}
\newcommand{\rik}{{\ri-\rk}}
\newcommand{\rjk}{{\rj-\rk}}
\newcommand{\rij}{{\ri-\rj}}
\newcommand{\rN}{{\r_1,...,\r_N}}
\newcommand{\rv}{\r}
\newcommand{\RR}{{\vec R}}

\newcommand{\sign}{\mathop{\rm sign}\nolimits}
\renewcommand{\kappa}{\varkappa}
\newcommand{\tr}{\mathop{\rm tr}\nolimits}

\newcommand{\arcctg}{\mathop{\rm arccot}\nolimits}

\newcommand{\arctg}{\mathop{\rm arctan}\nolimits}

\newcommand{\ch}{\mathop{\rm cosh}\nolimits}
\newcommand{\ctg}{\mathop{\rm cotan}\nolimits}
\newcommand{\cth}{\mathop{\rm cotanh}\nolimits}
\newcommand{\tg}{\mathop{\rm tan}\nolimits}
\newcommand{\sh}{\mathop{\rm sinh}\nolimits}
\renewcommand{\th}{\mathop{\rm tanh}\nolimits}

\addtocounter{secnumdepth}{3}
\addtocounter{tocdepth}{3}
\hypersetup{pdfpagemode=FullScreen}

\begin{document}
\bibliographystyle{hep}

\thispagestyle{empty}
{\bf
\Large
\centerline{Universit\`a degli Studi di Trento}
\vspace{1.0cm}
\centerline{Facolt\`a di Scienze Matematiche Fisiche e Naturali}
\vspace{1.5cm}
\large
\centerline{Tesi di Dottorato di Ricerca in Fisica}
}
\vspace{4.0cm}

\Huge
{\bf
\centerline{QUANTUM MONTE CARLO}
\vspace{0.5cm}
\centerline{STUDY}
\vspace{0.5cm}
\centerline{OF ULTRACOLD GASES}
}

\vspace{5.0cm}

\normalsize
$$
\begin{array}{p{6cm}p{3cm}p{6cm}}
\bf Relatori:&&\bf Candidato:\\
{\bf\large Dr. Stefano Giorgini}&&{\bf\large G.E. Astrakharchik}\\
{\bf\large Prof. Lev P. Pitaevskii}&&
\end{array}
$$

\vspace{2cm}

\centerline{\bf Dottorato di Ricerca in Fisica, XVII Ciclo}
\vspace{0.2cm}
\centerline{\bf 15 Dicembre 2004}

\pagenumbering{roman}
\setcounter{page}{1}

\tableofcontents

\pagenumbering{arabic}
\setcounter{page}{1}

\newpage
\section*{Notation and abbreviations}
\addcontentsline{toc}{chapter}{Notation and abbreviations}
For the notation of other quantities either an unambiguous standard notation is
used, or the notation is given explicitly in the text. List of the special notation
used throughout the Dissertation:
\begin{table}[h!]
\begin{tabular}{lll}
{\it SYMBOL}&{\it MEANING}&{\it DEFINITION}\\
$a_{1D}$& one-dimensional $s$-wave scattering length&(\ref{defa1D})\\
$a_{3D}$& three-dimensional $s$-wave scattering length&(\ref{a3D})\\
$a_\perp$& oscillator length of the transverse confinement &$a_\perp=\sqrt{\hbar/m\omega_\perp}$\\
$a_z$& oscillator length of the longitudinal confinement &$a_z=\sqrt{\hbar/m\omega_z}$\\
$D$ [dimensionless] & number of dimensions&\\
$D$ & diffusion constant &$D = \hbar^2/2m$\\
$E$ & total energy of the system \\
$E^{loc}(\R)$& local energy of a walker $\R$& (\ref{Eloc})\\
$\E^{loc}(r)$ & Bijl-Jastrow component of a local energy & (\ref{e})\\
$\F(\R)$ & drift force&(\ref{DriftForce})\\
$\FF(r)$ & Bijl-Jastrow component of the drift force& (\ref{F})\\
$f_1(\r)$ & one-body Bijl-Jastrow term& see (\ref{Jastrow})\\
$f_2(r)$ & two-body Bijl-Jastrow term& see (\ref{Jastrow})\\
$g_1(r)$ & non-diagonal element of the OBDM & (\ref{g21})\\
$g_2(r)$ & pair-distribution function & (\ref{g2hom})\\
$g_3(0)$ & value at zero of the three-body correlation function & (\ref{g3})\\
$g_{1D}$ & one-dimensional coupling constant & (\ref{g1D})\\
$g_{3D}$ & three-dimensional coupling constant & (\ref{g3D})\\
$L$ & size of the system or side of the simulation box\\
$m$ & particle mass&\\
$n_{1D}$ & linear density& $n_{1D} = N/L$\\
$n_{3D}$ & (total) particle density& $n_{3D} = N/V$\\
$N$ & number of particles&\\
$\r_i$& coordinate of $i$-th particle&$\r_i = (x_i,y_i, z_i)$\\
$\R$ & a point in $D N$-dimensional phase space (a {\it walker})&$\R =\{\r_1,...,\r_N\}$\\
$R$ & range of the potential&(\ref{range})\\
$\Rm$ & (variational) matching distance\\
$R_z$ & size of the cloud in $z$-direction \\
$R_\perp$ & size of the cloud in the transverse direction \\
$u(r)$& exponentiation of the Bijl-Jastrov term& (\ref{u2})\\
$V_{ext}(\r)$ & external potential\\
$V_{int}(|\rij|)$ & pair-interaction potential\\
$\lambda$& anisotropy parameter ({\it aspect ratio})&$\lambda=\omega_z/\omega_\perp$\\
$\mu$ [units of energy] & chemical potential&\\
$\mu$ [units of mass] & reduced mass &$\mu = m/2$\\
$\tau$ & imaginary time& $\tau = i t$\\
$\phi_0(\R)$ & ground state many body wave function \\
$\psi_T(\R)$ & trial many body wave function\\
$\omega_\perp$ & frequency of the transverse harmonic confinement& \\
$\omega_z$ & frequency of the longitudinal harmonic confinement& \\
\end{tabular}
\end{table}

List of used abbreviations:

\begin{table}[h!]
\begin{tabular}{ll}
BCS & Bardeen Cooper Schriffer\\
BEC & Bose-Einstein condensation\\
BJ & Bijl-Jastrow\\
DMC & Diffusion Monte Carlo\\
FN-MC & Fixed Node Monte Carlo\\
JS & Jastrow-Slater\\
GP & Gross-Pitaevskii\\
HR & hard rod\\
HS & hard sphere\\
LDA & local density approximation\\
LL & Lieb-Liniger\\
OBDM & one-body density matrix\\
TG & Tonks-Girardeau\\
SR & short range\\
SS & soft sphere\\
SW & square well\\
QMC & Quantum Monte Carlo\\
VMC & Variational Monte Carlo\\
\end{tabular}
\end{table}

\chapter*{Introduction}
\addcontentsline{toc}{chapter}{Introduction}
Although proposed by Einstein \cite{Einstein24, Einstein25} for an ideal quantum gas
a long time ago Bose-Einstein condensation (BEC) almost remained only a mathematical
artifact. After many years of intense experimental activity in 1995 BEC was observed
in alkali vapours in a remarkable series of experiments \cite{Anderson95,Davis95}.
Since that time there has been an explosion of experimental and theoretical interest
worldwide in the study of dilute Bose gases. The Bose condensate, a macroscopically
occupied quantum wave, exhibits peculiar properties and often is referred to as a
new state of matter.

One of the directions where very important achievements were made in the years
passed from the first realization of the BEC in gases, is the the development
techniques of cooling quantum gases to extremely low temperatures and of trapping
methods allowing for the realization of low-dimensional geometries (for example,
\cite{Gorlitz01,Schreck01,Greiner01,Moritz03,Tolra04,Stoferle04}). This combination
leads to highly non-trivial effects, like the {\it fermionization} of bosons which
may happen in one-dimensional quantum system. The progress in cooling methods has
led to the possibility of observing both Bose and Fermi systems at temperatures much
smaller than the degeneration temperature. The development of the techniques of
diagnostics allows to get a quantitative description of the system under
investigation: the size of the cloud, release energy, the momentum distribution,
structure factor and frequencies of collective excitations are available in many
experiments.

The most widely used approach for the description of quantum degenerate bosonic
system is the mean-field Gross-Pitaevskii (GP) theory\cite{Gross61,Pitaevskii61}. In
this approach all particles are considered to be in the same quantum state described
by the condensate wave function, which evolves in time according to the
Gross-Pitaevskii equation. The mean-field approach has proven very useful as it is
mathematically much easier to solve the equation for one particle in an effective
field of other particles, than to solve the full many body problem. The GP approach
holds when the depletion of the condensate is negligible or, more generally, when
the correlation length is much larger than the interparticle distance.
One-dimensional gases in a regime of strong quantum correlations where the above
condition fails, have already been realized.

The problem of solving the many-body Schr\"odinger equation and finding
multidimensional averages integrating out $3N$ degrees of freedom is very
complicated. The Monte Carlo methods are indispensable tools in the calculation of
multidimensional integrals (see, for example\cite{Ceperley95,Casulleras95,Moroni95})
and have been shown to be highly useful in the investigation of quantum systems
(see, for example, \cite{Astrakharchik02a,Gori-Giorgi04}).
We are most of all interested in the quantum properties of the system at zero
temperature.
The diffusion Monte Carlo method is the best suited for this type of study.

\begin{figure}
\begin{center}
\includegraphics*[width=0.4\columnwidth,angle=0]{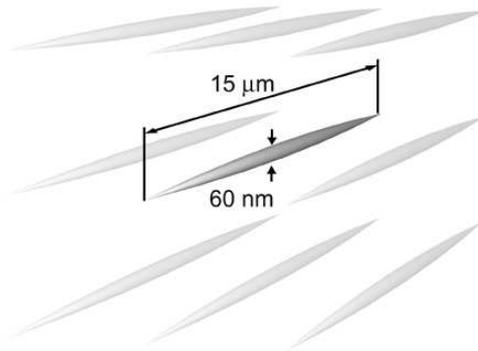}
\caption{An example of realization of a quasi one dimensional bosonic system
(taken from \cite{Moritz03}). Two counterpropagating laser beams create a tight two
dimensional optical lattice. In the transversal direction the gas is in the ground
state of the confining potential. Excitation of the next levels is highly suppressed
due to the low temperature $k_BT/\hbar\omega_\perp<6\cdot10^{-3}$ and low value of the
available one-dimensional energy $\mu/\hbar\omega_\perp-1<0.1$.}
\label{FigZurich}
\end{center}
\end{figure}

The confining potentials (magnetic trapping, optical trapping, etc.) can be well
described by harmonic potentials. If the frequencies of the confinement are equal in
all three directions ({\it i.e.} the trap is spherical) the sample of the gas inside
is three-dimensional. If, instead, the trap is made tighter in two directions, the
shape of the gas cloud becomes elongated, and in the limit $\omega_\perp\gg\omega_z$
the system becomes effectively one-dimensional (see Fig.~\ref{FigZurich}). The
crossover of a trapped gas from three- to one- dimensional behaviour is highly
interesting and we have studied it using a Quantum Monte Carlo
method\cite{Astrakharchik02b}.

The reduced dimensionality enhances the effect of interactions and the properties of
a one-dimensional system can be very different from the ones of a three-dimensional
gas. The phenomenon of Bose condensation is absent in a one-dimensional homogeneous
system. Furthermore, the behavior of repulsive bosons is very peculiar in one
dimensional system: in the limit of low density the particles get completely
reflected in the process of two-body collisions (limit of impenetrable particles)
and the interaction between particles plays a role of an effective Pauli principle.
In this Tonks-Girardeu limit fermionization of bose particles happens. The wave
function of strongly interacting bosons can be mapped onto a wave function of ideal
spinless fermions\cite{Girardeau60}. The system of bosons acquires many fermionic
properties: the energy, pair distribution function, static structure factor, etc.
are the same. In the low-density regime beyond mean-field effects are important and
they can not be accounted for by the Gross-Pitaevskii approach, which is valid
instead in the opposite regime of large densities. In the regime of intermediate
densities (recently realized in experiments\cite{Tolra04}) both methods are not
applicable. The system with contact repulsive interaction (Lieb-Liniger gas) is
exactly solvable. Many of its properties are known exactly: the ground-state energy
\cite{Lieb63}, value of the pair distribution function at zero
distance\cite{Gangardt03}, long- and short-range expansions of the
OBDM\cite{Olshanii03}. Still the complete description of correlation functions in
the Lieb-Liniger gas (also value at zero of the three-body correlation function
which was measured in the experiment\cite{Tolra04}) is unknown. The DMC is well
suited to study this problem\cite{Astrakharchik03}.

In a one-dimensional system with attractive contact interactions a two-body bound
state appears for any strength of the $\delta$-potential
While the Pauli exclusion principle prohibits fermions from occupying a state with
the same quantum numbers, bosons are free to populate the lowest state. An exact
result for a system of bosons\cite{McGuire64} shows that the ground state is a
soliton-like state with large negative energy. In a system of two-component fermions
two particles with different spin can form a bound state (particles with parallel
spins can not interact with contact potential), but other particles (or pairs) have
to stay apart. The exact solution\cite{Gaudin67,Krivnov75} shows formation of dimers
which form a gas-like state. In the dilute regime the internal structure of the
composite dimer can be neglected and the system behaves as a Tonks-Girardeau gas of
bosons with twice the mass of the atoms\cite{Astrakharchik04b}.

For a gas of 1D bosons we propose to obtain a large attractive interaction in this
special way: start with a gas of repulsive bosons, increase the strength of the
interactions using the Feshbach resonance up to Tonks-Girardeau regime
($a_{1D}\to-0$) and then change the sign of the interaction ($a_{1D}\to +0$). The
new state (``super-Tonks'') will have correlations which are even stronger than in
the Tonks-Girardeau gas. The super-Tonks gas is a metastable state which has
analogies with a gas of hard-rods of size $a_{1D}$. The super-Tonks is a metastable
state. A very important question is to find out if the super-Tonks gas is stable
and, thus, can be realized in an experiment. We use the variational Monte Carlo
method to investigate this problem\cite{Astrakharchik04d}. Interaction effects in
quasi-one-dimensional systems can be studied in experiments by exciting
``breathing'' mode oscillations. The local density approximation can be used to
obtain the density profile of a trapped system of bosons or fermions. We solve the
local density approximation for a quite general class of equations of state
analytically. Using the sum rule approach we extract the oscillation frequencies
numerically for all densities and analytically in the limits where the expansion of
the equation of state is known.

A peculiar property of a low temperature system is the possibility of being
superfluid. One of the most important predictions of Landau theory of superfluidity
is the existence of a finite critical velocity. If a body moves in a superfluid at
$T=0$ with velocity $V$ less then $v_c$, the motion is dissipationless. At $V>v_c$ a
drag force arises because elementary excitations are created. Recently, existence of
a critical velocity for the superfluid motion in a Bose-Einstein condensed gas was
confirmed in various experiments. For example, at MIT a trapped condensate was
stirred by a laser beam and the dissipated energy was
measured\cite{Raman99,Onofrio00}. According to Landau if one imagines to move a
small body through the system and there is no normal part, no dissipation will
happen if the speed is smaller than the speed of sound. We calculate the effect of a
small impurity moving through a condensate which is described by the
Gross-Pitaevskii equation\cite{Astrakharchik02c}. We want to find an answer to a question which is rather
complicated.
%
We know that in the large density mean-field regime that the system should be
superfluid. On the other side, in the Tonks-Girardeau regime the system is mapped to
the fermions, which are not superfluid.

An important question concerns effective interactions in 1D, {\it i.e.} how the
one-dimensional effective coupling constant is related to the three dimensional
$s$-wave scattering length. A solution for the problem of two-body scattering on a
pseudopotential in a waveguide was found by Olshanii\cite{Olshanii98} and shows a
resonant behavior in the regime $a_{3D}\sim a_\perp$ due to virtual excitations of
transverse modes confinement levels. Since in normal experimental conditions
$a_{3D}\ll a_\perp$, a resonant scattering in the vicinity of a Feshbach resonance
should be used in order to enter this regime. In experiments a possible way to
fulfill this condition is to make use of the Feshbach resonance.
An important question is to prove the existence of the confinement induced resonance
in a many body system. We consider a resonant scattering on a smooth attractive
potential of very small range and use Fixed-Node Monte Carlo to study the problem of
quasi-one-dimensional Bose gases with large scattering
length\cite{Astrakharchik04a,Astrakharchik04c}.

The use of Feshbach resonance allows one to vary the interaction strength in a
controlled way and tune the scattering length essentially to any arbitrary value.
Recent experiments on two-component ultracold atomic Fermi gases near a Feshbach
resonance have opened the possibility of investigating the crossover from a
Bose-Einstein condensate (BEC) to a Bardeen-Cooper-Schrieffer (BCS) superfluid. For
positive values of the $s$-wave scattering length $a_{3D}$, atoms with different
spins are observed to pair into bound molecules which, at low enough temperature,
form a Bose condensate~\cite{Jochim03,Greiner03,Zwierlein03}. The molecular BEC
state is adiabatically converted into an ultracold Fermi gas with $a<0$ and
$k_F|a|\ll 1$~\cite{Bartenstein04,Bourdel04}, where standard BCS theory is expected
to apply. In the crossover region the value of $|a_{3D}|$ can be orders of magnitude
larger than the inverse Fermi wave vector $k_F^{-1}$ and one enters a new
strongly-correlated regime known as unitary
limit~\cite{O'Hara02,Bartenstein04b,Bourdel04}. In dilute systems, for which the
effective range of the interaction $R_0$ is much smaller than the mean interparticle
distance, $k_FR_0\ll 1$, the unitary regime is believed to be universal. In this
regime, the only relevant energy scale should be given by the energy of the
noninteracting Fermi gas. The unitary regime presents a challenge for many-body
theoretical approaches because there is not any obvious small parameter to construct
a well-posed theory. Quantum Monte Carlo techniques are the best suited tools for
treating strongly-correlated systems. We use Fixed-Node Monte Carlo method to obtain
for the first time the equation of state covering all regimes (BEC, unitary,
BCS)\cite{Astrakharchik04e}. The equation of state can be tested in experiments by
measuring the frequencies of collective oscillations. We also investigate the
behavior of correlation functions.

The structure of the Dissertation is as follows.

In the Chapter~\ref{secTools} we introduce the analytical approaches and
approximations used in the subsequent Chapters. Chapter~\ref{secQMC} explains in
details the Quantum Monte Carlo methods used in the study. In Chapter~\ref{sec3D1D}
we consider a system of bosons in an anisotropic trap and study the transition from
a three dimensional behaviour to a quasi one dimensional one as the trap is made
very elongated. We study the properties of a quasi-one-dimensional Bose gas with
resonant scattering in Chapter~\ref{sec1DBG}. The system of $\delta$-interacting
bosons in the case of repulsive interactions (Lieb-Liniger gas) is investigated in
Chapter~\ref{secLL} and in the case of attractive interactions in
Chapter~\ref{secST}. The motion of an impurity as a test for superfluidity is
considered in Chapter~\ref{secdelta}. In the next two chapters we consider systems
of two component fermions in a quasi one dimensional system
(Chapter~\ref{sec1Dfermions}) and in a three-dimensional uniform system
(Chapter~\ref{secBECBCS}). Conclusions are drawn in the last Chapter
(Chapter~\ref{secConclusions}).

\chapter{Tools\label{secTools}}
\section{Introduction}
This Chapter is intended to introduce tools for the subsequent Chapters. Here we
define the quantities (correlation functions, static structure factor, etc.) that
later are used to describe the properties of quantum systems. We explain the analytical methods (Gross-Pitaevskii approach) and
approximations (local density approximation, pseudopotential approximation) used
in our study.
We review the 2-body scattering problem in three- and one-dimensional systems as it
gives insight into the many-body physics and is relevant for the implementation of
the Monte Carlo techniques. Most of the content of the Chapter is standart and is
presented for the completeness of the discussion. Only in several sections some new
results are obtained (Secs.~\ref{secLDA},\ref{secLuttinger}).

The structure of the Chapter is as follows.

In Section~\ref{secIntrCorr} we introduce quantities which characterize a quantum
system and can be accessed in experiments. We start by considering the
representations of the first and second quantization (Secs.~\ref{secCorr2nd},
\ref{secCorr1st}). A special attention is paid to the relation between mean averages
and correlation functions. The calculation of the correlation functions can be
largely simplified in a homogeneous system (Sec.~\ref{secg2hom}), although the case
of trapped systems is also considered (Sec.~\ref{secg2trap}). The momentum
distribution and static structure factor are introduced in Sec.~\ref{secnk}.

The scattering theory is addressed in Section~\ref{secIntrScat}.
The scope of our study is general and we consider the theory
in a three-dimensional system (Sec.~\ref{sec3D}), as well as in a one-dimensional
system (Sec.~\ref{sec1D}). The scattering problem is solved for a number of
potentials that appear in different models. The scattering solutions are used to
construct the trial wave function (see Chapter~\ref{secQMC}) and make comparison of
$N$-body and 2-body physics (see Chapter~\ref{secLL}). We discuss scattering on a
$\delta$-potential ({\it pseudopotential}) in a one dimensional system
(Sec.~\ref{secLLsc}), where the problem is well posed, and also in a three
dimensional system (Sec.~\ref{secPseudo}), where instead a regularization procedure
is needed. We relate the coupling constant to the $s$-wave scattering length (in 1D
and 3D) for the scattering on the pseudopotential, which is a highly important
theoretical tool widely used throughout this dissertation. In the conclusion of
Section~\ref{secIntrScat} we consider the case of resonant scattering, when the
scattering length can be much larger than the range of the potential.

A dilute quantum system of repulsive bosons shows very peculiar properties in 1D.
Fermionization of the bosonic system takes place (Tonks-Girardeau
gas\cite{Girardeau60}), and the particles behave as if they were ideal fermions. We
address some of the properties of an ideal Fermi gas in Section.~\ref{secTGHR}.
The Fermi momentum and Fermi energy of an ideal 1D Fermi gas (Sec.~\ref{secETGgas})
provide an important physical scale not only in the TG regime but in the whole range
of densities. The ground-state energy of a gas of impenetrable particles (hard-rod
gas) is calculated in Sec.~\ref{secEHR}. The properties of a gas of hard-rods are
important in the proposed relation of such a system to a short-range attractive
potential (super-Tonks) gas (see Chapter~\ref{secST}). Also the HR gas equation of
state is related to the expansion of the energy of a Lieb-Liniger gas in the regime
of strong correlations and this expansion is relevant for the estimation of the
properties of correlation functions in this regime.

In this dissertation the Monte Carlo results are systematically compared to the
predictions of the mean-field Gross-Pitaevskii approach (when GP equation is
applicable). In Section~\ref{secVarGPE} the GP equation is derived from a energy
functional, which later is used to study the properties of a condensate disturbed by
an impurity (see Chapter~\ref{secdelta}). In a similar way the GP equations in
restricted geometries (cigar- and disk- shaped condensates) are derived in
Sec.~\ref{secGP1D2D}. In this approach virtual excitations in the tight direction
are neglected and the resulting expression of the coupling constants is to be
compared with the one of an exact solution of a two-body scattering problem in 1D
obtained by Olshanii \cite{Olshanii98}.

If the equation of state of the homogeneous system is known, the local density
approximation allows one to estimate the properties of a system in the presence of
an external confinement. The general idea of this method is explained in
Sec.~\ref{secLDAgen} and the characteristic parameters in three- and one-dimensional
systems are discussed. We propose an exact solutions of the LDA problem for a
``perturbative'' equation of state both in one dimension (Sec.~\ref{secLDA1D}) and
in three dimensions (Sec.~\ref{secLDA3D}). The obtained formulas are applied to
bosonic systems (see Chapter~\ref{secST}) as well as fermionic systems (see
Chapter~\ref{sec1Dfermions}). In particular the LDA method together with the sum
rule approach (in 1D) and scaling approach (in 1D) can be used to estimate the
frequencies of collective excitations. Expansions for those frequencies are obtained
and later are compared to the result of the numerical results obtained using LDA
(see Figs.~\ref{figHR4}, \ref{fig1Df4}). Finally in Section~\ref{secSkLDA} we
consider the LDA applied to the Tonks-Girardeau gas and calculate the static
structure factor in a trapped system.

This introductory Chapter is concluded with a newly proposed derivation of the
dynamic form factor, pair distribution function and the one-body density matrix of a
weakly interacting bosonic gas in 1D. The Haldane description \cite{Haldane81} of
this system is corrected in order to replace the phononic excitation spectrum with
the more precise Bogoliubov spectrum. In this way we eliminate logarithmic
divergences present in the problem and estimate the prefactors of the long-range
asymptotics. In particular the coefficient of the decay of the OBDM
(Sec.~\ref{secPopov}) is compared to the exact DMC result (see Sec.~\ref{secLL}) and
is found to be extremely accurate (less than $0.3\%$ error).

\newpage
\section{Correlation functions and related quantities\label{secIntrCorr}}

\subsection{Correlation functions: second quantization form\label{secCorr2nd}}

Quantum description of identical particles can be conveniently done in terms of the
creation and annihilation field operators. The operator $\hat\Psi^\dagger(\r)$ puts
a particle into a point $\r$, while $\hat\Psi(\r)$ destroys a particle at the same
point. Field operators can be conveniently presented in terms of creation $\hat a_k$
(annihilation $\hat a^\dagger$) operator that puts (destroys) a particle in a single
particle orbital $\varphi_\k(\r)$:
\begin{eqnarray}
\left\{
\begin{array}{lll}
\hat\Psi^\dagger(\r) &=& \sum\limits_\k \varphi^*_\k(\r)\,\hat a_\k^\dagger\\
\hat\Psi(\r) &=& \sum\limits_\k \varphi_\k(\r)\,\hat a_\k
\end{array},
\right.
\label{Psi}
\end{eqnarray}
In a uniform gas occupying a volume $V$ single particle orbitals $\varphi_\k(\r)$
are plain waves $\varphi_\k(\r) = \frac{1}{\sqrt{V}}\,e^{i\k\r}$. In a system of
bosons operators (\ref{Psi}) commute $[\Psi(\r),\Psi^\dagger(\r')]=\delta(\r-\r')$,
$[\Psi(\r),\Psi(\r')]=0$ and anticommute in a system of fermions.

Before giving the definition of the correlation functions in terms of the field
operators (\ref{Psi}), let us discuss how the correlation functions come naturally
from the calculation of the mean values of operators. We shall start with a very
general form of a Hamiltonian consisting of one- and two- body operators
\begin{eqnarray}
\hat H = \hat F^{(1)}+\hat F^{(2)},
\end{eqnarray}
where the one-body operator $\hat F^{(1)}$ is a sum of operators $\hat f^{(1)}_i$
each acting only on one particle:
\begin{eqnarray}
\hat F^{(1)} &=& \sum\limits_{i=1}^N \hat f^{(1)}(\r_i)\\
\hat F^{(2)} &=& \frac{1}{2}\sum\limits_{i\ne j}^N \hat f^{(2)}(\r_i,\r_j)
\end{eqnarray}

For example, it can be an external potential $f^{(1)}(\r) = V_{ext}(\r)$ (and, thus,
the operator is diagonal the coordinate representation) or it can be the kinetic
energy $f^{(1)}(p)=p^2/2m$ (the operator is diagonal in the momentum
representation). A commonly used two-body operator is a particle-particle
interaction is usually defined in the coordinate representation $f^{(2)}(\r_1,\r_2)
= V_{int}(\r_1,\r_2)$.

In the second quantization representation the one-body $\hat F^{(1)}$ and two-body
$\hat F^{(2)}$ operators are conveniently expressed in terms of the field operators
(\ref{Psi}).
%
\begin{eqnarray}
\hat F^{(1)} &=&
\intint\hat\Psi^\dagger(\r)f^{(1)}(\r,\r')\hat\Psi(\r')\,\dr\dr'\\
\hat F^{(2)} &=&
\frac{1}{2}\intint\hat\Psi^\dagger(\r)\hat\Psi^\dagger(\r')
f^{(2)}(\r,\r')\hat\Psi(\r')\hat\Psi(\r)\,\dr\dr'
\label{thF2}
\end{eqnarray}

Here we assume that the one-body operator can be either local
$\langle\r|f^{(1)}|\r'\rangle = f^{(1)}(\r)\delta(\r-\r')$ (like in the case of an
external field), either non local (like in the case of the kinetic energy), so, in
general, we have two arguments $f^{(1)}=f^{(1)}(\r,\r')$. Instead, for the two-body
term we always assume that it is local ({\it i.e.} it has a form similar to the
particle-particle interaction energy) $\langle\r_1,\r_2|f^{(1)}|\r_1',\r_2'\rangle
= f^{(2)}(\r_1,\r_2) \delta(\r_1-\r_1')\delta(\r_2-\r_2')$, so in (\ref{thF2}) we
have only two arguments instead of four.

The quantum averages of $\hat F^{(1)}$ and $\hat F^{(2)}$ can be extracted from
$\hat f^{(1)}$ and $\hat f^{(2)}$ if the {\it correlation functions} are
known\footnote{At zero temperature the expectation value $\langle...\rangle$ is
taken with respect to the ground state of the system.}:
\begin{eqnarray}
\label{F1}
\langle\hat F^{(1)}\rangle &=&\intint\hat f^{(1)}(\r,\r')G_1(\r,\r')\,\dr\dr'\\
\langle\hat F^{(2)}\rangle &=&\frac{1}{2}\intint\hat f^{(2)}(\r,\r')G_2(\r,\r')\,\dr\dr'
\label{F2}
\end{eqnarray}

Here $G_1(\r,\r')$ and $G_2(\r,\r')$ are the {\it non normalized correlation
functions} defined as
\begin{eqnarray}
\label{G1}
G_1(\r,\r')&=&\langle\hat\Psi^\dagger(\r)\hat\Psi(\r')\rangle\\
G_2(\r,\r')&=&\langle\hat\Psi^\dagger(\r)\hat\Psi^\dagger(\r')\hat\Psi(\r')\hat\Psi(\r)\rangle
\label{G2}
\end{eqnarray}

The function $G_1(\r,\r')$ characterizes correlations existing between values of the
field in two different points $\r$ and $\r'$. The total phase does not enter in the
definition, but instead the relative phase between two points is important. The
diagonal term $\r=\r'$ of (\ref{G1}) gives the density of the system
$n(\r)=\langle\hat\Psi^\dagger(\r)\hat\Psi(\r)\rangle=G_1(\r,\r)$, so the trace of
the matrix $G_1$ gives the total number of particles $\tr G_1 = \int G_1(\r,\r)\,\dr
= N$. The function $G_2(\r,\r')$ characterizes the density correlations between
points $\r$ and $\r'$.

It is convenient to introduce dimensionless versions of functions (\ref{G1}) and
(\ref{G2}):
\begin{eqnarray}
\label{g1}
g_1(\r,\r') &=& \frac{G_1(\r,\r')}{\sqrt{G_1(\r,\r)}\sqrt{G_1(\r',\r')}}\\
g_2(\r,\r') &=& \frac{G_2(\r,\r')}{G_1(\r,\r)G_1(\r',\r')}
\label{g2}
\end{eqnarray}

The function (\ref{g1}) is limited to the range $[0,1]$ and can be understood as the
probability to destroy a particle at $\r$ and put it at $\r'$. It is always
possible to put a particle to the place where it was, so $g_1(\r,\r)=1$.
The non-diagonal long range asymptotic vanishes in trapped systems and also in
homogeneous systems in the absence of Bose-Einstein condensation
$g_1(\r,\r')\to 0, |\r,\r'|\to\infty$.

A more detailed introduction to the analytic properties of the correlation functions
can be found, for example, in \cite{Glauber63,Naraschewski99,Gangardt03b}

\subsection{Correlation functions: first quantization form\label{secCorr1st}}

The meaning of the correlation functions (\ref{g1},\ref{g2}) is best of all
understood in terms of the field operators as discussed in the previous Section.
Instead for the implementation of the Monte-Carlo technique it is necessary to
express the correlation functions in terms of the wave function $\psi(\R)$ of the
system. The easiest way to do SO is to find an expression of the operator average in
form similar to (\ref{F1},\ref{F2}).

The mean value of a one-body operator in the first quantization Is written as
\begin{eqnarray}
\nonumber
\langle F^{(1)} \rangle
= \frac{\int \psi^*(\R) F^{(1)}(\R) \psi(\R)\,\dR}{\int |\psi(\R)|^2\,\dR}=
\frac{\sum\limits_{i=1}^N\int
\psi^*(\r_1, ..., \r_N) f^{(1)}(\r_i) \psi(\r_1, ...., \r_N)\,\dR}{|\psi(\R)|^2\,\dR}
= \\= \frac{N \intint F^{(1)}(\r_1,\r_1')
|\psi(\r_1, ..., \r_N)|^2\,\dR}{\int|\psi(\R)|^2\,\dR}
=\intint f^{(1)}(\r,\r') G_1(\r, \r')\,\dr\dr',
\label{F1mean}
\end{eqnarray}
where $G_1(\r,\r')$ stands for
\begin{eqnarray}
G_1(\r, \r') =
\frac{N \int \psi^*(\r, \r_2, ..., \r_N) \psi^*(\r', \r_2, ..., \r_N)\,\dr_2 ... \dr_N}
{\int \psi^*(\r_1, ..., \r_N) \psi^*(\r_1, ..., \r_N)\,\dr_1 ... \dr_N}
\end{eqnarray}

The expression for the two-body correlation function (\ref{G2}) can be deduced from
the average of a two-body operator (\ref{F2}):
\begin{eqnarray}
\langle F^{(2)} \rangle
= \frac{\int \psi^*(\R) F^{(2)}(\R) \psi(\R)\,\dR}{\int |\psi(\R)|^2\,\dR}=
\frac{\frac{1}{2}\sum\limits_{i\ne j}^N\int
\psi^*(\r_1, ..., \r_N) f^{(2)}(\r_i, \r_j) \psi(\r_1, ...., \r_N)\,\dR}{|\psi(\R)|^2\,dR}
= \\= \frac{N(N-1) \int f^{(2)}(\r_1, \r_2)
|\psi(\r_1, ..., \r_N)|^2\,\dR}{2\int|\psi^*(\R)|^2\,\dR}
=\frac{1}{2}\intint f^{(2)}(\r_1, \r_2) g_2(\r_1, \r_2)\,\dr_1\dr_2,
\label{secQnt}
\end{eqnarray}
where the first quantization expression for the two body correlation function is
\begin{eqnarray}
G_2(\r', \r'') =
\frac{N(N-1) \int |\psi(\r', \r'', \r_3,..., \r_N)|^2\,\dr_3 ... \dr_N}
{\int|\psi(\r_1, ..., \r_N)|^2\,\dr_1 ... \dr_N}
\end{eqnarray}

\subsection{Homogeneous system\label{secg2hom}}

Many simplifications can be done in a homogeneous system due to the presence of the
translational symmetry. The correlation functions discussed above depend only on the
relative distance between two coordinates $|\r-\r'|$. The diagonal element of $G_1$
is simply a constant $G_1(\r,\r) = n$. The non-diagonal element of the normalized
one-body density matrix (in following we will address it as OBDM) is written as
\begin{eqnarray}
g_1(r) =
\frac{N}{n}\frac{\int \psi^*(\r, \r_2, ..., \r_N) \psi(0, \r_2, ..., \r_N)\,\dr_2 ... \dr_N}
{\int |\psi(\r_1, ..., \r_N)|^2\,\dr_1 ... \dr_N}
\label{g21}
\label{g1hom}
\end{eqnarray}

The normalized two-body density matrix (pair distribution function) is then given by
\begin{eqnarray}
g_2(r) =
\frac{N(N-1) \int |\psi(\r, 0, \r_3,..., \r_N)|^2\,\dr_3 ... \dr_N}
{n^2\int|\psi(\r_1, ..., \r_N)|^2\,\dr_1 ... \dr_N}
\label{g2hom}
\end{eqnarray}

At the zero temperature some of the properties of the pair distribution can be
easily understood. At large distances the correlation between particles becomes
weaker and weaker and we can approximate the field operator
$\hat{\Psi}(r)=\sqrt{\hat{n}(r)}e^{i~\hat{\phi}(r)}\approx $ $\sqrt{\hat{n}(r)},
r\rightarrow \infty$ and at zero temperature one has $g_2(r)\rightarrow
1-\frac{1}{N}$ and $g_2(r)$ approaches unity in the thermodynamic limit. On the
contrary at short distances particles ``feel'' each other and the value at zero can
be very different from the value in the bulk. In case of impenetrable particles, two
particles are not allowed to overlap, thus $g_2(0)=0$. For purely repulsive
interaction $g_2(0)<1$ and for purely attractive $g_2(0)>1$.

In the average of a two-body operator (\ref{F2}) it is possible to integrate out the
dummy variable and get a simple expression
\begin{eqnarray}
\left\langle F^{(2)}\right\rangle =\frac{n^{2}}{2}\int\!\!\!\!\int
f^{(2)}(\r_1-\r_2)~g_2(\r_1,\r_2)\,\dr_1\dr_2=\frac{Nn}{2}\int
f^{(2)}(r)~g_{2}(r)\,dr
\end{eqnarray}

For a particular case of a contact potential $V_{int}(r) = g \delta(r)$ the
potential energy is directly related to the value of the pair distribution function
at zero:
\begin{eqnarray}
\frac{E_{int}}{N} = \frac{1}{2}g n g_2(0)
\label{meanEcontact}
\end{eqnarray}

We will also give definition of the {\it three-body density matrix}\footnote{It is
interesting to note that the normalization factor $N!/(N-m)!$ of the $m$-body
correlation function comes from the number of ways to make groups of $m$-particles
out of $N$ particles, but at the same time can be found from the properties of the
field operators $\hat\Psi^\dagger|N\rangle=\sqrt{N}|N-1\rangle$,
$\hat\Psi|N\rangle=\sqrt{N+1}|N+1\rangle$}

\begin{eqnarray}
g_3(0) = \frac{N(N-1)(N-2)}{n^3}
\frac{\int|\psi(0, 0, 0, \r_4, ..., \r_N)|^2\, \dr_4... \dr_N}
{\int|\psi(\r_1, ..., \r_N)|^2\, \dr_1... \dr_N}
\label{g3}
\end{eqnarray}

Its value at zero gives the probability of finding three particles in the same
point.

\subsection{Momentum distribution and static structure factor\label{secnk}}

In terms of the field operator (\ref{Psi}) the momentum distribution $n_k$ is given as
\begin{eqnarray}
n_\k = \langle\hat\Psi^\dagger_\k\hat\Psi_\k\rangle,
\label{n_k}
\end{eqnarray}
The field operator in momentum space $\hat\Psi_\k$ is related to the $\hat\Psi(\r)$
by the Fourier transform
\begin{eqnarray}
\left\{
\begin{array}{lll}
\hat\Psi_\k &= &\int e^{-i\k\r} \hat\Psi(\r) \frac{\dr}{\sqrt{2\pi}}\\
\hat\Psi(\r) &= &\int e^{i\k\r} \hat\Psi_\k \frac{\dk}{\sqrt{2\pi}}
\end{array}
\right.
\label{Psi relation}
\end{eqnarray}

Substitution of (\ref{Psi relation}) into (\ref{n_k}) gives following expression for
the momentum distribution
\begin{eqnarray}
n_\k = \frac{1}{2\pi} \intint e^{i\k\vec s} G_1\left(\RR+\frac{\vec s}{2},\vec R-\frac{\vec s}{2}\right)\,d\RR d\vec s
\end{eqnarray}

Note, that the dependence on $\k$ enters through the relative distance, so the center
of the mass motion can be integrated out independently of $k$. This procedure is
used in the DMC (see Sec.\ref{secQuantities}). In the homogeneous system one has
\begin{eqnarray}
n_k = n\int e^{ikr} g_1(r)\,dr
\label{nk}
\end{eqnarray}

At zero temperature the dynamic structure factor $S(k,\omega)$ is related to the
$k$-component of the density operator
\begin{eqnarray}
\rho_k =
\int e^{-ikr}n(r)\,dr
\label{rhok}
\end{eqnarray}
in the following way
\begin{eqnarray}
S(k,\omega) = \sum\limits_n |\langle n |\hat\rho^\dagger_k-
\langle\hat\rho^\dagger_k\rangle|0\rangle|^2\delta(\hbar\omega-\hbar\omega_n)
\end{eqnarray}

It characterizes the scattering cross-section of inelastic reactions where the
scattering probe transfers momentum $\hbar k$ and energy $\hbar\omega$ to the
system. By integrating out the $\omega$ dependence we obtain the static structure
factor
\begin{eqnarray}
S(k) = \frac{\hbar}{N} \int_0^\infty S(k,\omega)\,d\omega
=\frac{1}{N}(\langle\rho_k\rho_{-k}\rangle - |\langle\rho_k\rangle|^2)
\label{Skmom}
\end{eqnarray}

This expression is used in QMC calculations (refer to Sec.~\ref{secSkDMC}). Another
useful representation can be obtained from Eqs.~(\ref{G2},\ref{rhok}) and
commutation relations for the field operator $\hat\Psi(r)$. It relates the static
structure factor to the two-body density matrix
\begin{eqnarray}
S(k) = 1+\frac{1}{N} \intint e^{ik(r_2-r_1)}(G_2(r_1,r_2)-n(r_1)n(r_2))\,dr_1dr_2
\label{Sknonuni}
\end{eqnarray}

In a homogeneous system the two-body density matrix depends only on the relative
distance $r = r_1-r_2$ and the static structure factor $S(r)$ is directly related to
the pair distribution function (\ref{g2})
\begin{eqnarray}
S(k) = 1 + n\int e^{ikr} (g_2(r)-1)\,dr
\label{Fourier Sk}
\end{eqnarray}

\subsection{Trapped system\label{secg2trap}}

The presence of an external harmonic confinement removes the translational
invariance. We will restrict ourselves to one dimensional case as the relevant to
the study presented in Chapter~\ref{sec1DBG}. The one-body density matrix (\ref{g1})
$g_1(z_1, z_2)$ depends on both arguments:
\begin{eqnarray}
n \left(\frac{z'+z''}{2}\right) g_1(z',z'') =
\frac{N\int \psi^*(z',..., z_N) \psi(z'',..., z_N)\,dz_2...dz_N}{\int |\psi(z_1,..., z_N)|^2\,dz_1...dz_N},
\label{g1trap}
\end{eqnarray}
where $n(z)$ is the density profile.

The momentum distribution of a trapped system is obtained from the OBDM by the
Fourier transform with the respect to the relative distance
\begin{eqnarray}
n(k) = \int\!\!\!\!\int g_1 \left(Z+\frac{z'}{2}, Z-\frac{z'}{2}\right) n(Z)~e^{ikz'}\,dZdz',
\end{eqnarray}


\section{The scattering problem\label{secIntrScat}}
\subsection{Introduction}

Apart from the rare cases when the exact wave function of the system is known (like
exactly solvable TG (\ref{wf1Dtonks}) and HR (\ref{wfHR}) gases) the construction of
the trial wave function is commonly done by matching the solution of a two-body
problem with a decay to a constant. We will pay much attention to the problem of a
two-body scattering both in three- and in one-dimensional geometries.

\subsection{Three-dimensional scattering problem\label{sec3D}}
\subsubsection{General approach\label{sec3Dsc}}

In this section we will formulate the generic scattering problem in a
three-dimensional space. At low density the interaction between particles of a gas
is well described by binary collisions.

Consider a collision of two particles having coordinates $r_1$ and $r_2$, masses
$m_1$ and $m_2$. The aim is to find the stationary solution $f_{12}(\r_1,\r_2)$ of the
Schr\"odinger equation:
\begin{eqnarray}
\left(-\frac{\hbar^2}{2m_1}\Delta_{\r_1}-\frac{\hbar^2}{2m_2}\Delta_{\r_2}
+V_{int}(|\r_1-\r_2|) \right)f_{12}(\r_1,\r_2) = \E_{12}f_{12}(\r_1,\r_2)
\label{scat0}
\end{eqnarray}

In absence of an external confinement the problem is translationary 
invariant and the center of mass moves with a constant velocity. The solution of
problem gets separated in the center of the mass frame. The Schr\"odinger equation
for the movement of the center of mass $\Rc=(m_1\r_1+m_2\r_2)/M$ is trivial:
\begin{eqnarray}
-\frac{\hbar^2}{2M}\Delta_\Rc f_\Rc(\Rc)= \E_\Rc f_\Rc(\Rc),
\label{scatR}
\end{eqnarray}
here $M=m_1+m_2$ is the total mass. The solution of Eq.~\ref{scatR} is a free
wave\footnote{The normalization of the scattering solution is of no interest to us,
so in following we will always omit the normalization factor.} $f_\Rc(\Rc) =
\exp\{i\vec k_\Rc \Rc\}$ with $\vec k_\Rc$ being the initial wavenumber of the
system and $\E_\Rc=\hbar^2k_\Rc^2/2M$.

The equation for the relative coordinate $\r=\r_1-\r_2$ involves the interaction
potential:
\begin{eqnarray}
\left(-\frac{\hbar^2}{2\mu}\Delta_{\r}+V_{int}(|\r|)\right) f(\r)= \E f(\r),
\label{scatr}
\end{eqnarray}
where
\begin{eqnarray}
\mu = \frac{m_1m_2}{m_1+m_2}
\label{mu}
\end{eqnarray}
is the reduced mass. Once solutions of Eqs.~\ref{scatR}-\ref{scatr} are known
the solution of the scattering problem (\ref{scat0}) is given by
\begin{eqnarray}
\left\{
\begin{array}{cll}
f_{12}(\r_1,\r_2) &=&f_\Rc(\Rc) f(\r)\\
\E_{12}&=&\E_\Rc+\E
\end{array}
\right.
\label{split}
\end{eqnarray}

In order to proceed further we will assume that the energy of the incident particle
$E$ is small and the solution has a spherical symmetry $f(\r) = f(|\r|)\equiv f(r)$.
In this case the Laplacian gets simplified $\Delta = \frac{\partial^2}{\partial
r^2}+\frac{2}{r}\frac{\partial}{\partial r}$ and the Eq.~\ref{scatr} is
conveniently rewritten by introducing function $g(r)$
\begin{eqnarray}
\label{u(r)}
u(r)&=&\frac{f(r)}{r}\\
u(0)&=&0
\label{u(0)}
\end{eqnarray}
in such a way that its form reminds a one-dimensional Schr\"odinger equation:
\begin{eqnarray}
-\frac{\hbar^2}{2\mu}u''(r)+V_{int}(r)u(r) = \E u(r)
\label{scatu}
\label{master}
\end{eqnarray}

The solution of this equation in a general form can be written as
\begin{eqnarray}
u(r) = \sin(kr+\delta(k)),
\label{usol}
\end{eqnarray}
where
\begin{eqnarray}
\hbar k=\sqrt{2mE}
\label{ISk}
\end{eqnarray}
is the momentum of the incident particle and $\delta(k)$ is the scattering phase.

The scattering at low energy (which describes well a binary collisions in a dilute
gas) has a special interest as it becomes universal and can be described in terms of
one parameter, the {\it $s$-wave scattering length} $a_{3D}$:
\begin{eqnarray}
a_{3D} = -\lim\limits_{k\to 0}\frac{\delta(k)}{k}
\label{a3D}
\end{eqnarray}

In the asymptotic limit of slow particles $k\to 0$ the scattering solution
(\ref{usol}) can be expanded
\begin{eqnarray}
f(r)\to const \left(1-\frac{a_{3D}}{r}\right)
\label{frlim}
\end{eqnarray}
and has the node at a distance equal to $a_{3D}$. It gives an equivalent definition
of the three-dimensional scattering length as a position of the first node of the
positive energy scattered solution in the low-momentum limit.

In the next several sections we will solve the problem of the scattering on a
hard-sphere potential (\ref{secHS}) and a soft-sphere potential (\ref{secSS}). We
will find explicit expressions for the scattered functions, which are of a great
importance, as in many cases can provide a physical insight into properties of a
many body problem. Indeed, at a certain conditions the correlation functions can be
related to the scattered function $f(r)$. Another point is that the two-body
Bijl-Jastrow term $f_2(r)$ (\ref{Jastrow}) in the construction of the trial
wave function is very often taken in a form of $f(r)$. Thus such calculations are
very important for the implementation of the Quantum Monte Carlo methods.

We will also find expressions for the scattering length $a_{3D}$ in terms of the
height (or depth) of the potential $V_0$:
\begin{eqnarray}
V_0 = \max\limits_{r}|V_{int}(r)|
\label{Vo}
\end{eqnarray}
and the {\it range of the potential} $R$, which in this Dissertation will be
understood as a characteristic distance on which the potential acts. In other
words the potential can be neglected for distances much larger than $R$:
\begin{eqnarray}
R = \min\limits_{r} \{V(|r|) \approx 0\}
\label{range}
\end{eqnarray}

\subsubsection{Scattering on a hard sphere potential\label{secHS}}

As pointed out in Sec.~\ref{sec3Dsc}, in the the limit of low energy collisions the
information about the interaction potential enters in the terms of only one
parameter, the $s$-wave scattering length and scattering on all potentials having
the same scattering length is the same (the scattering becomes {\it universal}).
This allows us to choose as simple potential as one can think of. If we consider the
scattering on a repulsive potential, then the easiest choice is the {hard sphere}
(HS) potential:
\begin{equation}
V^{HS}(r)=
\left\{
\begin{array}{cc}
+\infty, & r<a_{3D}\\
0,& r\ge a_{3D}
\end{array}\right.
\label{HS}
\end{equation}

This potential has only one parameter, which we name $a_{3D}$ in the definition
(\ref{HS}). Obviously it has the meaning of the range of the potential
(\ref{range}). At the same time it has meaning of the scattering length, as
introduced in (\ref{a3D}). It will come out naturally from the solution of the
scattering problem.

The Schr\"odinger equation (\ref{scatu}) becomes ($\mu=m/2$)
\begin{eqnarray}
-\frac{\hbar^2}{m}u''(r)+V^{HS}(r)u(r) = \E u(r)
\end{eqnarray}

A particle can not penetrate the hard core of the potential and the solution
vanishes for distances smaller than the size of the hard sphere\footnote{Note that
therefore the energy is purely kinetic and the interaction potential does not enter
in an explicit way, instead it sets the boundary condition on the solution.}:
\begin{eqnarray}
\left\{
{\begin{array}{ll}
\displaystyle u(r) = 0,& |r| < a_{3D}\\
\displaystyle u''(r) - k^2 u(r) = 0,&|r| \ge a_{3D}
\end{array}}
\right.
\label{eqHS}
\end{eqnarray}

The solution of the differential equation (\ref{eqHS}) can be easily found.
Together with (\ref{u(r)}) we obtain:
\begin{eqnarray}
f(r)=
\left\{
{\begin{array}{ll}
\displaystyle 0,& |r| < a_{3D}\\
\displaystyle A\sin(k(r-a_{3D}))\,/r,&|r| \ge a_{3D}
\end{array}}
\right.,
\label{fHS}
\end{eqnarray}
where $A$ is an arbitrary constant and $k$ is given by (\ref{ISk}). The phase shift
is linear in the wave vector of the incident particle $\delta(k) = -ka_{3D}$ and
from (\ref{a3D}) we prove that the range of the potential (\ref{HS}) has indeed
meaning of the three-dimensional scattering length as stated in the beginning of
this section.

\subsubsection{Scattering on a soft sphere potential\label{secSS}}

In order to test the universality assumption and if the details of the potential are
important it is useful to have a potential, where the range of the potential $R$ can
be varied while keeping the $s$-wave scattering length constant. In the case of the
hard-sphere (Sec.~\ref{secHS}) both distances are the same. The easiest way to
modify the hard sphere potential (\ref{HS}) in such a way that it has desired
properties is to make the height of the potential finite. The resulting potential is
called the {\it soft-sphere} (SS) potential:
\begin{eqnarray}
V^{SS}(r) =
\left\{
{\begin{array}{ll}
V_0,& r < R\\
0, & r \ge R
\end{array}}
\right.
\label{SS}
\end{eqnarray}
where $V_0$ is positive.

The Schr\"odinger equation (\ref{master}) for a pair of particles in the center of
mass system is given by
\begin{eqnarray}
\left\{
{\begin{array}{ll}
u''(r)+(k^2-\kappa^2)u(r) = 0,& r<R\\
u''(r)+k^2u(r) = 0, &r\ge R
\end{array}}
\right.,
\label{EQuSS}
\end{eqnarray}
where we express the energy of the incident particle in terms of the wave number
$k^2 = m\E/\hbar^2$ and introduce a characteristic wave number related to the
height of the potential:
\begin{eqnarray}
\kappa^2 = mV_0/\hbar^2
\label{kappaSS}
\end{eqnarray}

We are interested in scattering at small energy, so $\E<V_0$. For convenience we
introduce $\K^2=\kappa^2-k^2$, where $\K$ is real. The second equation out of the
pair (\ref{EQuSS}) has a free wave solution which extends with the same amplitude to
large distances, although the first equation has a decaying solution expressed in
terms of the hyperbolic sinus:
\begin{eqnarray}
u(r) =
\left\{
{\begin{array}{ll}
A\sh(\K r+\delta_1),& r<R\\
B\sin(kr+\delta), &r\ge R
\end{array}}
\right.
\label{uSS}
\end{eqnarray}

The phase $\delta_1$ must be equal to zero in order to obtain a solution which is
not divergent at $r=0$ (see condition~\ref{u(0)}). We impose continuity of solution
and its derivative in the point $R$:
\begin{eqnarray}
\left\{
{\begin{array}{lll}
A\sh(\K R) &=& B\sin(kR+\delta)\\
A\K\ch(\K R) &=& Bk\cos(kR+\delta)\\
\end{array}}
\right.
\label{A2}
\end{eqnarray}

Condition of the continuity of the logarithmic derivative $\K\cth(\K R)=
k\ctg(kR+\delta)$ fixes the phase $\delta(k)$ of the solution:
\begin{eqnarray}
\delta(k) = \arctg\left(\frac{k}{\K}\th \K R\right)-kR
\label{deltaSS}
\end{eqnarray}
This defines the relation between constants $A$ and $B$:
\begin{eqnarray}
A^2 = \frac{B^2}{\sh^2 kR+\left(\frac{\K}{k}\cos kR\right)^2}
\end{eqnarray}

By taking limit of low energy in (\ref{deltaSS}) and using the definition
(\ref{a3D}) one obtains the expression for the $s$-wave scattering length
for the scattering on the SS potential:
\begin{eqnarray}
a_{3D} = R\left[1-\frac{\th\kappa R}{\kappa R}\right]
\label{aSS}
\end{eqnarray}

If in the case of the hard core potential (\ref{HS}) the potential energy is absent,
it is no longer so here. This makes it reasonable to use a pair of potentials SS-HS
in order to test the universality of the $s$-wave description (see, {\it e.g.}, study done
in Chapter.~\ref{sec3D1D}).

\subsection{One-dimensional scattering problem\label{sec1D}}
\subsubsection{General approach\label{sec1Dsc}}

We already have explored some aspects of the scattering problem in three-dimensions
in Sec.~\ref{sec3D}. Here we will consider the problem of a one-dimensional
scattering.

The scattering solution in a uniform system separates in center of the mass frame,
as the property (\ref{split}) is valid also in a 1D case. Thus in the following we
will skip the trivial solution for the movement of the center of the mass and we
will address the most interesting part due to solution for the relative coordinate
$z = z_1-z_2$. The one dimensional Schr\"odinger equation for the relative motion is
written as
\begin{eqnarray}
-\frac{\hbar^2}{2\mu}f''(z) + V_{int}(z) f(z) = \E f(z),
\label{Scr1Dsc}
\end{eqnarray}
where the reduced mass $\mu$ is given by (\ref{mu}). We will always consider
scattering with a positive energy, even if the interaction potential itself might
be attractive. Then the scattering energy can be written as ${\cal E} =
\hbar^2 k^2/2\mu$, where $k$ is real. The equation (\ref{Scr1Dsc}) becomes
\begin{eqnarray}
f''(z) + \left(k^2-\frac{2\mu V_{int}(z)}{\hbar^2}\right) f(z) = 0
\label{Schr1D}
\end{eqnarray}

Its general solution can be written as\footnote{In the three-dimensional system we
look for solutions with spherical symmetry. In a one-dimensional system it is
equivalent to searching even solutions.}
\begin{eqnarray}
f(z) = \cos(k|z|+\Delta(k))
\label{f1D}
\end{eqnarray}

The one-dimensional scattering length is defined as the derivative of the phase
$\Delta(k)$ in the limit of low-energy scattering\footnote{The textbook definition
for the three-dimensional scattering length (\ref{a3D}) can be recasted in a similar
form $a_{3D} = \lim\limits_{k\to 0}\partial\delta/\partial k$. We prefer to have a
definition in terms of a derivative, as it does not cause any ambiguity in the
choice a free particle phase. In three dimensions the phase of sinus (\ref{usol}) in
absence of the scattering potential is fixed to zero due to the condition
(\ref{u(0)}), which is no longer so in $1D$ case, as it should be fixed to $\pi/2$.
Instead the definition (\ref{defa1D}) takes into account the difference between the
phase in presence of scatterer and in its absence. See also footnote
on p.~\pageref{p17}.}
\begin{eqnarray}
a_{1D} = -\lim\limits_{k\to 0}\frac{\partial \Delta(k)}{\partial k}
\label{defa1D}
\end{eqnarray}

\subsubsection{Scattering on a pseudopotential\label{secLLsc}}

In a one-dimensional system the contact $\delta$-potential is a ``good'' potential
and the problem of a scattering on it is solved in a standard manner, as described
in Sec.~\ref{sec1Dsc} without any special tricks. The situation is different in
three-dimensions where the $\delta$-potential has to be regularized (refer to
Sec.\ref{secPseudo}) in order to avoid a possible divergence which can be caused by
the behavior of a symmetric solution (\ref{u(r)}).

The $\delta$-pseudopotential turns out to be highly useful theoretical tool. Indeed
the commonly used Gross-Pitaevskii equation corresponds to pseudopotential
interaction $V_{int}(z)=g_{1D}\delta(z)$. A system of particles with
$\delta$-pseudopotential interaction (\ref{LL}) is one of few exactly solvable one
dimensional quantum systems.

The Schr\"odinger equation (\ref{Schr1D}) of the scattering on a pseudopotential
\begin{eqnarray}
-\frac{\hbar^2}{2\mu}f''(z) + g_{1D}\delta(z) f(z) = \E f(z),
\label{wf1DLLf"}
\end{eqnarray}

In the region $|z|>0$ it takes form of a free particle propagation $f''(z) +
k^2f(z) = 0$ with the even solution given by
\begin{eqnarray}
f(z) = \cos(k|z|+\Delta)
\label{fLL0}
\end{eqnarray}

We are left with the only point $z=0$, where the scattering potential is nonzero
$V_{int}(r)\ne 0$. The infinite strength of the $\delta$-potential makes the first
derivative of the potential be discontinuous. Indeed, the proper boundary condition
can be obtained by integrating the equation (\ref{wf1DLLf"}) from infinitesimally
small $-\varepsilon$ up to $+\varepsilon$. The integral of the continuous function
$f_2(z)$ is proportional to $\varepsilon$ and vanishes in the limit $\varepsilon\to
0$. Instead the $\delta$-function extracts the value of the function in zero and one
obtains the relation
\begin{eqnarray}
f'(\varepsilon) - f'(-\varepsilon) = \frac{2\mu g_{1D}}{\hbar^2} f(0)
\label{Lieb cont}
\end{eqnarray}

This boundary condition for the solution (\ref{fLL0}) provides a relation between
the scattering phase $\Delta$ and the momentum $k$ of an incident particle
\begin{eqnarray}
\Delta(k) = - \arcctg\frac{\hbar^2 k}{\mu\,g_{1D}}
\label{deltaLL}
\end{eqnarray}

Taking the limit of the low energy scattering from (\ref{defa1D}) one obtains the
value of the scattering length
\begin{eqnarray}
a_{1D} = -\frac{\hbar^2}{\mu g_{1D}}
\end{eqnarray}

This expression can be read the other around: for equal mass particles $\mu = m/2$
the strength of the potential $g_{1D}$ in a one-dimensional homogeneous system is
related to the value of the one-dimensional coupling constant as
\begin{eqnarray}
g_{1D} = -\frac{2\hbar^2}{ma_{1D}}
\label{g1D}
\end{eqnarray}

It is interesting to note, that the sign in the relation of the scattering length to
the coupling constant is opposite to the one of a three dimensional system. In $3D$
positive scattering length corresponds to repulsion and negative one to attraction.
Another difference is that the 3D coupling constant is directly proportional to the
scattering length, although $g_{1D}$ is inversely proportional to $a_{1D}$.

In terms of $a_{1D}$ the phase (\ref{deltaLL}) becomes
\begin{eqnarray}
\Delta(k) = \arcctg ka_{1D}
\label{LLphase}
\end{eqnarray}
The scattering solution (\ref{fLL0}) gets written as:
\begin{eqnarray}
f(z) = \cos(k|z|+\arcctg ka_{1D})
\label{fLL}
\end{eqnarray}

In the low energy limit $k\to 0$ the phase (\ref{LLphase}) can be expanded
$\Delta(k) = \pi/2-ka_{1D}+{\cal O}(k^3)$ 
and the scattering solution becomes
simply $f(z) = k(z-a_{1D})$. One sees that the one-dimensional scattering length
coincides with the position of the first node of the analytic continuation of the
low-energy solution\footnote{It turns out that this property is general and can be
used as an alternative to (\ref{defa1D}) definition of the one-dimensional
scattering length.\label{p17}}.

\subsubsection{Scattering on a 1D square well potential\label{sec1DSW}}

In this section we will consider scattering on a one-dimensional square well. The
potential is similar to the one of the soft sphere with the difference that now the
potential is attractive:
\begin{eqnarray}
V^{SW}(z) = - V_0\,\Theta(R^2-z^2),
\end{eqnarray}
where $R$ is the range of the potential. In the region $|z| < R$ the kinetic energy
of the slow particle can be neglected
\begin{eqnarray}
f''(z) + V_0 f(z) = 0
\end{eqnarray}

All solutions can be decomposed into a sum of even and odd solutions distinguished
by the boundary condition at zero which can be either $f(0) = 0$ or $f'(0) = 0$. We
choose the state with the minimal energy, {\it i.e.} $f'(0) = 0$, which leads to the
solution of the form
\begin{eqnarray}
f(z) = A\cos(\sqrt{V_0}\,z),\qquad |z|<R
\end{eqnarray}

In the other region $|z|>R$ the interaction potential is absent and the solution is
a plain wave
\begin{eqnarray}
f(z) = B\sin(kz+\delta_0),\qquad |z|<R
\end{eqnarray}

The scattering phase can be defined from the continuity condition of the logarithmic
derivative at the matching distance $R$. This condition reads as
\begin{eqnarray}
\frac{f'(R)}{f(R)}= -\sqrt{V_0} \tg(\sqrt V_0 R) = k \ctg (kR+\delta_0)
\label{Swdelta}
\end{eqnarray}

Eq. (\ref{Swdelta}) fixes the dependence of the phase on the wave number of the
scattering particle:
\begin{eqnarray}
\Delta(k) = -\arcctg\frac{\sqrt V_0 \tg(\sqrt V_0 R)}{k}-kR
\end{eqnarray}

Finally, from (\ref{defa1D}) we obtain the expression for the scattering length on
the 1D square well potential:
\begin{eqnarray}
a_{1D} =  R\left(1+\frac{\ctg(\sqrt V_0 R)}{\sqrt V_0 R}\right)
\label{a1DSW}
\end{eqnarray}

\subsubsection{Scattering on a hard-rod potential\label{secHR}}

The hard-rod potential is a one-dimension version of the hard core potential, which
in $3D$ correspond to a hard sphere (\ref{HS}). The HR potential is defined by its
radius $|a_{1D}|$
\begin{eqnarray}
V^{HR}(z)=
\left\{
\begin{array}{cc}
+\infty, & |z|<|a_{1D}|\\
0,& |z|\ge |a_{1D}|
\end{array}\right.
\label{HR}
\end{eqnarray}

The scattering phase in the solution (\ref{f1D}) is fixed by the condition that the
function vanishes at the HR radius $\Delta = -k|a_{1D}|-\pi/2$. From (\ref{defa1D})
immediately follows that the radius defined as (\ref{HR}) coincides with the value
of the one dimensional scattering length. Again, as in Sec.~\ref{secHS} we have a
hard core potential, for which its radius, the scattering length and the range of
the potential are completely the same.

The scattering solution on a hard rod potential reads as
\begin{eqnarray}
f(z) = \sin(k(|z|-|a_{1D}|))
\label{fHR}
\end{eqnarray}

\subsection{Pseudopotential}
\subsubsection{The pseudopotential method\label{secPseudo}}

As it was discussed above, in Secs.~\ref{sec3Dsc}-\ref{secSS}, scattering on
different short-ranged potentials in the low-energy limit is universal, {\it i.e.} depends
essentially on one parameter, the scattering length and the particular shape of the
potential is of no large importance. Thus it is very useful to relate scattering on
all those potentials to a scattering on a simple $\delta$-potential. In other words
instead of considering a particular shape of the interaction potential, we give the
description it terms of a free scattering solution at $|\r|>0$ with an appropriate
boundary condition at $r=0$, which takes properly into account the scattering length
and, thus, the interaction potential.

In one dimensional case the application of this scheme is straight as the
Schr\"odinger equation for two particles can be directly solved, as it is explained
in the Sec.~\ref{secLLsc}. In three dimensions the situation is more complicated as
the behavior of the solution (\ref{u(r)}) is not compatible with scattering on a
$\delta$-potential and special adjustments should be made.

Let us revise the solution of the Schr\"odinger equation in the limit of low energy
scattering. From the definition of the three-dimensional scattering length
(\ref{a3D}) it follows that the scattering solution vanishes at the distance $r =
a_{3D}$. Thus we define the scattering function of the pseuodopotential in such a
way that it satisfies the free scattering equation in the region $r>0$:
\begin{eqnarray}
(\Delta+k^2) f(r) = 0,\qquad r>0
\label{ffree}
\end{eqnarray}

We will use the expression (\ref{frlim}) to approach the $r\to 0$ limit:
\begin{eqnarray}
f(r) \to \chi \left(1-\frac{a_{3D}}{r}\right)
\label{fchi}
\end{eqnarray}
where the constant $\chi$ can be related to the scattering length by multiplying
(\ref{fchi}) by $r$ and differentiating
\begin{eqnarray}
\chi = \lim\limits_{r\to 0} \frac{\partial}{\partial r}(rf(r))
\end{eqnarray}

We can now modify Eq.~\ref{ffree} in such a way that it satisfies the correct
boundary condition (\ref{fchi}). By inserting (\ref{fchi}) into (\ref{ffree}) we
obtain\footnote{We used property $\Delta (1/r) = -4\pi\delta(r)$, which can be
easily obtained from the solution $f(\r) = -\frac{1}{4\pi}
\int\frac{\rho(\r')}{|\r-\r'|}\dr'$ to the Poisson equation
$\Delta f(\r) = \rho(\r)$ substituting the point charge $\rho(\r) =\delta(\r)$.}
\begin{eqnarray}
(\Delta+k^2)\, \chi \left(1-\frac{a_{3D}}{r}\right) = -4\pi \delta(r) \frac{\partial}{\partial r}(rf(r))
\end{eqnarray}
The operator $\delta(\r)\frac{\partial}{\partial r}(r\cdot)$ is called the {\it
pseudopotential}. Going back to energy units we obtain the relation between the
strength of the pseudopotential $g_{3D}$ ({\it coupling constant}) and the
three-dimensional scattering length
\begin{eqnarray}
\label{pseudopotential}
V_{int} &=& g_{3D} \delta(\r)\frac{\partial}{\partial r}(r\cdot)\\
g_{3D} &=& \frac{4\pi\hbar^2}{m}a_{3D},
\label{g3D}
\end{eqnarray}
where we considered the case of equal-mass particles $\mu = m/2$.

The pseudopotential (\ref{pseudopotential},\ref{g3D}) was used by Olshanii
\cite{Olshanii98} to solve quasi one dimensional scattering problem in a tight
harmonic transverse confinement.

Finally, the wave function $f(r)$ satisfies the equation\footnote{Additional
literature on the topic of pseudopotential description can be found in classical
articles \cite{Fermi36},\cite{Huang57} and in books \cite{Blatt52},p.74,
\cite{Huang87}.}:
\begin{eqnarray}
\left(-\frac{\hbar^2}{2m}\Delta_1
-\frac{\hbar^2}{2m}\Delta_2  +
\frac{4\pi\hbar^2}{m} \delta(r_{12}) \frac{\partial}{\partial r_{12}}
(r_{12}~\cdot~)\right) f(\r_1,\r_2) = \E f(\r_1,\r_2),
\end{eqnarray}

\subsection{Resonance scattering}

In the previous sections we considered situation, when the scattering happens on the
lowest energy level. In this case the $s$-wave scattering length $a$ for any finite
strength potential is smaller than the range of the potential $R$ (see, {\it e.g.}
Secs.~\ref{secSS},\ref{sec1DSW}) and equals to $R$ in the case of the infinite
strength potential (Secs.~\ref{secHS},\ref{secHR}). The pseudopotential description
(Secs.~\ref{secLLsc},\ref{secPseudo}) falls into a different class of problem used
at a small density, where the exact type of the potential is not important and it is
substituted by the boundary condition at $r=0$. In this sense the range of the
pseudopotential is zero $R=0$ and we have opposite condition
\begin{eqnarray}
|a|\gg R
\label{agtR}
\end{eqnarray}

A physical realization of $3D$ scattering satisfying the condition (\ref{agtR}) can
be achieved in the case of a {\it resonant scattering}. In this Section we will
describe scattering on the first exited state of attractive potentials supporting a
bound state in the case when the position of the excited state is close to
zero-energy continuum level.

\subsubsection{Scattering on a square-well potential\label{secSW}}

Let us consider an attractive version of the soft sphere potential (\ref{SS}):
\begin{eqnarray}
V^{SW}(r) =
\left\{
{\begin{array}{ll}
-V_0,& r < R\\
0, & r \ge R
\end{array}}
\right.
\label{SW}
\end{eqnarray}

Interaction (\ref{SW}) is called a {\it square-well} potential, with $V_0$
(positive) being its depth and $R$ being its range. The Schr\"odinger equation
(\ref{master}) for a pair of particles in the center of mass system is given by
\begin{eqnarray}
\left\{
{\begin{array}{ll}
u''(r)+(k^2+\kappa^2)\,u(r) = 0,& r<R\\
u''(r)+k^2u(r) = 0, &r\ge R
\end{array}}
\right.,
\label{EQuSW}
\end{eqnarray}
where, as usual, $k^2 = m\E/\hbar^2$ and
\begin{eqnarray}
\kappa^2 = -mV_0/\hbar^2>0
\label{kappaSW}
\end{eqnarray}

We are interested in finding solutions with positive energies, as that are the
solutions corresponding to a scattered state, instead solutions with negative energy
are localized. On the opposite to the situation described in Sec.~\ref{secSW}, the
interaction potential is always lower than the value of the scattering energy
$V_{int}(r)<\E$. For convenience we introduce $\K^2=\kappa^2+k^2>0$. In both regions
the solution is a free-wave like:
\begin{eqnarray}
u(r) =
\left\{
{\begin{array}{ll}
A\sin(\K r+\delta_1),& r<R\\
B\sin(kr+\delta), &r\ge R
\end{array}}
\right.
\label{uSW}
\end{eqnarray}

The condition (\ref{u(0)}) immediately fixes the phase $\delta_1 = 0$. The matching
equations for the function and its derivative read as
\begin{eqnarray}
\left\{
{\begin{array}{lll}
A\sin(\K R) &=& B\sin(kR+\delta)\\
A\K\cos(\K R) &=& Bk\cos(kR+\delta)\\
\end{array}}
\right.
\label{A2SW}
\end{eqnarray}

Condition of the continuity of the logarithmic derivative $\K\ctg(\K R)=
k\ctg(kR+\delta)$ fixes the phase $\delta(k)$ of the solution
\begin{eqnarray}
\delta(k) = \arctg\left(\frac{k}{\K}\tg\K R\right)-kR
\label{deltaSW}
\end{eqnarray}
This builds the relation between constants $A$ and $B$:
\begin{eqnarray}
A^2 = \frac{B^2}{\sin^2 kR+\left(\frac{\K}{k}\cos kR\right)^2}
\end{eqnarray}

By taking limit of low energy in (\ref{deltaSW}) and using the definition
(\ref{a3D}) one obtains the expression for the $s$-wave scattering length:

\begin{eqnarray}
a_{3D} = R\left[1-\frac{\tg\kappa R}{\kappa R}\right]
\label{aSW}
\end{eqnarray}

The dependence of the scattering length of the scattering on a soft sphere potential
(Eq.~\ref{aSS}) looks similar to (\ref{aSW}) with the only difference that the
trigonometric tangent is substituted with the hyperbolic one. The difference is
crucial. Indeed, as $0<\th (x)/x\le 1$, the scattering length on the SS potential is
always smaller than the range of the potential. Instead, the term $\tg(x)/x$ is
unbound. When the scattering happens at {\it resonant } momentum $\kappa R =
\pi/2+\Delta(\kappa)$ with small detuning $|\Delta(\kappa)|\ll 1$, the scattering
length becomes extremely large and changes its sign.

The square well potential is attractive and in principle can have the bound state
solution with energy $E_b = -\hbar^2k_b^2/m<0$. In outer region $r>R$ the solution
(\ref{uSW}) gets modified and decays exponentially fast. The condition of the
continuity of the logarithmic derivative in the limit $k\to 0$ is $\kappa\tg\kappa R
= k_b\th k_bR$. This condition can not be satisfied before crossing the resonance,
as inequality $\tg x > \th x$ holds for arguments $x<0<\pi/2$. Instead immediately
after the resonance position $\Delta(\kappa)>0$ a shallow bound state appears in the
system.

\subsubsection{Scattering on a modified P\"oschl-Teller potential\label{secMorse}}

The potential (\ref{SW}) considered in the previous section might be inconvenient in
some cases, as it produces large gradients of the solution at its border $r\approx
R$ due to the abrupt change of its value from $-V_0$ to zero. This can be avoided
by using, for example, the modified Po\"schl-Teller potential
\begin{eqnarray}
V(r) = -\frac{V_0}{\ch^2 (r/R)}
=-\frac{\hbar^2}{2mR^2}\frac{\lambda(\lambda-1)}{\ch^2 (r/R)},
\label{Morse}
\end{eqnarray}
where $V_0$ is the depth of the potential and $R$ is its range.

The problem of three-dimensional scattering on this potential can be solved
analytically (see, {\it e.g.} \cite{Flugge71}) and the dependence of the $s$-wave scattering
length on the depth of the potential well can be found explicitly:
\begin{eqnarray}
\frac{a_{3D}}{R}=\frac{\pi}{2}\ctg\frac{\pi\lambda}{2}+\gamma+\Psi(\lambda),
\end{eqnarray}
where $\gamma = 0.5772...$ is the Euler's constant and $\Psi$ is the Digamma
function. This dependence is expressed in the Fig.~\ref{figSc2}.

\section{Energy of the TG and HR gas\label{secTGHR}}
\subsection{Energy of the Tonks-Girardeau gas\label{secETGgas}}

In the very dilute $1D$ regime, when the one-dimensional gas parameter becomes
extremely small $n_{1D}|a_{1D}|\ll 1$, the $1D$ system of bosons can be mapped onto
$1D$ system of fermions \cite{Girardeau60}. In a fermionic system the number of
fermions is given by the volume of the fermi sphere (the bosons are mapped onto
spinless fermions). In a one-dimensional system this volume degenerates to $2k_F$:
\begin{equation}
N = L\int\limits_{-k_F}^{k_F}\frac{dk}{2\pi} = \frac{1}{\pi}k_FL
\end{equation}

We obtain that the relation of the fermi wave number $k_F$ to the density $n_{1D}$
is linear
%
\begin{equation}
k_F = \pi n_{1D}
\label{kTG}
\end{equation}

The value of $k_F$ fixes the scale for the correlation functions. The static
structure factor (\ref{Sk tonks}) completely changes its behavior at $k=2k_F$. The
value of $k_F$ fixes period of oscillations in the pair distribution function
(\ref{g2 tonks}). Being the only spatial length scale in a homogeneous system,
$1/k_F$ fixes at the same time value of the healing length $\xi$, and consequently
the border at which starts the asymptotic power law decay of the one-body density
matrix.

The chemical potential equals to the fermi energy (this is the definition of the
fermi energy):
\begin{equation}
\mu_F = \frac{\pi^2 \hbar^2}{2m} n^2_{1D}
\label{muTG}
\end{equation}

The energy is obtained by integration of the chemical potential. The energy per
particle turns out to be equal to
\begin{equation}
E_F = \frac{\pi^2 \hbar^2}{6m} n^2_{1D}
\label{ETG}
\end{equation}

\subsection{Hard-rod gas\label{secEHR}}

Let us consider a gas of $N$ hard rod bosons of size $a_{1D}$\footnote{As discussed
in Sec.~\ref{secHR}, the size of a hard-rod equals to the one-dimensional scattering
length on HR potential (\ref{HR}).}. The energy of the hard-rode gas is easily
obtained from the expression for the energy of the Tonks-Girardeau gas (\ref{ETG})
by subtracting the excluded volume $n \to N/(L-Na_{1D})$
\cite{Girardeau60,Krotscheck99}
\begin{equation}
\frac{E_{HR}}{N} = \frac{\pi^2\hbar^2n_{1D}^2}{6m}\frac{1}{(1-n_{1D}a_{1D})^2}
\label{EHR}
\end{equation}

The chemical potential is the derivative of the energy with respect to number of
particles
\begin{eqnarray}
\mu_{HR} = \frac{\pi^2\hbar^2n_{1D}^2}{2m}\frac{(1-a_{1D}n_{1D}/3)}{(1-a_{1D}n_{1D})^3},
\end{eqnarray}

If the density is small $n_{1D}a_{1D}\ll 1$, one is allowed to make an expansion of
(\ref{EHR}) in terms of the small parameter:
\begin{eqnarray}
\frac{E}{N} = \frac{\pi^2\hbar^2n_{1D}^2}{6m}+\frac{\pi^2\hbar^2n_{1D}^3a_{1D}}{3m}
\label{Eexclud}
\end{eqnarray}

It is interesting to note, while the ``excluded volume'' term was derived for
$a_{1D}>0$, it still provides the leading correction to the TG energy (\ref{ETG}) in
the Lieb-Liniger Hamiltonian (\ref{LL}), {\it i.e.} for $a_{1D}<0$. The point is that it
describes the interaction energy, which is absent in a TG gas (see argumentation
done on page~\pageref{HRdiscussion}). The equation of state in LL model can be found
exactly by solving the integral equations (\ref{LLintegraleqs}-\ref{LLe}). An iterative solution in the considered
region $n_{1D}|a_{1D}|\ll 1$ provides a way for the calculation of the expansion
\begin{eqnarray}
e(n|a_{1D}|) = \frac{\pi^2}{3} - \frac{2}{3}\pi^2 n_{1D}|a_{1D}|,
\label{exclvol}
\end{eqnarray}
where we adopt standard for LL equations notation (\ref{LLnotation}). This formula
is consistent with (\ref{Eexclud}) and can be obtained by solving recursively
the Lieb-Liniger integral equations (\ref{LLintegraleqs}-\ref{LLe}).

\section{Gross Pitaevskii Equation\label{secGPE}}
\subsection{Variational derivation of the GPE\label{secVarGPE}}

Let us consider $N$ identical bosons in an external potential $V_{ext}$. For $T\ll
T_c$ all particle stay in the ground state of the Hamiltonian:
\begin{eqnarray}
\hat H = \sum\limits_{i=1}^{N} \left[\frac{\hat p_i^2}{2m}+V_{ext}(\r_i)\right]
+\frac{1}{2}\sum\limits_{i\ne j}^N V_{int}(\rij),
\label{Hgp}
\end{eqnarray}

At low temperatures, namely when the de Broglie wavelength $\lambda_T$ becomes much
larger than the range of $V_{int}(r_{ij})$, only s-wave scattering between pairs of
bosons remains significant, and we can approximate $V_{int}(r_{ij})$ by a
pseudopotential (\ref{g3D}).

Generally, the ground state of $\hat H$ cannot be determined exactly. In the absence
of interactions however, it is a 
product state: all the bosons are in the ground state of the single particle
Hamiltonian. In the presence of {\it weak} interactions, one still can approximate
the ground state of $\hat H$ by a product state:
\begin{eqnarray}
|\phi_0\rangle =|\psi(1)\rangle ... |\psi(N)\rangle,
\label{Fock}
\end{eqnarray}
where all bosons are in the same state $|\psi\rangle$\footnote{It is important to
note that $|\psi\rangle$ is not a wave function and in this sense the derived below
GPE (\ref{GPE}) is not a ``non linear Schr\"odinger equation''. In particular its
time evolution is driven by the chemical potential $\mu$ instead of the energy of
the system $E$, as it happens for the solution of the Schr\"odinger equation.}.

Obviously, $|\phi_0\rangle$ is symmetric with the respect to exchange of particles
and has the correct symmetry for a system of bosons. Contrary to the non-interacting
case, $|\psi\rangle$ is no longer the ground state of the single particle
Hamiltonian, but has to be determined by minimizing the energy:
\begin{eqnarray}
E = \frac{\langle\phi_0|\hat H|\phi_0\rangle}{\langle\phi_0|\phi_0\rangle}
\end{eqnarray}

Let us calculate the value of (\ref{Hgp}) averaged over the Fock state (\ref{Fock}). In the
coordinate representation the external potential energy becomes:
%
\begin{eqnarray*}
\langle\phi_0|\!\sum\limits_{i=1}^N V_{ext}(\r_i)|\phi_0\rangle
\!=\!\int\!\!\psi^*(\r_N) ...\psi^*(\r_1)
\sum\limits_{i=1}^N V_{ext}(\r_i)
\psi(\r_1)...\psi(\r_N)\;\dR
\!=\!N\!\!\int\!\!\psi^*(\r) V_{ext}(\r) \psi(\r)\;\dr
\end{eqnarray*}

For the interaction between the particles we obtain:
%
\begin{eqnarray*}
\langle\phi_0|\sum\limits_{i\ne j}^N \frac{1}{2}V_{int}(r_{ij})|\phi_0\rangle
=\!\!\int\!\!\psi^*(\r_N) ...\psi^*(\r_1)
\frac{1}{2}\sum\limits_{i\ne j}^N V_{int}(r_{ij})
\psi(\r_1)...\psi(\r_N)\;\dR =\\
= \frac{1}{2}\sum\limits_{i\ne j}^N\!\!
\intint\!\!\psi^*(\r_i)\psi^*(\r_j)V_{int}(r_{ij})\psi(\r_i)\psi(\r_j)\;\dr_i\dr_j\!\!
=\!\!\frac{N(N-1)}{2}\!\!\!\intint\psi^*(\r)\psi^*(\r')V_{int}(|\r\!-\!\r'|)\psi(\r)\psi(\r')\;\dr\dr'
\end{eqnarray*}

Thus we obtain the expression of the total Hamiltonian in the first quantization
(see, also, (\ref{secQnt}))
\begin{eqnarray}
\langle\hat H\rangle\!\!
=\!\!N\!\!\int\!\! \psi^*\!(\r)\!\left(\!\!-\frac{\hbar^2\triangle}{2m}+
V_{ext}\!\!\right)\!\psi(\r)\;\dr
\!+\!\frac{N(N-1)}{2}\!\!\!\intint\!\!\!\psi^*\!(\r)\psi^*\!(\r')
V_{int}(|\r-\r'|)\psi(\r)\psi(\r')\;\dr\dr'
\label{H1st}
\end{eqnarray}

We now look for the minimum of the energy $\langle\phi_0|\hat H|\phi_0\rangle$
keeping the normalization fixed $\langle\phi_0|\phi_0\rangle = 1$. Because $\psi$
in general is
a complex number, we can consider the variations $\delta\psi$ and
$\delta\psi^*$ as independent. Using the method of Lagrange multipliers, the
approximate ground state $|\phi_0\rangle$ has to satisfy:
\begin{eqnarray}
\delta\left[\langle\phi_0|\hat H|\phi_0\rangle\right]-\mu\delta\langle\phi_0|\hat H|\phi_0\rangle = 0,
\label{Lagrange}
\end{eqnarray}
where $\mu$ is the Lagrange multiplier associated with the constraint
$\langle\phi_0|\phi_0\rangle = 1$.

Inserting the expression (\ref{H1st}) in equation (\ref{Lagrange}) and
setting to zero the linear term $\delta\psi^*$ we yield:
\begin{eqnarray}
\left(-\frac{\hbar^2}{2m}\triangle+V_{ext}\right)\psi(\r)
+(N-1)\left(\int V_{int}(|\r-\r'|)|\psi(\r')|^2\,\dr'\right)\psi(\r) = \mu\psi(\r)
\end{eqnarray}

Now we use that the properties of the $s$-wave scattering at the discussed conditions
can be described by using the pseudopotential (\ref{g3D}) and, finally, obtain
\begin{eqnarray}
\left(-\frac{\hbar^2}{2m}\triangle+V_{ext}\right)\psi(\r)+(N-1)g_{3D}|\psi(\r)|^2\psi(\r) = \mu\psi(\r)
\label{GPEN-1}
\end{eqnarray}

This is the Gross-Pitaevskii equation \cite{Gross61,Pitaevskii61}. It has a
straightforward interpretation: each boson evolves in the external potential
$V_{ext}$ and in the {\it mean-field} potential produced by the other $N-1$ bosons.

Let us clarify the meaning of the parameter $\mu$, which was introduced formally as
a Lagrange multiplier. Multiplying GP equation (\ref{GPEN-1}) by $\psi^*(r)$ and by
carrying out an integrating over $r$ we have:
\begin{eqnarray}
\mu = \int \psi^*(\r)\left(-\frac{\hbar^2\triangle}{2m}+V_{ext}\right)\psi(\r)\,\dr
+(N-1)\int \psi^*(\r)\psi^*(\r')V_{int}(|\r-\r'|)\psi(\r)\psi(\r')\,\dr\dr'
\end{eqnarray}

A direct comparison to (\ref{H1st}) shows that $\mu = \frac{d}{d N}
\langle\phi_0|\hat H|\phi_0\rangle$ (number of considered particles
is large) and thus $\mu$ has a physical meaning of the chemical potential.

An alternative way is to normalize the wave function to the number of particles in
the system $\langle\phi_0|\phi_0\rangle = N$. In this normalization GPE reads as
($N\gg 1$):

\begin{eqnarray}
\left(-\frac{\hbar^2}{2m}\triangle+V_{ext}\right)\psi(\r)+g_{3D}|\psi(\r)|^2\psi(\r) = \mu\psi(\r)
\label{GPE}
\end{eqnarray}

\subsection{Coupling constant in quasi one- and two- dimensional systems\label{secGP1D2D}}

The mean-field relation of the coupling constant in 1D, $g_{1D}$ and in 2D,
$g_{2D}$, to the three dimensional scattering length $a_{3D}$ in restricted
geometries can be found by repeating the derivation given in Sec.~\ref{secVarGPE}
while assuming that the order parameter $\psi$ can be factorized. We start from the
energy functional (\ref{H1st})
\begin{eqnarray}
E[\psi] = \int \left(\frac{\hbar^2}{2m} |\nabla \psi|^2+V_{ext}|\psi|^2 + \frac{g_{3D}}{2} |\psi|^4 \right) {\bf dr},
\label{energy functional}
\end{eqnarray}
where, according to (\ref{g3D}), $g_{3D} = 4\pi\hbar^2a/m$ is the three dimensional
coupling constant. The variational procedure
\begin{eqnarray}
i\hbar \frac{\partial \psi}{\partial t} = \frac{\delta E}{\delta \psi^*}
\label{var procedure}
\end{eqnarray}
gives time-dependent Gross-Pitaevskii equation
\begin{eqnarray}
i\hbar \frac{\partial\psi(\r,t)}{\partial t}=\left(-\frac{\hbar^2\triangle}{2m}+ g_{3D}|\psi(\r,t)|^2\right) \psi(\r,t)
\label{GPEt}
\end{eqnarray}

In the presence of an external confinement along one direction (disk-shaped
condensate) $V_{ext}(\r) = m\omega^2x^2/2$ we assume a gaussian {\it ansatz} for the
wave function $\psi({\bf r},t) = \psi_{osc}(x)\varphi(y,z,t)$ with $\psi_{osc}(x)
= \pi^{-1/4} a_{osc}^{-1/2} \exp\left(-x^2/2a_{osc}^2\right)$ being ground state
wave function of a harmonic oscillator. The integration over $x$ in (\ref{energy
functional}) can be easily done by using following properties of the gaussian
function $\psi_{osc}$:
\begin{enumerate}
\item Normalization properties
\begin{eqnarray}
\int \psi^2_{osc}(x)\, dx = 1,\qquad
\int \psi^4_{osc}(x)\, dx = \frac{1}{\sqrt{2\pi}a_{osc}}
\end{eqnarray}
\item The function $\psi_{osc}$ is a stationary solution of a one-dimensional
Schr\"odinger equation in a trap
\begin{eqnarray}
\left(-\frac{\hbar^2}{2m}\frac{\partial^2}{\partial x^2}+
\frac{m\omega^2 x^2}{2}\right) \psi_{osc}(x)
= \frac{\hbar\omega}{2} \psi_{osc}(x)
\end{eqnarray}
\end{enumerate}

Integrating out $x$ from the GP energy functional (\ref{energy functional}) and
doing the variational procedure (\ref{var procedure}) we obtain the Gross-Pitaevskii
equation in a quasi two dimensional system
\begin{eqnarray}
i\hbar \frac{\partial \varphi(y, z, t)}{\partial t}
=\left(-\frac{\hbar^2}{2m}\left(\frac{\partial^2}{\partial y^2}+\frac{\partial^2}{\partial z^2}\right)
+ g_{2D} |\varphi(y,z,t)|^2+\frac{\hbar \omega}{2}\right)\varphi(y,z,t),
\end{eqnarray}
where the two dimensional coupling constant is given by
\begin{eqnarray}
g_{2D} = \frac{g_{3D}}{\sqrt{2\pi}a_{osc}}= \frac{2\sqrt{2\pi}\hbar^2a}{ma_{osc}}
\label{g2D}
\end{eqnarray}

If the external potential restricts the motion in two dimensions ({\it i.e.} in a
cigar-shaped condensate) and the confinement is so strong that no excitations in the
radial direction are possible, the wave function gets factorized in the following
way: $\psi(\r,t) = \psi_{osc}(x)\psi_{osc}(y)\phi(z)$. The explicit integration in
(\ref{energy functional}) over $x$ and $y$ leads to one-dimensional Gross-Pitaevskii
equation
\begin{eqnarray}
i\hbar \frac{\partial \varphi(z, t)}{\partial t}
=\left(-\frac{\hbar^2}{2m}\frac{\partial^2}{\partial z^2}
+ g_{1D} |\varphi(z,t)|^2+\hbar\omega\right) \varphi(z,t)
\label{GPE1D}
\end{eqnarray}

Here $g_{1D}$ denotes effective one-dimensional coupling constant
\begin{eqnarray}
g_{1D} = \frac{g_{3D}}{2\pi a^2_{osc}} = \frac{2\hbar^2a}{ma_{osc}^2}
\label{g1DMF}
\end{eqnarray}

Comparing it with the definition of the 1D coupling constant $g_{1D} = -
2\hbar^2/(ma_{1D})$ (\ref{g1D}) we find the mean-field relation of one-dimensional
scattering length $a_{1D}$ to the three-dimensional scattering length $a$ and
oscillator length $a_{osc}$:
\begin{eqnarray}
a_{1D} = -\frac{a_{osc}^2}{a}
\label{a1DMF}
\end{eqnarray}

%

\section{Local Density Approximation\label{secLDA}}
It happens often, that properties of a homogeneous system are well known ({\it e.g.} the
homogeneous model is exactly solvable, or numerical calculation has been done), but
the properties of the system in an external field are not known. If number of
particles is large enough one can refer to the {\it local density approximation} in
order to obtain the desired properties.

\subsection{General method\label{secLDAgen}}

In the local density approximation one assumes that the chemical potential $\mu$ is
given by sum of the local chemical potential $\mu_{loc}$, which is the chemical
potential of the uniform system, and the external field:
\begin{equation}
\mu = \mu_{\hom}(n(\r)) + V_{ext}(\r)
\label{LDA}
\end{equation}

The local chemical potential $\mu_{\hom}$ is defined by the equation of state in
absence of the external field and accounts for the interaction between particles and
partially for the kinetic energy.

The value of the chemical potential $\mu$ is fixed by the normalization condition
\begin{eqnarray}
N = \int n(\r)\,\dr,
\label{LDAN}
\end{eqnarray}
where the density profile is obtained by inverting the density dependence of the
local chemical potential $n = \mu_{\hom}^{-1}$.

Once the chemical potential $\mu$ is known a lot of useful information can be
inferred: the density profile, energy, size of the cloud, density moments $\langle
r^2\rangle$, etc.

In the following we will always consider a harmonic external confinement:
\begin{eqnarray}
V_{ext}(\r) = \frac{1}{2}m\omega_x x^2+\frac{1}{2}m\omega_y y^2+\frac{1}{2}m\omega_z z^2
\end{eqnarray}

The normalization condition (\ref{LDAN}) becomes:
\begin{eqnarray}
N= \int\!\!\!\!\int\!\!\!\!\int\mu_{\hom}^{-1}\left[\mu-\frac{1}{2}m\omega_x x^2-\frac{1}{2}m\omega_y y^2-\frac{1}{2}m\omega_z
z^2\right]\,dx dy dz
\label{LDANharm}
\end{eqnarray}

The sizes of the cloud in three directions $R_x,R_y,R_z$ is fixed by the value of
the chemical potential and corresponding frequencies of the harmonic confinement
through relation:
\begin{eqnarray}
\mu = \frac{1}{2}m\omega_x R_x^2=\frac{1}{2}m\omega_y R_y^2=\frac{1}{2}m\omega_z R_z^2
\label{LDAR}
\end{eqnarray}

We express the distances in the trap in units of the size of the cloud: $\tilde r =
(x/R_x,y/R_y,z/R_z)$ and in front of the integral (\ref{LDANharm}) we have the
geometrical average $R_xR_yR_z = R^3$ appearing. It means that the trap frequencies
(even if the trap is not spherical) enter only through combination $\omega_{ho} =
(\omega_x\omega_y\omega_z)^{1/3}$ and the oscillator lengths correspondingly through
parameter $a_{ho} =\sqrt{\hbar/m\omega}$. Now the integral is to be taken inside a
sphere of radius $1$ and is symmetric in respect to $\tilde r$. It follows
immediately, that the normalization condition (\ref{LDANharm}) in general can be
written as
\begin{eqnarray}
\Delta_{3D}^3 =
\tilde\mu^{3/2}
\int_0^1 a^3\mu_{\hom}^{-1}\left[\frac{\hbar^2}{ma^2}
\tilde\mu\Delta_{3D}^2(1-\tilde r^2)\right]4\pi\tilde r^2\,d\tilde r,
\label{LDARdim}
\end{eqnarray}
here the dimensionless chemical potential $\tilde\mu$ is obtained by choosing
$\frac{N^{1/3}}{2}\hbar\omega_{ho}$ as the unit of energy in the trap, the density
in a homogeneous system $\mu_{\hom}^{-1}$ is measured in units of $a^{-3}$, where $a$
is a length scale convenient for the homogeneous system (for example it can be equal
to the $s$-wave scattering length $a_{3D}$), chemical potential ({\it i.e.} the argument
of the inverse function $\mu_{\hom}^{-1}$) is measured in units of $\hbar^2/ma^2$,
and, finally, the characteristic parameter $\Delta_{3D}$ is defined as
\begin{eqnarray}
\Delta_{3D} = N^{1/6}\frac{a}{a_{ho}}
\label{LDADelta}
\label{Delta3D}
\end{eqnarray}

From the Eq.~\ref{LDARdim}, which is basically a dimensionless version of
Eq.~\ref{LDANharm} we discover there is a scaling in terms of the characteristic
parameter $\Delta_{3D}$. In other words systems having different number of particles
and oscillator frequencies will have absolutely the same density profile and other
LDA properties (once expressed in the correct units as discussed above) if they have
equal values of parameter (\ref{LDADelta}).

A similar procedure can be carried in a one-dimensional case (we choose the $z$
axis), where the normalization condition reads as
\begin{eqnarray}
N = \int\mu_{\hom}^{-1}\left[\mu-\frac{1}{2}m\omega_zz^2\right]\,dz
\label{LDANharmZ}
\end{eqnarray}

Its dimensionless form is obtained by measuring the energies in the trap in units
of $\frac{1}{2}N\hbar\omega_z$
\begin{eqnarray}
\Delta_{1D} =\tilde\mu^{1/2}\int_{-1}^1 a\mu_{\hom}^{-1}\left[\frac{\hbar^2}{ma^2}
\tilde\mu\Delta_{1D}^2(1-\tilde z^2)\right]\,d\tilde z,
\label{LDA1Ddimensionless}
\end{eqnarray}
and the one-dimensional characteristic parameter is related to the number of
particles as
\begin{eqnarray}
\Delta_{1D} = \frac{N^{1/2}a}{a_z}
\label{Delta1D}
\end{eqnarray}

\subsection{Exact solution for 1D ``perturbative'' equation of state\label{secLDA1D}}

We will start from very general equation of state of a homogeneous system which can
be found in any type of first-order perturbation theory. In the zeroth approximation
one has\footnote{This approximation is called {\it polytropic}.}\footnote{Many
theories produces results that fall into the class of equations of state described
by formula (\ref{polytropic}). For example GP theory, ideal fermi gas, TG gas.}:
\begin{eqnarray}
\mu_{\hom}^{(0)}=C_{1}(na)^{\gamma_1}\frac{\hbar ^{2}}{ma^{2}},
\label{polytropic}
\end{eqnarray}
here $a$ is unit of length, $C_{1}$ is a numerical coefficient of the
leading term in the chemical potential and $\gamma _{1}$ is the power of the
dependence on the gas parameter $na.$ The next term of perturbation in
general can be written as
\begin{equation}
\mu_{\hom}^{(1)}=C_{1}(na)^{\gamma_1}(1+C_{2}(na)^{\gamma _{2}}+...)%
\frac{\hbar ^{2}}{ma^{2}},
\label{muhom}
\end{equation}
where $C_{2}(na)^{\gamma _{2}}\ll 1.$ We will use local density approximation
(Sec.~\ref{secLDA}) in order to obtain properties of trapped system. The equation
(\ref{LDA}) can be inverted by using (\ref{muhom}) to obtain the density profile
$n(z)$:
\begin{equation}
n(z)a={\left( \frac{1}{{C_{1}}}\frac{\mu}{\hbar ^{2}/ma^{2}\,}
\left(1-\frac{z^{2}}{R^{2}}\right) \right)}^{\frac{1}{{{\gamma}_{1}}}}-
\frac{{C_{2}}}{{{\gamma}_{1}}}{\left( \frac{1}{{C_{1}}}\frac{\mu}
{\hbar^2/ma^{2}\,}\left( 1-\frac{z^{2}}{R^{2}}\right) \right)}^{\frac{{1}+
{{\gamma}_{2}}}{{{\gamma}_{1}}}},
\label{LDA1Dz2}
\end{equation}
here size of the cloud $R$ is related to the chemical potential $\mu
=\frac{1}{2}m\omega ^{2}R^{2}$ (\ref{LDAR}).

The value of the chemical potential is fixed by the normalization condition
(\ref{LDAN}). It is convenient to make use of the integral equality \cite{Gradstein80}
\begin{equation}
\int\limits_{-1}^{1}(1-x^{2})^{\alpha}\,dx=\frac{\sqrt{\pi}\Gamma (\alpha
+1)}{\Gamma (\alpha +\frac{3}{2})},\qquad\alpha >-1
\end{equation}

Thus we have restriction on the polytropic indices
$\gamma_1>-1,\frac{\gamma_1+\gamma_2}{\gamma_1}>-1$. If those conditions are
satisfied, then the leading contribution to the chemical potential is given by
\begin{eqnarray}
\frac{\mu ^{(0)}}{\hbar ^{2}/ma^{2}}={\left( \frac{C_{1}^{\frac{1}{\gamma
_{1}}}\,\,\Gamma (\frac{1}\gamma_1+\frac{3}{2})}{\sqrt{2\,\pi}\Gamma (
\frac{1}\gamma_1+1)}{\Delta_{1D}}^{2}\right)}^{\frac{2\,\gamma _{1}}{
2+\gamma _{1}}},
\label{LDAmu1D0}
\end{eqnarray}
where $\Delta_{1D}$ is the characteristic parameter of a one-dimensional trapped
gas defined by (\ref{Delta1D}).

In the next order of accuracy the chemical potential is given by
\begin{eqnarray}
\frac{\mu ^{(1)}}{\hbar ^{2}/ma^{2}}=\frac{\mu ^{(0)}}{\hbar ^{2}/ma^{2}}+%
\frac{\,\sqrt{8\,\pi}\,C_{1}^{-\frac{1+{{\gamma}_{2}}}{{{\gamma}_{1}}}}C{%
_{2}}\,\,\,\,}{\left( 2+{{\gamma}_{1}}\right) {\Delta_{1D}}^{2}\,\,\,\,}\,\frac{%
\Gamma (1+\frac{1+{{\gamma}_{2}}}{{{\gamma}_{1}}})}{\Gamma (\frac{3}{2}+%
\frac{1+{{\gamma}_{2}}}{{{\gamma}_{1}}})}\left( \frac{\mu ^{(0)}}{\hbar
^{2}/ma^{2}}\right) ^{\frac{3}{2}+\frac{{1}+\,{{\gamma}_{2}}}{\,{{\gamma}%
_{1}}}}
\end{eqnarray}

The mean square displacement $\left\langle z^{2}\right\rangle =\frac{1}{N}
\int\limits_{-R}^{R}z^{2}n(z)\,\,dz$ is directly related to the
potential energy of the oscillator confinement and is given by
\begin{eqnarray}
\frac{\left\langle z^{2}\right\rangle}{R^{2}}=\frac{\,{{\gamma}_{1}}}{2+3\,%
{{\gamma}_{1}}}\left( 1+\frac{\sqrt{\pi}\,{C_{2}}\,{{\gamma}_{2}}}{2^{1+%
\frac{2}{{{\gamma}_{1}}}}}\frac{\ \Gamma (2+\frac{2}{{{\gamma}_{1}}}%
)\,\Gamma (1+\frac{1+{{\gamma}_{2}}}{{{\gamma}_{1}}})\,}{\Gamma {(\frac{1}{%
{{\gamma}_{1}}})}^{2}\Gamma (\frac{5}{2}+\frac{1+{{\gamma}_{2}}}{{{\gamma}%
_{1}}})}\,{\left( \frac{1}{\,{C_{1}}}\frac{\mu}{{\hbar}^{2}/m~{a}^{2}}%
\right)}^{\frac{{{\gamma}_{2}}}{{{\gamma}_{1}}}}\right)
\end{eqnarray}

The frequencies of the collective oscillations can be predicted within LDA. The
frequency of the breathing mode is inferred from the derivative of the mean square
displacement $\Omega_{z}^{2}=-2\left\langle z^{2}\right\rangle \left/ \frac{\partial
\left\langle z^{2}\right\rangle}{\partial \omega ^{2}}\right.$
\cite{Menotti02} and equals to
\begin{eqnarray}
\frac{\Omega _{z}^{2}}{\omega _{z}^{2}}=\left( 2+{{\gamma}_{1}}\right) +%
\frac{\sqrt{\pi}{C_{2}}\,{{\gamma}_{2}}\,\left( {{\gamma}_{1}}+{{\gamma}%
_{2}}\right) \left( \frac{3}{2}+\frac{1}{{{\gamma}_{1}}}\right) \Gamma (1+%
\frac{1+{{\gamma}_{2}}}{{{\gamma}_{1}}})\Gamma (2+\frac{2}{{{\gamma}_{1}}}%
)\,}{2^{1+\frac{2}{{{\gamma}_{1}}}}\Gamma (1+\frac{1}{{{\gamma}_{1}}}%
)\,\Gamma (\frac{1}{{{\gamma}_{1}}})\,\Gamma (\frac{5}{2}+\frac{1+{{\gamma}%
_{2}}}{{{\gamma}_{1}}})}{\left(\frac{\,\Gamma (\frac{3}{2}+\frac{1}{{{%
\gamma}_{1}}}){\Delta_{1D}}^{2}}{\sqrt{2\,\pi {{C}_{1}}}\Gamma (1+\frac{1}{{{%
\gamma}_{1}}})}\right)}^{\frac{2\,{{\gamma}_{2}}}{2+{{\gamma}_{1}}}}
\label{1Dfreq}
\end{eqnarray}
The obtained formula is very general and gives an insight to many interesting cases
where the perturbation theory can be developed. In the table
(\ref{tableFrequencies}) we summarize some of the examples.

\begin{table}[ht!]
\centering
\begin{tabular}{|l|l|l|l|r|l|}
\hline
Limit & $C_1$ & $\gamma_1$ & $C_2$ & $\gamma_2$ & $\Omega_z^2/\omega_z^2$\\ \hline
Lieb-Liniger: weak interaction & $2\pi^2$ & $1$ & $-\sqrt{2}/\pi$ & -1/2& $\displaystyle3+\frac{5(9{\pi )}^{1/3}\,}{32\,\sqrt{2}}/\,{\Delta_{1D} }^{2/3}$ \\ \hline
Lieb-Liniger: strong interaction & $\pi^2/2$ & 2 & -8/3 & 1 & $\displaystyle4-\frac{128\sqrt{2}}{15\pi ^{2}}{\Delta_{1D} }$ \\ \hline
Attractive Fermi gas: strong interaction &$\pi ^{2}/32$ & 2 & 2/3 & 1 & $\displaystyle4+\frac{64\sqrt{2}}{15\pi ^{2}}{\Delta_{1D} }$ \\ \hline
Attractive Fermi gas: weak interaction & $\pi ^{2}/8$ & 2 &$ -8/\pi ^{2}$ & -1&$\displaystyle 4+\frac{32}{3\pi ^{2}}/{\Delta_{1D}}$ \\ \hline
Repulsive Fermi gas: strong interaction &$ \pi ^{2}/2$ & 2 & $-8\ln (2)/3$ & 1& $\displaystyle4-\frac{128\sqrt{2}\ln 2}{15\pi ^{2}}{\Delta_{1D}}$ \\ \hline
Repulsive Fermi gas: weak interaction &$\pi^2/8$ & 2 &$8/\pi^2$ & -1 &$\displaystyle4-\frac{32}{3\pi ^{2}}/{\Delta_{1D}}$ \\ \hline
Gas of Hard-Rods & $\pi^2/2$ & 2 & 8/3 & 1 & $\displaystyle 4+\frac{128\sqrt{2}}{15\pi ^{2}}{\Delta_{1D}}$ \\ \hline
\end{tabular}
\caption{Summary for some of one-dimensional models where the expansion of the
equation of state is known. The first column labels the considered model. The
coefficients of the expansion are given in columns 2-5. The last column gives the
predictions for the oscillation frequencies calculated calculated as (\ref{1Dfreq}).
The parameter $\Delta_{1D}$ is defined by (\ref{LDADelta}). Note that the presence
of a term in the chemical potential independent of the density (for example, binding
energy of a molecule) does not modify the frequencies of oscillations and is
ignored.}
\label{tableFrequencies}
\end{table}

\subsection{Exact solution for 3D ``perturbative'' equation of state\label{secLDA3D}}

In this Section we will develop theory in three-dimensions for the ``perturbative''
equation of state which we define as:
\begin{eqnarray}
\mu_{\hom}^{(0)}=C_{1}(na^3)^{\gamma_1}(1+C_2(na^3)^{\gamma_2})\frac{\hbar^2}{ma^2},
\label{polytropic3D}
\end{eqnarray}
where the $|C_2(na^3)^{\gamma_2}|\ll 1)$ is the perturbative term.

Within the local density approximation we obtain the chemical potential in a
trapped system. The leading term is given by
\begin{eqnarray}
\frac{\mu^{(0)}}{N^{1/3}\hbar\omega_{ho}}=\frac{1}{\Delta_{3D}^2}
\left(\frac{C_1^{1/\gamma_1}\Delta_{3D}^6}{(2\pi)^{3/2}}
\frac{\Gamma(\frac{5}{2} + \frac{1}{\gamma_1})}
{\Gamma(1 + \frac{1}{\gamma_1})}\right)}^{\frac{2\gamma_1}{3\gamma_1 + 2},
\label{mu3D0}
\end{eqnarray}
where $\Delta_{3D}$ is the characteristic combination (\ref{Delta3D}). The next correction
to (\ref{mu3D0}) is given by
\begin{eqnarray}
\frac{\mu^{(1)}}{\mu^{(0)}}=
1 + \frac{2C_2}{2 + 3\gamma_1}
\frac{\Gamma(\frac{5}{2} + \frac{1}{\gamma_1})
\Gamma(1 + \frac{1 + \gamma_2}{\gamma_1})}{\Gamma(1 + \frac{1}{\gamma_1})
\Gamma(\frac{5}{2} + \frac{1 + \gamma_2}{\gamma_1})}
{\left(\frac{\Delta_{3D}^6\Gamma(\frac{5}{2} +\frac{1}{\gamma_1})}{{(2\pi C_1)}^{3/2}
\Gamma(1 + \frac{1}{\gamma_1})}\right)}^{\frac{2\,\gamma_2}{3\gamma_1+2}}
\label{mu3D1}
\end{eqnarray}

The density profile is given by
\begin{eqnarray}
n(r)a^3 = \left(\frac1{C_1}\frac\mu{\hbar^2/ma^2}\left(1-\frac{r^2}{R^2}\right)\right)^\frac1{\gamma_1}
-\frac{C_2}{\gamma_1}\left(\frac{1}{C_1}\frac{\mu}{\hbar^2/ma^2}\left(1-\frac{r^2}{R^2}\right)\right)^\frac{1+\gamma_2}{\gamma_1},
\end{eqnarray}
where the chemical potential is given by (\ref{mu3D0}-\ref{mu3D1}), the size of the
condensate $R$ is defined in (\ref{LDAR}) while the relation between different units
of energy is provided by $N^{1/3}\hbar\omega_{ho} = \Delta_{3D}^2 \hbar^2/ma^2$.

We give an explicit expression for the density in the center of the trap. The
leading term is:
\begin{eqnarray}
n^{(0)}(0)a^3 = {\left(\frac{\Gamma(\frac{5}{2}+\frac{1}{\gamma_1})}
{{\left(2\pi{C_1}\right)}^{\frac{3}{2}}\Gamma(1+\frac{1}{\gamma_1})
{{\Delta_{3D}}}^{\frac{4}{\gamma_1}}}\right)}^{\frac{2}{2+3\gamma_1}}{\Delta_{3D}}^{\frac{4}{\gamma_1}}
\end{eqnarray}

The next term is:
\begin{eqnarray}
n^{(1)}(0)=n^{(0)}(0)
-\frac{C_2}{\gamma_1}
\left(1-\frac{2}{(2+3\gamma_1)}
\frac{\Gamma(1+\frac{1+\gamma_2}{\gamma_1})
\Gamma(\frac{5}{2}+\frac{1}{\gamma_1})}{\Gamma(1+\frac{1}{\gamma_1})
\Gamma(\frac{5}{2}+\frac{1+\gamma_2}{\gamma_1})}\right)
{\left(\frac{{\Delta_{3D}}^6\Gamma(\frac{2+5\gamma_1}{2\gamma_1})}{{\left(2\pi{C_1}
\right)}^{\frac{3}{2}}\Gamma(\frac{1+\gamma_1}{\gamma_1})}\right)}^{\frac{2\left(1+\gamma_2\right)}{2+3\gamma_1}}
\end{eqnarray}

The scaling approach allows calculation of the frequencies of the collective
oscillations. The frequency of the breathing mode is\footnote{S. Stringari,
unpublished.} $\Omega^2/\omega_{ho}^{2}=3/2~\Xi-1$ in a spherical trap,
$\Omega_z^2/\omega_z^2=3-2/\Xi$ and $\Omega_\perp^2/\omega_\perp^2=\Xi$ in a very
elongated trap $\omega_z\ll\omega_\perp$. The parameter $\Xi$ defining oscillation
frequencies is given by
\begin{eqnarray}
\Xi = 2(1+\gamma_1)
+\frac{4C_2(\gamma_1+\gamma_2)\gamma_2}{1+\gamma_1+\gamma_2}
\frac{\Gamma(\frac{7}{2}+\frac{1}{\gamma_1})\Gamma(1+\frac{1+\gamma_2}{\gamma_1})}
{\Gamma(1+\frac{1}{\gamma_1})\Gamma(\frac{7}{2}+\frac{1+\gamma_2}{\gamma_1})}
{\left(\frac{\Delta_{3D}^6}{{(2\pi C_1)}^{\frac{3}{2}}}
\frac{\Gamma(\frac{5}{2}+\frac{1}{\gamma_1})}{\Gamma(1+\frac{1}{\gamma_1})}\right)}^{\frac{2\gamma_2}{2 + 3 \gamma_1}}
\end{eqnarray}

The ``expansion'' equation of states (\ref{muhom},\ref{polytropic3D}) can be
naturally applied to the problems, where it is possible to construct a perturbation
theory. Another possible application of the discussed above method is to consider
the parameters $C_1,C_2,\gamma_1,\gamma_2$ as variational and fix them by fitting to
an equation of state, where an exact solution to the LDA problem is not known. An
arbitrary equation of state can be expanded as (\ref{muhom},\ref{polytropic3D}) in
any point, for example, by demanding that the first three derivatives of the
function and the function itself coincide with the ones calculated from the
(\ref{muhom},\ref{polytropic3D}). The four conditions of the continuity fixes four
parameters.

\subsection{Static structure factor of a trapped Tonks-Girardeau gas\label{secSkLDA}}

The chemical potential of the Tonks-Girardeu gas is known due to fermion-bosonic
mapping \cite{Girardeau60}. It equals to the fermi energy of a one-dimensional
spinless fermi gas and is given by the formula (\ref{muTG}). The dependency on the
density is simple and lies within class of functions (\ref{polytropic}) for which
the LDA problem was solved in Sec.~\ref{secLDA1D}. The TG gas is described by the
subsequent set of parameters: $C_1 = \pi^2/2,\gamma_1 = 2,C_2=0$.

\label{CalogeroRef}
The value of the chemical potential of the TG gas in a trap is immediately found
from Eq.~\ref{LDAmu1D0} and equals to $\mu = N\hbar\omega_z$. Its integration with
the respect of the number of particles gives the total energy in the LDA\footnote{It
is interesting to note that the result (\ref{ETGLDA}) of an {\it approximate}
solution for particles of infinite repulsion coincides with an {\it exact} result
for particles interacting with $g/r^2$ interaction (Calogero-Sutherland model
\cite{Calogero69,Sutherland71}) in the limit $g\to 0$. Indeed, as shown in
\cite{Sutherland71} the energy of such a gas equals $E/N =
\frac{1}{2}\hbar\omega_z(1+\lambda(N-1))$, where $\lambda$ is related to the
strength of interaction $g=2\lambda(\lambda-1)$. There are two different ways of
taking the limit $g\to 0$: 1) $\lambda\to 0$, $E/N\to\frac{1}{2}\hbar\omega_z$ {\it i.e.}
this limit corresponds to non-interacting bosons all staying in the lowest state of
a harmonic oscillator 2) $\lambda\to 0$, $E/N\to\frac{1}{2}N\hbar\omega_z$, {\it i.e.}
this limit preserves the singularity of the interaction while makes the potential
energy vanishing.}:
\begin{equation}
\frac{E}{N}=N\frac{\hbar\omega_z}{2}
\label{ETGLDA}
\end{equation}

The density profile is a semicircle
\begin{equation}
n(z) = n_0\left(1-\frac{z^2}{R_z^2}\right)^{1/2},
\label{nTG}
\end{equation}
with system size given by (\ref{LDAR}) $R_z = \sqrt{2N}a_z$ and the density in the
center equal to $n_0a_z = \sqrt{2N}/\pi$. The mean square radius of the trapped
system is given then by
\begin{equation}
\sqrt{\langle z^2\rangle}=\sqrt{\frac{N}{2}}a_z
\label{zTGLDA}
\end{equation}

The static structure factor of a uniform system depends on value of momentum $k$ and
on the density ({\it i.e.} on the value of the fermi momentum $k_f$) as given by formula
(\ref{Sk tonks}). We approximate the static structure factor in a trap by averaging it
over the density profile (\ref{nTG}):
\begin{eqnarray}
S^{LDA}(k) = \frac{1}{2R_z}\int\limits_{-R}^R S(k,n(z))\,dz
=\frac{ka_z}{\sqrt{8N}}\arcsin\sqrt{1-\frac{(ka_z)^2}{8N}}
+1-\sqrt{1-\frac{(ka_z)^2}{8N}}
\label{SkTGLDA}
\end{eqnarray}

It is easy to check that it vanishes for small momenta
$S^{LDA}(k) \to 0, k\to 0$, while it saturates to unity at large values of momenta
$S^{LDA}(k) = 1,|k|>\sqrt{8N}/a_z$.

\section{Correlation functions in a Luttinger liquid\label{secLuttinger}}
\subsection{Stationary density-density correlation function\label{secg2stat}}

The long-range properties of a weakly interacting one-dimensional bosonic gas can be
calculated using the macroscopic representation of the field operator (\ref{Psi}):
$\hat\Psi(x) = \sqrt{\rho_0+\hat\rho^{\prime}(x)}e^{i\hat \varphi(x)}$, where
$\rho_0$ is the mean density\footnote{In this section we keep a different notation
for the linear density $\rho\equiv n_{1D} = N/L$} and $\hat\varphi(x)$ is the phase
operator. Those operators can be expressed in terms of quasiparticle creation and
annihilation operators (see., for example, \cite{Pitaevskii03} Eqs.(6.65-6.66), and
consider a one-dimensional system):
\begin{eqnarray}
\label{phi}
\hat \varphi &=& - i\sum_k
\sqrt{\frac{\pi}{\eta |k|L}}
(\hat b_ke^{ikx}-\hat b^\dagger_k e^{-ikx})\\
\hat \rho^{\prime}&=& \sum_k
\sqrt{\frac{\eta |k|}{4\pi L}}
(\hat b_ke^{ikx}+\hat b^\dagger_k e^{-ikx}),
\label{rho}
\end{eqnarray}
where we introduced an important parameter describing the interactions between
particles:
\begin{eqnarray}
\eta = \frac{2\pi\hbar\rho_0}{Mc}
\label{eta}
\end{eqnarray}

The operators (\ref{phi},\ref{rho}) satisfy the commutation rule $[\hat \varphi(x),
\hat\rho^{\prime}(x^{\prime})] = -i\delta(x-x^{\prime})$.

Our approach is applicable in a weakly interacting gas $\rho_0\to\infty$. Deep in this
regime the speed of sound has a square root dependence on the density $c =
\sqrt{g\rho_0/M}$ and the coefficient $\eta$ is large $\eta =
2\pi\hbar\sqrt{\rho_0/Mg}$. In the opposite regime of strong correlations $\rho\to
0$ (TG limit) the bosonic system of impenetrable particles is mapped onto a system
of non-interacting fermions \cite{Girardeau60} with the speed of sound given by the
fermi velocity $c_F = \pi\hbar\rho_0/M$ (\ref{kTG}) and is proportional to the
density. In this regime $\eta = 2$. By generalizing the definition of the fermi
velocity from the TG regime, where the fermionization of a bosonic system happens,
to an arbitary density we obtain a simple interpretation of the parameter
(\ref{eta}): $\eta = 2c_F/c$. The speed of sound in a system with a repulsive
contact potential is not larger than the fermi velocity, thus in LL system
(\ref{LL}) $\eta \ge 2$. The situation becomes different in a gas of hard-rods of
size $a_{1D}$. Presence of an excluded volume makes the available phase space be
effectively smaller $L\to L-Na_{1D}$, which in turn renormalizes the speed of sound (see
\ref{EHR}) and makes it be larger. 
In this special case of the super-Tonks gas (Sec.~\ref{secST}) the parameter
$\eta$ (\ref{eta}) can be smaller than $2$.

Following Haldane \cite{Haldane81} we introduce a new field $\hat\vartheta(x)$ such
that $\nabla\hat\vartheta(x) = \pi[\rho_0+\hat\rho^{\prime}(x)]$.
The operator $\hat\vartheta$ satisfy boundary conditions $\hat\vartheta(x+L) =
\hat\vartheta(x) +\pi N$ and increases monotonically by $\pi$ each
time $x$ passes the location of a particle. Particles are thus taken to be located
at the points where $\hat\vartheta(x)$ is a multiple of $\pi$, allowing the density
operator to be expressed as $\hat\rho(x) = \nabla\hat\vartheta(x)\{\sum_n
\delta[\hat\vartheta(x)-\pi\rho]\}$, or, equivalently,
\begin{eqnarray}
\hat\rho(x) = [\rho_0 + \hat\rho^{\prime}(x)] \sum\limits_{m=-\infty}^\infty
\exp[i2m\hat\vartheta(x)]
\end{eqnarray}

Integrating (\ref{rho}) we obtain an expression of this field in terms of creation
and annihilation operators:
\begin{eqnarray}
\hat\vartheta(x) = \vartheta_0 + \pi\rho_0 x -i\sum_k
\sqrt{\frac{\pi\eta}{4|k|L}}\sign k\,
(\hat b_k e^{ikx}-\hat b^\dagger_k e^{-ikx})
\label{theta}
\end{eqnarray}

We will start with calculation of asymptotics of the density-density correlation
function:
\begin{eqnarray}
\langle\hat\rho(x)\hat\rho(0)\rangle \approx
(\rho_0^2 + \langle\hat\rho^{\prime}(x)\hat\rho^{\prime}(0)\rangle)
\langle\sum\limits_{m,m^{\prime}}
\exp[i2(m\hat\vartheta(x)+m^{\prime}\hat\vartheta(0))]\rangle
\label{rho(x,0)}
\end{eqnarray}

First of all we calculate the contribution coming from density fluctuations
$\rho^{\prime}(x)$:
\begin{eqnarray}
\langle\hat\rho^{\prime}(x)\hat\rho^{\prime}(0)\rangle =
\langle\sum_{k,k^{\prime}}
\frac{\eta\sqrt{|kk^{\prime}|}}{4\pi L}
(\hat b_k e^{ikx}+\hat b^\dagger_k e^{-ikx}) (\hat b_{k^{\prime}}+\hat
b^\dagger_{k^{\prime}}) \rangle  \label{rho'(x,0)}
\end{eqnarray}

The creation and annihilation operators satisfy bosonic commutation relations
$[\hat b_k, \hat b^\dagger_{k^{\prime}}] = \delta_{k,k^{\prime}}$ and at
zero temperature excitations are absent $\langle \hat b^\dagger_k\hat
b_k\rangle = 0$, so in the averaging in Eq.~\ref{rho'(x,0)} we get non
zero result only for $\langle \hat b_k\hat b^\dagger_k\rangle = 1$, {\it i.e.}
\begin{eqnarray}
\langle\hat\rho^{\prime}(x)\hat\rho^{\prime}(0)\rangle
=\sum_{k} \frac{\eta |k|e^{ikx}}{4\pi L}
=\int_{-\infty}^\infty \frac{\eta |k|}{4\pi Mc}e^{ikx}\frac{dk}{2\pi}
=\frac{\eta}{4\pi^2} \left.\left(\frac{1}{x^2}-\frac{ik}{x}\right)e^{ikx}\right|_0^\infty
\end{eqnarray}

We consider contribution only from the lower limit of the integration $k = 0$
and, thus, obtain
\begin{eqnarray}
\langle\hat\rho^{\prime}(x)\hat\rho^{\prime}(0)\rangle =-\frac{\eta}{4\pi^2}x^{-2}
\label{luttNN0}
\end{eqnarray}

Now let us calculate the contribution of the phase fluctuations in~(\ref{rho(x,0)}):
\begin{eqnarray}
\nonumber
\langle\sum\limits_{m,m^{\prime}} \exp[i2(m\hat\vartheta(x)+m^{\prime}\hat%
\vartheta(0))] \rangle =~~~~~~~~~~~~~~~~~~~~~~~~~~~~~~~~~~~~~~~~~~~~~~~~~~
~~~~~~~~~~~~~~~~~~~~~~~~~~ \\
=\!\!\langle\!\sum\limits_{m,m^{\prime}}\!\!\exp\!\!\left\{\!i2(m\!+\!m^{\prime})\vartheta_0
\!+\!i2\pi m\rho_0 x\!+\!\sum_k\!\!
\sqrt{\frac{\pi\eta}{|k|L}}
\!\sign k\!\left(\hat b_k(m e^{ikx}\!+\!m^{\prime})-\hat b^\dagger_k(m
e^{-ikx}\!+\!m^{\prime}))\right)\!\! \right\}\!\rangle
\label{luttNN2}
\end{eqnarray}

An average of a phonon operator $\hat A$ is gaussian and satisfies
an equality $\langle\exp\{\hat A\}\rangle = \exp\{\langle\hat A^2\rangle/2\}$. In
this way we can pass from an average of an exponent to an exponent of averaged
quantities:
\begin{eqnarray}
\nonumber
\langle\sum\limits_{m,m^{\prime}} \exp[i2(m\hat\vartheta(x)+m^{\prime}\hat%
\vartheta(0))]\rangle =\sum\limits_{m,m^{\prime}} \exp\left\{
i2(m+m^{\prime})\vartheta_0 + i2\pi m\rho_0 x\right\}
~~~~~~~~~~~~~~~~~~~~~~~~~~~~~ \\
\exp\!\!\left\{\!\! \sum_k\!\!\frac{\pi\eta}{2|k|L} \langle (\hat
b_k (m e^{ikx}\!\!+\!m^{\prime})\!-\!\hat b^\dagger_k(m e^{-ikx}\!\!+\!m^{\prime})) (\hat
b_k (m e^{ikx}\!\!+\!m^{\prime})\!-\!\hat b^\dagger_k(m e^{-ikx}\!\!+\!m^{\prime})) \rangle
\!\!\right\}
\label{LuttNN}
\end{eqnarray}

At the zero temperature excitations are absent. This simplifies the calculation as
the averaging in (\ref{LuttNN}) gives simply $\langle...\rangle
=-(me^{-ikx}+m^{\prime})(me^{ikx}+m^{\prime}) =
-(m^2+m^{\prime 2}+2mm^{\prime}\cos kx)$. We substitute the summation in the exponent
of (\ref{LuttNN}) with integration over $k$:
\begin{eqnarray}
-\sum_k \frac{\pi\eta}{2|k|L}(m^2+m^{\prime 2}+2mm^{\prime}\cos kx)
=-2\int\limits_0^\infty\frac{\pi\eta}{2k}(m^2+m^{\prime 2}+2mm^{\prime}\cos kx)\frac{dk}{2\pi}
\label{exponent}
\end{eqnarray}

This integral has an infrared divergence unless $m^{\prime}= -m$, so we consider
only these terms. Now the integral converges at small $k$ and takes the leading
contribution in the interval $1/x<k<1/\xi$ where $\xi$ is minimal length at which
the hydrodinamic theory can be applied. In this region one can neglect the
contribution coming from the oscillating cosine term and one has
\begin{eqnarray}
-4m^2\int\limits_{1/x}^{1/\xi} \frac{\pi\eta}{2k}\frac{dk}{2\pi}
=-m^2\eta(\ln(1/\xi)-\ln(1/x)) =-\eta m^2\ln(x/\xi)
\label{crit exp}
\label{luttNN3}
\end{eqnarray}

Finally, collecting together (\ref{luttNN0},\ref{luttNN2},\ref{luttNN3}) we obtain
an expression for the stationary density-density correlation function
\begin{eqnarray}
\frac{\langle\hat\rho(x)\hat\rho(0)\rangle}{\rho_0^2}
= \left(1-\frac{\eta}{4\pi^2}(\rho_0x)^{-2}\right)
\left(1+ 2\sum\limits_{m=1}^\infty C_i \cos(2\pi m\rho_0 x)
\left(\frac{x}{\xi}\right)^{-\eta m^2}\right)
\label{g2(x)log}
\end{eqnarray}

\subsection{Time-dependent density-density correlation function\label{secg2time}}

In this Section we develop an approach which allows an estimation of correlations
between different moments of time. We substitute the stationary hydrodynamic
expressions of phase and density operators (\ref{phi},\ref{rho}) on time-dependent
hydrodynamic expressions (see, for example, \cite{Lifshitz80}, Eqs.(24.10)):
\begin{eqnarray}
\hat \varphi(x,t) &=& - i\sum_k \sqrt{\frac{\pi}{\eta |k|L}}
(\hat b_k e^{i(kx-|k|ct)}-\hat b^\dagger_k e^{-i(kx-|k|ct)}),  \label{phi(t)} \\
\hat \rho^{\prime}(x,t)& =& \sum_k \sqrt{\frac{\eta|k|}{4\pi L}} (\hat
b_k e^{i(kx-|k|ct)}+\hat b^\dagger_k e^{-i(kx-|k|ct)})  \label{rho(t)}
\end{eqnarray}

It is easy to note (see Eqs.~\ref{phi},\ref{rho}) that the time $t$ enters always
in the combination $(kx-|k|ct)$, which means that time-dependent solution can be
obtained from stationary solution by changing $kx\to kx-|k|ct $ in integrands and
carrying out integration again. Density fluctuations (\ref{rho'(x,0)}) are than
given by
\begin{eqnarray}
\langle\hat\rho^{\prime}(x,t)\hat\rho^{\prime}(0,0)\rangle
=\int_0^\infty\frac{\eta k}{4\pi}(e^{ik(x-ct)}+e^{ik(x+ct)})\frac{dk}{2\pi}
= - \frac{\eta}{8\pi^2} \left(\frac{1}{(x+ct)^2} + \frac{1}{(x-ct)^2}\right)
\end{eqnarray}

Here again we considered the contribution from the lower limit $k=0$.

The contribution from the phase fluctuations (\ref{luttNN2}) is calculated analogously to
(\ref{exponent}):
\begin{eqnarray}
-\eta m^2 \left[\int\limits_0^\infty \frac{[1-\cos k(x+ct)]}{k}dk +\int\limits_0^\infty \frac{[1-\cos k(x-ct)]}{k}dk\right]
\end{eqnarray}

The main contribution to integrals comes from momenta $1/(x+ct)<k<1/\xi$ in the
first integral and $1/(x-ct)<k<1/\xi$ in the second one. As we are interested in
description of asymptotically large distances condition $x>ct$ is always fulfilled.
In this conditions the integration gives
\begin{eqnarray}
-\frac{1}{2}\eta m^2\left[\ln\frac{x+ct}{\xi}+\ln\frac{x-ct}{\xi}\right]
=-\frac{1}{2}\eta m^2\ln\frac{x^2-c^2t^2}{\xi^2}
\end{eqnarray}

Thus we find that the asymptotic behavior of the time-dependent density-density
correlation function is given by
\begin{eqnarray}
\frac{\langle\hat\rho(x,t)\hat\rho(0,0)\rangle}{\rho_0^2}
= 1-\frac{\eta}{8\pi^2\rho_0^2} \left(\frac{1}{(x+ct)^2} + \frac{1}{(x-ct)^2}\right)
+2\sum\limits_{m=1}^\infty C_i\cos(2\pi m\rho_0 x)\left(\frac{x^2-c^2t^2}{\xi^2}\right)^{-\frac{1}{2}\eta m^2}
\label{g2(x,t)log}
\end{eqnarray}

\subsection{Calculation with non-logarithmic accuracy \label{secSkstat}}

The phonon dispersion $\omega = c|k|$ relation which was used in the derivation
above leads to infrared divergence in some of the integrals (\ref{exponent}) and was
resolved by truncation of the integral. This problem can be cured using a more
precise Bogoliubov dispersion law:
\begin{eqnarray}
\omega(k) = \sqrt{(kc)^2+\left(\frac{\hbar k^2}{2M}\right)^2}
\end{eqnarray}

It is easy to see that results for the new dispersion can be obtained by changing
formally
the speed of sound $c|k|\to c|k|\sqrt{1+(\hbar k/2Mc)^2}$
in definitions of hydrodynamic operators (\ref{phi(t)}-\ref{rho(t)}. This will lead
to a converging value of the integral (\ref{exponent}):
\begin{eqnarray}
-2\int\limits_0^\infty \frac{\eta\pi2m^2(1-\cos kx)}{2k\sqrt{1+(\hbar k/2Mc)^2}}\frac{dk}{2\pi}
=-\eta m^2\int\limits_0^\infty\frac{1-\cos z}{\sqrt{1+\varepsilon^2z^2}}\frac{dz}{z},
\label{luttInt}
\end{eqnarray}

Here we introduced the notation $z = xk$ and $\varepsilon = \hbar/2Mcx$.
Let us split the integral (\ref{luttInt}) in two parts:
\begin{eqnarray}
\int\limits_0^\infty\frac{1-\cos z}{\sqrt{1+\varepsilon^2z^2}} \frac{dz}{z}
\approx \int\limits_0^N (1-\cos z)\frac{dz}{z} +\int\limits_N^\infty \frac{dz%
}{z\sqrt{1+\varepsilon^2z^2}},
\end{eqnarray}
in such a way that $1\ll N \ll 1/\varepsilon$. The term $(1+\varepsilon^2z^2)$
can be neglected in the integration up to $N$ and oscillating term can
be neglected at larger distances. In order to proceed further we shift the
lower integration limit by short distance $\epsilon \to 0$. Then the first
integral becomes equal to $\ln N/\epsilon-\mathop{\rm Ci}\nolimits N+%
\mathop{\rm Ci}\nolimits\epsilon$. The second integral can be easily
calculated by using substitution $y^2=1+\varepsilon^2z^2$ and equals
$\frac{1}{2}\ln \frac{\sqrt{1+\varepsilon^2N^2}+1}{\sqrt{1+\varepsilon^2N^2}-1}$.
Collecting everything together
we obtain
\begin{eqnarray}
\int\limits_0^\infty\frac{1-\cos z}{\sqrt{1+\varepsilon^2z^2}} \frac{dz}{z}
\approx \gamma + \ln \frac{4Mcx}{\hbar}
= \gamma + \ln \frac{8\pi\rho_0 x}{\eta},
\end{eqnarray}
where $\gamma \approx 0.577$ is Euler's constant.

The static density-density correlation function (\ref{g2(x)log}) (more precisely its $m
\ne 0$ part) is equal to
\begin{eqnarray}
\frac{\langle\hat\rho(x)\hat\rho(0)\rangle}{\rho_0^2} =
1+2\sum\limits_{m=1}^\infty
\left(\frac{\eta}{8\pi C}\right)^{\eta m^2}
\frac{\cos(2\pi m\rho_0 x)}{(\rho_0 x)^{\eta m^2}},
\label{g2(x)}
\end{eqnarray}
where $C = e^\gamma \approx 1.781$.

The time-dependent result differ from the stationary case (\ref{g2(x)}) only by
substitution $x \to \sqrt{x^2-c^2t^2}$ in the denominator, as it was already shown
in the calculation with logarithmic accuracy (compare \ref{g2(x)log} and
\ref{g2(x,t)log}):
\begin{eqnarray}
\frac{\langle\hat\rho(x,t)\hat\rho(0,0)\rangle}{\rho_0^2} =
1+2\sum\limits_{m=1}^\infty \left(\frac{\eta}{8\pi C}\right)^{\eta m^2}
\frac{\cos(2\pi m\rho_0 x)}{(\rho_0 \sqrt{x^2-c^2t^2})^{m^2\eta}},
\label{g2(x,t)}
\end{eqnarray}

\subsection{Dynamic form factor}

The dynamic form factor is related to the time-dependent density-density correlation
function by the means of the Fourier transform:
\begin{eqnarray}
S(k,\omega) = \frac{\rho_0}{\hbar}\int\!\!\!\!\int e^{i(\omega t-kx)}
\left[\frac{\langle\hat\rho(x,t)\hat\rho(0,0)\rangle}{\rho_0^2}-1\right]\, dx\, dt
\label{def Skw}
\end{eqnarray}

%

The correlation function was calculated with non-logarithmic accuracy and is given
by formula (\ref{g2(x,t)}). The evaluation of direct Fourier transform
(\ref{def Skw}) would give us the expression for the dynamic form factor. It turns
out that it is easier to go other way around, {\it i.e.} guess the form of the $m$-th
component of $S(k,\omega)$
\begin{eqnarray}
S(k,\omega) = A (\omega^2-c^2(k-2mk_F)^2)^{\frac{m^2\eta}{2}-1},
\end{eqnarray}
make the inverse Fourier transform to go back from $(k,\omega)$ to $(x,t)$ and
comparing obtain result to (\ref{g2(x,t)}) fix the value of the constant $A$.
Here we use notation $k_F = \pi\rho_0$.
We start by doing the integration over momentum.
%
$S(x,\omega) = \int\limits_{-\infty}^\infty e^{ikx} S(k,\omega)\frac{dk}{2\pi}$.
We introduce notation $\triangle k = k+2mk_F$ the integral is limited to
the region $(-\omega/c,\omega/c)$:
\begin{eqnarray}
S(x,\omega) =
A c^{m^2\eta-2} e^{i2mk_Fx} \int\limits_{-\omega / c}^{\omega / c}
\cos(\triangle k x)
\left[\left(\frac{\omega}{c}\right)^2-(\triangle k)^2\right]^{\frac{m^2\eta}{2}-1}
\frac{d(\triangle k)}{2\pi}
\end{eqnarray}

As a reference we use formula 3.771(464/465) from Gradstein-Ryzhik book \cite{Gradstein80}:
\begin{eqnarray}
\nonumber
\int_{0}^{u}}(u^{2}-x^{2})^{\nu-{1\over2}}{
\cos (ax) dx
= {\sqrt{\pi}\over2}\left({2u\over a}\right)^{\nu}\Gamma
\left(\nu+{1\over2}\right)J_{\nu}(au),\quad
\left[a>0, u>0,\mbox{Re} \nu>-{1\over2}\right]
\end{eqnarray}

The substitution $u=\omega/c$, $a=x$, $\nu = (m^2\eta-1)/2$ gives an expression
in terms of $(x,\omega)$
\begin{eqnarray}
S(x,\omega) =
\frac{A e^{i2mk_Fx}}{2\sqrt\pi c}
\left(\frac{2\omega c}{x}\right)^{\frac{m^2\eta-1}{2}}
\Gamma\left(\frac{m^2\eta}{2}\right)
J_{\frac{m^2\eta-1}{2}}\left(\frac{\omega x}{c}\right),
\end{eqnarray}
where $J_n(x)$ is the Bessel function of the first kind.
The integration over the frequencies
$S(x,t) = \int\limits_{0}^\infty e^{-i\omega t} S(x,\omega) \frac{d\omega}{2\pi}$
can be done easily done by using the formula (6.699.5) from \cite{Gradstein80}
\begin{eqnarray}
\nonumber
\int_{0}^{\infty}x^{\nu}\cos (ax) J_{\nu}(bx)\,dx
=2^\nu \frac{b^\nu}{\sqrt{\pi}}\Gamma\left(\frac{1}{2}+\nu
\right)(b^2-a^2)^{-\nu-\frac{1}{2}}\quad
\left[0< a<b,\quad |\mbox{Re}\,\nu|< {1\over 2}\right]
\end{eqnarray}
and gives
\begin{eqnarray}
S(x,t) = A\frac{(2c)^{m^2\eta-1}}{2\pi^2}
\Gamma^2\left(\frac{m^2\eta}{2}\right)\frac{\cos(2mk_F)}{(x^2-c^2t^2)^{\frac{m^2\eta}{2}}}
\end{eqnarray}

Comparing this result with (\ref{g2(x,t)}) we fix so far unknown coefficient of the
proportionality to the value $A =
\frac{8\pi^2c\rho_0}{\hbar\Gamma^2\left(m^2\eta/2\right)}
\left(\frac{\hbar}{8Cmc^2}\right)^{m^2\eta}$
and, finally, obtain
\begin{eqnarray}
S(k, \omega) =
\!\!\sum\limits_{m=1}^\infty\!
\frac{8\pi^2\rho_0 c}{\Gamma^2\left(\frac{m^2\eta}{2}\right)\hbar}\!\!
\left(\frac{\hbar}{8CMc^2}\right)^{m^2\eta}\!\!
\left[
\frac{(\omega^2\!\!-\!\!c^2(k\!\!-\!\!2mk_F)^2)^{\frac{m^2\eta}{2}-1}
\!+\!(\omega^2\!\!-\!\!c^2(k\!\!+\!\!2mk_F)^2)^{\frac{m^2\eta}{2}-1}}{2}\right]
\label{Skw}
\end{eqnarray}
%

\subsection{Popov's coefficient\label{secPopov}}

The introduced above approach allows us to find the asymptotic behavior of the
one-body density matrix and estimate the coefficient of its decay. Within the first
order of accuracy we split the average as
\begin{eqnarray}
g_1(x) =\langle \sqrt{\hat\rho(x)\hat\rho(0)} e^{i(\hat\varphi(x)-\hat
\varphi(0))}\rangle \approx \langle \sqrt{\hat\rho(x)\hat\rho(0)}\rangle
\langle e^{i(\hat\varphi(x)-\hat\varphi(0))}\rangle
\label{g1lutt}
\end{eqnarray}

We first calculate the contribution coming from the phase fluctuations
\begin{eqnarray}
g_{1}^{phase}(x)=\langle e^{i(\hat{\varphi}(x)-\hat{\varphi}(0))}\rangle
=\left\langle \exp \left\{ \sum_{k}\sqrt{\frac{\pi}{\eta|k|L}}(
\hat{b}_{k}(e^{ikx}-1)-\hat{b}_{k}^{\dagger }(e^{-ikx}-1)\right\}
\right\rangle
\label{luttg1ph}
\end{eqnarray}

The average of the exponent can be further developed by using the relation
for the gaussian average $\left\langle \exp A\right\rangle =
\exp \left\langle \frac{1}{2}A^{2}\right\rangle $. At zero temperature
excitations are absent and the only nonzero average is $\left\langle \hat{b}%
_{k}\hat{b}_{k}^{\dagger }\right\rangle =1$. Thus we obtain
\begin{eqnarray}
g_{1}^{phase}(x)=\exp \left\{ -\frac{1}{2}\sum_{k}\frac{\pi}{\eta |k|L}
(e^{ikx}-1)(e^{-ikx}-1)\right\} =\exp \left\{ -\int\limits_{-\infty}^{\infty}
\frac{\pi(1-\cos kx)}{\eta|k|}\frac{dk}{2\pi }\right\}
\label{luttg1ph2}
\end{eqnarray}

At this point we substitute the phononic excitation spectrum with the proper
Bogoliubov dispersion. This can be done by changing
$\eta\rightarrow \eta/\sqrt{ 1+(\hbar k/2Mc)^{2}}$.
\begin{eqnarray}
g_{1}^{phase}(x)=\exp \left\{ -\frac{1}{\eta }\int\limits_{0}^{\infty }\frac{%
\sqrt{1+(\hbar k/2Mc)^{2}}(1-\cos kx)dk}{k}\right\} \label{g1phint}
\end{eqnarray}

Formally this integral is diverging. However, we will take use of the properties of the
$\delta$-function
\begin{eqnarray}
\int\limits_{-\infty }^{\infty }\cos \frac{kx}{2\pi }=\delta (x)
\label{deltaint}
\end{eqnarray}

As we are interested in long range asymptotical behavior we subtract
(\ref{deltaint}) from the exponent of (\ref{g1phint}) and consider a well-convergent
expression
\begin{eqnarray}
g_{1}^{phase}(x)=\exp \left\{ -\frac{1}{\eta }\int\limits_{0}^{\infty
}\left( \frac{\sqrt{1+(\hbar k/2Mc)^{2}}}{k}-1\right) (1-\cos kx)\right\} dk
\end{eqnarray}

Partial integration together with the notation $\varepsilon =\hbar/(2xMc)\ll 1$ and $z=kx$ gives%
\begin{eqnarray}
g_{1}^{phase}(x)=\exp \left\{ -\frac{1}{\eta }\int\limits_{0}^{\infty}
\frac{z-\sin z}{z^{2}\sqrt{1+\varepsilon ^{2}z^{2}}}\right\}\,dz
\end{eqnarray}

This equation can be calculated with non-logarithmic accuracy at $x\gg \xi $
by splitting the integral in three parts $1\ll N\ll 1/\varepsilon $

\begin{eqnarray}
\int\limits_{0}^{\infty }\frac{z-\sin z}{z^{2}\sqrt{1+\varepsilon ^{2}z^{2}}}
dz=\int\limits_{\lambda }^{N}\frac{1}{z}dz-\int\limits_{\lambda }^{N}\frac{
\sin z}{z^{2}}dz+\int\limits_{N}^{\infty }\frac{1}{z\sqrt{1+\varepsilon
^{2}z^{2}}}dz
=\gamma -1-\ln \frac{\hbar}{4xMc}
\end{eqnarray}

The calculation gives the result
\begin{eqnarray}
g_{1}^{phase}(x)
=\left( \frac{e^{1-\gamma }\eta }{8\pi \rho _{0}x}\right) ^{\frac{1}{\eta }}
\label{psipsi}
\end{eqnarray}

In order to take into account the density fluctuations we develop (\ref{g1lutt})
using Taylor expansion $\sqrt{1+x} = 1+x/2-x^2/8+{\cal O}(x^3)$, so $g_1^{\rho}(x)
=\langle \sqrt{\hat\rho(x)\hat\rho(0)}\rangle \approx
\rho_0+\frac{1}{4\rho_0}\langle
[\hat\rho^{\prime}(x)-\hat\rho^{\prime}(0)]\hat\rho^{\prime}(0)\rangle$. Writing the
density operator in terms of creation and annihilation operators (eq. \ref{rho}) we
obtain
\begin{eqnarray}
g_1^{\rho}(x) =\rho_0+ \frac{1}{4} \sum_k
\frac{\rho_0\hbar|k|}{2LMc(k)} \langle(\hat b_k (e^{ikx}\!-\!1)+\hat b^\dagger_k
(e^{-ikx}\!-\!1)) (\hat b_k + \hat b^\dagger_k) \rangle
=\rho_0+\frac{1}{4} \int\limits_0^\infty \frac{\rho_0\hbar k}{Mc(k)}
(\cos kx\!-\!1)
\frac{dk}{2\pi}
\nonumber
\end{eqnarray}

We substitute the speed of sound for the Bogoliubov dispersion relation $c(k) =
c\sqrt{1+(\hbar k/2Mc)^2}$ and express the integral in dimensionless units
$\varepsilon = \hbar/(2Mcx)$, $z = kx$
\begin{eqnarray}
\langle \sqrt{\hat\rho(x)\hat\rho(0)}\rangle =\rho_0+ \frac{\rho_0\hbar}{
8\pi Mc x^2} \int\limits_0^\infty \frac{(\cos z -1)z\,dz}{\sqrt{1+\varepsilon^2 z^2}}
\end{eqnarray}

The integral can be evaluated and expanded for small $\varepsilon$
\begin{eqnarray}
\int\limits_0^\infty\frac{(\cos z -1)z\,dz}{\sqrt{1+\varepsilon^2 z^2}} =
\frac{1}{\varepsilon^2}+ \frac{\pi}{2\varepsilon^2}\left(I_1\left(\frac{1}{
\varepsilon}\right)-L_{-1}\left(\frac{1}{\varepsilon}\right)\right) \approx
\frac{1}{\varepsilon^2}+ (-1-3\varepsilon^2 +{\cal O}(\varepsilon^4)),
\end{eqnarray}
where $I_1(z)$ is modified Bessel function of first kind and $L_{-1}(z)$ is
modified Struve function.

In terms of the parameter $\eta$ we have
\begin{eqnarray}
\langle \sqrt{\hat\rho(x)\hat\rho(0)}\rangle =\rho_0\left(1+\frac{1}{\eta}-\frac{\eta}{16\pi\rho_0^2x^2}\right)
\label{rhorho}
\end{eqnarray}

Combining together (\ref{psipsi}) and (\ref{rhorho}) we obtain finally the
expression for the coefficient of the long-range asymptotics
\begin{eqnarray}
g_1(x) = \rho_0\left(\frac{e^{1-\gamma}\eta}{8\pi}\right)^{\frac{1}{\eta}}
\left(1+\frac{1}{\eta}-\frac{\eta}{16\pi\rho_0^2x^2}\right)
\left(\rho_0 x\right)^{-\frac{1}{\eta}}
\label{Casympt}
\end{eqnarray}


In order to get an expression for the one-body density matrix at a finite
temperature, one should account for thermal quasi-particle excitations. The
long-range excitations are phonons and obey Bose-Einstein statistics $\langle \hat
b^\dagger_k \hat b\rangle = (\exp(\hbar k c/k_B T)-1)^{-1}$. We are interested at
the long-range behavior of the one-body density matrix, which corresponds to the
limit $k\to 0$. In this conditions one can do a Taylor expansion and get $\langle
\hat b^\dagger_k \hat b\rangle = k_B T/\hbar |k| c$. The calculation of the average
(\ref{luttg1ph}) leads to appearance of an additional term, which depends on the
temperature (compare with (\ref{luttg1ph2})):
\begin{eqnarray}
g_{1}^{phase}(x)=\exp \left\{-\int\limits_{-\infty}^{\infty}
\frac{\pi(1-\cos kx)(1+2k_B T/\hbar|k| c)}{\eta|k|}\frac{dk}{2\pi }\right\}
\end{eqnarray}

As we will show, the additional thermal suppression becomes dominating and will
change the asymptotic behaviour of $g_1(x)$ significantly. The effect of the
thermal phase fluctuations can be separated:
\begin{eqnarray}
g_{1}^{phase}(x)=
g_{1,T=0}^{phase}(x)
\exp \left\{-\int\limits_{-\infty}^{\infty}
\frac{\pi(1-\cos kx)2k_B T/\hbar|k| c}{\eta|k|}\frac{dk}{2\pi}\right\}
\label{luttg1T}
\end{eqnarray}
where the zero temperature part $g_{1,T=0}^{phase}(x)$ is readily given by the
formula (\ref{psipsi}). The integral in (\ref{luttg1T}) is well behaved and can be
easily calculated. It turns out to be proportional to $-|x|$ leading to exponential
decay of the thermal fluctuation part at large distances:
\begin{eqnarray}
g_{1}^{phase}(x)
=\left( \frac{e^{1-\gamma }\eta }{8\pi\rho_0x}\right)^{\frac{1}{\eta}}
\exp\left(-\frac{|x|}{\xi_T}\right)
\end{eqnarray}

The characteristic thermal decay length $\xi_T$ is inversely proportional to the
temperature:
\begin{eqnarray}
\xi_T = \frac{2\hbar^2\rho_0}{mk_BT}=4\pi\rho_0\lambda^2,
\end{eqnarray}
where the de Broglie thermal length is defined in the usual way $\lambda =
\hbar/\sqrt{2\pi m k_B T}$.

\chapter{Quantum Monte Carlo technique\label{secQMC}}
\section{Introduction}
%
%
%

Quantum Monte Carlo methods (QMC) are very powerful tools for the investigation of
quantum many body systems (for a review see, for example,
\cite{Ceperley95},\cite{Guardiola98}). The usage of QMC techniques provides deep insight
into understanding of the physical problem. It allows one to accomplish the {\it ab
initio} calculation and, starting from a microscopic model (commonly a model
Hamiltonian), earn knowledge of the macroscopic behavior of the system. Often it
turns out that this approach is the only accessible tool for studying sophisticated
problems, as in order to have a model, which can be solved analytically in exact
way, one usually has to make severe assumptions, which can be relaxed in QMC. In
many cases it is possible to construct analytically a perturbation theory, then its
applicability is restricted by smallness of the perturbation parameter and also in
cases like that QMC methods can be used to avoid the restrictions. The QMC
techniques solve the many-body Schr\"odinger equation for the ground state and for
excited states at zero temperature. Similar to other MC approaches, these techniques
are based on stochastic numerical algorithms, which are powerful when one is
treating systems with many degrees of freedom.

We are interested in exploring the quantum properties of systems. The quantum
effects manifest the most at the lowest temperatures, when the system stays in the
ground state. Thus we choose the Diffusion Monte Carlo method to address the
problem. This method is {\it exact}\footnote{Of course, as this method is a
statistical one and all outputs are obtained within statistical errors which can be
decreased by making a longer series of measurements.} for calculation a ground state
energy of a bosonic system

In order to study a fermionic system we use Fixed-Note Monte Carlo technique (FN-MC),
which is a modification of the DMC method. In general this approach gives an upper
bound to the ground state energy, but with a good choice of the trial wave function
the difference can be significantly minimized.

In this chapter we will start from the Variational Monte Carlo method which is
applicable both for bosons and fermions. Then we will discuss bosonic Diffusion
Monte Carlo method and fermionic Fixed-Node Monte Carlo method. We will address in
details construction of the trial wave functions and, next, will discuss the
implementation of the measurements of the quantities of interest.

\section{Variational Monte Carlo \label{secVMC}}
\subsection{Variational principle}

The simplest of the Quantum Monte Carlo methods is the {\it variational} method
(VMC). The idea of this method is to use an approximate wave function $\psi_T$ ({\it
variational} or {\it trial wave function}) and then by sampling the probability
distribution
\begin{eqnarray}
p(\R)~=~|\psi_T(\R)|^2
\label{p(r)}
\end{eqnarray}
calculate averages of physical quantities. It is easy to show that the average
\begin{eqnarray}
E_T = \frac{\langle\psi_T|\hat H|\psi_T\rangle}{\langle\psi_T|\psi_T\rangle} \ge E_0
\label{Et}
\end{eqnarray}
gives an upper bound to the ground-state energy. By minimizing the variational
energy with respect to the external parameters one can optimize the wave function
within the given class of wave functions considered.

Importantly, the variational principle also applies to excited states. For a trial
wave function $\psi_T$ with a given symmetry, the variational estimate provides an
upper bound to the energy of the lowest excited state of the Hamiltonian $\hat H$
with that symmetry.

\subsection{Applications}

If the wave function of the ground state is known exactly, then the VMC sampling will
provide {\it exact} ground state properties. For example, in a system of hard-rods
one has a knowledge of the ground state wave function and energy, but the correlation
functions are not known. In this case the VMC calculation allows to complete the
description of the system. In a similar manner if one knows exactly an eigenstate
wave function, then the properties of the system can be obtained by the VMC sampling
in an exact way.

The VMC can be effectively used for studying the metastable states. In particular
one can make estimations of the critical parameters leading to the system collapse
(see Sec.~\ref{secHRstability}).

The VMC is used to optimize the variational parameters before doing DMC or FN-MC
calculations, as the efficiency of the latter depends significantly on the quality
of the trial wave function.

\subsection{Implementation}

It is convenient to work in the coordinate representation, as this representation is
the most natural for writing the interaction potential and external potential. In a
system of $N$ particles in $D$ dimensions the distribution function depends on $D N$
variables $p(\R) = p(\rv_1, ..., \rv_N)$. An average value of an operator $\hat A$
is then given by a $D N$-multidimensional integral
\begin{eqnarray}
\langle A\rangle = \frac{\int...\int A(\rv_1,...,\rv_N)p(\rv_1,...,\rv_N))
\vec{dr_1}...\vec{dr_N}}{\int...\int p(\rv_1,...,\rv_N))
\vec{dr_1}...\vec{dr_N}}
\label{A}
\end{eqnarray}

Even for a few particles $N\approx 10$ the structure of the integral becomes too
difficult for implementation usual discretization methods of integration and instead
one can use stochastic {Monte Carlo} methods. The idea of the method is to generate
a set of states ({\it chain}) $\R_1,...,\R_M$ with the probability distribution
$p(\R)$ and approximate the $\langle A\rangle$ as the average
\begin{eqnarray}
\langle A\rangle \approx \frac{1}{M}\sum\limits_{i=1}^M A(\R_i)
\end{eqnarray}

Such a chain where the next configuration $\R'$ depends only on the previous
configuration $\R$ (Markov chain) can be generated by the Metropolis algorithm
\cite{Metropolis53}: the new configuration is accepted with the probability $P(\R
\rightarrow \R')$ given by the rule
\begin{eqnarray}
P(\R \rightarrow \R') =
\left\{
\begin{tabular}{ccc}
$1$, & if $p(\R')\ge p(\R)$ \\
$p(\R')/p(\R)$, & if $p(\R') < p(\R)$ \\
\end{tabular}
\right.
\label{Metr}
\end{eqnarray}

In a quantum system the probability distribution is given by the square of the
wave function module as Eq.~\ref{p(r)}. The specific construction of wave functions
will be discussed in Sec.~\ref{secWF}.

The efficiency significantly depends on the type of the trial moves. A natural way
to generate a new configuration is to move all particles $\r_i' = \r_i+\vec\xi_i,
i=\overline{1,N}$, here $\vec\xi_i$ is a random shift delimited to range
$|\vec\xi_i|<\Xi$. If the amplitude of the shift $\Xi$ is too large the acceptance
rate becomes too small, if instead $\Xi$ is very small, almost all moves are
accepted, but the generated configurations are strongly correlated. The acceptance
rate of about $50\%$ in general provides a good choice.

The efficiency of a variational calculation can be highly improved by doing a
complex move consisting of separate moves of one particle at a time. Each
independent move allow a larger displacement at the fixed acceptance rate. Indeed,
the amplitude of each individual move can be an order of magnitude larger and
consequently leading to a faster convergence of the sampling.

\section{Diffusion Monte Carlo\label{secDMC}}

The Diffusion Monte Carlo method (DMC) can be successfully applied to the
investigation of bosonic systems at low temperatures. It is based on solving the
Schr\"odinger equation in the imaginary time and allows calculation of the exact (in
statistical sense) value of the ground state energy.

\subsection{Schr\"odinger equation}

The evolution of a quantum system is described by Schr\"odinger equation
\begin{eqnarray}
i\hbar \frac{\partial}{\partial t} \varphi (\R,t) = \hat H \varphi (\R,t),
\label{Schrodinger}
\end{eqnarray}
Instead of considering the time-evolution we will look for the ground state
properties. That can be done by introducing the {\it imaginary time} $\tau = -i
t/\hbar$. We rewrite the Schr\"odinger equation and introduce a constant energy
shift $E$, whose meaning will become clearer later:
\begin{eqnarray}
-\frac{\partial}{\partial \tau} \varphi (\R,\tau) = (\hat H -E) \varphi (\R,\tau),
\label{shifted Schrodinger}
\end{eqnarray}

The formal solution $\psi (\R,\tau) = e^{-(\hat H - E)\tau} \psi (\R,0)$ can be
expanded in eigenstate functions of the Hamiltonian $\hat H\phi_n~=~E_n\phi_n$,
where we order the eigenumbers in an increasing order $E_0 < E_1 < ...$
\begin{eqnarray}
\psi (\R,\tau) = \sum\limits_{n=0}^\infty c_n \phi_n(\R,\tau) =
\sum\limits_{n=0}^\infty c_n \phi_n(\R,0) e^{-(E_n - E)\,\tau}
\label{sum}
\end{eqnarray}
The amplitudes of the components change with time, either increasing or decreasing
depending on the sign of $(E_n-E)$. At large times the term that corresponds to the
projection on the ground state dominates the sum. In other words all excited states
decay exponentially fast and only contribution from ground state survives
\begin{eqnarray}
\psi (\R,\tau) \to c_0~\phi_0(\R,0)~e^{-(E_0 - E)\,\tau},\qquad \mbox{when~~~} \tau \to \infty
\label{psi limit}
\end{eqnarray}

In the long time limit the wave function remains finite only if $E$ is equal to
$E_0$. The ground state energy $E_0$ will be estimated in a different way
(Sec.~\ref{secEo}), but the fact that its estimation used for the energy shift value
$E$ leads to a stable normalization of $\psi(\R,\tau)$ proves in a different way
that the estimator is correct (for the implementation see Eq.~\ref{branching}).

The Hamiltonian of a system of $N$ particles interacting via pair-wise potential
$V_{int}$ and subjected to an external field $V_{ext}$ in a most general form can be
written as
\begin{eqnarray}
\hat H = -\frac{\hbar^2}{2m} \sum\limits_{i=1}^N \Delta_i
+ \sum\limits_{i<j}^N V_{int}(|\vec r_i -\vec r_j|)
+ \sum\limits_{i=1}^N V_{ext}(\vec r_i),
\label{Hgeneral}
\end{eqnarray}

The Schr\"odinger equation (\ref{shifted Schrodinger}) reads as\footnote{The
subscript $\R$ of the differential operator indicates that the derivative
has to be taken for every component of $\R$.}
\begin{eqnarray}
-\frac{\partial}{\partial \tau} \psi (\R,\tau) =-D \Delta_\R \psi (\R,\tau)
+ V(\R) \psi (\R,\tau) - E \psi (\R,\tau),
\label{generalShr}
\end{eqnarray}
where we introduced the notation $D = \hbar^2 /2m$ and terms depending only on
particle coordinates are denoted as $V(\R) =\sum\limits_{i<j}^N V_{int}(|\vec
r_i -\vec r_j|) + \sum\limits_{i=1}^N V_{ext}(\vec r_i)$.

The efficiency of the method can be significantly improved if additional information
on the wave function is used. The idea is to
approximate the true wave function $\psi(\R,\tau)$ by a trial one $\psi_T(\R)$ and let
the algorithm correct the guess done. This approach is called
{\it importance sampling} and consists in solving the
Schr\"odinger equation for the modified wave function
\begin{eqnarray}
f(\R,\tau) = \psi_T (\R) \psi (\R,\tau)
\label{f}
\end{eqnarray}

Another reason for using the product of wave functions as the probability
distribution instead of sampling $\psi$ is that the average over the latter is ill
defined $\langle A\rangle = \int A\,\psi\,\dR/\int\psi\,\dR$, on the contrary the
average over the product of wave functions has the meaning of the mixed estimator
$\langle A\rangle = \int \psi_T A\,\psi\,\dR/
\int\psi_T\psi\,\dR$. From (\ref{generalShr}) it follows that
the distribution function $f$ satisfies the equation
\begin{eqnarray}
-\frac{\partial}{\partial \tau} f (\R,\tau) = -D\Delta_\R f (\R,\tau)
+ D\nabla_\R (\F f(\R,\tau)) + (E^{loc}(\R) - E) f(\R,\tau),
\end{eqnarray}
here $E^{loc}$ denotes the {\it local energy} which is the average of the
Hamiltonian with respect to trial wave function\footnote{for details of the
calculation refer to Sec.~\ref{secQuantities}}
\begin{eqnarray}
E^{loc}(\R) = \psi^{-1}_T(\R) \hat H \psi_T(\R)
\label{local energy}
\label{Eloc}
\end{eqnarray}
and $\F$ is the {\it drift force} which is proportional to the gradient of the trial
wave function and points in the direction of the maximal increase of
$\psi_T$\footnote{The definition (\ref{DriftForce}) of the drift force can be
understood in a classical analogy. In a classical system the probability
distribution has a Boltzman form. The coordinate part of the distribution function
is related to the potential energy $p(\R) = const \exp(-U(\R))$ (we put the
fictitious temperature to one). The classical force is an antigradient of the
potential energy $\F = - \nabla_\R U(\R) = \nabla_\R \ln p(\R)$. If we approximate a
quantum probability distribution by the square of a trial wave function (\ref{p(r)}),
then the force equals exactly to (\ref{DriftForce}).}

\begin{eqnarray}
\F = \frac{2}{\psi_T(\R)}\nabla_\R \psi_T(\R)
\label{DriftForce}
\end{eqnarray}

\subsection{Green's function\label{secGreen}}

The formal solution of the Schr\"odinger equation written in coordinate space is
given by
\begin{eqnarray}
\langle\R|f(\tau)\rangle =
\sum\limits_{\R'} \langle\R|e^{-(\hat H-E)\tau}| \R'\rangle
\langle\R'|f(0)\rangle,
\end{eqnarray}
or, expressed in terms of the {\it Green's function} $G(\R,\R',\tau) =
\langle\R|e^{-(\hat H-E)t}|\R'\rangle$, the above equation reads as
\begin{eqnarray}
f(\R,\tau) = \int G(\R,\R',\tau) f(\R',0)\,{\bf dR'}
\label{Green}
\end{eqnarray}

In other words, the differential Schr\"odinger equation (\ref{Schrodinger})
corresponds to the integral equation (\ref{Green}), which can be integrated with
help of Monte Carlo methods. Although the Green's function $G(\R',\R,\tau)$ is not
known, it can be approximated at small times $\tau$, and then equation (\ref{Green})
can be solved step by step
\begin{eqnarray}
f(\R,\tau + \Delta\tau) = \int G(\R,\R',\Delta\tau) f(\R',\tau)\,{\bf dR'}
\label{Green approx}
\end{eqnarray}

The asymptotic solution for large times can be obtained by propagating $f(\R,\tau)$ for a large number
of time steps $\Delta \tau$.
\begin{eqnarray}
f(\R,\tau)\to\psi_T(\R)\phi_0(\R),\quad\tau\to\infty
\label{f limit}
\end{eqnarray}

For further convenience let us split the Hamiltonian into three pieces
\begin{eqnarray}
\label{H summands}
\hat H = \hat H_1 + \hat H_2 + \hat H_3,
\end{eqnarray}
where
\begin{eqnarray}
\begin{array}{lcl}
\hat H_1&=&-D \Delta,\\
\hat H_2&=&D((\nabla_\R \F) + \F \nabla_\R)),\\
\hat H_3&=&E^{loc}(\R) - E
\end{array}
\end{eqnarray}

Let us introduce the corresponding Green's functions:
\begin{eqnarray}
G_i(\R,\R',\tau) = \langle\R|e^{-\hat H_i\tau}|\R'\rangle,\qquad i = 1,2,3
\label{GreenCoord}
\end{eqnarray}

The exponent of a sum of two operators (on the contrary to an exponent of
$c$-numbers) in general can not be written as a product of two exponents. The exact
relation takes into account the non-commutativity $\exp\{-(\hat A+\hat B)\tau+[\hat
A,\hat B]\tau^2/2\}=\exp\{-\hat A\tau\}\exp\{-\hat B\tau\}$. The primitive
approximation consists in neglecting the noncommutativity
\begin{eqnarray}
e^{-\hat H\tau} = e^{-\hat H_1\tau}e^{-\hat H_2\tau}e^{-\hat H_3\tau}+O(\tau^2)
\label{t2 approximation}
\end{eqnarray}

This formula, rewritten in the coordinate representation, gives the expression for
the Green's function
\begin{eqnarray}
\nonumber
G(\R,\R',\tau) = \intint
G_1(\R,{\bf R_1},\tau) G_2({\bf R_1},{\bf R_2},\tau) G_3({\bf R_2},\R',\tau)\,
{\bf dR}_1 {\bf dR}_2
\end{eqnarray}

From Eq.~\ref{GreenCoord} we find that the Green's function should satisfy Bloch
differential equation:
\begin{eqnarray}
\left\{
{\begin{array}{rcll}
\displaystyle -\frac{\partial}{\partial \tau} G_i(\R,\R',\tau) &=&
\displaystyle \hat H_i\,G_i(\R,\R',\tau),
&\displaystyle i = 1,2,3\\
\displaystyle G_i(\R,\R',0)& =&\displaystyle\delta(\R-\R')&\\
\end{array}}
\right.
\label{G}
\end{eqnarray}

The equation for the kinetic term has the form
\begin{eqnarray}
-\frac{\partial G_1(\R,\R',\tau)}{\partial \tau} =-D\Delta G_1(\R,\R',\tau)
\end{eqnarray}

This is the diffusion equation with diffusion constant $D = \hbar^2/2m$. It can be
conveniently solved in a momentum representation where the kinetic energy operator
is diagonal. Going back to the coordinate representation one finds that the solution
is a Gaussian
\begin{eqnarray}
G_1(\R,\R',\tau) = (4 \pi D\tau)^{-3N/2}\exp\left\{-\frac{(\R-\R')^2}{4D\tau}\right\}
\label{GF1}
\end{eqnarray}

The equation for the drift force term is
\begin{eqnarray}
-\frac{\partial G_2(\R,\R',\tau)}{\partial \tau} =-D \nabla_\R (\F G_2(\R,\R',\tau))
\end{eqnarray}
and its solution is
\begin{eqnarray}
G_2(\R,\R',\tau) = \delta(\R-\R(\tau)),
\label{GF2}
\end{eqnarray}
here $\R(\tau)$ is the solution of the classical equation of motion
\begin{eqnarray}
\left\{
{\begin{array}{rcl}
\displaystyle\frac{{\bf dR}(\tau)}{d\tau}&=&\displaystyle D F(\R(\tau)),\\
\displaystyle\R(0)&=&\displaystyle\R'
\end{array}}
\right.
\label{importance sampling}
\end{eqnarray}

The last equation from (\ref{G}) has a trivial solution, which describes the {\it
branching} term
\begin{eqnarray}
G_3(\R,\R',\tau) = \exp\{(E-E^{loc}(\R))\,\tau\}~\delta(\R-\R')
\label{GF3}
\end{eqnarray}

\subsection{Primitive algorithm}

If the wave function of the system $f(\R,\tau)$ is real and positive, as it
happens in case of ground state of a bose system, it can be treated as population
density distribution\footnote[1]{The formula (\ref{Walkers}) should be understood
in the statistical sense, the average of any value $A$ over the l.h.s. and r.h.s
distributions are equal to each other in the limit when size of the population $N_W$
tends to infinity $\int A(\R) f(\R,\tau)\,{\bf dR} =
\lim\limits_{N_W\to\infty} \int A(\R)
\sum\limits_{i=1}^{N_W} C \delta ({\bf R - R_i}(\tau))\,{\bf dR}$}
(the algorithm for fermions, where the wave function has nodes will be discussed in
Sec.~\ref{secFNMC})
\begin{eqnarray}
f(\R,\tau) = \sum\limits_{i=1}^{N_W} C \delta ({\bf R - R_i}(\tau)),
\label{Walkers}
\end{eqnarray}
here $C$ is a positive constant, ${\bf R_i}(\tau)$ are coordinates of a population
element (so called {\it walker}) in $3N$-dimensional configuration space, $f({\bf
R},\tau)\,{\bf dR}$ gives the probability to find a walker at time $\tau$ in
the vicinity ${\bf dR}$ of a point $\R$.

Let us now interpret the action of the each of the three terms of the Hamiltonian
(\ref{H summands}) on the population distribution or, being the same, the action of
the corresponding Green's functions (\ref{GF1}, \ref{GF2}, \ref{GF3}).
In terms of Markov chains the Green's function is the
$G(\R,\R',\tau)$ is the transition matrix which determines the evolution of
the distribution (see Eq.~\ref{Green approx}).

The first term means the {\it diffusion} of each of the walkers in the configuration
space
\begin{eqnarray}
{\bf R^{(1)}}(t+\Delta\tau) = \R(\tau) + {\bf \chi},
\label{R1}
\end{eqnarray}
here ${\bf \chi}$ is a random value having a gaussian distribution $\exp\{-\chi^2/(4
D \Delta\tau)\}$.

The second term describes the action of the drift force, which guides the walkers to
places in the configuration space, where the trial wave function is maximal. This is
the way how importance sampling acts in the algorithm
\begin{eqnarray}
{\bf R^{(2)}}(t+\Delta\tau) = \R(\tau) +  D F(\R)\Delta\tau
\label{R2}
\end{eqnarray}

The Green's functions of steps (\ref{GF1}),(\ref{GF2}) are normalized to unity $\int
G(\R,\R',\tau)\,\dR = 1$. The normalization of wave function $f$ is then conserved,
which means that the number of walkers remains constant.

The third term is the {\it branching}
\begin{eqnarray}
f^{(3)}(\R, \tau+\Delta\tau) = \exp\{-(E^{loc}(\R)-E)\,\Delta\tau\}~f(\R,\tau)
\label{third step}
\label{branching}
\end{eqnarray}

The corresponding Green's function $G_3(\R,\R',\tau)$ (\ref{GF3}) is no longer
normalized. That means that the weight of a walker $\R$ changes on this step, thus
walkers with lower local energy have larger weights and walkers with larger local
energy have smaller weights. On this step each walker is to be duplicated
$b=\exp\{-(E^{loc}(\R)-E)\,\Delta\tau\}$ times. In general the number $b$ is not an
integer. A possible solution is to throw a random number $\xi \in (0,1)$ and
duplicate the walker $[b+\xi]$ times, where the brackets $[\cdot]$ stand for the
integer part of a number.

Now it is clear that by adjusting the value of $E$ one can control the size of the
population and keep it within the desired range. If the value of $E$ is taken to be
equal the estimator $E_0$ (\ref{Eloc}) averaged over the population and the
population size does not change, it means that $E_0$ is equal to the ground state
energy (see, also, Eq.~\ref{psi limit}).

The branching is an essential part of the DMC algorithm as it ``corrects'' the trial
wave function. Indeed, the first two steps (\ref{GF1}), (\ref{GF2}) alone without
(\ref{GF3}) are equivalent to sampling the trial wave function and provide the same
result as the variational calculation (Sec.~\ref{secVMC}). If the trial wave function
is an exact eigenfunction of the Hamiltonian, the local energy equals to the
corresponding eigennumber and is independent of $\R$, thus the branching becomes
irrelevant as it acts in the same way on all walkers.

\subsection{Higher-order algorithm}

The primitive approximation (\ref{t2 approximation}) for the Green's function has a
first-order of accuracy, {\it i.e.} the resulting energy has a linear dependence on the
timestep. It means that measurements with different time-steps are necessary in
order to make extrapolation to zero timestep $\Delta\tau\to 0$ as the dependence on
the timestep is very strong. Instead one can consider a higher order approximations.
One of the possible second-order expansions is
\begin{eqnarray}
e^{-\hat H\tau} =e^{-\hat H_3\tau/2}e^{-\hat H_2\tau/2}e^{-\hat H_1\tau}e^{-\hat H_2\tau/2}e^{-\hat H_3\tau/2}+O(\tau^3)
\label{t3 approximation}
\end{eqnarray}

This expansion leads to a quadratic dependence of the energy on the timestep which
makes this approximation very useful. The point is that each calculation has an
intrinsic statistical error, which depends on the length of the calculation and on
the variance of the measured quantity. Once the desired level of accuracy is chosen
one can make a study of the dependence on the timestep and adjust it to the maximal
value, which still gives an error smaller than the desired statistical error. Using
this timestep one can avoid the extrapolation procedure at all.

Here is the summary of the higher order scheme used in the calculations. One step
which propagates the system in the imaginary time from $\tau$ to $\tau+\Delta\tau$.
The walker is moved from position $\R_{i-1}$ to position $\R_{i}$ and is replicated
with the weight calculated during the branching\\
$~~~~~~~~$1) Gaussian jump (\ref{GF1}):\\
$~~~~~~~~~~~~~~~~~~~~~~~\R = \R_{i-1} + \chi,\qquad f(\chi) = \exp\{-\R^2/4\Delta\tau\}$\\
$~~~~~~~~$2) Drift force (\ref{GF2}):\\
$~~~~~~~~~~~~~~~~~~~~~~~\R' = \R + \F(\R)\,\Delta\tau/2$\\
$~~~~~~~~~~~~~~~~~~~~~~~\R'' = \R + (\F(\R)+\F(\R')\,\Delta\tau/4$\\
$~~~~~~~~~~~~~~~~~~~~~~~\R''' = \R + \F(\R'')\,\Delta\tau$\\
$~~~~~~~~$3) Branching (\ref{GF3}):\\
$~~~~~~~~~~~~~~~~~~~~~~~\R'''\to\R_i$

\section{Fixed-node Diffusion Monte Carlo method\label{fndmc}}


The FN-DMC method~\cite{Reynolds82} modifies the DMC method to allow an approximate
treatment of excited states of many-body systems. The idea of the FN-DMC method is
to treat excited states by ``enforcing'' the positive definiteness of the
probability distribution $f(\R,\tau)=\psi_T(\R)\psi(\R,\tau)$. The function
$f(\R,\tau)$ is positive definite everywhere in configuration space, and can hence
be interpreted as a probability distribution, if $\psi_T(\R)$ and $\psi(\R,\tau)$
change sign together, and thus share the same (high-dimensional) nodal surface. To
ensure positive definiteness of $f(\R,\tau)$, the trial wave function $\psi_T(\R)$
imposes a nodal constraint, which is fixed during the calculation. Within this
constraint, the function $f(\R,\tau)$ is propagated (following a scheme very similar
to that outlined in Sec.~\ref{secDMC}), and reaches an asymptotic distribution for
large $\tau\to\infty$, $f(\R,\tau)=\psi_T(\R)\,\psi(\R,\tau)$. In the FN-DMC method,
$\psi_T(\R)$ is an approximation to the exact excited eigenfunction of the many-body
Schr\"odinger equation (and not the groundstate wave function as in the DMC method).
It can be proven that, due to the nodal constraint, the fixed-node energy is a
variational upper bound to the exact eigenenergy for a given
symmetry~\cite{Reynolds82}. In particular, if the nodal surface of $\psi_T(\R)$ were
exact, then $\psi(\R,\tau)$ would be exact. Thus, the FN-DMC energy depends
crucially on a good parameterization of the many-body nodal surface.

Thus, in a FN-DMC simulation the function $f({\bf R},\tau)=\psi_T({\bf R})\Psi({\bf
R},\tau)$, where $\Psi({\bf R},\tau)$ denotes the wave function of the system and
$\psi_T({\bf R})$ is a trial function used for importance sampling, is evolved in
imaginary time according to the Schr\"odinger equation
\begin{eqnarray}
-\frac{\partial f({\bf R},\tau)}{\partial\tau}= &-& D\nabla_{\bf R}^2 f({\bf R},\tau) + D \nabla_{\bf R}[{\bf F}({\bf R})
f({\bf R},\tau)]
+ [E_L({\bf R})-E_{ref}]f({\bf R},\tau)
\label{FNDMC}
\end{eqnarray}

In the above equation ${\bf R}=({\bf r}_1,...,{\bf r}_N)$, $E_L({\bf R})=
\psi_T({\bf R})^{-1}H\psi_T({\bf R})$ denotes the local energy, ${\bf F}({\bf R})=2\psi_T({\bf R})^{-1}\nabla_{\bf R}
\psi_T({\bf R})$ is the quantum drift force, $D=\hbar^2/(2m)$ plays the role of an effective diffusion constant, and $E_{ref}$
is a reference energy introduced to stabilize the numerics. The energy and other
observables of the state of the system are calculated from averages over the
asymtpotic distribution function $f({\bf R},\tau\to\infty)$. To ensure positive
definiteness of the probability distribution $f$ for fermions, the nodal structure
of $\psi_T$ is imposed as a constraint during the calculation. It can be proved
that, due to this nodal constraint, the calculated energy is an upper bound to the
eigenenergy for a given symmetry~\cite{Reynolds82}. In particular, if the nodal
surface of $\psi_T$ were exact, the fixed-node energy would also be exact.

Construction of the trial wave function, as well as evaluation of the energy,
is described in Sec.~\ref{secFNMC}.


\section{Construction of trial wave functions: system of Bosons\label{secWF}}
\subsection{Introduction}

This Section is aimed to provide technical details on the construction of trial
bosonic wave functions. First of all we will discuss the general Bijl-Jastrow trial
wave function and than we will explain in details the construction of trial
wave functions used to obtain properties of the bosonic systems. The idea of this
section is to make a reference for the technical part of the calculations, in order
to leave the subsequent discussion free for physical discussions.

\subsection{Bijl-Jastrow wave function}

The bosonic function must be symmetric with respect to exchange of two particles.
The most natural way to construct the trial wave function of a system of Bosons is
to consider a product of one-body and two-body terms (we neglect three-body
and higher terms):
\begin{eqnarray}
\Psi(\rN) =\prod\limits_{i=1}^N f_1(\ri)\prod\limits_{j<k}^N f_2(|\rjk|)
\label{Jastrow}
\end{eqnarray}

This construction is called Bijl-Jastrow trial wave function. The one-body term
$f_1(\r)$ accounts for the external potential and, commonly, has the same structure
as the solution of an ideal system in the same external potential. The interaction
between particles is accounted by the two-body Bijl-Jastrow term $f_2(r)$ which must
go to a unity (uncorrelated value) at large distances. If the periodic boundary
conditions are used the restriction on the two-body term is stronger, the function
must go to the unity already at the half size of the simulation box. This condition
ensures that the particles do not interact with their own images and no artificial
correlations are introduced.

Once the exact type of the Bijl-Jastrow terms is chosen, one should also calculate
the first and second derivatives in order to implement the QMC method. Actually the
algorithm can be optimized by noticing (see Sec.~\ref{secQuant}) that the trial
wave function always comes in one of the three combinations:

1) the logarithm of the Bijl-Jastrow term (is necessary for the Metropolis algorithm
in the variational calculation and calculations of the non-local quantities, {\it e.g.}
the one-body density matrix)
\begin{eqnarray}
u(r) = \ln f(r)
\label{u2}
\end{eqnarray}

2) the logarithmic derivative of the Bijl-Jastrow term (is needed for the
calculation of the drift force (\ref{DriftForce}))
\begin{eqnarray}
{\cal F}(r) = \frac{f'(r)}{f(r)}
\label{F}
\end{eqnarray}

3) the second derivative enters only in the calculation of the kinetic part of the
local energy. The following combination is relevant:
\begin{eqnarray}
\E^{loc}(r) =-\frac{f''(r)}{f(r)}+\left(\frac{f'(r)}{f(r)}\right)^2
+ \frac{mV_{int}(r)}{\hbar^2}-\frac{D-1}{r}\frac{f'(r)}{f(r)},
\label{e}
\end{eqnarray}
where $D$ is number of dimensions.

\subsection{One-body Bijl-Jastrow term in an anisotropic trap}

Let the external field be an anisotropic trap with the aspect ratio $\lambda$:
$V_{ext}(\r)=\frac{1}{2}m\omega_\perp^2(x^2+y^2+\lambda^2z^2)$. We choose the
one-body Bijl-Jastrow term (\ref{Jastrow}) in form of a Gaussian with the widths
$\alpha$ and $\beta$ being variational parameters:
\begin{eqnarray}
f_1(\r) =\exp\{-\alpha(x^2+y^2)-\beta\lambda z^2\}
\label{f1}
\end{eqnarray}

Then the one body contribution to the drift force is
\begin{eqnarray}
\vec F_1(\ri) = -\left(2\alpha x, 2\alpha y, 2\beta\lambda z\right)
\end{eqnarray}

The local energy is
\begin{eqnarray}
E^{loc}(\R) = N(2\alpha+\beta\lambda)
+\sum\limits_{j<k}^N\E^{loc}_2(|\rjk|)
-\frac{1}{2}\sum\limits_{i=1}^N |\vec F_i(\rN)|^2
+\sum\limits_{i=1}^N\frac{x_i^2+y_i^2+\lambda^2z_i^2}{2}
\end{eqnarray}
where we used oscillator units: energy is measured in units of $\hbar\omega_\perp$
and the distances in units of the oscillator length $a_\perp$.

\subsection{One-dimensional wave functions\label{secwf1D}}

In this section we will discuss construction of a wave functions which are used to
solve one-dimensional problems. Apart from the case of the Tonks-Girardeau and gas
of hard rods, where the wave function is known exactly
(Secs.~\ref{wf1Dtonks},\ref{wf1DHR}), the wave function are constructed in a
Bijl-Jastrow form based on two-body solutions found in the Secs.~\ref{sec1D} (the
same method is used also in the construction of three-dimensional wave functions
Secs.~\ref{secwf3D}).

\subsubsection{Tonks-Giradeau wave function\label{wf1Dtonks}}

As was first shown by Girardeau\cite{Girardeau60}, the wave function of the
Tonks-Girardeau gas is equal to the absolute value of the wave function of $1D$ ideal
fermions. The interaction potential corresponds to $\delta$-function with infinite
strength, or in other words the TG model describes impenetrable particles of a zero
size. The component of an {\it exact wave function} is given by
\begin{eqnarray}
f_2(z) = |\sin(\pi z /L)|
\label{wf2Tonks}
\end{eqnarray}

Strictly speaking the wave function (\ref{wf2Tonks}) does not fall into the class of
Bijl-Jastrow functions (\ref{Jastrow}) as the term $f_2(z)$ does not go to a
constant even in the large-range limit, but always experience oscillations. At the
same time it does not cause problems in our calculations as it turns out that the
scattering energy $\E = \pi^2\hbar^2/mL^2$ of the exact solution
\cite{Girardeau60} corresponds to lowest energy of one particle of reduced mass in a
box with zero boundary conditions and the $f_2(z)$ goes to one in a smooth way at
the maximal allowed distance $z = L/2$.

The drift force contribution (\ref{F}) is given by
\begin{eqnarray}
\FF(z) = \sqrt{\E} \ctg\sqrt{\E}z
\end{eqnarray}

and the $1D$ local energy (\ref{e}) equals to
\begin{eqnarray}
\EE_2(z) = \E(1+\ctg^2\sqrt{\E}z)
\end{eqnarray}

\subsubsection{Hard-rod wave function (exact)\label{wf1DHR}}

In this section we will discuss the wave function of the hard-rod gas, {\it i.e.} the
one-dimensional gas of impenetrable particles (\ref{HR}) of radius $a_{1D}$. Already
in his original work\cite{Girardeau60}, Girardeau noted that the {\it exact} ground
state wave function of a hard-rod system can be obtained from the wave function of the
TG gas (Sec~\ref{wf1Dtonks}) by subtracting the excluded volume. That can be done by
the transformation
\begin{eqnarray}
z_i' = z_i - ia_{1D},\qquad i=\overline{1,N}
\label{HRtransformation}
\end{eqnarray}

This exact wave function is used for calculation of correlation properties in the
super-Tonks regime.

\subsubsection{Hard-rod wave function (approximate)\label{wf1DHRapprox}}

The transformation (\ref{HRtransformation}) makes the implementation of the
calculation quite sophisticated. Instead one can construct an approximate
wave function in the same spirit as it will be done in the subsequent sections, {\it i.e.}
by using the exact solution for the two particle scattering problem (\ref{Scr1Dsc}).

Using the solution (\ref{fHR}) we propose
\begin{eqnarray}
f_2(z) =
\left\{
{\begin{array}{ll}
\displaystyle 0,& |z| \le |a_{1D}|\\
\displaystyle \left|\frac{A}{z} \sin(\sqrt{\E}(|z|-|a_{1D}|))\right|,& |z| > |a_{1D}|\\
\end{array}}
\right.
\label{wfHR}
\end{eqnarray}

The advantage of this wave function is that it is solution of a two-body problem, so
the interaction energy is always constant, which is much easier to sample
numerically. The force and local energy can be easily obtained from $1D$ hard-sphere
wave function. The difference is that $\E$ is a variational parameter here.

The drift force contribution (\ref{F}) is given by
\begin{eqnarray}
\FF(z) =
\left\{
{\begin{array}{ll}
\displaystyle 0,& |z| \le |a_{1D}|\\
\displaystyle \sqrt{\E} \ctg\sqrt{\E}(|z|-|a_{1D}|), & |z| > |a_{1D}|\\
\end{array}}
\right.
\end{eqnarray}

The $1D$ local energy (\ref{e}) is
\begin{eqnarray}
\E^{loc}_2(z) =
\left\{
{\begin{array}{ll}
\displaystyle 0,& |z| \le |a_{1D}|\\
\displaystyle E(1+\ctg^2\sqrt{E}(|z|-|a_{1D}|)), & |z| > |a_{1D}|\\
\end{array}}
\right.
\end{eqnarray}

\subsubsection{Wave function of the Lieb Liniger gas\label{secLLwf}}

In this section we will describe the construction of the wave function whis is used
to solve the Lieb-Liniger equation in presence of an external confinement (refer to
Sec.~\ref{secLLtrap}). The aim is to obtain a wave function suitable for the
description of the gas in a wide range of the density, starting from the
Tonks-Girardeau and up to the Gross-Pitaevskii regimes. The important point is that
TG wave function always has nodes, although in GP regime the nodes are absent.

At the distances $|z|>0$ the interaction potential is absent and the solutions are
are simple sinus and cosine functions. We want to choose a solution which goes to
one at the matching distance $\Rm$, which is treated as a variational parameter,
and is a solution of a two-body problem (\ref{fLL0}) at a smaller distances. Also we
want to have a symmetry in sign reversing. The solution that satisfies those
conditions is
\begin{eqnarray}
f_2(z)
=\left\{
{\begin{array}{ll}
\displaystyle 1, & z < -\Rm\\
\displaystyle  \cos k(z-\Rm), & -\Rm \le z < 0\\
\displaystyle  \cos k(z+\Rm), & 0 \le z < \Rm\\
\displaystyle 1, & \Rm \le z\\
\end{array}}
\right.
\label{wfLL}
\end{eqnarray}

At the matching points $\pm \Rm$ the derivative is automatically equal to zero and
the function matches smoothly to a constant. The phase $\Delta(k) = k\Rm$ is
related to the scattering length $a_{1D}$ by the boundary condition
(\ref{LLphase}), which we will write as\footnote{Note that for a repulsive gas
$a_{1D}<0$ while for attractive $a_{1D}>0$.}
\begin{eqnarray}
ka_{1D}\,\tg k\Rm = 1
\label{Lieb k}
\end{eqnarray}

Once $k$ is obtained by solving numerically this equation the drift force
contribution (\ref{F}) can be calculated from formula
\begin{eqnarray}
\FF(z)
=\left\{
{\begin{array}{ll}
\displaystyle -k \tg k(z-\Rm), & 0 \le |z| < \Rm\\
\displaystyle 0, & |z|\ge \Rm \\
\end{array}}
\right.
\end{eqnarray}

The energy contribution (\ref{e}) is then described by
\begin{eqnarray}
\EE(z) =
\left\{
{\begin{array}{ll}
\displaystyle k^2(1+ \tg^2(k(z-\Rm)), & 0 \le |z| < \Rm\\
\displaystyle 0, & |z|\ge \Rm \\
\end{array}}
\right.
\end{eqnarray}

\subsubsection{Phonon trial wave function ($\delta$-potential)\label{secWFphonon}}

Here we shall construct a trial wave function which at distances short is a two-body
solution of one-dimensional $\delta$-function scattering and has ``phonon'' like
behavior at large distances (see \cite{Reatto67}).

The trial wave function is chosen in the following form
\begin{eqnarray}
f_2(z) =
\left\{
{\begin{array}{ll}
A \cos k(z-B), & z < R\\
|\sin^\alpha (\pi z/L)|, & z \ge R
\end{array}}
\right.
\label{wfphonon}
\end{eqnarray}

There are five undefined parameters $A,B,k,\alpha,R$ and four continuity equations.
One parameter is left free. We will chose matching distance $R$ as the guiding
parameter.

\begin{itemize}
\item[1)] Continuity condition at zero is the same as in the Lieb-Liniger trial wave
function (\ref{Lieb cont}) and is given by the formula (\ref{Lieb k})
\begin{eqnarray}
ka_{1D} \tg kB = 1
\end{eqnarray}

This equation fixes the value of the phase shift $kB$.

\item[2)] Continuity condition at the matching point
\begin{itemize}
\item[a)] continuity of the wave function:
\begin{eqnarray}
A \cos k(R-B) = \sin^\alpha (\pi R/L)
\label{wfbAcont}
\end{eqnarray}

\item[b)] continuity of the first derivative, which together with the condition a)
means continuity of the logarithmic derivative:
\begin{eqnarray}
-k \tg k(R-B) = \alpha\frac{\pi}{L} \ctg (\pi R/L)
\end{eqnarray}

\item[c)] continuity of the second derivative or, together with a) and b)
means continuity of the local energy:
\begin{eqnarray}
-k^2 = \alpha
\left(\frac{\pi}{L}\right)^2 [(\alpha-1)\ctg^2(\pi R/L)-1]
\end{eqnarray}
\end{itemize}
\end{itemize}

One can prove that following relation holds $kL/\pi\sin 2\pi R/L + \sin
2k(R-B)=0$. Another useful relation is $\tg k(R-B) = (ka_{1D}\sin kR+\cos kR)
/(ka_{1D}\cos kR-\sin kR)$.

The value of the scattering momenta is a solution of the equation
\begin{eqnarray}
\frac{(\sin kR-ka_{1D}\cos kR)(\cos kR + ka_{1D}\sin kR)}
{k((ka_{1D})^2+1)}=\frac{L}{2\pi} \sin \frac{2\pi R}{L}
\label{kk}
\end{eqnarray}

The maximal value of the l.h.s. is reached at $k=0$ and equals to $R-a_{1D}$. For
matching distance much smaller than $L$ the sinus function on the r.h.s. can be
expanded and the condition for the existence of the solution is $a_{1D}<0$ which is
always fulfilled for the repulsive gas.

All other parameters can be found from the following formulae:
\begin{eqnarray}
\left\{
\begin{array}{lll}
B &=& \frac{1}{k} \arcctg ka\\
\alpha &=& 1+\tg\frac{\pi R}{L}\left(\frac{kL}{\pi} \ctg k(R-B)+\tg\frac{\pi R}{L}\right)\\
A &=& \frac{\sin^\alpha (\pi R/L)}{\cos(k(R-B))}\\
\end{array}
\right.
\label{AAA}
\label{BBB}
\end{eqnarray}

The contribution to the energy is given by
\begin{eqnarray}
-\frac{f''}{f}+\left(\frac{f'}{f}\right)^2 =
\left\{
{\begin{array}{ll}
k^2[1+\tg^2 k(z-B)], & z < R\\
\alpha \left(\frac{\pi}{L}\right)^2
\left[1+\ctg^2 \frac{\pi x}{L}\right], & z \ge R
\end{array}}
\right.
\end{eqnarray}

\subsubsection{Super-Tonks trial wave function (attractive $\delta$-potential)}

In this section we will describe the construction of the trial wave function $\psi_T$
used fir the investigation of the super-Tonks system (see
Chapter~\ref{secHRstability}).

We use Bijl-Jastrow construction (\ref{Jastrow}) with the two-body term which is
chosen to be similar to (\ref{wfLL})
\begin{equation}
f_2(z)=\left\{
\begin{array}{cr}
\cos[k(|z|-\Rm)],&\quad |z| \le \Rm\\
1,   &  |z|>\Rm
\end{array}\right.
\label{wfST}
\end{equation}

The cut-off length $\Rm$ is a variational parameter, while the wave vector $k$ (for
a given $\Rm$) is chosen in a such way that the boundary condition imposed by the
$\delta$-function potential (\ref{Lieb k}) at $z=0$ is satisfied:
$-k\tg(k\Rm)=1/a_{1D}$. For distances smaller than the cut-off length, $|z|\le\Rm$,
the above wave function corresponds to the exact solution with positive energy of
the two-body problem with the interaction potential $g_{1D}\delta(z)$ (see,
formula~\ref{fLL0}).

\subsubsection{Scattering on the resonance state of a Bose gas}

In the case of the Hamiltonian (\ref{eq1Dham}) we use Gaussian construction for the
one-body Bijl-Jastrow term
\begin{equation}
f_1(z) = \exp\left\{-\frac{z^2}{2\alpha_z^2}\right\}
\label{VMC4}
\end{equation}
where the Gaussian width $\alpha_z$ is treated as a variational parameter. The
two-body correlation term $f_2(z)$ (\ref{Jastrow}) is chosen as
\begin{equation}
f_2(z) =
\left\{
\begin{array}{cl}
\cos[k_z(|z|-\bar{Z})],& |z| \le \bar{Z}   \\
1,   &|z|>\bar{Z} \;.
\end{array}
\right.
\label{VMC5}
\end{equation}

The cut-off length $\bar{Z}$ is fixed at $\bar{Z}=500 a_{1D}$, while the wave vector
$k_z$ is chosen such that the boundary condition at $z=0$ imposed by the
$\delta$-function potential is satisfied: $-k_z\tg(k_z \bar{Z})=1/a_{1D}$. For
negative $a_{1D}$ ($g_{1D}>0$) the correlation function, Eq.~\ref{VMC5}, is
positive everywhere. For positive $a_{1D}$ ($g_{1D}<0$), in contrast, $f_2(z)$
changes sign at $|z|=a_{1D}$. The parameterization given by Eq.~\ref{VMC5} is used
in our stability analysis performed within a VMC framework (see Sec.~\ref{secScV})
and in our DMC calculations for $g_{1D}> 0$. To perform the FN-DMC calculations for
negative $g_{1D}$, we need to construct a trial wave function that is positive
definite everywhere. In the FN-DMC calculations, we thus use an alternative
parameterization, which imposes the constraint $f_2=0$ for $a_{1D}\le z$,
\begin{equation}
f_2(z)=
\left\{
\begin{array}{cl}
0,&   z \le a_{1D}   \\
\cos[k_z(|z|-\bar{Z})],&  a_{1D} < z \le \bar{Z}   \\
1,& z>\bar{Z}
\end{array}\right.
\label{VMC6}
\end{equation}

\subsection{Three-dimensional wave functions\label{secwf3D}}
\subsubsection{Hard sphere trial wave function}

The problem of scattering on a hard sphere potential (\ref{HS}) was studied in
Sec.~\ref{secHS}. In dilute systems for small interparticle distance $r$ the two
body Bijl Jastrow term $f_2(r)$ is well approximated by the solution $f(r)$
(\ref{fHS}), {\it i.e.} by the wave function of a pair of particles in vacuum. At large
distances the pair wave function asymptotically goes to a constant value, as the
particles become uncorrelated.

Taking these facts into account we introduce the trial function in the following
way\cite{Giorgini99} (here we introduce dimensionless notation by measuring the
distance $r$ in units of the hard sphere radius $a_{3D}$ and energy $E$ in units of
$\hbar^2/(ma^2_{3D})$)
\begin{eqnarray}
f_2(r) =
\left\{
{\begin{array}{ll}
\displaystyle \frac{A\sin(\sqrt{2E}(r-1))}{r},& |r| \le \Rm\\
\displaystyle 1- B\exp\left\{-\frac{r}{\alpha}\,\right\},& |r| > \Rm\\
\end{array}}
\right.
\label{f_PP}
\end{eqnarray}

The request the function be smooth at the matching point $\Rm$, {\it i.e.}
\begin{itemize}
\item[1)] the function $f_2(r)$ itself must be continuous:
\begin{eqnarray}
\frac{A\sin(\sqrt{2E}(\Rm-1))}{\Rm} = 1 - B
\exp\left\{-\frac{\Rm}{\alpha}\,\right\}
\end{eqnarray}

\item[2)] derivative $f_2'(r)$ must be continuous
\begin{eqnarray}
\frac{A\sqrt{2E}\cos(\sqrt{2E}(\Rm-1))}{\Rm}
-\frac{A\sin(\sqrt{E}(\Rm-1))}{\Rm^2}
=\frac{B}{\alpha}\exp\left\{-\frac{\Rm}{\alpha}\,\right\}
\end{eqnarray}

\item[3)] the local energy $f_2(r)^{-1}(-\hbar^2\Delta_1/2m-\hbar^2\Delta_1/2m+
V_{int}(\rij))f_2(r)$ must be continuous
\begin{eqnarray}
2E =
\frac{\left(\displaystyle\frac{1}{\alpha^2}-\frac{2}{\Rm\alpha}\right)
B \exp\left(-\displaystyle\frac{\Rm}{\alpha}\,\right)
}{1-B \exp\left(-\displaystyle\frac{\Rm}{\alpha}\,\right)}
\end{eqnarray}
\end{itemize}

The solution of this system is
\begin{eqnarray}
\left\{
{\begin{array}{l}
A =\displaystyle\frac{R}{\sin(u(1-1/R))}\frac{\xi^2-2\xi}{\xi^2-2\xi+u^2},\\
B =\displaystyle\frac{u^2 \exp(\xi)}{\xi^2-2\xi+ u^2},
\end{array}}
\right.
\label{f_PP 1}
\end{eqnarray}
where we used the notation $u = \sqrt{2E}R$ and $\xi = R/\alpha$. The value of
$\xi$ is obtained from the equation
\begin{eqnarray}
1-\frac{1}{R} = \frac{1}{u} \arctg\frac{u(\xi-2)}{u^2+\xi-2}
\label{f_PP 2}
\end{eqnarray}

There are three conditions for the determination of five unknown parameters,
consequently two parameters are left free. The usual way to define them is minimize
the variational energy in Variational Monte Carlo which yields an optimized trial
wave function.

\subsubsection{Soft sphere trial wave function\label{secwfSS}}

At short distances the scattering solution $f(r)$ (\ref{uSS}) is expected
to proved a good approximation for the two-body Bijl-Jastrow term $f_2(r)$
(\ref{Jastrow}) in a gas with the soft sphere interaction potential (\ref{SS}). At
the larger distances $f_2(r)$ should saturate to a constant in a smooth way. We
choose an exponential type of decay:
\begin{eqnarray}
f_2(r) =
\left\{
{\begin{array}{ll}
\displaystyle
\frac{A\sh(\K r)}{r},& r<R\\
\displaystyle
\frac{B\sin(kr+\delta)}{r}, &R\le r<\Rm\\
\displaystyle
1-C \exp\left(-\frac{r}{\alpha}\right),&\Rm\le \frac L2\\
\end{array}}
\right.
\label{wfSS}
\end{eqnarray}

The description of a continuous matching at the point $R$ is described in
Sec.~\ref{secSS}. Now we shall discuss the matching procedure at the point $\Rm$.
As usual we have three matching conditions:
\newcommand{\sinf}{\sin(k\Rm+\delta)}
\newcommand{\cosf}{\cos(k\Rm+\delta)}
\newcommand{\ctgf}{\ctg(k\Rm+\delta)}
\begin{enumerate}
\item Continuity of the function $f_2(r)$:
\begin{eqnarray}
f(\Rm) = \frac{B\sinf}{\Rm} = 1-C \exp\left(-\frac{\Rm}{\alpha}\right)
\label{wf cont}
\end{eqnarray}
\item  Continuity of the logarithmic derivative $f_2'(r)/f_2(r)$ which fixes the value
of the parameter $C$:

\begin{eqnarray}
C = \exp(\Rm/\alpha)
\frac{k\ctgf-1/\Rm}{k\ctgf-1/\Rm+1/\alpha}
\label{C}
\end{eqnarray}

Substitution of (\ref{C}) into (\ref{wf cont}) fixes value of $B$
\begin{eqnarray}
B = \frac{\Rm/\alpha}{\sinf}
\frac{1}{k\ctgf-1/\Rm+1/\alpha}
\label{B}
\end{eqnarray}
\item The kinetic energy must be continuous at $r = \Rm$. This condition yields
\begin{eqnarray}
-\left(\frac{f''(\Rm)}{f(\Rm)}+\frac{2}{\Rm}\frac{f'(\Rm)}{f(\Rm)}\right)
= k^2
\label{kinetic e}
\end{eqnarray}
\end{enumerate}

After some mathematics the following procedure is obtained:
\begin{enumerate}
\item By choosing the value of the scattering length $a_{3D}$ and the range of the
potential $R$ define the value of $\kappa$ ({\it i.e.} height of the potential $V_0$ as
related by (\ref{kappaSS})) by solving the transcendental equation (\ref{aSS})
\item Introduce $x = k \Rm$, $y = \Rm/\alpha$ and $\bar\delta = \delta/x$.
Both equations (\ref{deltaSS}) and
\begin{eqnarray}
1+\bar\delta = \frac{1}{x}\arctg\left(\frac{x(y-2)}{x^2+y-2}\right)
\label{1+bardelta}
\end{eqnarray}
has to be satisfied in order to loop to determine $x = k\Rm$ and $\delta$. This
can be done using iterative procedure:
\begin{enumerate}
\item Choose the value for the $\bar\delta_a^{(i)}$ (the first time it is
initialized with $\bar\delta_a^{(0)} = -a/\Rm$) and obtain the value of $x^{(i)}$ as
the solution of Eq.~\ref{1+bardelta}
\item Fix the scattering momentum $k^{(i)} = x^{(i)}/\Rm$ and obtain the phase $\delta_b^{(i)}$ as a solution of (\ref{deltaSS})
\item Iterate $(\bar\delta_a^{(i+1)} = (\delta_a^{(i)}+\delta_b^{(i)})/2$ until
both numbers converge to the same value by repeating steps (a-c)
\end{enumerate}
\item Then the the constants (A,B,C) are given by formulae
\begin{enumerate}
\item $C = \exp(y)x^2/(x^2+y^2-2y)$
\item $B = \Rm/\sin(x+\delta)$
\item $A = B\sin(k\Rm+\delta)/\sh(k\Rm)$
\end{enumerate}
\end{enumerate}

Once all parameters are fixed, the two-body term $f_2(r)$ is given by (\ref{wfSS}),
the drift force contribution (\ref{F}) is given by
\begin{eqnarray}
\FF(r) =
\left\{
\begin{array}{lll}
\displaystyle \sqrt{V-\E}\frac{\cth(\sqrt{V-\E}\,r)}{r}-\frac{1}{r},& |r|\le R\\
\displaystyle \sqrt{\E}\ctg(\sqrt{\E}r+\delta)-\frac{1}{r},& R \le |r| \le \Rm\\
\displaystyle \frac{C}{\alpha}\left(\exp\left(\frac{r}{\alpha}-C\right)\right)^{-1},& |r| > \Rm\\
\end{array}
\right.
\end{eqnarray}

Local energy (\ref{e}) equals to
\begin{eqnarray}
\EE(r) =
\left\{
\begin{array}{ll}
\displaystyle \E + (V-\E)\left(\frac{\cth(\sqrt{V-\E}\,r)}{r}\right)^2,& |r|<R\\
\displaystyle \E(1+\ctg^2(\sqrt{E}r+\delta),& R \le |r| \le \Rm\\
\displaystyle \frac{C}{\alpha}\left[\frac{1}{\alpha}
\left(1+\frac{C}{\exp\left(r/\alpha\right)-C}\right)-\frac{2}{r}\right]
\frac{1}{\exp\left(r/\alpha\right)-C},& |r| > \Rm\\
\end{array}
\right.
\end{eqnarray}

\subsubsection{Trial wave function of 3D zero range potential}

We construct a wave function for a zero range potential in a three-dimensional case.
The correct scattering length $a$ is imposed on the trial wave function $f(r)$ by
corresponding boundary condition at zero distance.
\begin{eqnarray}
\left.\frac{(rf(r))'}{rf(r)}\right|_{r=0} = - \frac{1}{a}
\end{eqnarray}

We choose the trial wave function in the form
\begin{eqnarray}
f(r) = \frac{A}{r} \sin (kr+B)
\end{eqnarray}
for $r<R$ and $f(r) = 1$ otherwise.

The boundary condition at zero gives the constraint:
\begin{eqnarray}
\tg B = -ka
\label{BB}
\end{eqnarray}

We impose continuity of the derivative at the matching distance which gives us
another condition on the parameters of the trial wave function
\begin{eqnarray}
\tg\left(kR+B\right) = kR
\end{eqnarray}

The constant $B$ can be easily eliminated providing an equation which fixes the
momentum $k$:
\begin{eqnarray}
\frac{\tg(kR-\arctg kR)}{kR} = \frac{a}{R}
\end{eqnarray}

Ones it is solved, the equation (\ref{BB}) fixes the value of the $B$. Finally the
value of $A$ is fixed by continuity of the wave function itself
\begin{eqnarray}
A = \frac{L/2}{\sin(kL/2+B)}
\end{eqnarray}

The drift force is given by
\begin{eqnarray}
\FF(r) = k\ctg (kr+B)-\frac{1}{r}
\end{eqnarray}

The Bijl-Jastrow contribution to the 3D local energy depends on the distance $r$ as
\begin{eqnarray}
\EE(r) = k^2 + \left(k\ctg (kr+B)-\frac{1}{r}\right)^2
\end{eqnarray}

\subsubsection{Scattering on the resonance state of a Bose gas}

To describe the lowest-lying gas-like state of the Hamiltonian~(\ref{eq_h3D}), we
use for the one-body Bijl-Jastrow term an {\it ansatz} similar to (\ref{f1}):
\begin{eqnarray}
f_1(\r) = \exp\left\{-\frac{x^2+y^2}{2\alpha_\rho^2}-\frac{z^2}{2\alpha_z^2}\right\}
\end{eqnarray}
Here, $\alpha_z$ and $\alpha_\rho$ determine the Gaussian width of $\psi_T$ in the
longitudinal and transverse direction, respectively. These variational parameters
$\alpha_z$ and $\alpha_{\rho}$ are optimized in the course of the VMC calculation by
minimizing the energy expectation value. The two-body correlation factor $f_2(r)$
(\ref{Jastrow}) is chosen to reproduce closely the scattering behavior of two bosons
at low energies. For the hard-sphere potential (\ref{HS}), we take
\begin{equation}
f_2(\r)=
\left\{
\begin{array}{cl}
0,& |\r| \le a_{3D}\\
1-{a_{3D}}/{|\r|},&  |\r|>a_{3D}
\end{array}
\right.
\label{VMC2}
\end{equation}

The constraint $f_2=0$ for $r \le a_{3D}$ accounts for the boundary condition
imposed by the hard-sphere potential, it is exact even for the many-body system. For
the short-range potential (\ref{Morse}), we use instead
\begin{equation}
f_2(\r)=
\left\{
\begin{array}{cl}
0,& \frac{x^2 + y^2}{a^2}+\frac{z^2}{b^2} \le 1\\
1-1/\sqrt{\frac{x^2 + y^2}{a^2}+\frac{z^2}{b^2}},& \frac{x^2 + y^2}{a^2}+\frac{z^2}{b^2}>1
\end{array}
\right.
\label{VMC3}
\end{equation}
where $a$ and $b$ denote the lengths of the semi-axes of an ellipse. For two
particles under highly-elongated confinement, the nodal surface is to a good
approximation ellipticly shaped as will be discussed in Sec.~\ref{secScIVsubI}.
Thus, the parameters $a$ and $b$ are determined by fitting the elliptical surface to
the nodal surface obtained by solving the Schr\"odinger equation for $N=2$,
Eqs.~\ref{eq_h3Dn2} and \ref{eq_se3Dn2}, by performing a B-spline basis set
calculation. In contrast to $V^{HS}$, the constraint $f_2=0$ in Eq.~\ref{VMC3}
parameterizes the many-body nodal surface for $V^{SR}$ only approximately. We expect
that our parameterization leads to an accurate description of quasi-1D Bose gases if
the average distance between particles is much larger than the semi-axes of the
ellipse. The trial wave functions discussed here in the context of our VMC
calculations also enter our FN-DMC calculations.

\section{Construction of trial wave functions: system of Fermions\label{secFNMC}}
\subsection{Trial wave function in the BCS limit}

In the construction of the trial function the antisymmitrization is included through
the Slater determinant ${\cal D}{\bf(R)}$
\begin{eqnarray}
\psi_T(\R) = {\cal D}{\bf(R)}\prod\limits_{i=1}^Nf_1(\ri)
\prod\limits_{j<k}^Nf_2(|\rjk|)
\label{D}
\end{eqnarray}

Thus at the variational trial move one has to calculate the ratio of two
determinants in addition to usual one- and two- body correlation terms present in
the bosonic VMC algorithm (compare with (\ref{Jastrow}) and see Sec.~\ref{secVMC}).
An element of the Slater matrix is given by ${\cal D}_{i\alpha} =
\varphi_\alpha(\ri)$, where $\varphi_\alpha(\r)$ is a single particle orbital. In
further latin indices will always refer to particle number and the greek indices to
orbital number. During a trial move in which position of only one particle get
changed, just one row of the Slater matrix changes. This means that instead of a
direct calculation of the Slater determinant a more efficient method can be used.
Before doing the trial move one should calculate the inverse matrix ${\cal\overline
D}$ such that ${\cal D\overline D} = I$ or in terms of the matrix elements
\begin{eqnarray}
\sum\limits_{\alpha=1}^N{\cal D}_{i\alpha}{\cal \overline D}_{j\alpha} = \delta_{ij}
\end{eqnarray}

If we denote the matrix with coordinate of the $i^{th}$ particle changed as ${\cal
D'}$ then the ratio of interest becomes
\begin{eqnarray}
\frac{|{\cal D'}|^2}{|{\cal D}|^2} =
\frac{|{\cal D\overline DD'}|^2}{|{\cal D}|^2} =
\frac{|{\cal D}|^2|{\cal \overline DD'}|^2}{|{\cal D}|^2} =
|{\cal D'\overline D}|^2
\end{eqnarray}

The matrix ${\cal D'\overline D}$ is almost diagonal. Indeed, only $i^{th}$ row is
different from the one of a unitary matrix. It means that the determinant of such a
matrix equals to the $i^{th}$ element of this row, {\it i.e.}
\begin{eqnarray}
q = \frac{|{\cal D'}|}{|{\cal D}|} =
\sum\limits_{\alpha=1}^N\varphi_\alpha(\ri'){\cal \overline D}_{i\alpha}
\label{q}
\end{eqnarray}

After the move is accepted the inverse matrix must be updated. There is a fast way
of doing it. Instead of direct inversion of the determinant matrix one can use $q$
from eq.(\ref{q}):
\begin{eqnarray}
{\cal \overline D}_{j\alpha}
= \left\{
{\begin{array}{ll}
{\cal \overline D}_{j\alpha}/q,&j = i\\
{\cal \overline D}_{j\alpha} -
{\cal \overline D}_{i\alpha}
\sum\limits_{\beta=1}^N\frac{\varphi_\beta(\ri'){\cal \overline D}_{j\beta}}{q}
,&j \ne i\\
\end{array}}
\right.
\end{eqnarray}

Differentiating the trial wave function (\ref{D}) one finds the expression for the
kinetic energy. It is equal to
\begin{eqnarray}
T^{loc}(\R) = \frac{\hbar^2}{2m}\left\{
\sum\limits_{i=1}^N\E^{i}_{\cal D}+
2\sum\limits_{j<k}^N\E^{loc}_2(|\rjk|)
-\sum\limits_{i=1}^N |\vec F_i(\R)|^2\right\},
\label{T fermions}
\end{eqnarray}
where the two-body contribution to the local energy is the same as in the bosonic
case
\begin{eqnarray}
\E^{loc}_2(r) = 
-\frac{f_2''(r)}{f_2(r)} 
-\frac{(D-1)}{r} 
\frac{f_2'(r)}{f_2(r)} 
+ \left(\frac{f_2'(r)}{f_2(r)}\right)^2
\end{eqnarray}
and there is an additional term coming from the determinant part of the trial
wave function.
\begin{eqnarray}
\E^{i}_{\cal D}(\vec r) = -\frac{\Delta_i {\cal D}(\R)}{{\cal D}(\R)}
+\left(\frac{\nabla_i {\cal D}(\R)}{{\cal D}(\R)}\right)^2
\end{eqnarray}

The drift force (\ref{DriftForce}) appearing in (\ref{T fermions}) is given by
\begin{eqnarray}
\vec F_i(\R) =
\frac{\nabla_i {\cal D}(\R)}{{\cal D}(\R)}\frac{\ri}{r_i}
+\sum\limits_{k\ne
i}^N\frac{f_2'(|\rik|)}{f_2(|\rik|)}\frac{\rik}{|\rik|}
\end{eqnarray}

The derivatives of the determinant are related to the derivatives of the orbitals in
an easy way (see formula (\ref{q}))
\begin{eqnarray}
\frac{\nabla_i {\cal D}(\R)}{{\cal D}(\R)}
=\sum\limits_{\alpha=1}^N {\cal \overline D}_{i\alpha} \nabla_i \varphi_\alpha(\ri)
\end{eqnarray}

\subsection{Kinetic energy}

We always use coordinate representation for the wave functions in our calculations,
thus the calculation of the potential energy, which is diagonal in this
representation, is trivial. Instead calculation of the kinetic energy demands
knowledge of wave function derivatives. Let us calculate the first and second
derivatives of the fermion wave function:
\begin{eqnarray}
\vec\nabla_{\ri}\Psi(\R)
=\Psi(\R)\left(
\frac{\vec\nabla_{\ri} \D}{\D}
+\sum\limits_{j} \frac{\vec\nabla_{\ri} f_2(|\rij|)}{f_2(|\rij|)}
\right)
\end{eqnarray}

\begin{eqnarray}
\Delta_{\ri}\Psi(\R)
=\Psi(\R)
\left(
\left[
\frac{\vec\nabla_{\ri} \D}{\D}
+\sum\limits_{j} \frac{\vec\nabla_{\ri} f_2(|\rij|)}{f_2(|\rij|)}
\right]^2+
{\cal E}^{loc}_i + {\cal D}^{loc}_i
\right)
\end{eqnarray}

The local energy is defined as
\begin{eqnarray}
{\cal E}^{loc}_i
&=&\sum\limits_{j}\left[
\frac{\Delta_{\ri}f_2(|\rij|)}{f_2(|\rij|)}-\left(\frac{\vec\nabla_{\ri}f_2(|\rij|)}{f_2(|\rij|)}\right)^2\right]
\\
{\cal D}^{loc}_i &=&
\frac{\Delta_{\ri}\D}{\D}-\left(\frac{\vec\nabla_{\ri}\D}{\D}\right)^2
\end{eqnarray}

The second derivative of the determinant is calculated explicitly, while the second
derivative of the two-body Jastrow term is calculated by assuming spherical
symmetry:
\begin{eqnarray}
\Delta_{\ri}\D& =&
\left(\frac{\partial^2}{\partial x_i^2}
+\frac{\partial^2}{\partial y_i^2}
+\frac{\partial^2}{\partial z_i^2}
\right)\D\\
\Delta f_2(r) & =&
f''_2(r)+\frac{2}{r}f'(r)
\end{eqnarray}

\subsection{Calculation of the tail energy}

A simulation of a homogeneous system is done by considering a finite box of size
$L$. One restricts interaction between the particles to a distance of $L/2$. Larger
distances should be avoided in order not to have a double counting of a same
particle which would leave to artificial correlation. Thus one introduces a cut-off
at $L/2$ and a proper calculation of the energy is necessesary.

The situation is different for a Bijl-Jastrow construction of the wave function and a
Slater determinant. We will consider a generalization of the wave function containing
a product of both terms. The energy per particle in the thermodynamic limit
$N\rightarrow \infty $ is given by the integral of the interaction energy from the
cut-off length $L/2$ to infinity.
\begin{eqnarray}
E_{pot}^{tail}=\sum\limits_i\sum\limits_{
j<i,|\rij|>L/2
}
V(|\rij|)\rightarrow n\int\limits_{L/2}^{\infty }V(r)\,d^{3}r
\end{eqnarray}

In the thermodynamic limit $N\rightarrow \infty $ the Jastrow force becomes zero as
the summation on $j$ is approximated by a symmetric uniform distribution of
particles outside a sphere of $L/2$ radius. So, the tail of a kinetic energy for a
Jastrow wave function is
\begin{eqnarray}
E_{J}^{tail}=\frac{\hbar ^{2}n}{m}\int\limits_{L/2}^{\infty }{\cal E}
^{loc}(r)d^{3}r=\frac{\hbar ^{2}n}{m}\int\limits_{L/2}^{\infty }\left[-
\frac{f^{\prime \prime }(r)}{f(r)}-\frac{2}{r}\frac{f^{\prime }(r)}{f(r)}
+\left( \frac{f^{\prime }(r)}{f(r)}\right) ^{2}\right] d^{3}r
\end{eqnarray}

On the opposite, there is no similar cancellation due to the symmetry in the Slater
term, but instead due to linearity the square of the first derivative is exactly
cancelled by the force squared term, thus
\begin{eqnarray}
E_{Det}^{tail}=\frac{\hbar ^{2}n}{m}\int\limits_{L/2}^{\infty }\left[ -\frac{
g^{\prime \prime }(r)}{g(r)}-\frac{2}{r}\frac{g^{\prime }(r)}{g(r)}\right]d^{3}r
\end{eqnarray}

\subsection{Bijl-Jastrow term (square well trial wave function)}

Now let us specify the Bijl-Jastrow term which will take care of the interactions
between spin up and spin down particles. We consider an attractive interaction
potential which supports a bound state. Thus we can describe resonant scattering
with very large scattering lengths $a_{3D}$. It also means that for the unit of
length it is preferable to take instead the range of potential $R$ instead of
$a_{3D}$ which can be even diverging (the unitary regime).

We consider scattering on the square well (SW) potential (\ref{SW}). The scattering
problem was studied in Sec.~\ref{secSW}. Here we only summarize the construction of
the Bijl-Jastrow term:
\begin{itemize}
\item[1)] the equation for the scattering momentum is
\begin{eqnarray}
\frac{1}{k}\left[\arctg\frac{kL}{2}-\arctg\left(\frac{k}{\sqrt{\kappa^2+k^2}}
\tg\sqrt{\kappa^2+k^2} R\right)\right] = \frac{L}{2}-R
\end{eqnarray}
\item[2)] the shift phase $\delta$ is defined as
\begin{eqnarray}
\delta = \arctg\left(\frac{kL}{2}\right)-\frac{kL}{2}
\end{eqnarray}
\item[3)] normalization factor $B$
\begin{eqnarray}
B = \frac{L/2}{\sin(kL/2+\delta)}
\end{eqnarray}
\item[4)] normalization factor $A$
\begin{eqnarray}
A = B\frac{\sin(k R+\delta)}{\sin(\sqrt{\kappa^2+k^2} R)}
\end{eqnarray}
\end{itemize}

The Bijl-Jastrow contribution to the force and the local energy are given by
following expressions:
\begin{eqnarray}
\frac{f'(r)}{f(r)} =\left\{
{\begin{array}{ll}
\K\ctg(\K r)-\frac{1}{r},& r<R\\
k\ctg(kr+\delta)-\frac{1}{r}, &r\ge R
\end{array}}
\right.
\end{eqnarray}
\begin{eqnarray}
E_{loc}^{3D} =\left\{
{\begin{array}{ll}
\K^2-(\K\ctg(\K r)-\frac{1}{r})^2,& r<R\\
k^2-(k\ctg(kr+\delta)-\frac{1}{r})^2, &r\ge R
\end{array}}
\right.
\end{eqnarray}

\subsection{Trial wave function: zero energy scattering state}

On the BCS side of the resonance the scattering length is negative $a<0$.
Here the attractive square well potential well of strength $V_{0}=\hbar
^{2}\varkappa ^{2}/m$. There is no bound state anymore and instead one has a
solution with positive energy ${\cal E}=\hbar ^{2}k^{2}/m$. The scattering solution
is
\begin{eqnarray}
f(r)=\left\{
\begin{array}{cc}
\frac{A}{r}\sin {\cal K}r, & r<R \\
\frac{B}{r}\sin (kr+\delta ), & r>R
\end{array}
\right.
\end{eqnarray}

The scattering length $a$ phase is related to the phase of the scatters
wave. In the low-energy limit the phase is simply $\delta =-ka$. In this
limit ${\cal K}\varkappa $ and we have simple solution%
\begin{eqnarray}
f(r)=\left\{
\begin{array}{cc}
\frac{A}{r}\sin kr, & r<R \\
B(1+\frac{|a|}{r}), & r>R
\end{array}
\right.
\end{eqnarray}

\begin{eqnarray}
f^{\prime }(r)=\left\{
\begin{array}{cc}
\frac{A}{r}\sin kr\left( \varkappa \mathop{\rm ctg}\nolimits-\frac{1}{r}
\right), & r<R \\
-\frac{B|a|}{r^{2}}, & r>R
\end{array}
\right.
\end{eqnarray}

\subsubsection{Matching to a constant}

The wave function constructed in the previous section is long-ranged as it decays
only as $1/r$ at large distances. This leads to overestimation of the correlations
in deep BCS limit. For example at $a=-5R$ the variational energy is 8 times larger
than the energy of a fermi gas. The fixed-node MC algorithm corrects this behavior
and the resulting energy is close to the energy of a fermi gas.
\begin{eqnarray}
f(r)=\left\{
\begin{array}{cc}
\frac{A}{r}\sin \varkappa r, & r<R \\
B(1+\frac{|a|}{r}), & R<r<R_m \\
1+C\exp (-k_{m}r), & R_m<r
\end{array}
\right.
\end{eqnarray}

The matching conditions fix values of the constants $A,B,C$:
\begin{eqnarray}
\left\{
\begin{array}{lll}
A&=&\frac{BR(1+|a|/R)}{\sin \varkappa R}\\
B&=&\left( 1+\frac{|a|}{R_{m}}\left( 1{\bf -}\frac{1}{R_{m}k_{m}}\right)\right)^{-1}\\
C&=&\frac{B|a|\exp (k_{m}R_{m})}{R_{m}^{2}k_{m}}
\end{array}
\right.
\end{eqnarray}

\section{Measured quantities\label{secQuant}\label{secQuantities}}
\subsection{Local energy\label{secEo}}

The most general form of a Hamiltonian of a system of $N$ interacting
bosons in an external field is (\ref{Hgeneral}):
\begin{eqnarray}
\hat H = -\frac{\hbar^2}{2m}\sum\limits_{i=1}^N \Delta_{\ri}
+ \sum\limits_{i=1}^N V_{ext}(\ri) + \sum\limits_{j<k}^N V_{int}(|\rjk|),
\label{Hlocenergy}
\label{H}
\end{eqnarray}
where $m$ is mass of a particle, $V_{ext}(\vec r)$ is the external field,
$V_{int}(|\r|)$ is particle-particle interaction potential. Given the many-body
wave function $\Psi(\rN)$ the local energy is defined according to (\ref{Eloc}):
\begin{eqnarray}
E^{loc}(\rN) = \frac{\hat H \Psi(\rN)}{\Psi(\rN)}
\end{eqnarray}

Operator of the external field and particle-particle interaction are diagonal in
this representation and are calculated trivially as a summation over particles and
pairs of the second and third terms of (\ref{Hlocenergy}). Calculation of the
kinetic energy, first term of (\ref{Hlocenergy}) is more tricky, as the Laplacian
operator is not diagonal.

\subsubsection{Local kinetic energy and the drift force}

In this section we will find the expression of the local kinetic energy
\begin{eqnarray}
T^{loc}(\rN) = -\frac{\hbar^2}{2m}\frac{\Delta\Psi(\rN)}{\Psi(\rN)}
\end{eqnarray}

Let us calculate the second derivative in two steps, as the first derivative is
important for the calculation of the drift force. We consider the Bijl-Jastrow form
(\ref{Jastrow}) of the trial wave function and will express the final results in
terms of one- and two- body Bijl-Jastrow terms $f_1$ and $f_2$. The gradient of the
many-body trial wave function is given by
\begin{eqnarray}
\vec\nabla_{\ri}\Psi(\rN)=\Psi(\rN)\left(\frac{f_1'(\ri)}{f_1(\ri)}\frac{\ri}{r_i}
+\sum\limits_{k\ne i}^N\frac{f_2'(|\rik|)}{f_2(|\rik|)}\frac{\rik}{|\rik|}\right)
\end{eqnarray}

The full expression for the Laplacian is
\begin{eqnarray}
\nonumber
\Delta_{\ri}\Psi(\rN)\qquad=\qquad\Psi(\rN)\left(
\frac{f_1'(\ri)}{f_1(\ri)}\frac{\ri}{r_i}+\sum\limits_{k\ne
i}^N\frac{f_2'(|\rik|)}{f_2(|\rik|)}\frac{\rik}{|\rik|}
\right)^2+\\+\Psi(\rN)\left(
\frac{f_1''(\ri)}{f_1(\ri)}-\left(\frac{f_1'(\ri)}{f_1(\ri)}\right)^2
+\sum\limits_{k\ne i}^N\left[
\frac{f_2''(|\rik|)}{f_2(|\rik|)}-\left(\frac{f_2'(|\rik|)}{f_2(|\rik|)}\right)^2
\right]\right)
\end{eqnarray}

The kinetic energy can be written in a compact form
\begin{eqnarray}
T^{loc}(\rN) = \frac{\hbar^2}{2m}\left\{\sum\limits_{i=1}^N\E^{loc}_1(\ri)+
2\sum\limits_{j<k}^N\E^{loc}_2(|\rjk|)-\sum\limits_{i=1}^N |\vec F_i(\rN)|^2\right\},
\end{eqnarray}
where we introduced notation for the one- and two- body contribution to the local
energy (see, also, (\ref{e}))
\begin{eqnarray}
\E^{loc}_1(\vec r)&=& -\frac{f_1''(\vec r)}{f_1(\vec r)}+\left(\frac{f_1'(\vec r)}{f_1(\vec r)}\right)^2\\
\E^{loc}_2(r)&=& -\frac{f_2''(r)}{f_2(r)}+\left(\frac{f_2'(r)}{f_2(r)}\right)^2
\end{eqnarray}
and introduced the drift force (see (\ref{F}))
\begin{eqnarray}
\vec F_i(\rN) =
\frac{f_1'(\ri)}{f_1(\ri)}\frac{\ri}{r_i}
+\sum\limits_{k\ne
i}^N\frac{f_2'(|\rik|)}{f_2(|\rik|)}\frac{\rik}{|\rik|}
\end{eqnarray}

\subsubsection{Exponentiation}

It is convenient (see Eq.\ref{u2}) to do the exponentiation of the one- and two-
body terms $u_1(\vec r) = \ln f_1(\vec r)$, $u_2(r) = \ln f_2(r)$. The point is that
numerically a better precision is achieved by working with numbers of the same
order. The formula for the kinetic energy becomes simpler
\begin{eqnarray}
T^{loc}(\rN) = -\frac{\hbar^2}{2m}\left\{
\sum\limits_{i=1}^N u_1''(\ri)+
2\sum\limits_{j<k}^Nu_2''(|\rjk|) +\sum\limits_{i=1}^N |\vec
F_i(\rN)|^2\right\}
\end{eqnarray}
with
\begin{eqnarray}
\vec F_i(\rN) = u_1'(\ri)\frac{\ri}{r_i} +\sum\limits_{k\ne i}^N u_2''(|\rik|)\frac{\rik}{|\rik|}
\end{eqnarray}

%

\subsection{Static structure factor\label{secSkDMC}}

It is natural to give the definition of the static structure factor $S(\k)$ in the
momentum space as the correlation function of the momentum distribution between
elements $-\k$ and $\k$ (\ref{Skmom}):
\begin{equation}
N S(\k) = \langle\rho_{-\k}\rho_\k\rangle - |\langle\rho_\k\rangle|^2,
\label{Skdef}
\end{equation}

Using the properties of the Fourier component $\rho_{-\k} = (\rho_\k)^*$ it can be
rewritten in a different way
\begin{equation}
N S(\k) = \langle|\rho_\k|^2\rangle - |\langle \rho_\k\rangle|^2
\end{equation}

In the Diffusion Monte Carlo calculation the density distribution is approximated by
the density of walkers (see (\ref{Walkers}))
\begin{equation}
n(\r) = \sum\limits_{i=1}^N \delta(\r-\r_i)
\end{equation}

With the means of the Fourier transform we express it in the momentum space
\begin{equation}
\rho_\k = \int e^{i\k\r} n(\r) d\r = \sum\limits_{i=1}^N e^{i\k\r_i}
=\sum\limits_{i=1}^N \cos\k\r_i +i \sum\limits_{i=1}^N \sin\k\r_i
\end{equation}
and obtain a simple expression for the static structure factor
\begin{equation}
N S(\k) =
\left<
\left(\sum\limits_{i=1}^N \cos \k\r_i\right)^2+
\left(\sum\limits_{i=1}^N \sin \k\r_i\right)^2
\right>
-\left|\left<\sum\limits_{i=1}^N \cos \k\r_i\right>\right|^2
-\left|\left<\sum\limits_{i=1}^N \sin \k\r_i\right>\right|^2
\label{Sk}
\end{equation}

In a trapped system there are no restrictions on the value of momentum $\k$,
although, naturally, the momentum distribution vanishes for $k\ll 1/R$, where $R$ is
the size of the system. Instead, if periodic boundary conditions are used, the value
of momenta is quantized and is dependent on the size of the box
\begin{equation}
k_{n_{x,y,z}} = \frac{2\pi}{L} n_{x,y,z}
\end{equation}

At the same in a homogeneous system the two last terms in (\ref{Sk}) are vanishing.

\subsection{One body density matrix in a homogeneous system}

The one body density matrix (OBDM) $g_1$ of a homogeneous system described by the
many body wave function $\psi(\r_1, ..., \r_N)$ according to (\ref{g1hom}) is equal
to
\begin{eqnarray}
g_1(|\r~'-\r~''|)=N \frac{\int...\int\psi^*(\r\,', \r_2, ..., \r_N)
\psi(\r\,'', \r_2, ..., \r_N)\,d\r_2 ...d\r_N}{\int...\int|\psi(\r_1, ..., \r_N)|^2\,d\r_1 ... d\r_N}
\label{OBDM}
\end{eqnarray}

Since in DMC calculation does not sample directly the ground-state probability
distribution $\phi_0^2$, but instead the mixed probability $\psi_T\phi_0$ (\ref{f})
one obtains the {\it mixed} one-body density matrix as the output
\begin{eqnarray}
g_1^{mixed}(\r) =
N \frac{\int...\int \psi^*_T(\r_1+\r, \r_2,..., \r_N) \phi_0 (\r_1, \r_2, ..., \r_N)\,d\r_2 ... d\r_N}
{\int...\int \psi^*_T(\r_1, ...,\r_N)\phi_0(\r_1, ...,\r_N)\,d\r_1 ... d\r_N},
\label{OBDMhom}
\end{eqnarray}

This formula can be rewritten in a way convenient for the Monte Carlo sampling:
\begin{eqnarray}
g_1^{mixed}(r) =
n\frac{\int...\int[\psi^*_T(\r_1+\r, \r_2,..., \r_N)(\psi^*_T(\r_1,\r_2,...,\r_N))^{-1}]
f(\r_1, ..., \r_N)d\r_1 ... d\r_N}{\int...\int f(\r_1, ..., \r_N)d\r_1 ... d\r_N},
\label{OBDM mixed}
\end{eqnarray}
where we have used the asymptotic formula (\ref{f limit}) and have taken into
account that in a homogeneous system $g_2$ depends only on the module of the
relative distance between two particles. If the trial wave function is chosen as a
product of pair functions (\ref{Jastrow}) then using the notation (\ref{u2}) $u(|\r_i
-\r_j|) = \ln f_2(|\r_i -\r_j|)$) and $f_1 \equiv 0$ one has $\psi_T(\r_1, ...,\r_N)
= \prod\limits_{i<j}\exp\{u(|\r_i-\r_j|)\}$. Then the ratio of the trial wave function
appearing in (\ref{OBDM mixed}) becomes
\begin{eqnarray}
\frac{\psi_T(\r_1+\r, ...,\r_N)}{\psi_T(\r_1, ...,\r_N)}
= \exp\left\{\sum\limits_{j>1} \mu(|\r_1+\r-\r_j|)-\mu(|\r_1-\r_j|)\right\}
\end{eqnarray}

In order to gain better statistics it is convenient to average over all possible
pairs of particles
\begin{eqnarray}
g_1^{mixed}(r) =
\frac{1}{N} \sum\limits_{i=1}^N\frac{\psi_T(\r_1, ..., \r_i+\r, ...,\r_N)}{\psi_T(\r_1, ...,\r_N)}
=\frac{1}{N} \sum\limits_{i=1}^N\exp\left\{\sum\limits_{j \neq i}^N
u(|\r_i+\r-\r_j|)-u(|\r_i-\r_j|)\right\}
\end{eqnarray}

The asymptotic limit of the OBDM gives the condensate density
\begin{eqnarray}
\lim\limits_{r\to\infty} g_1(r) = \frac{N_0}{V}
\end{eqnarray}
and the condensate fraction is obtained by the calculating the asymptotic ratio
\begin{eqnarray}
\lim\limits_{r\to\infty} \frac{g_1(r)}{n} = \frac{N_0}{N}
\end{eqnarray}

\subsection{One body density matrix in a harmonic trap}

While in a homogeneous system the OBDM depends only on the relative distance, for a
system in external potential it is no longer true (\ref{g1trap}). Instead one define the OBDM in a
convenient way by integrating out the center of the mass motion.
\begin{eqnarray}
\overline g_1(\r) =
\int g_1\left(\RR+\frac{\r}{2}, \RR-\frac{\r}{2}\right)\,d\RR,
\label{DMCg2trap}
\end{eqnarray}

Here the standard notation for the center of the mass variables is used
$\RR=(\r_1+\r_2)/2$, $\r = \r_1-\r_2$. The point in the definition (\ref{DMCg2trap})
is that the momentum distribution can be obtained by the Fourier transform with
respect to $\r$
\begin{eqnarray}
n(\vec k) = \int\overline g_1(\r) e^{i\vec k\r}\,d\r
\end{eqnarray}

For practical purposes it is convenient to change the notation
\begin{eqnarray}
\left\{
\begin{array}{lll}
\RR &=& \r_1\\
\r &=& \r_1-\r_2
\end{array}
\right.
\end{eqnarray}
Using this notation the mixed OBDM becomes
\begin{eqnarray}
g_1^{mixed} (\RR, \r) =
N \frac{\int...\int \psi^*_T(\RR+\r, \r_2,..., \r_N) \phi_0 (\RR, \r_2, ..., \r_N)\,d\r_2 ... d\r_N}
{\int...\int \psi^*_T(\r_1, ...,\r_N)\phi_0(\r_1, ...,\r_N)\,d\r_1 ... d\r_N},
\label{OBDM trap}
\end{eqnarray}
which reminds us the expression for the OBDM of a homogeneous system (\ref{g21}).

The function $\overline g_1$ can be measured in the QMC simulation
\begin{eqnarray}
\overline g_1^{mixed}(\r) =
\frac{\int...\int
[\psi^*_T(\RR+\r, \r_2,..., \r_N)(\psi^*_T(\RR,\r_2,...,\r_N))^{-1}]
f(\RR, \r_2, ..., \r_N)d\RR d\r_2 ... d\r_N}
{\int...\int f(\r_1, ..., \r_N)d\r_1 ... d\r_N},
\end{eqnarray}
by taking an average of the following quantity
\begin{eqnarray}
\frac{\psi_T(\RR+\r, ...,\r_N)}{\psi_T(\RR, ...,\r_N)} =
\exp\left\{u_1(\RR+\r)-u_1(\RR)+
\sum\limits_{j>1} u_2(|\r_1+\r-\r_j|)-u_2(|\r_1-\r_j|)\right\},
\end{eqnarray}
where $u_1(\r)$ stands for the one-body exponent in the Jastrow-Bijl wave function
(see Eq.~\ref{u2}), which for harmonic confinement is taken to be equal to
(\ref{f1}).

\subsection{Pair distribution}

The pair distribution function (TBDM) in a homogeneous system is given by the
formula (\ref{g2hom})
\begin{eqnarray}
g_2(|\r_2-\r_1|) = \frac{N(N-1)}{n^2}\frac{\int|\psi(\R)|^2\dr_3...\dr_N}{\int|\psi(\R)|^2\,\dR}
\end{eqnarray}

Let us explain now how this formula is implemented in Monte Carlo calculation. We
make summation over all pairs of particles:
\begin{eqnarray}
g_2(r)
=\frac{N(N-1)}{n^2 L}
\frac{\int\delta(\r_1-\r_2-\r)|\psi(\R)|^2\,\dR}{\int|\psi(\R)|^2\,\dR}
=\frac{2}{nN}
\frac{\int\sum\limits_{i<j}\delta(r_{ij}-r)|\psi(\R)|^2\,\dR}{\int|\psi(\R)|^2\,\dR}
\end{eqnarray}

If we do a discretization of the coordinate with spacing $h$ and introduce function
$\vartheta_h(z)$ which is one if $z<h$ and zero otherwise, do summation over 
absolute value (the distribution is obviously symmetric) we obtain following
expressions:
\begin{itemize}

\item[1)] In one dimensional system:
\begin{eqnarray}
g^{1D}_2(r)=\left\langle \frac{2}{2hnN}\sum\limits_{i<j}\vartheta_h(|r_{ij}-r|)|
\right\rangle
\end{eqnarray}

In a uncorrelated system $\vartheta_h(|z|) = 2h/L$ is constant and $g_2(z)=1-1/N$.

\item[2)] In two-dimensional system distance $z$ enters explicitly in the
expression of the pair distribution function leading to larger
numerical variance at small distances
\begin{eqnarray}
g^{2D}_2(z)=
\left\langle \frac{2}{2\pi z hnN}\sum\limits_{i<j}\vartheta_h(|z_{ij}-z|)|\right\rangle
\end{eqnarray}

\item[3)] In a three-dimensional system the corresponding expression is
\begin{eqnarray}
g^{3D}_2(z)=
\left\langle \frac{2}{4\pi z^2 hnN}\sum\limits_{i<j}\vartheta_h(|z_{ij}-z|)|\right\rangle
\end{eqnarray}
\end{itemize}

\subsection{Pure estimators and extrapolation technique\label{secExtrapolation}}

As a result of the VMC calculation one obtains a {\it variational} esimator for a
quantity (let it be described by an operator $\hat A$), which corresponds to an
average over the trial wave function $\psi_T$:
\begin{eqnarray}
\langle \hat A \rangle_{var.} =
\frac{\langle\psi_T|\hat A|\psi_T\rangle}{\langle\psi_T|\psi_T\rangle}
\end{eqnarray}

Instead, the DMC method asymptotically provides a more precise {\it mixed}
estimator, which we denote as
\begin{eqnarray}
\langle \hat A \rangle_{mix.} =
\frac{\langle\phi_0|\hat A|\psi_T\rangle}{\langle\phi_0|\psi_T\rangle}
\end{eqnarray}

Still, this type of average can differ from the ``pure'' ground state average,
which corresponds to the true quantum-mechanical equilibrium value at a zero
temperature
\begin{eqnarray}
\langle \hat A \rangle_{pure} = \frac{\langle\phi_0|\hat A|\phi_0\rangle}{\langle\phi_0|\phi_0\rangle}
\end{eqnarray}

The DMC method gives an exact result for the energy, as the mixed average of the
local energy $E_{loc} = \psi_T^{-1}\hat H\psi_T$ coincides with the pure estimator.
This can be easily seen by noticing that when $\langle \phi_0$ acts on $\hat H$, it
gives exactly the ground state energy.

We will show that averages of local operators can be calculated in a ``pure'' way.
This means that the pair distribution function, radial distribution, size of the
condensate can be found essentially exactly. We assume that $\langle \R |\hat A
|\R'\rangle = A(\R) \langle \R |\R'\rangle$. The ``pure'' average can be related to
the mixed one in the following way
\begin{eqnarray}
\langle \hat A \rangle_{pure}
=
\frac{\langle\phi_0|A(\R)\frac{\phi_0(\R)}{\psi_T(\R)}|\psi_T\rangle}
{\langle\phi_0|\frac{\phi_0(\R)}{\psi_T(\R)}|\psi_T\rangle}
=
\frac{\langle  A(\R) P(\R) \rangle_{mix}}{\langle P(\R) \rangle_{mix}},
\end{eqnarray}
where $P(\R)$ is defined as
\begin{eqnarray}
P(\R) = \frac{\phi_0(\R)}{\psi_T(\R)} \langle \phi_0|\psi_T\rangle
\end{eqnarray}
and gives the number of descendants of a walker $\R$ for large times
$\tau\to\infty$. Practically it is enough to wait a sufficiently large, but a finite
time $T$. One makes measurements of a local quantity for all of the walkers, but
calculates the average after the time $T$, so that each walker was replicated
according to the weight $P(\R)$.

One of important examples of a non-local quantity is the non-diagonal element of the
one-body density matrix (see (\ref{OBDM mixed})). This quantity deserves a special
attention, so we will explain an extrapolation technique, which can be applied for
finding averages of a non-local operators.

Let us denote the difference between the trial wave function and ground-state wave
function as $\delta \psi$
\begin{eqnarray}
\phi_0 = \psi_T + \delta \psi
\end{eqnarray}

Then the ground-state average can be written as
\begin{eqnarray}
\langle\hat A\rangle_{pure} =
\langle\phi_0|\hat A|\phi_0\rangle =
\langle\psi_T|\hat A|\psi_T\rangle + 2\langle\phi_0|\hat A|\delta\psi\rangle
+ \langle\delta\psi|\hat A|\delta\psi\rangle
\end{eqnarray}

If $\delta \psi$ is small the second order term
$\langle\delta\psi|\hat A|\delta\psi\rangle$ can be neglected.
After substitution
$\langle\phi_0|\hat A|\delta\psi\rangle =
\langle\psi_T|\hat A|\phi_0\rangle -
\langle\psi_T|\hat A|\psi_T\rangle$
the extrapolation formula becomes
\begin{eqnarray}
\langle\hat A\rangle_{pure} \approx 2 \langle\hat A\rangle_{mix.} - \langle\hat A\rangle_{var.}
\label{extrapolation}
\end{eqnarray}

It is possible to write another extrapolation formula of the same order of
accuracy:
\begin{eqnarray}
\langle\hat A\rangle_{pure} \approx \frac{\langle\hat A\rangle^2_{mix.}}{\langle\hat
A\rangle_{var.}}
\label{extrapolation2}
\end{eqnarray}

Of course, if the extrapolation technique is applicable, formulae
(\ref{extrapolation}) and (\ref{extrapolation2}) give the same result. The second
formula is preferable for extrapolation of a non-negative quantity ({\it e.g.} the
OBDM), if the function can be very close to zero, as (\ref{extrapolation2})
preserves the sign of the function.

In the end of this Section we will mention that it can be proven that the
measurement of the superfluid density in DMC method is a pure estimator
and to a large extent is not biased by the chose of a trial wave
function \cite{Astrakharchik01,Astrakharchik02b}.

\chapter{3D-1D crossover of a trapped Bose gas\label{sec3D1D}}
\section{Introduction}

The study of trapped Bose systems in low dimensions is currently attracting a lot of
interest. In particular, 1D systems are expected to exhibit remarkable properties
which are far from the mean-field description and are not present in 2D and 3D. The
peculiarity of 1D physics consists in the role played by fluctuations, which destroy
long-range order even at zero temperature \cite{Schwartz77,Haldane81}, and in the
occurrence of characteristic effects due to correlations such as the fermionization
of the gas in the Tonks-Girardeau regime \cite{Girardeau60}. Recent experiments with
highly anisotropic, quasi-one-dimensional traps have shown first evidences of 1D
features in the aspect ratio and energy of the released
cloud\cite{Gorlitz01,Schreck01} as well as in the coherence properties of
condensates with fluctuating phase\cite{Dettmer01}. From a theoretical viewpoint,
the emergence of 1D effects in the properties of binary atomic collisions, by
increasing the confinement in the transverse direction, has been pointed out by
Olshanii\cite{Olshanii98}. In the case of harmonically trapped gases, the occurrence
of various regimes possessing true or quasi-condensate and the possibility of
entering the Tonks-Girardeau gas regime of impenetrable bosons has been discussed
in\cite{Petrov00}.

The ground-state properties and excitation spectrum of a homogeneous 1D system of
bosons interacting through a repulsive contact potential have been calculated
exactly by Lieb and Liniger long time ago\cite{Lieb63,Lieb63b}. For a fixed
interaction strength the Lieb-Liniger equation of state reproduces in the high
density regime the mean-field result obtained using the Bogoliubov model and in the
opposite limit of low density coincides with the ground-state of impenetrable bosons
\cite{Girardeau60}. For 1D systems in harmonic traps, the exact many-body
ground-state wave function in the Tonks-Girardeau regime has been recently
calculated\cite{Girardeau01}, and the equation of state interpolating between the
mean-field and the Tonks-Girardeau regime has been obtained within the local density
approximation in\cite{Dunjko01}. Methods based on local density approximation in the
longitudinal direction and on the Gross-Pitaevskii equation for the transverse
direction have been recently employed to predict the frequency of the collective
excitations \cite{Menotti02} and the ground-state energy in the 3D-1D cross-over as
well as in the 1D mean field - Tonks-Girardeau gas cross-over \cite{Das02}.

In this chapter we present exact Diffusion Monte-Carlo (see Sec.~\ref{secDMC})
results for the 3D-1D cross-over in harmonically trapped Bose gases. As a function
of the anisotropy parameter of the trap we calculate the ground-state properties of
the system and for highly anisotropic traps we point out the occurrence of important
beyond mean-field effects including the fermionization of the gas.

\section{Theory}
\subsection{Model Hamiltonian}

In order to describe a cold bosonic gas in a trap we use the model Hamiltonian of
type (\ref{H})
\begin{equation}
\hat H=- \frac{\hbar^2}{2m}\sum_{i=1}^N\Delta_i+\sum_{i<j}V_{int}(|\ri-\rj|)+\sum_{i=1}^N V_{ext}(\ri) \;,
\label{ham}
\end{equation}

Our system consists of $N$ spinless bosons of mass $m$ interacting through the
two-body interatomic potential $V_{int}(r)$ and is subject to the external field
which is taken to be harmonic and axially symmetric:
\begin{eqnarray}
V_{ext}(\r)=m(\omega_\perp^2 r_\perp^2 + \omega_z^2z^2)/2,
\label{Vext}
\end{eqnarray}
where $z$ is the axial coordinate, $r_\perp$ is the radial transverse coordinate and
$\omega_z$, $\omega_\perp$ are the corresponding oscillator frequencies.

For the interatomic potential we use two different repulsive model potentials: the
hard-sphere (HS) potential (\ref{HS}) and the soft-sphere (SS) potential
(\ref{SS}). In the case of the HS potential the $s$-wave scattering length
coincides with the radius of the sphere and in the case of the SS potential is
given by (\ref{aSS}). For finite $V_0$ one always has $R>a_{3D}$, while for
$V_0\to+\infty$ the SS potential coincides with the HS one with $R=a_{3D}$. The
height $V_0$ of the potential is fixed by the value of the range $R$ in units of
the scattering length, for which we choose $R=5a_{3D}$. It is worth noticing that
the HS and the SS model with $R=5a_{3D}$ represent two extreme cases for a
repulsive interatomic potential. In the HS case, the energy is entirely kinetic,
while for the SS potential $a\simeq(m/\hbar^2)\int_0^{\infty}V(r)r^2 dr$, according
to Born approximation, and the energy is almost all potential. By comparing the
results of the two model potentials we can investigate to what extent the
ground-state properties of the system depend only on the $s$-wave scattering length
and not on the details of the potential.

\subsection{Relevant parameters and DMC approach}

The relevant parameters of the problem are the number of particles $N$, the ratio
$a_{3D}/a_\perp$ of the scattering length to the transverse harmonic oscillator length
$a_\perp=\sqrt{\hbar/m\omega_\perp}$ and the anisotropy parameter
$\lambda=\omega_z/\omega_\perp$. For a given set of parameters we solve exactly,
using the Diffusion Monte-Carlo method (Sec.~\ref{secDMC}), the many-body
Schr\"odinger equation (\ref{Schrodinger}) for the ground state and we calculate the
energy per particle and the mean square radii of the cloud in the axial and radial
directions. Importance sampling is used through the Bijl-Jastrow trial wave function
(\ref{Jastrow}). For the one-body term, which accounts for the external confinement,
we use a simple gaussian {\it ansatz} (\ref{f1})
$f_1(r_\perp,z)=\exp\{-\alpha_\perp r_\perp^2-\alpha_z z^2\}$, with $\alpha_\perp$
and $\alpha_z$ optimized variational parameters. The two-body term $f_2(r)$ accounts
instead for the particle-particle interaction and is chosen using the same technique
employed in Ref. \cite{Giorgini99} for a homogeneous system. Of course, since DMC is an
exact method, the precise choice of $\psi_T(\R)$ is to a large extent unimportant
and the results obtained are not biased by the choice of the trial
wave function\footnote{Mean square radii are calculated using the pure estimator technique developed in \cite{Casulleras95}}.

\subsection{Mean-field approach}

The DMC results are compared with the predictions of mean-field theory which are
obtained from the stationary Gross-Pitaevskii (GP) equation (\ref{GPE1D})
\begin{equation}
\left(-\frac{\hbar^2}{2m}\Delta+V_{ext}(\r)+g(N-1)|\Phi(\r)|^2\right)\Phi(\r)=\mu\Phi(\r) \;,
\label{GP}
\end{equation}
where $\Phi(\r)$ is the order parameter normalized to unity: $\int|\Phi(\r)|^2\dr=1$
and $g=4\pi\hbar^2a_{3D}/m$ is the coupling constant (\ref{g3D}). Further, finite
size effects have been taken into account in the GP equation by the factor $N-1$ in
the interaction term \cite{Esry97} (see, also, \ref{GPEN-1}). In the case of
anisotropic traps with $\lambda<1$, the GP equation (\ref{GP}) is expected to
provide a correct description of the system if the transverse confinement is weak
$a_{3D}/a_\perp\ll 1$, and if the mean separation distance between particles is much
smaller than the healing length $1/n_{3D}^{1/3}\ll\xi$, where $\xi=1/\sqrt{8\pi
n_{3D}a_{3D}}$ and $n_{3D}$ is the central density of the cloud. In terms of the
linear density along $z$, $n_{1D}(z)=2\pi\int_0^\infty r_\perp
n_{3D}(r_\perp,z)\,dr_\perp$, this latter condition reads $1/n_{1D}\ll
a_\perp^2/a_{3D}$. If the mean separation distance between particles in the
longitudinal direction becomes much larger than the effective 1D scattering length
given by $a_\perp^2/a_{3D}$ \cite{Petrov00}, the mean-field approximation breaks
down because of the lack of off diagonal long range order.

The system enters the 1D regime when the motion in the radial direction becomes
frozen. In this regime the radial density profile of the cloud is fixed by the
harmonic oscillator ground state, resulting in a mean square radius which coincides
with the transverse oscillator length $\sqrt{\langle r_\perp^2\rangle}=a_\perp$.
Further, the energy per particle is dominated by the trapping potential and one has
the condition $E/N-\hbar\omega_\perp\ll\hbar\omega_\perp$.

\subsection{1D system: local density approximation}

If the discretization of levels in the longitudinal direction can be neglected, {\it i.e.}
if $E/N-\hbar\omega_\perp\gg\hbar\omega_z$, the 1D system can be described within
the local density approximation (LDA). In this case, the chemical potential of the
system is calculated through the local equilibrium equation (\ref{LDA}) which we
write separating in an explicit way the dominant contribution of the transverse
confinement $\hbar\omega_\perp$:
\begin{equation}
\mu=\hbar\omega_\perp + \mu_{hom}(n_{1D}(z))+\frac{m}{2}\omega_z^2z^2,
\label{LDA3D1D}
\end{equation}
Here $\mu_{hom}(n_{1D})$ is the chemical potential corresponding to a homogeneous
1D system of density $n_{1D}$. If the ratio $a_{3D}/a_{\perp}\ll 1$, the local chemical
potential can be obtained from the Lieb-Liniger (LL) equation (\ref{LL}) of state with the
effective 1D coupling constant $g_{1D}=g_{3D}/(2\pi a_{\perp}^2)$ (\ref{g1DMF}).
One finds: $\mu_{hom}=\partial[n_{1D}\epsilon_{LL} (n_{1D})]/\partial n_{1D}$,
where $\epsilon_{LL}$ is the LL energy per particle.
The LDA problem in one-dimension was already studied in Sec.~\ref{secLDA1D}. The
chemical potential $\mu$ as a function of $N$ and trap parameters is given by
formula (\ref{LDAmu1D0}). The ground-state energy of the system with a given number of
particles can then be calculated through direct integration of $\mu(N)$.

If $n_{1D}a_\perp^2/a_{3D}\gg 1$, the system is weakly interacting and the LL
equation of state coincides with the mean-field prediction:
$\epsilon_{LL}=g_{1D}n_{1D}/2$. In the notation of Sec.~\ref{secLDA1D} it
corresponds to $C_1 = 2, \gamma_1 =1,C_2=0$. From formula (\ref{LDAmu1D0}) one finds
the following results for the energy per particle
\begin{equation}
\frac{E}{N}-\hbar\omega_\perp=\frac{3}{10}\left(3N\lambda\frac{a_{3D}}{a_\perp}\right)^{2/3}\hbar\omega_\perp \;,
\label{1DMF1}
\end{equation}
and exploiting formula (\ref{LDA1Dz2}) one obtains an expression for the mean square
radius of the cloud in the longitudinal direction
\begin{equation}
\sqrt{\langle z^2\rangle}=\left(3N\lambda\frac{a_{3D}}{a_\perp}\right)^{1/3}\frac{a_\perp}{\sqrt{5}\lambda} \;.
\label{1DMF2}
\end{equation}

In the opposite limit, $n_{1D}a_\perp^2/a\ll 1$, the system enters the
Tonks-Girardeau regime and the LL equation of state has the Fermi-like behavior
(\ref{ETG}) $\epsilon_{LL}=\pi^2\hbar^2n_{1D}^2/6m$. The energy per particle and the
mean square radius of the trapped system are easily extracted from the results for a
purely one-dimensional system (\ref{ETGLDA}),(\ref{zTGLDA})
\begin{equation}
\frac{E}{N}-\hbar\omega_\perp=\frac{N\lambda}{2}\hbar\omega_\perp \;,
\;\;\;\;\;\;
\sqrt{\langle z^2\rangle}=\sqrt{\frac{N}{2\lambda}}a_\perp \;.
\label{1DTG}
\end{equation}

In terms of the parameters of the system, the two regimes can be identified by
comparing the corresponding energies. The mean-field energy becomes favorable if
$N\lambda a_\perp^2/a^2_{3D}\gg 1$, whereas the Tonks-Girardeau gas is preferred if
the condition $N\lambda a_\perp^2/a^2_{3D}\ll 1$ is satisfied\footnote{This
characteristic for LDA parameter is  universal, for example systems with different
number of particles behave similarly if the combination
$N\lambda a_\perp^2/a^2_{3D}$ is the same. It becomes clear from
from equation (\ref{LDA1Ddimensionless}) noting that $N\lambda a_\perp^2/a^2_{3D}=
\Delta_1^2$.}.

\subsection{1D system: beyond local density approximation\label{secLLtrap}}

In order to account for effects beyond local density approximation we have also
applied the DMC method to a system of $N$ particles interacting through the
Lieb-Liniger Hamiltonian in the presence of harmonic confinement
\begin{eqnarray}
\hat H^{trap}_{LL}= N\hbar\omega_\perp - \frac{\hbar^2}{2m}\sum_{i=1}^N\frac{\partial^2}{\partial z_i^2}
+g_{1D}\sum_{i<j}\delta(z_i-z_j)+\sum_{i=1}^N \frac{m\omega_z^2 z_i^2}{2} \;.
\label{hamLL}
\end{eqnarray}

When the number of particles is large, the properties of the ground state of the
Hamiltonian (\ref{hamLL}) coincide with the ones obtained from the LL equation of
state within LDA. However, for small systems one expects deviations and the DMC
method provides us with a powerful tool. The relevant parameters are the same as for
the 3D simulation with the Hamiltonian (\ref{ham}): the number of particles $N$, the
ratio $a_{3D}/a_\perp$ fixing the strength of the contact potential $g_{1D}$ and the
anisotropy parameter $\lambda$ fixing the strength of the longitudinal confinement
in units of $\hbar\omega_\perp$.

The importance sampling is realized through the Bijl-Jastrow trial wave function
(\ref{Jastrow}). The one-body term is of gaussian form (\ref{f1})
$f_1(z)=\exp\{-\alpha_z z^2\}$. The two-body term $f_2(z)$ is given by the formula
(\ref{wfLL}). The construction of the the discussed in details in
Sec.~\ref{secLLwf}. There are two variational parameters: gaussian width $\alpha_z$
and the matching distance $\Rm$. We fix them by minimizing the variational energy.

We notice that our choice of the two-body Jastrow factor reproduces both the weakly
interacting and the Tonks-Girardeau regime. In fact, if $k\Rm$ is small,
$f_2(z)\simeq 1, z=0$ and the contact potential is almost transparent. On the
other hand, if $k\Rm$ approaches $\pi/2$, $f_2(z)\to 0,z=0$ and the
contact potential behaves as an impenetrable barrier.

\section{Results}
\subsection{Small system, medium scattering length}

\begin{figure}[ht!]
\begin{center}
\includegraphics*[angle=-90,width=0.6\columnwidth]{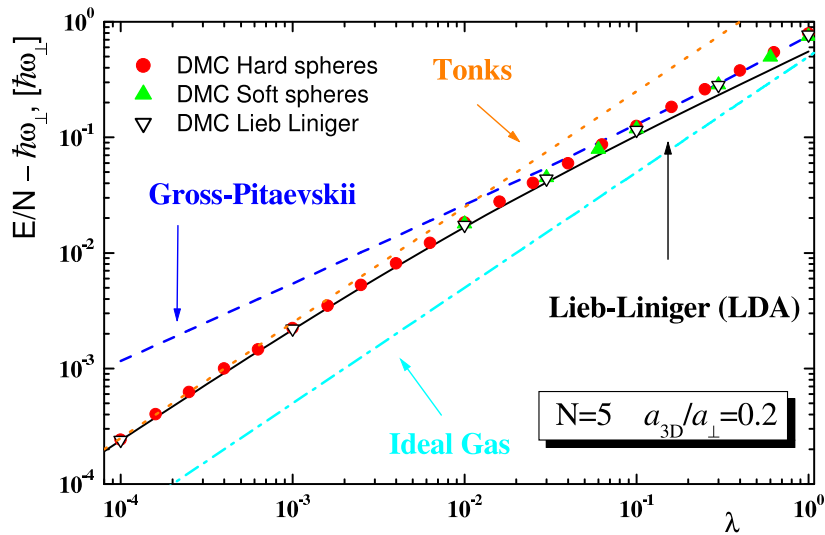}
\caption{Energy per particle as a function of $\lambda$. DMC results: HS potential (solid circles), SS potential (solid
triangles), LL Hamiltonian (\ref{hamLL}) (open triangles).
Dashed line: GP equation (\ref{GP}), solid line: LL equation of states in LDA,
dotted line: TG gas, dot-dashed line: non-interacting gas. Error bars are smaller
than the size of the symbols.}
\label{Fig3D1D1}
\end{center}\vspace{-6 mm}
\end{figure}

\begin{figure}[ht!]
\begin{center}
\includegraphics*[angle=-90,width=0.6\columnwidth]{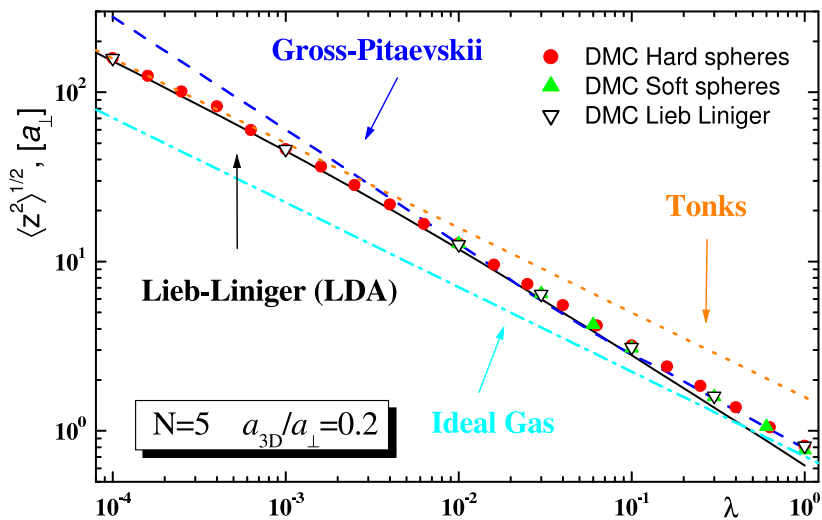}
\caption{Mean square radius along $z$ as a function of $\lambda$.
Dashed line: GP equation (\ref{GP}), solid line: LL equation of states in LDA,
dotted line: TG gas, dot-dashed line: non-interacting gas. Error bars are smaller
than the size of the symbols.}
\label{Fig3D1D2}
\end{center}\vspace{-6 mm}
\end{figure}

We first consider a system of very few particles ($N=5$) and we consider different
values of the ratio $a_{3D}/a_\perp$. Figs.~\ref{Fig3D1D1}-\ref{Fig3D1D2} refer to
$a_{3D}/a_\perp=0.2$, and we present results for the energy per particle and the
mean square radius of the cloud in the longitudinal direction as a function of the
anisotropy parameter $\lambda=\omega_z/\omega_\perp$. Results from the GP equation
(\ref{GP}) and from the Lieb-Liniger equation of state in LDA are also shown. We
find that the HS and SS potential give practically the same results even for the
largest values of $\lambda$, showing that for these parameters we are well within
the universal regime where the details of the potential are irrelevant. For large
values of $\lambda$ the DMC results agree well with the predictions of GP equation.
By decreasing $\lambda$ beyond mean-field effects become visible and both the energy
per particle and the mean square radius approach the LL result when $N\lambda
a_\perp^2/a^2\sim 1$, corresponding to $\lambda\sim 10^{-2}$. Finally, for the
smallest values of the anisotropy parameter ($\lambda\sim 10^{-4}$) we find clear
evidence of the Tonks-Girardeau gas behavior both in the energy and in the shape of
the cloud. It is worth stressing that beyond mean-field effects occurring in the
small $\lambda$ regime can be only obtained by using DMC. A Variational Monte-Carlo
(VMC) calculation based on the trial wave function $\psi_T(\R)$ described above,
would yield results in good agreement with mean-field over the whole range of values
of $\lambda$. DMC results using the Lieb-Liniger Hamiltonian $H_{LL}$ of Eq.
(\ref{hamLL}) are also shown and coincide with the results of the 3D Hamiltonian
(\ref{ham}). This shows that the 3D interatomic potential is correctly described by
the 1D $\delta$-potential even for the largest values of $\lambda$. In fact, due to
the small number of particles, the density profile of the cloud in the transverse
direction is correctly described by the harmonic oscillator ground-state
wave function (see Fig.~\ref{Fig3D1D9}). The 1D character of the system is also
evident from Fig.~\ref{Fig3D1D1} which shows that $E/N-\hbar\omega_\perp$ is always
smaller than the transverse confining energy. Deviations of DMC results from the LL
equation of state arise because of finite size effects. These effects become less
and less important as $\lambda$ decreases and one enters the regime
$(E/N-\hbar\omega_\perp)/\hbar\omega_\perp \gg\lambda$ where LDA applies. In terms
of the mean square radius of the cloud (see Fig.~\ref{Fig3D1D2}), the condition of
applicability of LDA requires $\langle z^2\rangle^{1/2}$ much larger than the
corresponding ideal gas (IG) value.

\subsection{Small system, small scattering length}

\begin{figure}[ht!]
\begin{center}
\includegraphics*[angle=-90,width=0.6\columnwidth]{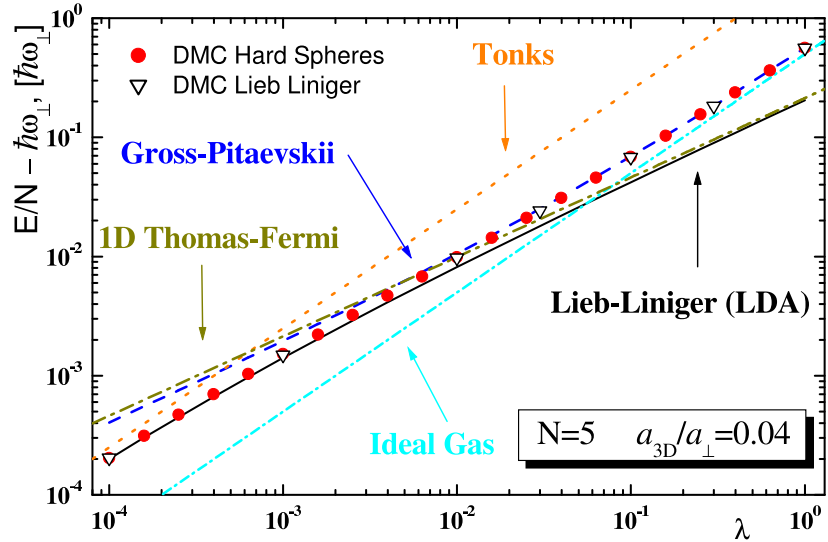}
\caption{Energy per particle as a function of $\lambda$.
DMC results: HS potential (solid circles), LL Hamiltonian (\ref{hamLL}) (open triangles), 
Dashed line: GP equation (\ref{GP}), solid line: LL equation of states in LDA,
dotted line: TG gas, dot-dashed line: non-interacting gas.}
\label{Fig3D1D3}
\end{center}\vspace{-6 mm}
\end{figure}

\begin{figure}[ht!]
\begin{center}
\includegraphics*[angle=-90,width=0.6\columnwidth]{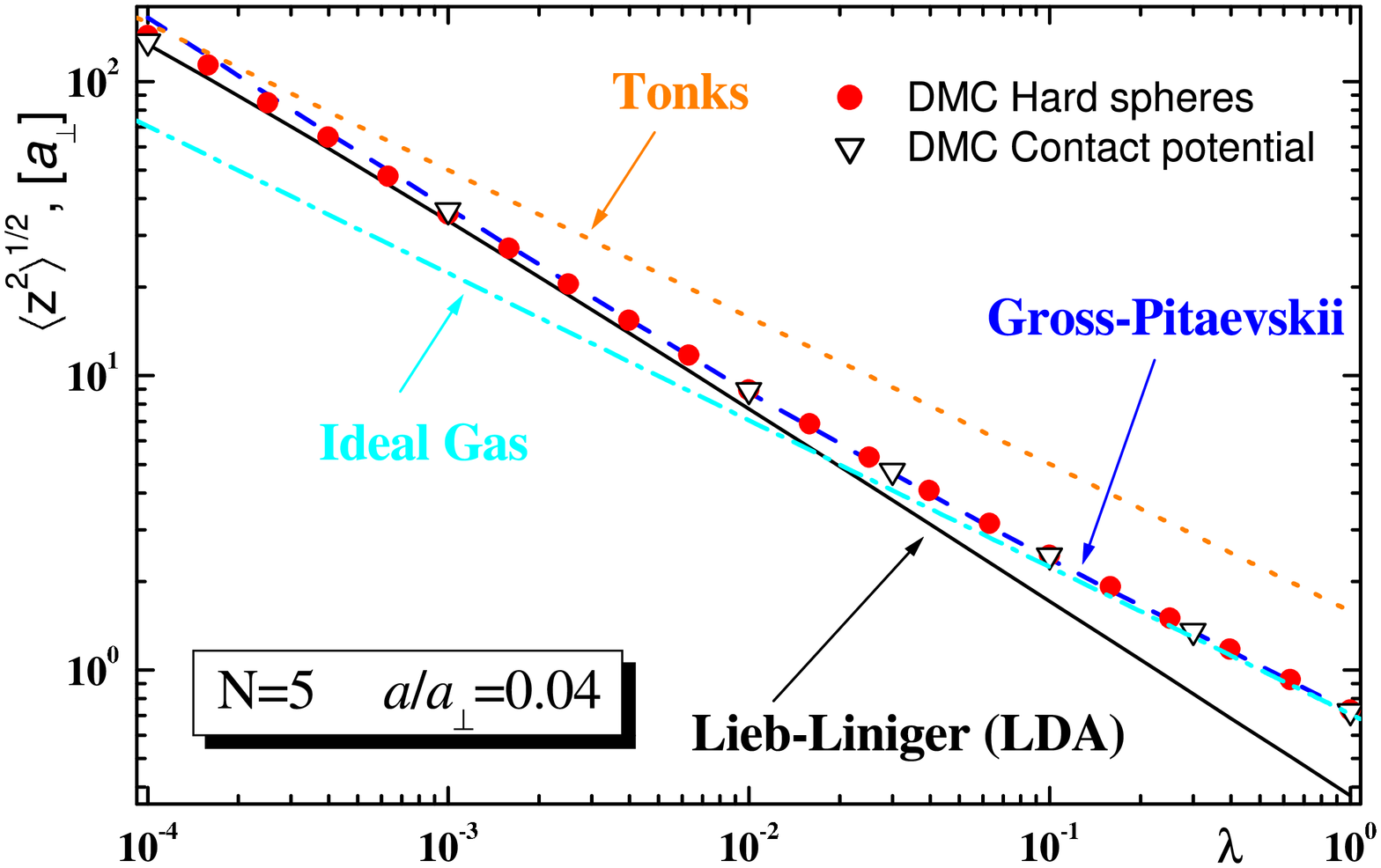}
\caption{Mean square radius along $z$ as a function of $\lambda$.
Dashed line: GP equation (\ref{GP}), solid line: LL equation of states in LDA,
dotted line: TG gas, dot-dashed line: non-interacting gas.}
\label{Fig3D1D4}
\end{center}\vspace{-6 mm}
\end{figure}

In Figs.~\ref{Fig3D1D3}-\ref{Fig3D1D4} we present results for $a_{3D}/a_\perp=0.04$,
corresponding to a less tight transverse confinement or, equivalently, to a smaller
scattering length. By decreasing $a_{3D}/a_\perp$ we enter more deeply in the universal
regime where the theory of pseudo-potentials applies and we find no difference
between HS and SS results. Further, as for the $a_{3D}/a_\perp=0.2$ case, we find no
difference between DMC results obtained starting from Hamiltonian (\ref{ham}) and
from the 1D Hamiltonian (\ref{hamLL}). Compared to Figs.~\ref{Fig3D1D1}-\ref{Fig3D1D2},
the cross-over between mean field and Lieb-Liniger occurs for smaller values of
$\lambda$. For the smallest values of $\lambda$ beyond mean-field effects become
evident, though one would need to decrease $\lambda$ even further to enter the
Tonks-Girardeau gas regime.

\subsection{Small system, large scattering length}

\begin{figure}[ht!]
\begin{center}
\includegraphics*[angle=-90,width=0.6\columnwidth]{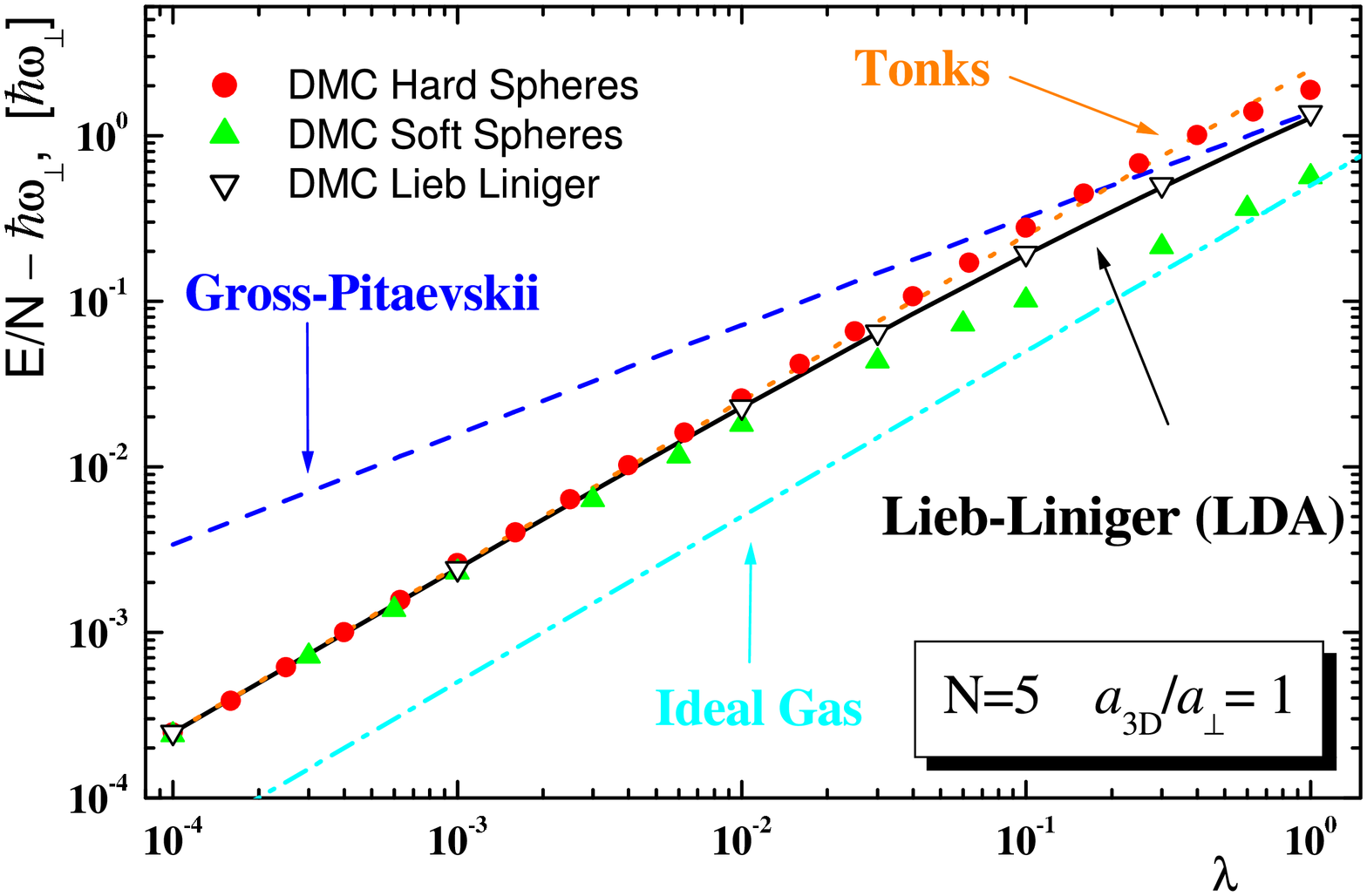}
\caption{Energy per particle as a function of $\lambda$. Dashed line:
GP equation (\ref{GP}), solid line: LL equation of states in LDA, dotted line:
TG gas, dot-dashed line: non-interacting gas.}
\label{Fig3D1D5}
\end{center}\vspace{-6 mm}
\end{figure}

\begin{figure}[ht!]
\begin{center}
\includegraphics*[angle=-90,width=0.6\columnwidth]{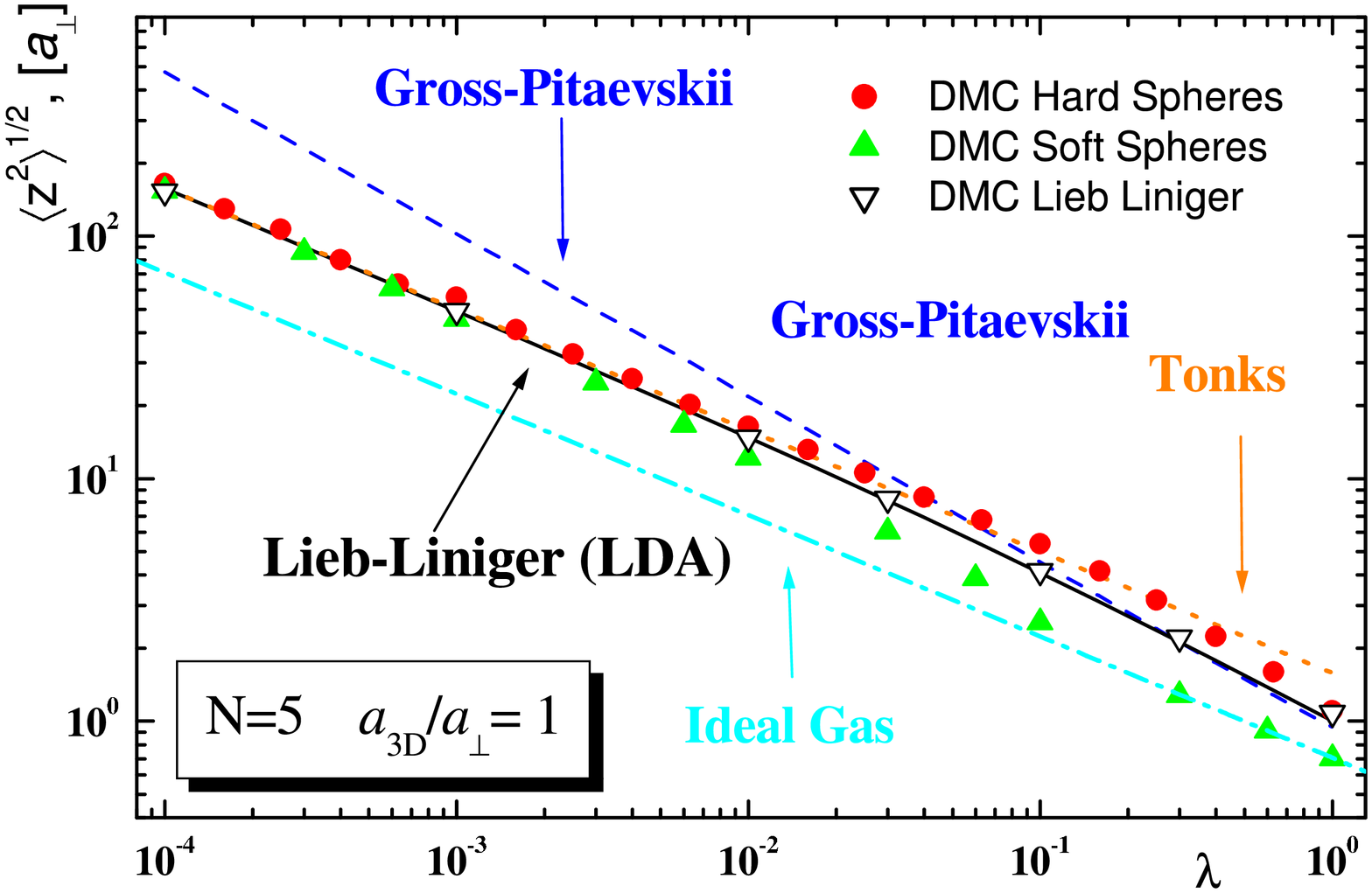}
\caption{Mean square radius along $z$ as a function of $\lambda$.
Dashed line: GP equation (\ref{GP}), solid line: LL equation of states in LDA,
dotted line: TG gas, dot-dashed line: non-interacting gas.}
\label{Fig3D1D6}
\end{center}\vspace{-6 mm}
\end{figure}

The results for $a_{3D}/a_\perp=1$ are shown in Figs.~\ref{Fig3D1D5}-\ref{Fig3D1D6}. In this
case the HS and SS potential give significantly different results in the large
$\lambda$ regime. For both potentials the mean-field description is inadequate. By
decreasing $\lambda$ the HS system enters the Tonks-Girardeau regime before
approaching the LL results, whereas the SS system crosses from the ideal gas (IG)
regime to the Tonks-Girardeau regime. For this value of $a_{3D}/a_\perp$ the DMC results
with the Hamiltonian (\ref{hamLL}) coincide exactly with the ones of the LL equation
of state in LDA due to the large coupling constant.

\subsection{Large system}

\begin{figure}[ht!]
\begin{center}
\includegraphics*[angle=-90,width=0.6\columnwidth]{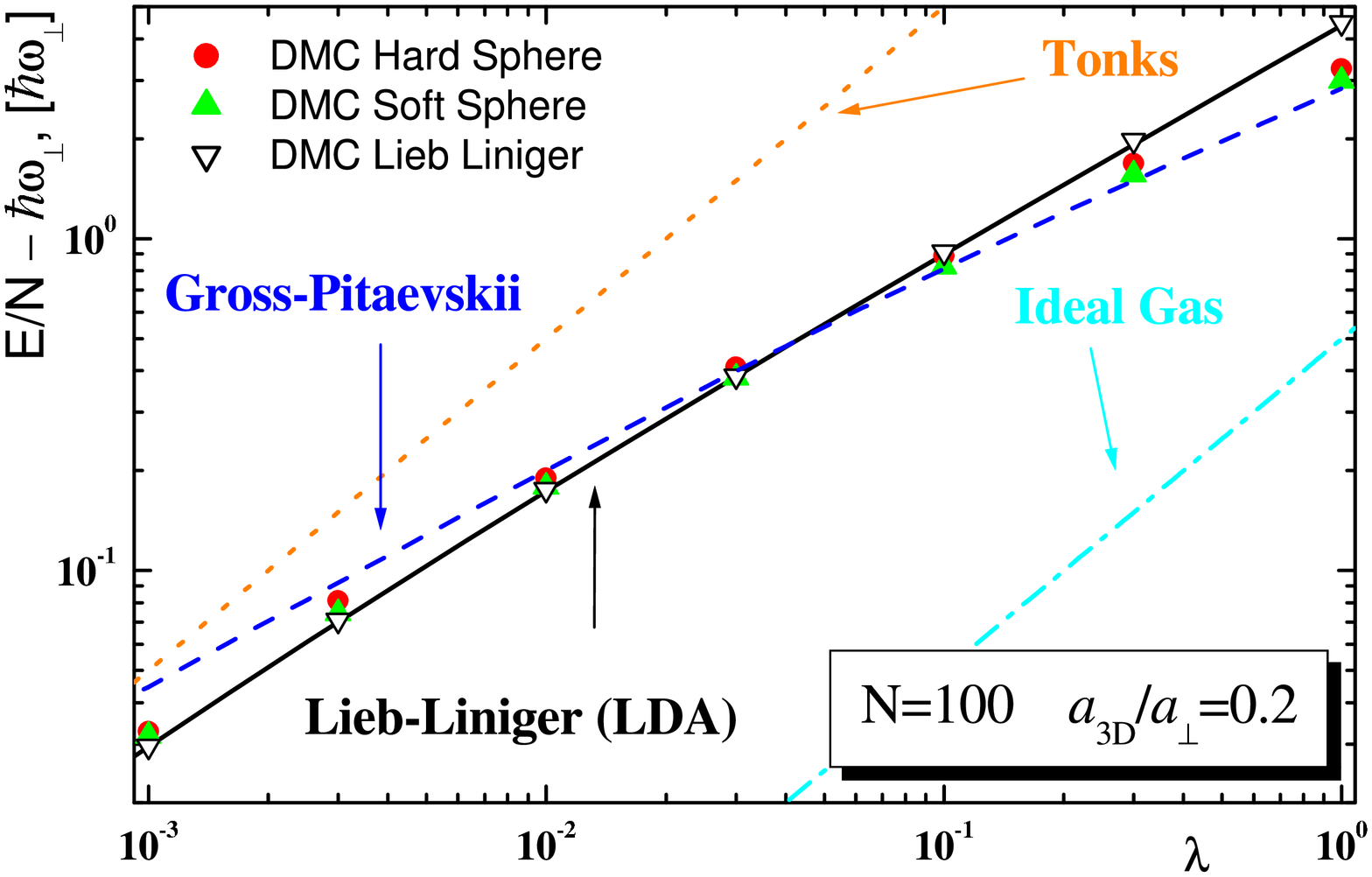}
\caption{Energy per particle as a function of $\lambda$.
Dashed line: GP equation (\ref{GP}), solid line: LL equation of states in LDA,
dotted line: TG gas, dot-dashed line: non-interacting gas.
}
\label{Fig3D1D7}
\end{center}\vspace{-6 mm}
\end{figure}

\begin{figure}[ht!]
\begin{center}
\includegraphics*[angle=-90,width=0.6\columnwidth]{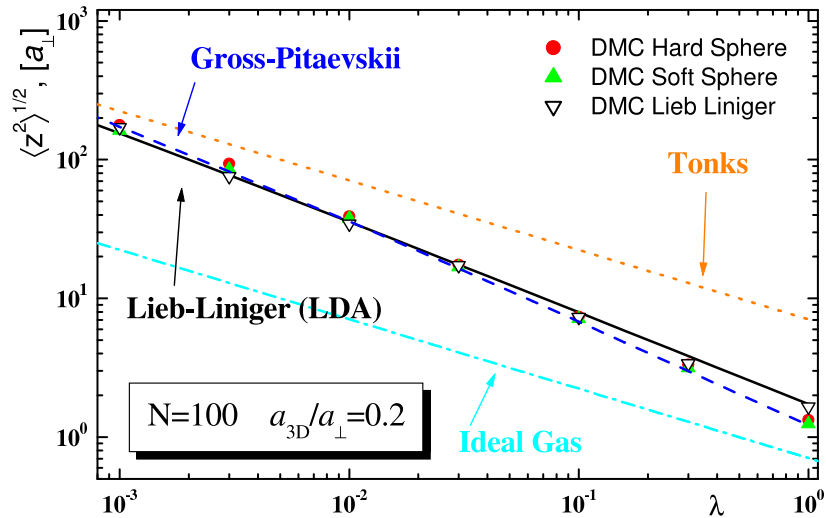}
\caption{Mean square radius along $z$ as a function of $\lambda$.
Dashed line: GP equation (\ref{GP}), solid line: LL equation of states in LDA,
dotted line: TG gas, dot-dashed line: non-interacting gas.}
\label{Fig3D1D8}
\end{center}\vspace{-6 mm}
\end{figure}

Figs.~\ref{Fig3D1D7}-\ref{Fig3D1D8} refer to a much larger system with $N=100$ and
$a_{3D}/a_\perp=0.2$. In this case, we see a clear cross-over from 3D mean field, at
large $\lambda$, to 1D LL at small $\lambda$. Important beyond mean-field effects
become evident in the energy per particle as $N\lambda a_\perp^2/a^2\sim 1$,
corresponding to $\lambda\sim 10^{-3}$. The Tonks-Girardeau regime would correspond
to even smaller values of $\lambda$ which are difficult to obtain in our simulation.
However, for the smallest values of $\lambda$ reported in Fig.~\ref{Fig3D1D7} we
find already very good agreement with the LL equation of state. One should notice
that small deviations from mean field are also visible for $\lambda\sim 1$, and are
due to high density corrections to the GP equation. The DMC results with the 1D
Hamiltonian (\ref{hamLL}) follow exactly the LDA prediction showing that the
deviations seen in Figs.~\ref{Fig3D1D1}-\ref{Fig3D1D2} are due to finite size
effects. In the cross-over region from the mean-field to the 1D LL regime, residual
3D effects are still present (see Fig.~\ref{Fig3D1D9}) and produce small deviations
from the LL equation of state.

\subsection{Radial size of the system}

\begin{figure}[ht!]
\begin{center}
\includegraphics*[width=0.6\columnwidth]{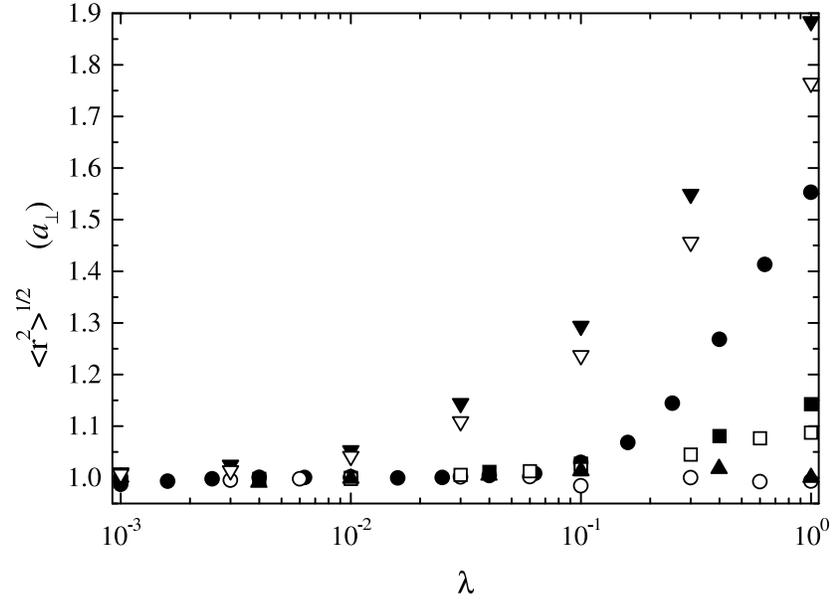}
\caption{Mean square radius in the radial direction as a function of $\lambda$. Solid symbols: HS potential;
open symbols: SS potential. Down triangles: $N=100$ and $a_{3D}/a_\perp=0.2$; circles: $N=5$ and $a_{3D}/a_\perp=1$;
squares: $N=5$ and $a_{3D}/a_\perp=0.2$; up triangles: $N=5$ and $a_{3D}/a_\perp=0.04$.
Error bars are smaller than the size of the symbols.}
\label{Fig3D1D9}
\end{center}
\end{figure}

Finally, in Fig.~\ref{Fig3D1D9}, we show results for the mean square radius in the
transverse direction. The cross-over from 3D to 1D is clearly visible in the case of
$N=100$, for both the HS and SS potential, and for the HS potential in the case of
$N=5$ and $a_{3D}/a_\perp=1$. For the system with $N=5$ and $a_{3D}/a_\perp=0.2$ we only see
small deviations from $\sqrt{\langle r_\perp^2\rangle}=a_\perp$ for the largest
values of $\lambda$. In the $a_{3D}/a_\perp=0.04$, as well as in the $a_{3D}/a_\perp=1$ case
with the SS potential, the transverse density profile is well described by the
harmonic oscillator wave function and we find $\sqrt{\langle r_\perp^2\rangle}\simeq
a_\perp$ over the whole range of values of $\lambda$. It is worth noticing that for
the $N=5$ system with SS potential, the largest deviations from $\sqrt{\langle
r_\perp^2\rangle}=a_\perp$ are achieved for $a_{3D}/a_\perp=0.2$, corresponding to a
transverse confinement $a_\perp=R$ where $R$ is the range of the SS potential.

\section{Conclusions}

In this chapter we present exact Quantum Monte Carlo results of the ground-state
energy and structure of a Bose gas confined in highly anisotropic harmonic traps.
Starting from a 3D Hamiltonian, where interparticle interactions are model by a
hard-sphere or a soft-sphere potential, we show that the system exhibits striking
features due to particle correlations. By reducing the anisotropy parameter
$\lambda$, while the number of particles $N$ and the ratio $a_{3D}/a_\perp$ of scattering
to transverse oscillator length are kept fixed, the system crosses from a regime
where mean-field theory applies to a regime which is well described by the 1D
Lieb-Liniger equation of state in local density approximation. In the cross-over
region both theories fail and one must resort to exact methods to account properly
for both finite size effects and residual 3D effects. For very small values of
$\lambda$ we find clear evidence, both in the energy per particle and in the
longitudinal size of the cloud, of the fermionization of the system in the
Tonks-Girardeau regime.

\chapter{Quasi 1D Bose gases with large scattering length\label{sec1DBG}}
\section{Introduction\label{secScI}}

Quasi-1D Bose gases have been realized in highly-elongated traps by tightly
confining the transverse motion of the atoms to their zero point oscillations
\cite{Gorlitz01,Schreck01,Greiner01}. As compared to the 3D case, the role of quantum fluctuations is
enhanced in 1D and these systems are predicted to exhibit peculiar properties,
which cannot be described using traditional mean-field theories, but require more
advanced many-body approaches. Particularly intriguing is the strong coupling
regime, where, due to repulsion between particles, the quasi-one dimensional Bose gas behaves as
if it consisted of fictitious spinless fermions (Tonks-Girardeau
gas\cite{Tonks36,Girardeau60,Olshanii98,Petrov00}). This regime has not been achieved yet,
but is one of the main focus areas of present experimental investigations in this
field \cite{Moritz03,Reichel03}. An interesting possibility to approach the strongly
correlated TG regime is provided by magnetic field induced atom-atom Feshbach
resonances \cite{Inouye98,Cornish00,Loftus02,O'Hara02,Bourdel03}. By utilizing this technique one can tune the $3D$
$s$-wave scattering length $a_{3D}$, and hence the strength of atom-atom
interactions, to essentially any value including zero and $\pm\infty$.

Degenerate quantum gases near a Feshbach resonance have recently received a great
deal of interest both experimentally and theoretically. At resonance
($|a_{3D}|\to\infty$) the 3D scattering cross-section $\sigma$ is fixed by the
unitary condition, $\sigma=4\pi/k^2$, where $k$ is the relative wave vector of the
two atoms. In this regime it is predicted that, if the effective range $R$ of the
atom-atom interaction potential is much smaller than the average interparticle
distance, the behavior of the gas is universal, {\it i.e.}, independent of the details of
the interatomic potential and independent of the actual value of $a_{3D}$
\cite{Heiselberg01,Cowell02}. This is known as the unitary regime \cite{Ho04}.
In the case of 3D Bose gases, this unitary regime can most likely not be realized
in experiments since three-body recombination is expected to set in when $a_{3D}$
becomes comparable to the average interparticle distance. Three-body recombination
leads to cluster formation and hence makes the gas-like state unstable. The
situation is different for Fermi gases, for which the unitary regime has already
been reached experimentally \cite{O'Hara02,Bourdel03}. In this case, the Fermi pressure
stabilizes the system even for large $|a_{3D}|$.

In quasi-one dimensional geometries a new length scale becomes relevant, namely,
the oscillator length $a_\perp=\sqrt{\hbar/(m\omega_\perp)}$ of the tightly confined
transverse motion, where $m$ is the mass of the atoms and $\omega_\perp$ is the
angular frequency of the harmonic trapping potential. For $|a_{3D}|\gg a_\perp$, the
gas is expected to exhibit a universal behavior if the effective range $R$ of the
atom-atom interaction potential is much smaller than $a_\perp$ and the mean
interparticle distance is much larger than $a_\perp$. It has been predicted that
three-body recombination processes are suppressed for strongly interacting 1D Bose
gases (see Eq.~\ref{g3 TG}). These studies raise the question whether the unitary
regime can be reached in Bose gases confined in highly-elongated traps, that is,
whether the quasi-one dimensional bosonic gas-like state is stable against cluster formation as
$a_{3D}\to\pm\infty$.

This chapter is devoted to the investigation of the properties of a quasi-one
dimensional Bose gas at zero temperature over a wide range of values of the $3D$
scattering length $a_{3D}$ using Quantum Monte Carlo techniques (see
Chapter~\ref{secQMC}). We find that the system 1) is well described by a 1D model
Hamiltonian with contact interactions and renormalized coupling constant
\cite{Olshanii98} for any value of $a_{3D}$, 2) reaches the regime of a TG gas for a
critical positive value of the 3D scattering length $a_{3D}$, 3) enters a unitary
regime for large values of $|a_{3D}|$, that is, for $|a_{3D}|
\rightarrow \infty$, where the properties of the quasi-one dimensional Bose gas become
independent of the actual value of $a_{3D}$ and are similar to those of a 1D gas of
hard-rods and 4) becomes unstable against cluster formation for a critical value of
the 1D gas parameter, or equivalently, for a critical negative value of the 3D
scattering length $a_{3D}$.

The structure of this Chapter is as follows. Section~\ref{secScII} discusses the
energetics of two bosons in quasi-one dimensional harmonic traps. to a 1D model Hamiltonian with
contact interactions and renormalized coupling constant~\cite{Olshanii98}. The
eigenenergies of the system are calculated by exact diagonalization of both the 3D
and the 1D Hamiltonian. We use these results for two particles to benchmark our
quantum MC calculations presented in Sec.~\ref{secScIV}. Section~\ref{secScIII}
discusses the relation between the 3D and the 1D Hamiltonian for $N$ bosons under
quasi-one dimensional confinement. Section~\ref{secScIV} presents
our MC results for $N=2$ and $N=10$ atoms in highly-elongated harmonic traps over a
wide range of values of the 3D scattering length $a_{3D}$. A comparison of the
energetics of the lowest-lying gas-like state for the 3D and the 1D Hamiltonian is
carried out.
In the $N=2$ case, we additionally compare with the essentially exact
results presented in Sec.~\ref{secScII}. In the $N=10$ case, we additionally compare with the
energy of the lowest-lying gas-like state of the 1D Hamiltonian calculated using
the local density approximation (LDA). Section~\ref{secScV} discusses the stability of the
lowest-lying gas-like state against cluster formation when $a_{3D}$ is negative
using the variational Monte Carlo (VMC) method. We provide a quantitative estimate
of the criticality condition. Finally, Sec.~\ref{secScVI} draws our conclusions.

\section{Two Bosons under quasi-one-dimensional confinement\label{secScII}}

Consider two interacting mass $m$ bosons with position vectors $\vec{r}_1$ and
$\vec{r}_2$, where $\vec{r}_i=(x_i,y_i,z_i)$, in a waveguide with harmonic
confinement in the radial direction. If we introduce the center of mass coordinate
$\vec{R}$ and the relative coordinate $\vec{r}=\vec{r}_2-\vec{r}_1$, the problem
separates. Since the solution to the center of mass Hamiltonian is given readily,
we only consider the internal Hamiltonian $H_{3D}^{int}$, which can be conveniently
written in terms of cylindrical coordinates $\vec{r}=(\rho,\phi,z)$,
\begin{eqnarray}
\hat H^{int}_{3D} = -\frac{\hbar^2}{2\mu}\Delta +
V_{int}(\vec{r}) + \frac{1}{2} \mu \omega_\perp^2 \rho^2, \label{eq_wgint3D}
\end{eqnarray}
where $\mu$ denotes the reduced two-body mass, $\mu=m/2$, and $V_{int}(\vec{r})$ denotes
the full 3D atom-atom interaction potential.

Considering a regularized zero-range pseudo-potential $V_{int}(\vec{r}) = 2 \pi \hbar^2
a_{3D}/\mu \delta(\vec{r})\frac{\partial}{\partial r}r$, where $a_{3D}$ denotes the
3D scattering length, Olshanii~\cite{Olshanii98} derives an effective 1D Hamiltonian,
\begin{eqnarray}
H^{int}_{1D}= - \frac{\hbar^2}{2 \mu} \frac{d^2}{d z^2}
+ g_{1D} \delta(z) + \hbar \omega_\perp,\label{eq_wgint1D}
\end{eqnarray}
and renormalized coupling constant $g_{1D}$,
\begin{eqnarray}
g_{1D}=\frac{2 \hbar^2 a_{3D}}{m a_\perp^2}
\left [ 1-|\zeta(1/2)|
\frac{a_{3D}}{\sqrt{2}a_\perp} \right]^{-1},
\label{eq_g1Dren}
\end{eqnarray}
which reproduce the low energy scattering solutions of the full 3D Hamiltonian,
Eq.~\ref{eq_wgint3D}. Here, $\zeta(\cdot)$ denotes the Riemann zeta function,
$\zeta(1/2) = -1.4604$. Alternatively, $g_{1D}$ can be expressed through the
effective 1D scattering length $a_{1D}$~\cite{Olshanii98},
\begin{eqnarray}
g_{1D}= -\frac{2\hbar^2}{ma_{1D}},
\label{eq_g1Da1D}
\end{eqnarray}
where
\begin{eqnarray}
a_{1D} = -a_\perp\left(\frac{a_\perp}{a_{3D}}-\frac{|\zeta(1/2)|}{\sqrt 2}\right)
\label{eq_a1Dren}
\end{eqnarray}

Olshanii's result shows that the waveguide gives rise to an effective interaction,
parameterized by the coupling constant $g_{1D}$, which can be tuned to any strength
by changing the ratio between the 3D scattering length $a_{3D}$ and the transverse
oscillator length $a_\perp$.

The renormalized coupling constant, Eq.~\ref{eq_g1Dren}, can be compared with the
unrenormalized coupling constant $g_{1D}^0$ (\ref{g1DMF}),
\begin{eqnarray}
g_{1D}^0=\frac{2\hbar^2 a_{3D}}{m a_\perp^2}, \label{eq_g1Dunren}
\end{eqnarray}

\begin{figure}
\vspace*{-1.0in}
\centerline{\includegraphics*[height=11cm,width=0.6\columnwidth]{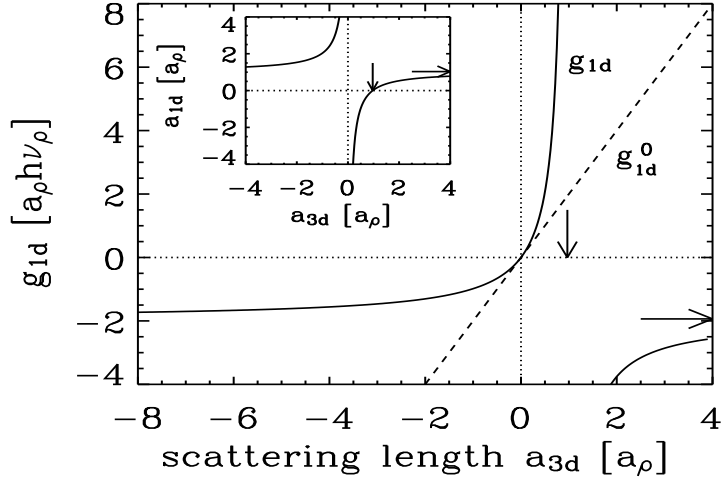}}
\vspace*{-.3in}
\caption{One-dimensional coupling constants $g_{1D}$ [Eq.~\protect\ref{eq_g1Dren},
solid line] and $g_{1D}^{0}$ [Eq.~\protect\ref{eq_g1Dunren}, dashed line] as a
function of the 3D scattering length $a_{3D}/a_\perp$. The vertical arrow
indicates the value of $a_{3D}$ for which $g_{1D}$ diverges,
$a_{3D}^{c}=0.9684a_\perp$. The horizontal arrow indicates the asymptotic value of
$g_{1D}$ as $|a_{3D}| \rightarrow \infty$, $g_{1D}=-1.9368a_\perp \hbar
\omega_\perp$. Inset: One-dimensional scattering length $a_{1D}$,
Eq.~\protect\ref{eq_a1Dren}, as a function of $a_{3D}/a_\perp$. The vertical
arrow indicates the value of $a_{3D}$ for which $a_{1D}$ goes through zero,
$a_{3D}^{c}=0.9684a_\perp$. The horizontal arrow indicates the asymptotic value of
$a_{1D}$ as $|a_{3D}| \rightarrow \infty$, $a_{1D}=1.0326a_\perp$. The angular
frequency $\omega_\perp$ determines the frequency $\nu_\perp$, $\omega_\perp=2
\pi \, \nu_\perp$ (also, $\hbar \omega_\perp=h \nu_\perp$). }
\label{figSc1}
\end{figure}

Figure~\ref{figSc1} shows the unrenormalized coupling constant $g_{1D}^0$ [dashed
line, Eq.~\ref{eq_g1Dunren}] together with the renormalized coupling constant
[solid line, Eq.~\ref{eq_g1Dren}]. For $|a_{3D}|\ll a_\perp$, the renormalized
coupling constant $g_{1D}$ is nearly identical to the unrenormalized coupling
constant $g_{1D}^0$. For large $|a_{3D}|$, however, the confinement induced
renormalization becomes important, and the effective 1D coupling constant $g_{1D}$,
Eq.~\ref{eq_g1Dren}, has to be used. At the critical value $a_{3D}^c=0.9684
a_\perp$ (indicated by a vertical arrow in Fig.~\ref{figSc1}), $g_{1D}$ diverges.
For $a_{3D}\to\pm\infty$, $g_{1D}$ reaches an asymptotic value,
$g_{1D}=-1.9368a_\perp \hbar \omega_\perp$ (indicated by a horizontal arrow in
Fig.~\ref{figSc1}). Finally, $g_{1D}$ is negative for all negative 3D scattering
lengths. The inset of Fig.~\ref{figSc1} shows the effective 1D scattering length
$a_{1D}$, Eq.~\ref{eq_a1Dren}, as a function of $a_{3D}$. For small positive
$a_{3D}$, $a_{1D}$ is negative and it changes sign at $a_{3D}=a_{3D}^c$ ($a_{1D}=0$
for $a_{3D}=a_{3D}^c$). Moreover, $a_{1D}$ reaches, just as $g_{1D}$, an asymptotic
value for $|a_{3D}| \rightarrow \infty$, $a_{1D}=1.0326a_\perp$ (indicated by a
horizontal arrow in the inset of Fig.~\ref{figSc1}). The renormalized 1D scattering
length $a_{1D}$ is positive for negative $a_{3D}$, and approaches $+\infty$ as
$a_{3D}\rightarrow -0$. Figure~\ref{figSc1} suggests that tuning the 3D scattering
length $a_{3D}$ to large values allows a universal quasi-one dimensional regime,
where $g_{1D}$ and $a_{1D}$ are independent of $a_{3D}$, to be entered.

The effective coupling constant $g_{1D}$, Eq.~\ref{eq_g1Dren}, has been derived
for a wave guide geometry, that is, with no axial confinement. However, it also
describes the scattering between two bosons confined to other quasi-one dimensional geometries.
Consider, {\it e.g.}, a Bose gas under harmonic confinement. If the confinement in the
axial direction is weak compared to that of the radial direction, the scattering
properties of each atom pair are expected to be described accurately by the
effective coupling constant $g_{1D}$ and the effective scattering length $a_{1D}$.

The internal motion of two bosons under highly-elongated confinement can be
described by the following 3D Hamiltonian
\begin{eqnarray}
\hat H^{int}_{3D} = -\frac{\hbar^2}{2 \mu}\Delta +
V_{int}(\vec{r}) +\frac{1}{2} \mu \left(\omega_\perp^2 \rho^2 + \omega_z^2\label{eq_h3Dn2}z^2 \right),
\end{eqnarray}
where $\omega_z$ denotes the angular frequency in the longitudinal direction,
$\omega_z= \lambda\,\omega_\perp$ ($\lambda$ denotes the aspect ratio, $\lambda \ll
1$). The eigenenergies $E_{3D}^{int}$ and eigenfunctions $\psi_{3D}^{int}$ of this
Hamiltonian satisfy the Schr\"odinger equation,
\begin{eqnarray}
\hat H^{int}_{3D} \psi_{3D}^{int}(\rho,z)=
E^{int}_{3D} \psi_{3D}^{int}(\rho,z).\label{eq_se3Dn2}
\end{eqnarray}

The corresponding 1D Hamiltonian reads
\begin{eqnarray}
H^{int}_{1D}= - \frac{\hbar^2}{2 \mu} \frac{d^2}{dz^2}
+ g_{1D} \delta(z) + \frac{1}{2}\mu\omega_z^2 z^2 + \hbar\omega_\perp
\label{eq1Dham}
\end{eqnarray}

The 1D eigenenergies $E^{int}_{1D}$ of the stationary Schr\"odinger equation,
\begin{eqnarray}
\hat H^{int}_{1D} \psi_{1D}^{int}(z)=E^{int}_{1D} \psi_{1D}^{int}(z),\label{eq_se1Dn2}
\end{eqnarray}
can be determined semi-analytically by solving the transcendental equation~\cite{Busch98},
\begin{eqnarray}
g_{1D} = 2 \sqrt{2}
\frac{\Gamma(\chi_z+1)}{\Gamma(\chi_z+1/2)} \;
\tg(\pi \chi_z) \;
\hbar \omega_z \, a_z, \label{eq_trans}
\end{eqnarray}
self consistently for $\chi_z$ (for a given $g_{1D}$). In the above equation,
$\chi_z$ is an effective (possibly non-integer) quantum number, which determines
the energy $E_z$,
\begin{eqnarray}
\chi_z=\frac{E_z}{2\hbar \omega_z}-\frac{1}{4}.\label{eq_nuz}
\end{eqnarray}

The energy $E_z$, in turn, determines the internal 1D eigenenergies $E_{1D}^{int}$,
\begin{eqnarray}
E_{1D}^{int} = \lambda E_z + \hbar \omega_\perp.\label{eq_e1Dint}
\end{eqnarray}

In Eq.~\ref{eq_trans}, $a_z$ denotes the characteristic oscillator length in the
axial direction, $a_z = \sqrt{\hbar/(m \omega_z)}$.

To compare the eigenenergies $E^{int}_{3D}$ and $E^{int}_{1D}$, we use, for the 3D
atom-atom interaction potential $V(r)$, a short-range (SR) modified
P\"oschl-Teller potential (\ref{Morse}) $V^{SR}(r)$ that can support two-body bound states,
\begin{eqnarray}
V^{SR}(r)=-\frac{V_0}{\ch^2(r/R)}
\label{eq_vsr}
\end{eqnarray}

\begin{figure}[tbp]
\vspace*{-1.0in}
\centerline{\includegraphics*[width=0.57\columnwidth]{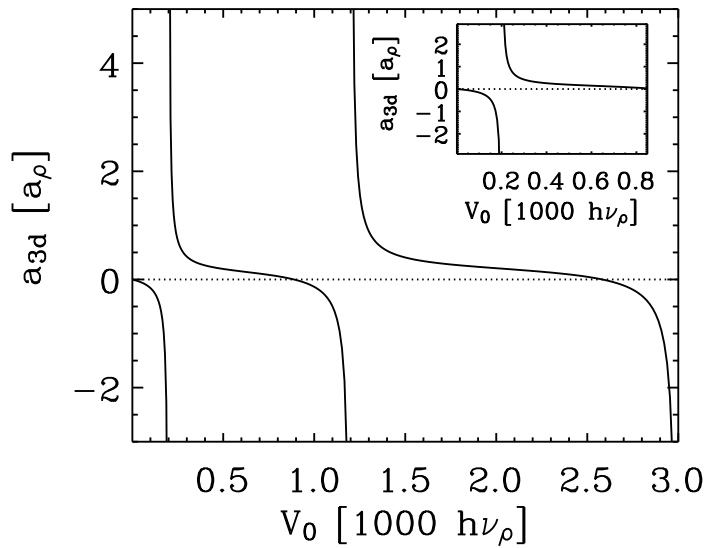}}
\vspace*{-.3in}
\caption{Three-dimensional $s$-wave scattering length $a_{3D}$ as a function
of the well depth $V_0$ for the short-range model potential $V^{SR}$,
Eq.~\protect\ref{eq_vsr}. Each time the 3D scattering length diverges a new
two-body $s$-wave bound state is created. Inset: Enlargement of the well depth
region used in our calculations.}
\label{figSc2}
\end{figure}

In the above equation, $V_0$ denotes the well depth, and $R$ the range of the
potential. In our calculations, $R$ is fixed at a value much smaller than the
transverse oscillator length, $R = 0.1 a_\perp$. To simulate the behavior of
$a_{3D}$ near a field-dependent Feshbach resonance, we vary the well depth $V_0$,
and consequently, the scattering length $a_{3D}$. It has been shown that such a
model describes many atom-atom scattering properties near a Feshbach resonance
properly~\cite{Tiesinga00}. Figure~\ref{figSc2} shows the dependence of the 3D
scattering length $a_{3D}$ on $V_0$. Importantly, $a_{3D}$ diverges for particular
values of the well depth $V_0$. At each of these divergencies, a new two-body
$s$-wave bound state is created. The inset of Fig.~\ref{figSc2} shows the range of
well depths $V_0$ used in our calculations.

\begin{figure}[tbp]
\vspace*{-.2in}
\centerline{\includegraphics*[width=0.8\columnwidth]{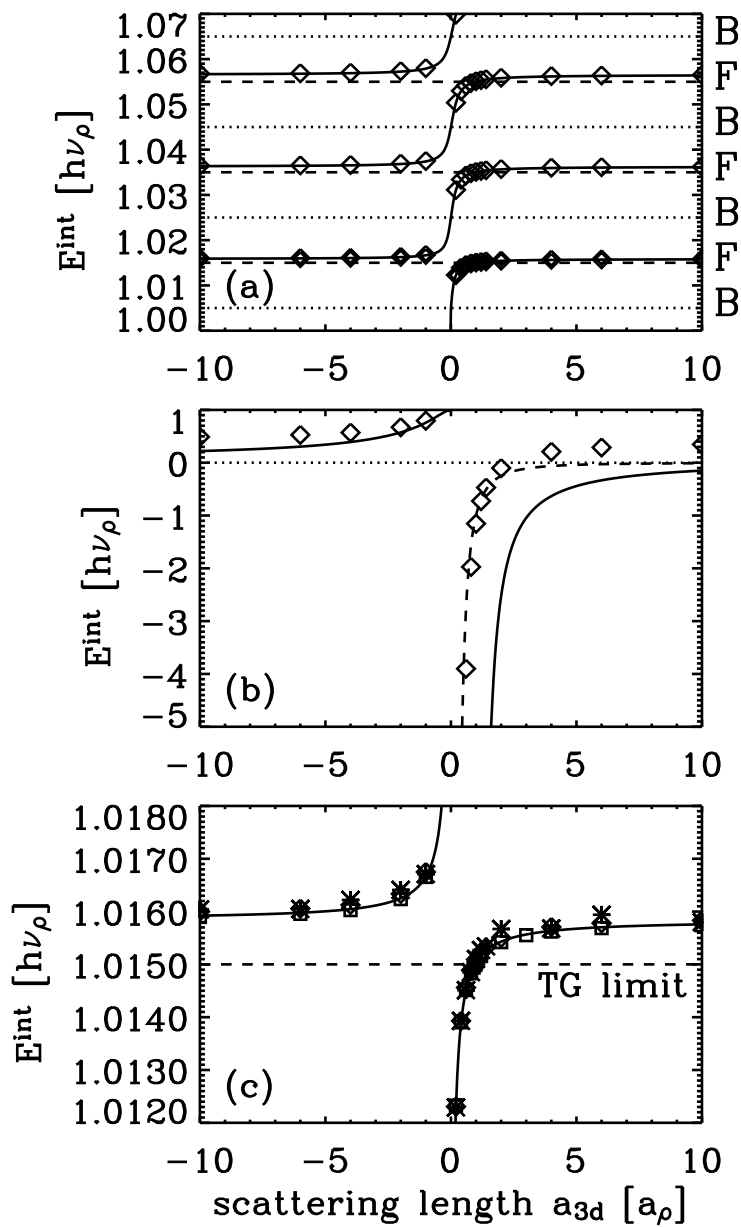}}
\caption{Internal eigenenergies $E^{int}$ as a function of the 3D scattering
length $a_{3D}/a_\perp$ for two bosons under highly-elongated confinement with
$\lambda=0.01$. (a) 3D $s$-wave eigenenergies $E^{int}_{3D}$ (diamonds) of gas-like
states obtained using the short-range model potential $V^{SR}$,
Eq.~\protect\ref{eq_vsr}, in a B-spline basis set calculation together with
internal 1D eigenenergies $E^{int}_{1D}$ (solid lines). Excellent agreement between
the 3D and 1D energies is found. Horizontal dotted lines show the lowest internal
eigenenergies for two non-interacting spin-polarized bosons, while horizontal
dashed lines show those for two non-interacting spin-polarized fermions (indicated
respectively by ``B'' and ``F'' on the right hand side). (b) Binding energy of
molecular-like bound states. In addition to the 3D and 1D energies [diamonds and
solid lines, respectively; see (a)], a dashed line shows the 3D binding energy
$-\hbar^2/(m a_{3D}^2)$. (c) Enlargement of the lowest-lying gas-like state. In
addition to the 3D and 1D energies shown in (a), asterisks show the 3D energies for
the interaction potential $V^{SR}$ calculated using the FN-MC technique, and
squares the 1D energies for the contact interaction potential calculated using the
FN-MC technique. The statistical uncertainty of the FN-MC energies is
smaller than the symbol size. Good agreement between the FN-MC energies
(asterisks and squares) and the energies calculated by alternative means (diamonds
and solid lines) is found.}
\label{figSc3}
\end{figure}

To benchmark our MC calculations (see Chapter~\ref{secQMC} and Sec.~\ref{secScIV}),
we solve the 3D Schr\"odinger equation, Eq.~\ref{eq_se3Dn2}, with $\lambda=0.01$
for various well depths $V_0$ using a two-dimensional B-spline basis in $\rho$ and
$z$. Figure~\ref{figSc3} shows the resulting 3D eigenenergies $E^{int}_{3D}$
(diamonds) as a function of the 3D scattering length $a_{3D}$. We distinguish
between two sets of states:

1) States with $E^{int}_{3D} \ge \hbar \omega_\perp$ are referred to as gas-like
states; their behavior is, to a good approximation, characterized by the 3D
scattering length $a_{3D}$, and is hence independent of the detailed shape of the
atom-atom potential. The energies of the gas-like states are shown in
Fig.~\ref{figSc3}(a).

2) States with $E^{int}_{3D}<\hbar \omega_\perp$ are referred to as molecular-like
bound states; their behavior depends on the detailed shape of the atom-atom
potential. The energies of these bound states are shown in Fig.~\ref{figSc3}(b). The
well depth $V_0$ of the short-range interaction potential $V^{SR}$ is chosen such
that $V^{SR}$ supports --- in the absence of the confining potential --- no
$s$-wave bound state for $a_{3D}<0$, and one $s$-wave bound state for $a_{3D}>0$.
Figure~\ref{figSc3}(b) shows that the bound state remains bound for $|a_{3D} |
\rightarrow \infty$ and for negative $a_{3D}$ if tight radial confinement is
present. In addition, a dashed line shows the 3D binding energy, $-\hbar^2/(m
a_{3D}^2)$, which accurately describes the highest-lying molecular bound state in
the absence of any external confinement if $a_{3D}$ is much larger than the range
$R$ of the potential $V^{SR}$.

The B-spline basis calculations yield not only the internal 3D eigenenergies
$E_{3D}^{int}$, but also the corresponding wave functions $\psi_{3D}^{int}$. The
nodal surface of the lowest-lying gas-like state, which is to a good approximation
an ellipse in the $\rho z$-plane, is a crucial ingredient of our many-body
calculations. Section~\ref{secScIV} discuss in detail how this nodal surface is used
to parametrize our trial wave function entering the MC calculations.

To compare the energy spectrum for $N=2$ of the effective 1D Hamiltonian with that
of the 3D Hamiltonian, Fig.~\ref{figSc3} additionally shows the 1D eigenenergies
$E_{1D}^{int}$ (solid lines) obtained by solving the Schr\"odinger equation for
$H_{1D}^{int}$, Eq.~\ref{eq1Dham}, semi-analytically [using the renormalized
coupling constant $g_{1D}$, Eq.~\ref{eq_g1Dren}]. Figure~\ref{figSc3}(a)
demonstrates excellent agreement between the 3D and the 1D internal energies for
all states with gas-like character. For positive $a_{3D}$, the effective 1D
Hamiltonian fails to reproduce the energy spectrum of the molecular-like bound
states of the 3D Hamiltonian accurately [see Fig.~\ref{figSc3}(b), and also
\cite{Bergeman03,Bolda03}].

Our main focus is in the lowest-lying energy level with gas-like character. This
energy branch is shown in Fig.~\ref{figSc3}(c) on an enlarged scale. A horizontal
dashed line shows the lowest internal 3D eigenenergy for two non-interacting
spin-polarized fermions (where the anti-symmetry of the wave function enters in the
$z$ coordinate). Our numerical calculations confirm~\cite{Bergeman03} that for
$a_{3D}=a_{3D}^c$ ($g_{1D} \rightarrow \infty$) the two boson system behaves as if
it consisted of two non-interacting spin-polarized fermions (TG gas). The energy
$E^{int}_{3D}$ is larger than that of two non-interacting fermions for
$a_{3D}>a_{3D}^c$, and approaches the first excited state energy of two
non-interacting bosons for $a_{3D} \rightarrow -0$ [indicated by a dotted line in
Fig.~\ref{figSc3}(a)].

For positive $g_{1D}$, the 1D Schr\"odinger equation, Eq.~\ref{eq_se1Dn2}, does
not support molecular-like bound states. Consequently, the wave function of the
lowest-lying gas-like state is positive definite everywhere. For negative $g_{1D}$,
however, one molecular-like two-body bound state exists. If $a_{1D}\ll a_z$ the
bound-state wave function is approximately given by the eigenstate
$\psi_{1D}^{int}$ of the 1D Hamiltonian without confinement,
Eq.~\ref{eq_wgint1D},
\begin{eqnarray}
\psi_{1D}^{int}(z)=\exp\left(-\frac{|z|}{a_{1D}}\right),
\end{eqnarray}
with eigenenergy $E_{1D}^{int}$,
\begin{eqnarray}
E_{1D}^{int}= -\frac{\hbar^2}{ma_{1D}^2} + \hbar \omega_\perp.
\label{eq_e1Dbinding}
\end{eqnarray}

For the highly-elongated trap with $\lambda=0.01$ shown in Fig.~\ref{figSc3}(b) and
positive $a_{1D}$ the above binding energy nearly coincides with the exact
eigenenergy of the molecular-like bound state obtained from the solution of the
transcendental equation~(\ref{eq_trans}) (solid line). The two-body binding energy,
Eq.~\ref{eq_e1Dbinding}, is largest for $a_{1D} \rightarrow +0$ ($g_{1D}
\rightarrow -\infty$); in this case, the molecular-like bound state wave function
is tightly localized around $z=0$, where $z=z_2-z_1$. Consider a system with
$a_{1D} \ll a_z$. For negative $g_{1D}$ (positive $a_{1D}$), the nodes along the
relative coordinate $z$ of the lowest-lying gas-like wave function (in this case,
the first excited state) are then approximately given by $\pm a_{1D}$. Thus,
imposing the boundary condition $\psi_{1D}^{int}=0$ at $|z|= a_{1D}$ and
restricting the configuration space to $z > a_{1D}$ allows one to obtain an
approximation to the eigenenergy of the first excited eigen state. Furthermore,
imposing the boundary condition $\psi_{1D}^{int}=0$ at $z=a_{1D}$ is identical to
solving the 1D Schr\"odinger equation for a hard-rod interaction potential
$V^{HR}(z)$ (\ref{HR}),
\begin{eqnarray}
V^{HR}(z) =
\left\{ \begin{array}{cll} \infty           & \mbox{ for } & z<a_{1D} \\
                                0            & \mbox{ for } & z \ge a_{1D}\;.
\end{array}\right.
\end{eqnarray}

For $N=2$, asterisks in Fig.~\ref{figSc3}(c) show the fixed-node diffusion Monte
Carlo (FN-MC) results obtained using the above fictitious hard-rod potential (see
Sec.~\ref{secScIVsubII}). Good agreement is found with the exact 1D eigenenergies
obtained from Eqs.~\ref{eq_trans}-\ref{eq_nuz}. For $N>2$ bosons, our 1D FN-MC
algorithm and our usage of the hard-rod equation of state both take advantage of a
reduction of configuration space similar to that discussed here for two bosons (see
Sec.~\ref{secScIII} and Chapter~\ref{secQMC}).

\section{$N$ bosons under quasi-one-dimensional confinement\label{secScIII}}

For tightly-confined trapped gases the 1D regime is reached if the transverse
motion of the atoms is frozen, with all the particles occupying the ground state of
the transverse harmonic oscillator. At zero temperature, this condition requires
that the energy per particle is dominated by the trapping potential,
$E/N=\hbar\omega_\perp+\epsilon$, where the excess energy $\epsilon$ is much
smaller than the separation between levels in the transverse direction,
$\epsilon\ll\hbar\omega_\perp$. In the following we consider situations where the
Bose gas is in the 1D regime for any value of the 3D scattering length $a_{3D}$.
For a fixed trap anisotropy parameter $\lambda$ and a fixed number of particles $N$
the above requirement is satisfied if $N\lambda\ll 1$. For $\lambda=0.01$ and
$N=10$ (as considered in Sec.~\ref{secScIVsubI}) this condition is fulfilled.

To compare the 3D and 1D energetics of a Bose gas, we consider the 3D and 1D
Hamiltonian describing $N$ spin-polarized bosons,
\begin{eqnarray}
\hat H_{3D}= \sum_{i=1}^N \left[-\frac{\hbar^2}{2m}\Delta_i
+ \frac{1}{2}m \left( \omega_\perp^2 \rho_i^2
+ \omega_z^2 z_i^2 \right) \right] + \sum_{i<j}^N V(r_{ij}),
\label{eq_h3D}
\end{eqnarray}
and
\begin{eqnarray}
\hat H_{1D}=\sum_{i=1}^N \left(-\frac{\hbar^2}{2m}
\frac{\partial^2}{\partial z_i^2} + \frac{1}{2} m \omega_z^2 z_i^2 \right)
+ g_{1D}\sum_{i<j}^N \delta(z_{ij}) + N \hbar\omega_\perp,
\label{eq_h1D}
\end{eqnarray}
respectively. The corresponding eigenenergies and eigenfunctions are given by
solving the Schr\"odinger equations,
\begin{eqnarray}
\hat H_{3D} \psi_{3D}(\vec{r}_1,...,\vec{r}_N)=E_{3D}\psi_{3D}(\vec{r}_1,...,\vec{r}_N)
\label{eq_se3D}
\end{eqnarray}
and
\begin{eqnarray}
\hat H_{1D} \psi_{1D}(z_1,...,z_N)=E_{1D} \psi_{1D}(z_1,...,z_N),
\label{eq_se1D}
\end{eqnarray}
respectively. In contrast to Sec.~\ref{secScII}, here we do not separate out the
center of mass motion since the MC calculations used to solve the 3D and 1D
many-body Schr\"odinger equations can be most conveniently implemented in Cartesian
coordinate space (see Sec.~\ref{secGreen}). In the following, we refer to
eigenstates of the confined Bose gas with energy greater than $N\hbar\omega_\perp$
as gas-like states, and to those with energy smaller than $N\hbar\omega_\perp$ as
cluster-like bound states.

Section~\ref{secScIVsubI} compares the energetics of the lowest-lying gas-like state
of the 3D Schr\"odinger equation, Eq.~\ref{eq_se3D}, obtained using the
short-range potential V$^{SR}$, Eq.~\ref{eq_vsr}, with that obtained using the
hard-sphere potential $V^{HS}$ Eq.~\ref{HS}. For $V^{HS}$, the $s$-wave scattering
length $a_{3D}$ coincides with the range of the potential (see Sec.~\ref{secHS}).
For $V^{SR}$, in contrast, $R$ determines the range of the potential, while the
scattering length $a_{3D}$ is determined by $R$ and $V_0$.
For $a_{3D}\ll a_\perp$, both potentials give nearly identical results for the
energetics of the lowest-lying gas-like state, which depend to a good approximation
only on the value of $a_{3D}$. For $a_{3D}\gtrsim a_\perp$, instead, deviations due
to the different effective ranges become visible and only $V^{SR}$ yields results,
which do not depend on the short-range details of the potential and which are
compatible with a 1D contact potential.

Section~\ref{secScIVsubI} also discusses the energetics of the 1D Hamiltonian,
Eq.~\ref{eq_h1D}. For small $|g_{1D}|$, the energetics of the many-body 1D
Hamiltonian are described well by a 1D mean-field equation with non-linearity. For
negative $g_{1D}$, the mean-field framework describes, for example, bright solitons
\cite{Carr00,Kanamoto03}, which have been observed experimentally \cite{Strecker02,Khaykovich02}. For
large $|g_{1D}|$, in contrast, the system is highly-correlated, and any mean-field
treatment will fail. Instead, a many-body description that incorporates higher
order correlations has to be used. In particular, the limit $|g_{1D}| \rightarrow
\infty$ corresponds to the strongly-interacting TG regime.

For infinitely strong particle interactions ($|g_{1D}| \rightarrow \infty$),
Girardeau shows~\cite{Girardeau60}, using the equivalence between the 1D
$\delta$-function potential and a ``1D hard-point potential'', that the energy
spectrum of the 1D Bose gas coincides with that of $N$ non-interacting
spin-polarized fermions. The lowest eigenenergy per particle of the trapped 1D Bose
gas, Eq.~\ref{eq_se1D}, is, in the TG limit, given by\footnote{See, also,
footnote on p.~\pageref{CalogeroRef}}
\begin{eqnarray}
\frac{E^{TG}_{1D}}{N}=\left(\frac{\lambda N}{2} + 1\right) \hbar \omega_\perp
\label{eq_etg}
\end{eqnarray}

The corresponding gas density is given by the sum of squares of single-particle
wave functions
\begin{eqnarray}
n^{TG}_{1D}(z)= \frac{1}{\sqrt{\pi} a_z} \sum_{k=0}^{N-1} \frac{1}{2^k k!}
H_k^2(z/a_z) \exp \left\{-(z/a_z)^2\right\},
\label{eq_nf}
\end{eqnarray}
with the normalization $\int_{-\infty}^\infty n^{TG}_{1D}(z)\,dz = N$. In
Eq.~\ref{eq_nf}, the $H_k$ denote Hermite polynomials, and $z$ denotes the
distance measured from the center of the trap. For large numbers of atoms, the
density expression in Eq.~\ref{eq_nf} can be calculated using the
LDA~\cite{Dunjko01},
\begin{eqnarray}
n_{1D}^{TG}(z)= \frac{\sqrt{2N}}{ \pi a_{z}}
\left(1-\frac{z^2}{2Na_z^2} \right)^{1/2} \, .\label{eq_ntg}
\end{eqnarray}

The above result cannot reproduce the oscillatory behavior of the exact density,
Eq.~\ref{eq_nf}, but it does describe the overall behavior properly (see
Sec.~\ref{secScV}).

To characterize the {\it inhomogeneous} 1D Bose gas further, we consider the
many-body Hamiltonian of the {\it homogeneous} 1D Bose gas,
\begin{eqnarray}
H_{1D}^{hom}=\sum_{i=1}^N-\frac{\hbar^2}{2m}\frac{\partial^2}{\partial z_i^2}
+ g_{1D}\sum_{i<j}^N \delta(z_{ij}) + N \hbar \omega_\perp
\label{eq_h1Dwg}
\end{eqnarray}

By introducing the energy shift $N \hbar \omega_\perp$, our classification of
gas-like states and cluster-like bound states introduced after Eq.~\ref{eq_se1D}
remains valid. For positive $g_{1D}$, $H_{1D}^{hom}$ corresponds to the
Lieb-Liniger (LL) Hamiltonian. The gas-like states of the LL Hamiltonian, including
its thermodynamic properties, have been studied in detail \cite{Lieb63,Lieb63b,Yang69}. The energy
per particle of the lowest-lying gas-like state, the ground state of the system, is
given by
%
\begin{eqnarray}
\frac{E_{1D}^{LL}(n_{1D})}{N}=\frac{\hbar^2}{2m} e(\gamma) n_{1D}^2,
\label{eq_ell}
\end{eqnarray}
where $n_{1D}$ denotes the density of the homogeneous system, and $e(\gamma)$ a
function of the dimensionless parameter $\gamma= 2/(n_{1D} |a_{1D}|)$.

We use the known properties of the LL Hamiltonian to determine properties of the
corresponding inhomogeneous system, Eq.~\ref{eq_h1D}, within the LDA. This
approximation provides a correct description of the trapped gas if the size of the
atomic cloud is much larger than the characteristic length scale $a_z$ of the
confinement in the longitudinal direction~\cite{Dunjko01}. Specifically, consider
the local equilibrium condition,
\begin{eqnarray}
\mu(N)= \hbar \omega_\perp + \mu_{local}[n_{1D}(z)]+\label{eq_mu}
\frac{1}{2}m \omega_z^2 z^2,
\end{eqnarray}
where $\mu_{local}(n_{1D})$ denotes the chemical potential of the homogeneous system
with density $n_{1D}$,
\begin{eqnarray}
\mu_{local}(n_{1D}) = \frac{\partial\left[
n_{1D} E_{1D}^{LL}(n_{1D})/N
\right]}{\partial n_{1D}}. \label{eq_mulocal}
\end{eqnarray}

The chemical potential $\mu(N)$, Eq.~\ref{eq_mu}, can be calculated using
Eq.~\ref{eq_mulocal} together with the normalization of the density,
$\int_{-\infty}^\infty n_{1D}(z)\,dz = N$. Integrating the chemical potential
$\mu(N)$ then determines the energy of the lowest-lying gas-like state of the
inhomogeneous $N$-particle system within the LDA. The LDA treatment is
computationally less demanding than solving the many-body Schr\"odinger equation,
Eq.~\ref{eq_se1D}, using MC techniques. By comparing with our full 1D many-body
results we establish the accuracy of the LDA (see Sec.~\ref{secScIVsubI}).

For negative $g_{1D}$, the Hamiltonian given in Eq.~\ref{eq_h1Dwg} supports
cluster-like bound states. The ground state energy and eigenfunction of the system
are~\cite{McGuire64}
\begin{eqnarray}
\frac{E_{1D}^{hom}}{N} =-\frac{\hbar^2}{6m a_{1D}^2} (N^2-1)+\hbar\omega_\perp,
\label{eq_e1Dwg}
\end{eqnarray}
and
\begin{eqnarray}
\psi_{1D}^{hom}(z_1,...,z_N)=
\prod_{i<j}^N\exp\left\{\frac{-|z_i-z_j|}{a_{1D}}\right\},
\label{eq_psi1Dwg}
\end{eqnarray}
respectively. The eigenstate given by Eq.~\ref{eq_psi1Dwg} depends only on the
relative coordinates $z_{ij}$, that is, it is independent of the center of mass of
the system. Adding a confinement potential [see Eq.~\ref{eq_h1D}] with $\omega_z$
such that $a_z \gg a_{1D}$ leaves the eigenenergy $E_{1D}^{hom}$ of this
cluster-like bound state to a good approximation unchanged, while the corresponding
wave function becomes localized at the center of the trap. This state describes a
bright soliton, whose energy can also be determined within a mean-field
framework~\cite{Kanamoto03}. An excited state of the many-body 1D Hamiltonian with
confinement corresponds, {\it e.g.}, to a state, where $N-1$ particles form a
cluster-like bound state, {\it i.e.}, a soliton with $N-1$ particles, and where one
particle approximately occupies the lowest harmonic oscillator state, {\it i.e.}, has
gas-like character. Similarly, molecular-like bound states can form with fewer
particles.

The above discussion implies that the lowest-lying gas-like state of the 1D
Hamiltonian with confinement, Eq.~\ref{eq_h1D} with negative $g_{1D}$,
corresponds to a highly-excited state. For dilute 1D systems with negative
$g_{1D}$, the nodal surface of this excited state can be well approximated by the
following nodal surface: $\psi_{1D}=0$ for $z_{ij}=a_{1D}$, where
$i,j=1,...,N$ and $i<j$. As in the two-body case, the many-body energy can then be calculated
approximately by restricting the configuration space to regions where the wave
function is positive. This corresponds to treating a gas of hard-rods of size
$a_{1D}$. In the low density limit, we expect that the lowest-lying gas-like state
of the 1D many-body Hamiltonian with $g_{1D}<0$ is well described by a system of
hard-rods of size $a_{1D}$.

In addition to treating the full 1D many-body Hamiltonian, we treat the
inhomogeneous system with negative $g_{1D}$ within the LDA. The equation of state
of the {\em{uniform}} hard-rod gas with density $n_{1D}$ is given
by (\ref{EHR}) \cite{Girardeau60}:
\begin{eqnarray}
\frac{E_{1D}^{HR}(n_{1D})}{N}=   \frac{\pi^2 \hbar^2 n_{1D}^2}
{6m \, (1-n_{1D}a_{1D})^2}  +  \hbar \omega_\perp .
\label{eq_e1Dhr}
\end{eqnarray}

We use this energy in the LDA treatment (see Eqs.~\ref{eq_ell} through
\ref{eq_mulocal} with $E_{1D}^{LL}$ replaced by $E_{1D}^{HR}$). The hard-rod
equation of state treated within the LDA provides a good description when $g_{1D}$
is negative, but $|g_{1D}|$ not too small (see Secs.~\ref{secScIVsubI} and
\ref{secScV}). To gain more insight, we determine the expansion for inhomogeneous
systems with $N \lambda \ll 1$ using the equation of state for the homogeneous
hard-rod gas,
\begin{eqnarray}
\frac{E_{1D}}{N} - \hbar\omega_\perp =
\hbar\omega_\perp \frac{N\lambda}{2}\left( 1 + \frac{128\sqrt{2}}{45\pi^2}
\sqrt{N\lambda}\frac{a_{1D}}{a_\perp} + ... \right) \;.
\label{eq_e1Dexpansion}
\end{eqnarray}

The first term corresponds to the energy per particle in the TG regime (see
Eq.~\ref{eq_etg}), the other terms can be considered as small corrections to the TG
energy. In the unitary limit, that is, for $a_{1D}/a_\perp=1.0326$, expression
(\ref{eq_e1Dexpansion}) becomes independent of $a_{3D}$, and depends only on
$N\lambda$. Similarly, the linear density in the center of the cloud, $z=0$, is to
lowest order given by the TG result, $n_{1D}^{TG}(0)=\sqrt{2N\lambda}/ (\pi
a_\perp)$ (see Eq.~\ref{eq_ntg}). Section~\ref{secScV} shows that the TG density
provides a good description of inhomogeneous 1D Bose gases over a fairly large range
of negative $g_{1D}$.

\section{Energetics of quasi-one-dimensional Bose gases\label{secScIV}}

Table~\ref{tabSc1} summarizes the techniques used to solve the 1D and 3D Hamiltonian,
respectively. This table is meant to guide the reader through our result sections.
Section~\ref{secScIVsubII} discusses our MC energies for two-particle systems,
while Sec.~\ref{secScIVsubI} discusses the energetics for larger systems,
calculated within various frameworks. Finally, Sec.~\ref{secScV} discusses the
stability of quasi-one dimensional Bose gases.

\begin{table}[ht!]
\center{
\begin{tabular}{|c|l|l|l|}
\hline
Hamiltonian & interaction & ~~~~technique & Section \\ \hline
\hline
$\hat H_{3D}$ & $V^{HS}$   & DMC     & \ref{secScIVsubII}\\
$\hat H_{3D}$ & $V^{SR}$   & FN-MC   & \ref{secScIVsubI}, \ref{secScIVsubII} \\ \hline
$\hat H_{1D}$ & $g_{1D}>0$ & DMC     & \ref{secScIVsubI}, \ref{secScIVsubII} \\
$\hat H_{1D}$ & $g_{1D}>0$ & LDA, LL & \ref{secScIVsubII}  \\
$\hat H_{1D}$ & $g_{1D}<0$ & FN-MC   & \ref{secScIVsubI}, \ref{secScIVsubII} \\
$\hat H_{1D}$ & $g_{1D}>0$ & LDA, hard-rod & \ref{secScIVsubII} \\
$\hat H_{1D}$ & $g_{1D}<0$ & VMC     & \ref{secScV} \\
\hline
\end{tabular}
}
\caption{Guide that summarizes the techniques used to solve the 3D and 1D Hamiltonian,
Eqs.~\protect\ref{eq_h3D} and \protect\ref{eq_h1D}, respectively. Column 2
specifies the atom-atom interactions of the many-body Hamiltonian, column 3 lists
the techniques used to solve the corresponding many-body Schr\"odinger equation,
and column 4 lists the sections that discuss the results obtained using this
approach.}
\label{tabSc1}
\end{table}

\subsection{Two-body system\label{secScIVsubI}}

Section~\ref{secScII} discusses the calculation of the energy spectrum related to
the internal motion of two bosons under highly-elongated confinement,
Eq.~\ref{eq_h3Dn2}, using a B-spline basis, and the eigenspectrum related to the
internal motion of two bosons under 1D confinement, Eq.~\ref{eq1Dham}, using
Eqs.~\ref{eq_trans} through (\ref{eq_e1Dint}). We now use these essentially exact
eigenenergies to benchmark our FN-MC calculations. Toward this end, we solve the 3D
Schr\"odinger equation, Eq.~\ref{eq_h3D}, and the 1D Schr\"odinger equation,
Eq.~\ref{eq_h1D}, for various interaction strengths for $N=2$ and $\lambda=0.01$
using FN-MC techniques. The resulting MC energies $E_{3D}$ and $E_{1D}$ include the
center of mass energy of $(1 + \lambda/2)\hbar \omega_\perp$. To compare with the
internal eigenenergies discussed in Sec.~\ref{secScII}, we subtract the center of
mass energy from the FN-MC energies.

\begin{figure}[tbp]
\vspace*{-1.0in}
\centerline{\includegraphics*[width=0.57\columnwidth]{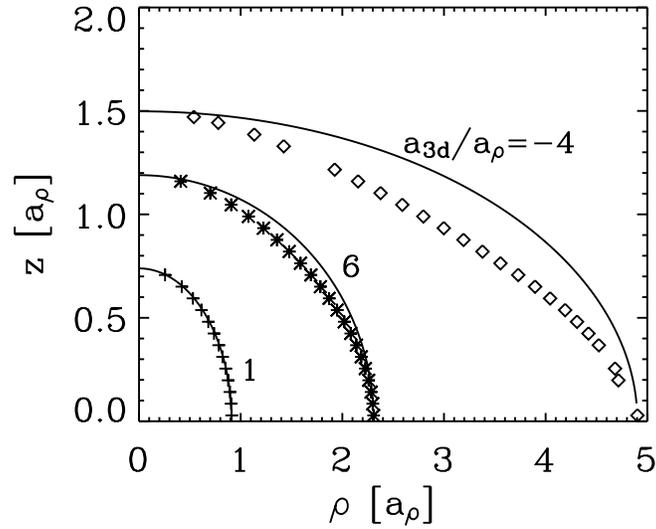}}
\vspace*{-.3in}
\caption{
Nodal surface of the trial wave function $\psi_T$ (solid lines,
Eq.~\protect\ref{VMC3}) together with the essentially exact nodal surface
calculated using a B-spline basis set (symbols) for $\lambda=0.01$, $N=2$, and
three different scattering lengths, $a_{3D}/a_\perp=1$ (pluses),
$a_{3D}/a_\perp=6$ (asterisks), and $a_{3D}/a_\perp=-4$ (diamonds). The nodal
surface is shown as a function of the internal coordinates $z$ and $\rho$. Good
agreement between the elliptical nodal surface (solid lines) and the essentially
exact nodal surface (symbols) is visible for $a_{3D}/a_\perp=1$ and $6$. Small
deviations are visible for $a_{3D}/a_\perp=-4$. }
\label{figSc5}
\end{figure}

For $N=2$, the lowest-lying gas-like state of the 3D Hamiltonian for the
short-range potential $V^{SR}$ is the first excited eigenstate. Consequently, we
solve the 3D Schr\"odinger equation by the FN-MC technique using the trial wave
function $\psi_T$ given by Eqs.~\ref{Jastrow} and \ref{VMC3}. Figure~\ref{figSc5}
shows the elliptical nodal surface of the trial wave function $\psi_T$ (solid
lines) together with the essentially exact nodal surface calculated using a
B-spline basis set (symbols; see also Sec.~\ref{secScII}) for $\lambda=0.01$ and
three different scattering lengths, $a_{3D}/a_\perp=1,6$, and $-4$. Notably, the
semi-axes $a$ along the $\rho$ coordinate is larger than that along the $z$
coordinate ($a/b >1$), ``opposing'' the shape of the confining potential, for which
the characteristic length along the $\rho$ coordinate is smaller than that along
the $z$ coordinate ($a_\perp/a_z<1$). Figure~\ref{figSc5} indicates good agreement
between the essentially exact nodal surface and the parameterization of the nodal
surface by an ellipse for $a_{3D}/a_\perp=1$ and $6$; some discrepancies become
apparent for negative $a_{3D}$. Since the FN-MC method results in the exact energy
if the nodal surface of $\psi_T$ coincides with the nodal surface of the exact
eigenfunction, comparing the FN-MC energies for two particles with those obtained
from a B-spline basis set calculation provides a direct measure of the quality of
the nodal surface of $\psi_T$. Figure~\ref{figSc3}(c) compares the internal 3D
energy of the lowest-lying gas-like state calculated using a B-spline basis
(diamonds, see Sec.~\ref{secScII}) with that calculated using the FN-MC technique
(asterisks). The agreement between these two sets of energies is --- within the
statistical uncertainty --- excellent for all scattering lengths $a_{3D}$
considered. We conclude that our parameterization of the two-body nodal surface,
Eq.~\ref{VMC3}, is accurate over the whole range of interaction strengths
$a_{3D}$ considered.

\begin{figure}[tbp]
\vspace*{-1.0in}
\centerline{\includegraphics*[width=0.55\columnwidth]{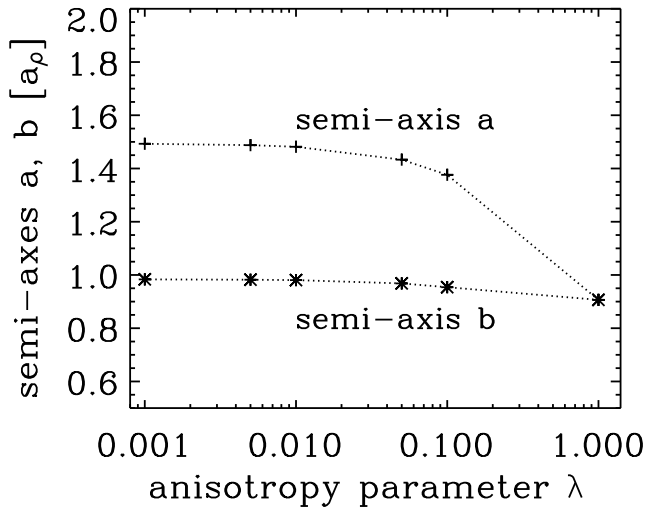}}
\vspace*{-.3in}
\caption{
Semi-axes $a$ (pluses) and $b$ (asterisks) obtained by fitting an ellipse (see
Eq.~\protect\ref{VMC3}) to the essentially exact nodal surface for two bosons
under cylindrical confinement, calculated using a B-spline basis set as a function
of the anisotropy parameter $\lambda$, for $a_{3D}/a_\perp=2$. Dotted lines are
shown to guide the eye. For $\lambda \le 0.01$, the nodal surface is nearly
independent of the anisotropy parameter $\lambda$.}
\label{figSc6}
\end{figure}

We expect that our parameterization of the two-body nodal surface is to a good
approximation independent of the confining potential in $z$ (for small enough
$\lambda$). In fact, we expect our nodal surface to closely resemble that of the
scattering wave function at low scattering energy of the 3D wave guide Hamiltonian
given by Eq.~\ref{eq_wgint3D}. To quantify this statement, Fig.~\ref{figSc6}
shows the semi-axes $a$ and $b$ (pluses and asterisks, respectively) obtained by
fitting an ellipse, see Eq.~\ref{VMC3}, to the nodal surface obtained by solving
the Schr\"odinger equation for the two-body Hamiltonian, Eq.~\ref{eq_h3Dn2},
using a B-spline basis for various aspect ratios $\lambda=0.001,...,1$, and fixed
scattering length, $a_{3D}=2a_\perp$ (similar results are found for other
scattering lengths). Indeed, the nodal surface for a given $a_{3D}/a_\perp$ is
nearly independent of the aspect ratio $\lambda$ for $\lambda \le 0.01$. These
findings for two particles imply that the parameterization of the nodal surface of
$\psi_T$ used in the FN-MC many-body calculations should be good as long as the
density along $z$ is small.

Next, consider the 1D Hamiltonian, Eq.~\ref{eq_h1D}, for $N=2$. For positive
$g_{1D}$, the lowest-lying gas-like state is the ground state of the two-body
system and we hence use the DMC technique with $\psi_T$ given by Eqs.~\ref{VMC4}
and \ref{VMC5}; for $g_{1D}<0$, the lowest-lying gas-like state is the first
excited state, and we instead use the FN-MC technique with $\psi_T$ given by
Eqs.~\ref{VMC4} and \ref{VMC6}. Figure~\ref{figSc3} shows the 1D energies of
the lowest-lying gas-like state calculated using Eqs.~\ref{eq_trans} through
\ref{eq_e1Dint} (solid line), together with those calculated by the FN-MC
technique (squares). We find excellent agreement between these two sets of 1D
energies.

The comparison for two bosons between the FN-MC energies and the energies
calculated by alternative means serves as a stringent test of our MC codes, since
these codes are implemented such that the number of particles enters simply as a
parameter.

\subsection{N-body system\label{secScIVsubII}}

This section presents our many-body study, which investigates the properties of
quasi-one dimensional Bose gases over a wide range of scattering lengths $a_{3D}$.
We focus specifically on three distinct regimes: 1) $0<a_{3D}<a_{3D}^c$ ($g_{1D}$
is positive), 2) $|a_{3D}| \rightarrow \infty$ ($g_{1D}$ and $a_{1D}$ are
independent of $a_{3D}$; unitary regime); and iii) $a_{3D} \rightarrow -0$
($a_{1D}$ is large and positive; onset of instability). We discuss the energetics
of quasi-one dimensional Bose gases for $N=10$. Our results presented here support
our earlier conclusions, which are based on a study conducted for a smaller system,
{\it i.e.}, for $N=5$~\cite{Astrakharchik04a}.

For small $\lambda$ ($\lambda=0.01$), the radial angular frequency $\omega_\perp$
dominates the eigenenergies of the 3D and of the 1D Schr\"odinger equation. The
shift of the eigenenergy of the lowest-lying gas-like state as a function of the
interaction strength is, however, set by the axial angular frequency $\omega_z$. To
emphasize the dependence of the eigenenergies of the lowest-lying gas-like state on
$\omega_z$, we report the energy per particle subtracting the constant offset
$\hbar \omega_\perp$, that is, we report the quantity $E/N-\hbar \omega_\perp$.

\begin{figure}[tbp]
\vspace*{-.2in}
\centerline{\includegraphics*[width=0.7\columnwidth]{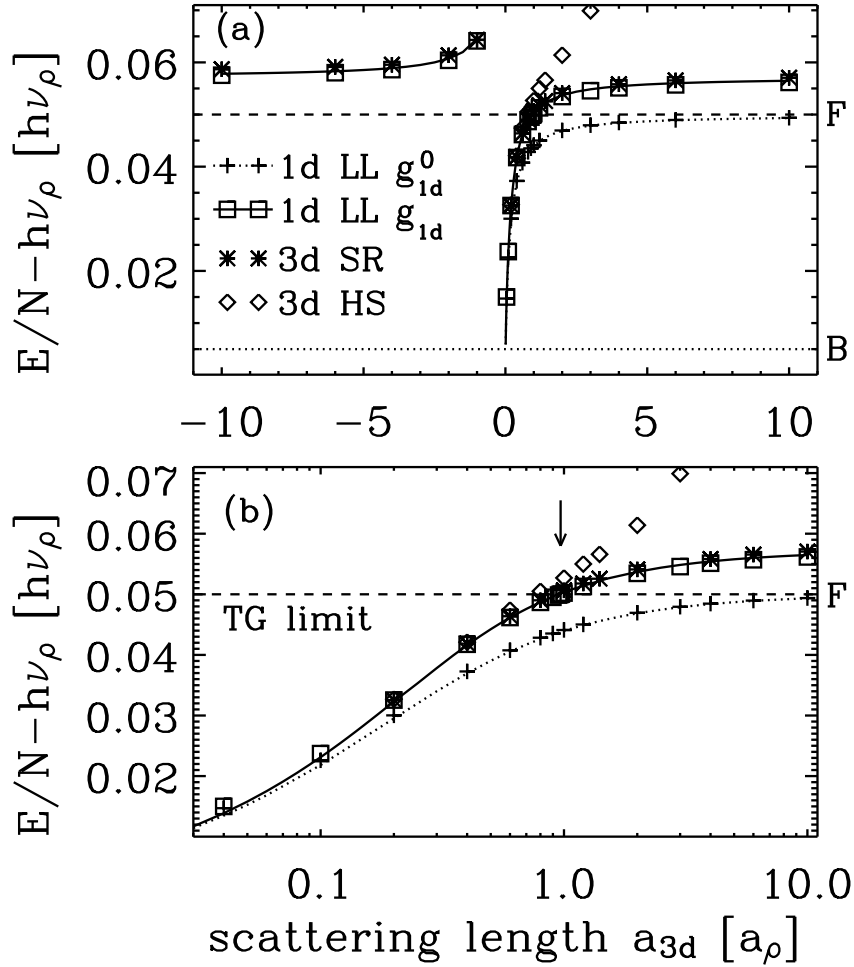}}
\vspace*{-.3in}
\caption{Three-dimensional FN-MC energy per particle, $E_{3D}/N-\hbar \omega_\perp$,
calculated using $V^{HS}$ (diamonds) and $V^{SR}$ (asterisks), respectively,
together with 1D FN-MC energy per particle, $E_{1D}/N-\hbar \omega_\perp$,
calculated using $g_{1D}$ [squares, Eq.~\protect\ref{eq_g1Dren}] and $g_{1D}^{0}$
[pluses, Eq.~\protect\ref{eq_g1Dunren}], respectively, as a function of $a_{3D}$
[(a) linear scale; (b) logarithmic scale] for $N=10$ and $\lambda=0.01$. The
statistical uncertainty of the FN-MC energies is smaller than the symbol size.
Dotted and solid lines show the 1D energy per particle calculated within the LDA
for $g_{1D}^0$, Eq.~\protect\ref{eq_g1Dunren} [using the LL equation of state]
and for $g_{1D}$, Eq.~\protect\ref{eq_g1Dren} [using the LL equation of state for
$g_{1D}>0$, and the hard-rod equation of state for $g_{1D}<0$], respectively. A
dotted horizontal line indicates the energy per particle of a non-interacting Bose
gas, and a dashed horizontal line indicates the TG energy per particle. A vertical
arrow the position where $g_{1D}$, Eq.~\protect\ref{eq_g1Dren}, diverges.}
\label{figSc4}
\end{figure}

Consider the lowest-lying gas-like state of the 3D Schr\"odinger equation.
Figure~\ref{figSc4} shows the 3D energy per particle, $E_{3D}/N-\hbar
\omega_\perp$, as a function of $a_{3D}$ for $N=10$ under quasi-one dimensional
confinement, $\lambda=0.01$, for the hard-sphere two-body potential $V^{HS}$
(diamonds) and the short-range potential $V^{SR}$ (asterisks). The energies for
$V^{HS}$ are calculated using the DMC method [with $\psi_T$ given by
Eqs.~\ref{Jastrow} and \ref{VMC2}], while those for $V^{SR}$ are calculated using
the FN-MC method [with $\psi_T$ given by Eqs.~\ref{Jastrow} and \ref{VMC3}]. For
small $a_{3D}/a_\perp$, the energies for these two two-body potentials agree within
the statistical uncertainty. For $a_{3D}\gtrsim a_\perp$, however, clear
discrepancies are visible. The DMC energies for $V^{SR}$ cross the TG energy per
particle (indicated by a dashed horizontal line),
$E/N-\hbar\omega_\perp=\hbar\omega_\perp \lambda N/2$, very close to the value
$a_{3D}^{c}=0.9684 a_\perp$ (indicated by a vertical arrow in
Fig.~\ref{figSc4}(b)), while the energies for $V^{HS}$ cross the TG energy per
particle at a somewhat smaller value of $a_{3D}$. 

For $a_{3D}>a_{3D}^c$, the energy for the short-range potential $V^{SR}$ of the
lowest-lying gas-like state increases slowly with increasing $a_{3D}$, and becomes
approximately constant for large values of $|a_{3D}|$. The limit $|a_{3D}|
\rightarrow \infty$ corresponds to the unitary regime (see below). Notably, the 3D
energy behaves smoothly as $a_{3D}$ diverges. The 3D energy slowly increases
further for increasing negative $a_{3D}$, and changes more rapidly as $a_{3D}
\rightarrow -0$. The $|a_{3D}| \rightarrow \infty$ regime and the $a_{3D}
\rightarrow -0$ regime are discussed in more detail below.

To compare our results obtained for the 3D Hamiltonian, $H_{3D}$, with those for the
1D Hamiltonian, $H_{1D}$, we also solve the Schr\"odinger equation for $H_{1D}$,
Eq.~\ref{eq_h1D}, for the lowest-lying gas-like state. For positive coupling
constants, $g_{1D}>0$, the lowest-lying gas-like state is the many-body ground
state, and we hence use the DMC method [with $\psi_T$ given by Eqs.~\ref{VMC4} and
\ref{VMC5}]. For $g_{1D}<0$, however, the 1D Hamiltonian supports cluster-like bound
states. In this case, the lowest-lying gas-like state corresponds to an excited
many-body state, and we hence solve the 1D Schr\"odinger equation by the FN-MC
method [with $\psi_T$ given by Eqs.~\ref{VMC4} and \ref{VMC6}].

Figure~\ref{figSc4} shows the resulting 1D energies per particle, $E_{1D}/N-\hbar
\omega_\perp$, for the renormalized coupling constant $g_{1D}$
[squares, Eq.~\ref{eq_g1Dren}], and the unrenormalized coupling constant
$g_{1D}^0$ [pluses, Eq.~\ref{eq_g1Dunren}], respectively. The 1D energies
calculated using the two different coupling constants agree well for small
$a_{3D}$, while clear discrepancies become apparent for $a_{3D} \gtrsim a_{3D}^c$.
In fact, the 1D energies calculated using the unrenormalized coupling constant
$g_{1D}^0$ approach the TG energy (dashed horizontal line) asymptotically for
$a_{3D}\rightarrow\infty$, but do not become larger than the TG energy. The 1D
energies calculated using the renormalized 1D coupling constant $g_{1D}$ agree well
with the 3D energies calculated using the short-range potential $V^{SR}$
(asterisks) up to very large values of the 3D scattering length $a_{3D}$. In
contrast, the 1D energies deviate clearly from the 3D energies calculated using the
hard-sphere potential $V^{HS}$ (diamonds) at large $a_{3D}$.

The 1D energies calculated using the renormalized coupling constant agree with the
3D energies calculated using the short-range potential $V^{SR}$ also for $|a_{3D}|
\rightarrow \infty$, that is, in the unitary regime, and for negative $a_{3D}$.
Small deviations between the 1D energies calculated using the renormalized 1D
coupling constant $g_{1D}$ and the 3D energies calculated using the short-range
potential $V^{SR}$ are visible; we attribute these to the finite range of $V^{SR}$.
The deviations should decrease with decreasing range $R$ of the short-range
potential $V^{SR}$. On the other hand, $R$ determines to first order the
energy-dependence of the scattering length $a_{3D}$. Thus, usage of an
energy-dependent coupling constant $g_{1D}$ should
also reduce the deviations between the 1D energies and the 3D energies calculated
using the short-range potential $V^{SR}$~\cite{Granger04}. Such an approach is,
however, beyond the scope of this paper.

We conclude that the renormalization of the effective 1D coupling constant $g_{1D}$
and of the 1D scattering length $a_{1D}$ are crucial to reproduce the results of
the 3D Hamiltonian $H_{3D}$ when $a_{3D}\gtrsim a_\perp$ and when $a_{3D}$ is
negative.

In addition to treating the 1D many-body Hamiltonian using the FN-MC technique, we
solve the 1D Schr\"odinger equation using the LL equation of state ($g_{1D}>0$) and
the hard-rod equation of state ($g_{1D}<0$) within the LDA (see
Sec.~\ref{secScIII}). These treatments are expected to be good when the size of the
cloud is much larger than the harmonic oscillator length $a_z$, where $a_z
=\sqrt{\hbar/m \omega_z}$, that is, when $a_{3D}$ is large and positive or when
$a_{3D}$ is negative.

Dotted lines in Fig.~\ref{figSc4} show the 1D energy per particle calculated within
the LDA for $g_{1D}^0$ (using the LL equation of state), while solid lines show the
1D energy per particle calculated within the LDA for $g_{1D}$,
Eq.~\protect\ref{eq_g1Dren} (using the LL equation of state for $g_{1D}>0$, and
the hard-rod equation of state for $g_{1D}<0$). Remarkably, the LDA energies nearly
coincide with the 1D many-body DMC energies calculated using the unrenormalized
coupling constant (pluses) and the renormalized coupling constant (squares),
respectively. Finite-size effects play a minor role only for $a_{3D}\ll a_\perp$.
Our calculations thus establish that a simple treatment, {\it i.e.}, a hard-rod equation
of state treated within the LDA, describes inhomogeneous quasi-one dimensional Bose
gases with negative coupling constant $g_{1D}$ well over a wide range of 3D
scattering lengths $a_{3D}$.

For $a_{3D} \rightarrow -0$, that is, for large $a_{1D}$, the hard-rod equation of
state treated within the LDA, cannot properly describe trapped quasi-one
dimensional Bose gases, which are expected to become unstable against formation of
cluster-like many-body bound states for $a_{1D} \approx 1/n_{1D}$. Thus,
Sec.~\ref{secScV} investigates the regime with negative $a_{3D}$ in more detail
within a many-body framework.

\section{Stability of quasi-one-dimensional Bose gases\label{secScV}}

This section discusses the stability of {\it inhomogeneous} quasi-one dimensional
Bose gases with negative $g_{1D}$, that is, with $a_{3D}>a_{3D}^c$ and $a_{3D}<0$,
against cluster formation. Section~\ref{secScIVsubI} shows that the FN-MC results
for the 1D Hamiltonian, Eq. (\ref{eq_h1D}), are in very good agreement with the
FN-MC results for the 3D Hamiltonian. Hence, we carry our analysis out within the
1D model Hamiltonian, Eq.~\ref{eq_h1D}; we believe that our final conclusions
also hold for the 3D Hamiltonian, Eq.~\ref{eq_h3D}. For the inhomogeneous 1D
Hamiltonian $H_{1D}$, Eq.~\ref{eq_h1D}, the lowest-lying gas-like state is a
highly-excited state (see Sec.~\ref{secScIII}). We now address the question whether
this state is stable quantitatively using the VMC method.

We solve the 1D many-body Schr\"odinger equation for the Hamiltonian $H_{1D}$,
Eq.~\ref{eq_h1D}, by the VMC method using the trial wave function $\psi_T$ given
by Eqs.~\ref{VMC4} and \ref{VMC5}. This many-body wave function has the same
nodal constraint as a system of $N$ hard-rods of size $a_{1D}$. However, contrary
to hard-rods, for interparticle distances smaller than $a_{1D}$ the amplitude of
the wave function increases as $|z|$ decreases. This effect arises from the
attractive nature of the 1D effective potential and gives rise in the many-body
framework to the formation of cluster-like bound states as the average
interparticle distance is reduced below a certain critical value.

\begin{figure}[tbp]
\vspace*{-1.0in}
\centerline{\includegraphics*[width=0.57\columnwidth]{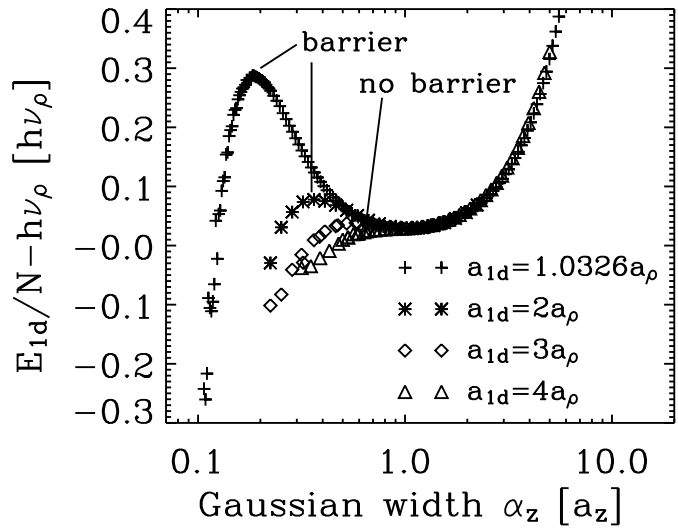}}
\vspace*{-.3in}
\caption{
VMC energy per particle, $E_{1D}/N-\hbar \omega_\perp$, as a function of the
variational parameter $\alpha_z$ for $N=5$, $\lambda=0.01$ and
$a_{1D}/a_\perp=1.0326$ (pluses), $2$ (asterisks), $3$ (diamonds) and $4$
(triangles). An energy barrier is present for $a_{1D}/a_\perp=1.0326$ and 2, but
not for $a_{1D}/a_\perp=4$. }
\label{figSc_stab1}
\end{figure}

\begin{figure}[tbp]
\vspace*{-1.0in}
\centerline{\includegraphics*[width=0.57\columnwidth]{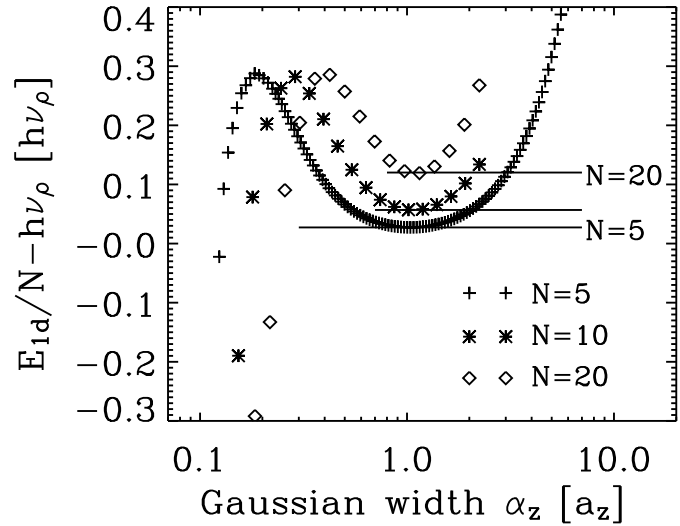}}
\vspace*{-.3in}
\caption{VMC energy per particle, $E_{1D}/N-\hbar \omega_\perp$, as a
function of the variational parameter $\alpha_z$ for $a_{1D}/a_\perp=1.0326$
(corresponding to the unitary regime), $\lambda=0.01$, and $N=5$ (pluses), $10$
(asterisks), and $20$ (diamonds). (The $N=5$ data are also shown in
Fig.~\protect\ref{figSc_stab1}.) The height of the energy barrier decreases with
increasing $N$. Horizontal solid lines show the corresponding energies for $N=5$,
$10$ and $20$ obtained using the FN-MC technique, which are in excellent agreement
with the VMC energy obtained for $\alpha_z=\alpha_{z,min}$. }
\label{figSc_stab2}
\end{figure}

Figure~\ref{figSc_stab1} shows the resulting VMC energy per particle, $E_{1D}/N-
\hbar \omega_\perp$, for $N=5$ and $\lambda=0.01$ as a function of the Gaussian
width $\alpha_z$ for four different values of $a_{1D}$. For
$a_{1D}/a_\perp=1.0326$ and $2$, Fig.~\ref{figSc_stab1} shows a local minimum at
$\alpha_{z,min} \approx a_{z}$. The minimum VMC energy nearly coincides with the
FN-MC energy (see also Fig.~\ref{figSc_stab2}), which suggests that our variational
wave function provides a highly accurate description of the quasi-one dimensional
many-body system. The energy barrier at $\alpha_z \approx 0.2 a_{z}$ decreases with
increasing $a_{1D}$, and disappears for $a_{1D}/a_\perp\approx 3$. We interpret
this vanishing of the energy barrier as an indication of instability~\cite{Bohn98}.
For small $a_{1D}$, the energy barrier separates the lowest-lying gas-like state
from cluster-like bound states. Hence, the gas-like state is stable against cluster
formation. For larger $a_{1D}$, this energy barrier disappears and the gas-like
state becomes unstable against cluster formation.

We stress that our stability analysis should not be confused with that carried out
for attractive inhomogeneous 3D systems at the level of mean-field Gross-Pitaevskii
theory~\cite{Baym96}. In fact, a mean-field type analysis of inhomogeneous 1D Bose
gases does not predict stability of gas-like states~\cite{Carr01}. In our analysis,
the emergence of local energy minima in configuration space is due to the structure
of the two-body correlation factor $f_2(z)$ entering the VMC trial wave function
$\psi_T$, Eqs.~\ref{VMC4} and \ref{VMC5}. It is a many-body effect that cannot be
described within a mean-field Gross-Pitaevskii framework.

To additionally investigate the dependence of stability on the number of particles,
Fig.~\ref{figSc_stab2} shows the VMC energy for $\lambda=0.01$ as a function of the
variational parameter $\alpha_z$ for different values of $N$, $N=5,10$ and $20$.
The scattering length $a_{1D}$ is fixed at the value corresponding to the unitary
regime, $a_{1D}=1.0326 a_\perp$. Figure~\ref{figSc_stab2} shows that the height of
the energy barrier decreases for increasing $N$. Figures~\ref{figSc_stab1} and
\ref{figSc_stab2} suggest that the stability of 1D Bose gases depends on $a_{1D}$
and $N$. To extract a functional dependence, we additionally perform variational
calculations for larger $N$ and different values of $\lambda$ and $a_{1D}$. We find
that the onset of instability of the lowest-lying gas-like state can be described
by the following criticality condition
\begin{equation}
\sqrt{N\lambda}\frac{a_{1D}}{a_\perp}\simeq 0.78,
\label{stability2}
\end{equation}
or, equivalently, by $\sqrt{N}{a_{1D}}/{a_z} \simeq 0.78$. Our 1D many-body
calculations thus suggest that the lowest-lying gas-like state is stable if
$\sqrt{N\lambda}a_{1D}/a_\perp\lesssim 0.78$, and that it is unstable if
$\sqrt{N\lambda}a_{1D}/a_\perp\gtrsim 0.78$. The stability condition, Eq.
(\ref{stability2}), implies that reducing the anisotropy parameter $\lambda$ should
allow stabilization of relatively large quasi-one dimensional Bose gases.

\begin{figure}[tbp]
\vspace*{-1.0in}
\centerline{\includegraphics*[width=0.55\columnwidth]{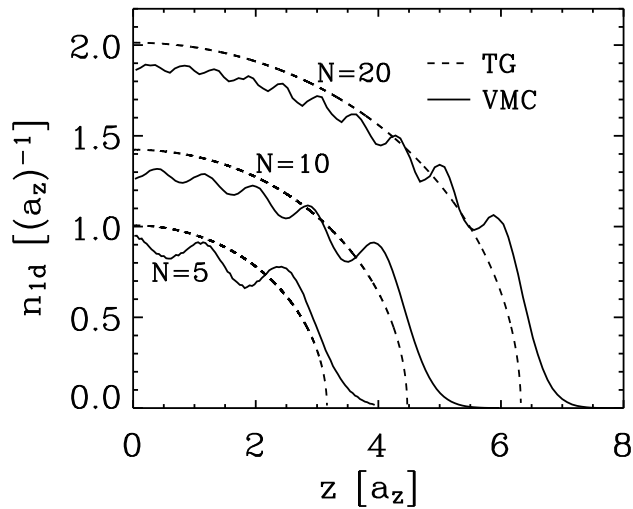}}
\vspace*{-.3in}
\caption{TG density [Eq.~\protect\ref{eq_ntg}, dashed lines]
as a function of $z$ together with VMC density (solid lines), obtained by solving
the 1D many-body Schr\"odinger equation, Eq.~\protect\ref{eq_se1D}, for $N=5$ and
$a_{1D}/a_\perp=3.6$, for $N=10$ and $a_{1D}/a_\perp=2.6$, and for $N=20$ and
$a_{1D}/a_\perp=1.8$. The TG density at the center of the trap, $z=0$, deviates
from the VMC density at the center of the trap by less than 10~\%.}
\label{figSc_stab3}
\end{figure}

To express the stability condition, Eq.~\ref{stability2}, in terms of the 1D gas
parameter $n_{1D} a_{1D}$, where $n_{1D}$ denotes the linear density at the trap
center, we approximate the density for negative $g_{1D}$ by the TG density,
Eq.~\ref{eq_ntg}. Figure~\ref{figSc_stab3} compares the TG density with that
obtained from the VMC calculations for $N=5, 10$ and 20 and values of
$a_{1D}/a_\perp$ close to the criticality condition, Eq.~\ref{stability2}. The
density at the center of the trap is described by the TG density to within 10~\%.
Since the TG density at the trap center is given by $\sqrt{2N}/(\pi a_z)$ (see
Eq.~\ref{eq_ntg}), the stability condition, Eq.~\ref{stability2}, expressed in
terms of the 1D gas parameter reads $n_{1D}a_{1D}\lesssim 0.35$.

\section{Conclusions\label{secScVI}}

This paper presents a thorough study of the properties of inhomogeneous,
harmonically-confined quasi-one dimensional Bose gases as a function of the 3D
scattering length $a_{3D}$. The behavior of confined Bose gases strongly depends on
the ratio of the harmonic oscillator length in the tight transverse direction,
$a_\perp$, to the interaction range $R$ and to the average interparticle distance
$1/n^{1/3}$, where $n$ denotes the 3D central density.

Quasi-1D bosonic gases have been realized experimentally in highly-elongated
harmonic traps. The strength of atom-atom interactions can be varied over a wide
range by tuning the value of the 3D $s$-wave scattering length $a_{3D}$ through
application of an external magnetic field in the proximity of a Feshbach resonance.
For $R\ll a_\perp$, the scattering length $a_{3D}$ determines to a good
approximation the effective 1D scattering length $a_{1D}$ and the effective 1D
coupling constant $g_{1D}$, which can be, just as the 3D coupling constant, tuned
to essentially any value including zero and $\pm\infty$. By exploiting Feshbach
resonance techniques, one should be able to achieve strongly-correlated quasi-one
dimensional systems. The strong coupling regime is achieved for $1/n^{1/3}\gg
a_\perp$, it includes the TG gas, where a system of interacting bosons behaves as if
it consisted of non-interacting spinless fermions, and the so-called unitary
regime, where the properties of the gas become independent of the actual value of
$a_{3D}$. In the unitary regime, the gas is dilute, that is, $nR^3\ll 1$, and at
the same time strongly-correlated, that is, $n|a_{3D}|^3\gg 1$.

The present analysis is carried out within various theoretical frameworks. We
obtain the 3D energetics of the lowest-lying gas-like state of the system using a
microscopic FN-MC approach, which accounts for all degrees of freedom explicitly.
The resulting energetics are then used to benchmark our 1D calculations. Full
microscopic 1D calculations for contact interactions with renormalized coupling
constant $g_{1D}$ result in energies that are in excellent agreement with the full
3D energies. This agreement implies that a properly chosen many-body 1D Hamiltonian
describes quasi-one dimensional Bose gases well.

We also consider the LL and the hard-rod equation of state of a 1D system treated
within the LDA. These approaches provide a good description of the energy of the
lowest-lying gas-like state for as few as five or ten particles. Finite size
effects are to a good approximation negligible. Our detailed microscopic studies
suggest that these LDA treatments provide a good description of quasi-one
dimensional Bose gases. In particular, we suggest a simple treatment of 1D systems
with negative $g_{1D}$ using the hard-rod equation of state.

Finally, we address the question of whether the lowest-lying gas-like state of
inhomogeneous quasi-one dimensional Bose gases is actually stable. We find,
utilizing a variational 1D many-body framework, that the lowest-lying gas-like
state is stable for negative coupling constants, up to a minimum critical value of
$|g_{1D}|$. Our numerical results suggest that the stability condition can be
expressed as $n_{1D} a_{1D} \simeq0.35$. Since our conclusions are derived from
variational 1D calculations, more thorough microscopic calculations are needed to
confirm our findings. We believe, however, that our findings will hold even in a 3D
framework or when three-body recombination effects are included explicitly.

While our study was performed for inhomogeneous quasi-one dimensional Bose gases,
many findings also apply to homogeneous quasi-one dimensional Bose gases.
Furthermore, the Fermi-Bose mapping~\cite{Girardeau60,Granger04,Cheon99,Girardeau03}, which
allows one to map an interacting 1D gas of spin-polarized fermions to an
interacting 1D gas of spin-polarized bosons, suggests that many of the results
presented here for quasi-one dimensional Bose gases may directly apply to quasi-one
dimensional Fermi gases.

\chapter{Ground state properties of a one-dimensional Bose gas\label{secLL}}
\section{Introduction}

Recent progress achieved in techniques of confining Bose condensates has lead to
experimental realizations of quasi-one dimensional
systems\cite{Gorlitz01,Schreck01,Dettmer01,Richard03,Moritz03,Tolra04}. The 1D regime is reached
in highly anisotropic traps, where the axial motion of the atoms is weakly confined
while the radial motion is frozen to zero point oscillations by the tight transverse
confinement. Another possible realization of a quasi one-dimensional system could be
a cold gas on a chip.
The possibility of an experimental observation revived interest in analytical
description of the properties of a one-dimensional bose gas. To first approximation
one parameter, the one-dimensional scattering length $a_{1D}$, is sufficient to
describe the interatomic potential, which in this case can be modeled by repulsive
$\delta$-function pseudopotential. Many properties of this integrable model like the
ground state energy\cite{Lieb63}, excitation spectrum\cite{Lieb63b}, thermodynamic
functions at finite temperature\cite{Yang69} were obtained exactly already in 60-ies
using the Bethe {\it ansatz} method. Gaudin in his book \cite{Gaudin83} devoted to
this powerful method writes that so far almost nothing is known about the
correlation functions, apart from the case of impenetrable bosons
\cite{Lenard64,Vaidya79,Jimbo80}. Lately active work was carried out in this
direction, there are recent calculations of short-range expansion of the one-body
density matrix\cite{Olshanii03}, the value at zero of the two-body correlation
function\cite{Gangardt03}. Still there are no exact calculations of the correlation
functions present in the literature.

We use Diffusion Monte Carlo method (Sec.~\ref{secDMC}), which is exact apart from
the statistical uncertainty, in order to address the problem of $\delta$-interacting
bosons in the ground state of a homogeneous system. We argue that the trial
wave function we propose provides a very good description of the ground state wave
function. As a benchmark test for our DMC calculation we recover the equation of
state which is known exactly\cite{Lieb63}. For the first time we find complete
description of the one-body density matrix and pair distribution function. We show
that our results are in agreement with known analytical predictions. We calculate
momentum distribution and static structure factor which are accessible in an
experiment. We calculate exactly the value of three-body correlation function which
is very important quantity as it governs rates of inelastic processes. Also we
address effects of the external confinement.

We present study of the correlation functions of a homogeneous system. We find the
one-body density matrix at all densities. We calculate the pair-distribution
function. Fourier transform relate those quantities to the momentum
distribution and the static structure factor, which are accessible expirementaly. We
calculate a momentum distribution in a trapped system. We expand our prediction on
homogeneous three-body correlation function, relevant for the estimation of
three-body collision rate, pair-distribution function and static structure factor in
traps. We discuss in details known analytical limits and propose a precise
expression for the decay coefficient of the one-body density matrix. We provide
relevant details of our Monte Carlo study. In particular it is argued that the trial
wave function we construct provides good description and variational energy is only
slightly higher than the exact one.

The structure of this chapter is as follows.
In section \ref{secLLH} we discuss a model used to describe a cold one-dimensional
gas and make a summary of known analytical expressions for correlation functions.
Section \ref{SMC} is devoted to a brief description of the Monte Carlo scheme used
for numerical solution of the Schr\"odinger equation and investigation of a finite
size errors that is relevant for infinite system simulation. The trial wave function
used for the importance sampling is discussed in detail. We present the result for a
homogeneous system in section \ref{SH}. One-body density matrix, pair distribution
function and three-body correlation function are calculated and compared to known
exact results and analytical expansions. The information about the momentum
distribution and static structure factor is extracted by means of the Fourier
transformation. In section \ref{ST} we discuss effects of the external trapping.
Modifications to the construction of the trial wave function are discussed. The
results for the pair distribution function and momentum distribution are presented.
Finally, in Section~\ref{SC} we draw our conclusions.

\section{Lieb-Liniger Hamiltonian\label{secLLH}}

A cold bosonic gas confined in a waveguide or in a very elongated
trap can be described in terms of a one-dimensional model if the
energy of the motion in the long longitudinal direction is
insufficient to excite the levels of transverse confinement.
Further, if the range of interparticle interaction potential is
much smaller than the characteristic length of the external
confinement, one parameter is sufficient to describe the
interaction potential, namely the one-dimensional $s$-wave
scattering length. In this case the particle-particle interactions
can be safely modeled by a $\delta$-pseudopotential. Such a system
is described by the homogeneous Lieb-Liniger Hamiltonian
\begin{eqnarray}
\hat H_{LL} = -\frac{\hbar^2}{2m}\sum\limits_{i=1}^N\frac{\partial^2}{\partial z^2_i}
+g_{1D}\sum\limits_{i<j}\delta(z_i-z_j),
\label{LL}
\end{eqnarray}
where the positive one-dimensional coupling constant is related to the
one-dimensional $s$-wave scattering length $g_{1D} = -2\hbar^2/ma_{1D}$ (\ref{g1D})
with $m$ being mass of an atom. In the presence of tight harmonic transverse
confinement (we denote the oscillator length as $a_\perp$, the one-dimensional
scattering length $a_{1D}$ was shown \cite{Olshanii98} to experience a resonant
behavior in terms of $a_{3D}$ due to virtual excitations of transverse oscillator
levels
\begin{eqnarray}
a_{1D} = -a_\perp\left(\frac{a_\perp}{a_{3D}}-1.0326\right)
\label{a1D}
\end{eqnarray}

By tuning the $a_{3D}$ by Feshbach resonance value of $a_{1D}$ can by widely varied.
Without using the Feshbach resonance one typically has $a_{3D}\ll a_\perp$ condition
fulfilled. In this situation the relation (\ref{a1D}) simplifies $a_{1D} =
-a_\perp^2/a_{3D}$ (compare with the mean-field result (\ref{a1DMF})).

\begin{figure}[ht!]
\begin{center}
\includegraphics[width=0.6\columnwidth]{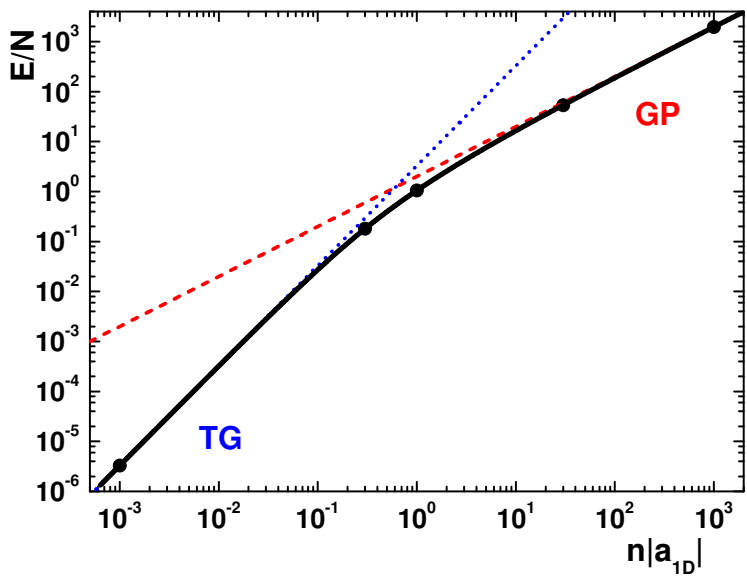}
\caption{Energy per particle: Bethe {\it ansatz} solution (solid line),
DMC (circles), GP limit (dashed line), TG limit (dotted line).
Energies are in units of $\hbar^2/(ma_{1D}^2)$.}
\label{Fig Energy}
\end{center}
\end{figure}

All properties in this model depend only on one parameter, the dimensionless density
$n_{1D}a_{1D}$. On the contrary to 3D case, where at low density the gas is weakly
interacting, in 1D system small values of the gas parameter $n_{1D}a_{1D}$ mean
strongly correlated system. This peculiarity of 1D system can be easily seen by
comparing the characteristic kinetic energy $\hbar^2n^{2/D}/2m$, with $D$ being
number of dimensions, to the mean-field interaction energy $g n$. The equation of
state of this model first was obtained Lieb and Liniger \cite{Lieb63} by using the
Bethe {\it ansatz} formulation. The energy of the system is conveniently expressed
as
\begin{eqnarray}
\frac{E}{N} = e(n_{1D}|a_{1D}|)\frac{\hbar^2n_{1D}^2}{2m},
\label{LLnotation}
\end{eqnarray}
where the function $e(n_{1D}|a_{1D}|)$ is obtained by solving a system of LL
integral equations. In the mean-field regime $n_{1D}a_{1D}\gg 1$ the energy per
particle is linear in the density $E^{GP}/N = g_{1D} n_{1D}/2$, although in the
strongly correlated regime the dependence is quadratic (\ref{ETG}): $E^{TG}/N =
\pi^2n_{1D}^2/6m$. Still an explicit general expression of the energy for an
arbitrary value of $n_{1D}a_{1D}$ is not known. The dependence of the energy on the
density resulting from numerical solution of the LL integral equations is plotted in
Fig.~\ref{Fig Energy}. On the same figure we present the energy obtained by a
different method of solving the Schr\"odinger equation, the Diffusion Monte Carlo
method (see section \ref{SMC}). Results of both methods are in perfect coincidence.

The chemical potential is defined as the derivative of the total energy
with respect to the number of particles $\mu = \partial E/\partial N$.
The healing length $\xi = \hbar/(\sqrt 2 mc)$ is related to the speed of
sound $c$, which in turn can be extracted from the chemical potential
$mc^2 = n_{1D}\frac{\partial}{\partial n_{1D}} \mu$.

These quantities can be obtained in an explicit way in the regime of strong
interactions $n_{1D}a_{1D}\ll 1$. In this limit the energy of incident particle is
not sufficient to tunnel through the particle-particle interaction potential. Two
particle can never be at the same point in space, which together with spatial
peculiarity of 1D system acts as an effective Fermi exclusion principle. Indeed, in
this limit the system of bosons acquires many fermi-like properties. There exist a
direct mapping of the wave function of strongly interacting bosons onto a
wave function of noninteracting fermions due to Girardeau \cite{Girardeau60}. We will
refer to this limit as the Tonks-Girardeau limit. The speed of sound in this gas is
related to the fermi momentum of the one-component fermi system at the same density
$c = p_F/m =
\pi n_{1D} \hbar/m$. The chemical potential equals to the fermi energy (\ref{muTG})
$\mu = \pi^2n_{1D}^2/2m$ (see $n_{1D}a_{1D}\ll 1$ limit in the Fig.~\ref{Fig Energy}).

Further, due to this mapping one knows the pair distribution function, which
exhibits Friedel-like oscillations
\begin{eqnarray}
g_2(z) = 1 - \frac{\sin^2 \pi n_{1D} z}{(\pi n_{1D} z)^2}
\label{g2 tonks}
\end{eqnarray}

The static structure factor of the TG gas is given by
\begin{equation}
S(k)=\left\{
\begin{array}{cc}
|k|/(2\pi n_{1D}),&|k|<2\pi n_{1D}\\
1,&|k|>2\pi n_{1D}\\
\end{array}
\right.
\label{Sk tonks}
\end{equation}

The one-body density matrix $g_1(z)$ was calculated in terms of series expansion at
small and large distances \cite{Lenard64,Vaidya79,Jimbo80}. Its slow long-range
decay
\begin{eqnarray}
g_1(z) = \frac{\sqrt{\pi e}2^{-1/3}A^{-6}}{\sqrt{zn}},
\quad n_{1D}|a_{1D}|\ll 1
\label{g1 TG}
\end{eqnarray}
leads to an infrared divergence in the momentum distribution $n(k) \propto
1/\sqrt{|k|}$.

Beyond the TG regime full expressions of the correlation functions are not known.
The long-range asymptotics ({\it i.e.} distances much larger than the healing length
$\xi$) can be obtained from the hydrodynamic theory of the low-energy phonon-like
excitations \cite{Reatto67,Schwartz77,Haldane81,Korepin93}. One finds following
power-law decay (\ref{Casympt})
\begin{eqnarray}
g_1(z) = \frac{C_{asympt}}{|z n_{1D}|^\alpha},
\label{g1 asypmt}
\end{eqnarray}
where $\alpha = mc/(2\pi\hbar n_{1D})$ and coefficient $C_{asympt}$ is given by
formula (\ref{g1 MF}). In the TG regime
$c = \pi\hbar n_{1D}/m$ and thus $\alpha=1/2$ as anticipated above. In the opposite
GP regime ($n_{1D}|a_{1D}|\gg 1$), the result is $\alpha =
1/(\pi\sqrt{2n_{1D}|a_{1Dd}|})$ which decreases as $n_{1D}|a_{1D}|$ increases. That
means that in the Lieb-Liniger theory $\alpha\le 1/2$. Instead in the super-Tonks
regime one deals with special situation $\alpha>1/2$. In the mean-field limit the
relation of the coefficient of proportionality in Eq.~\ref{g1 asypmt} to $\alpha$
was established by Popov \cite{Popov80} Of course power-law decay of the
non-diagonal element of the one-body density matrix
(\ref{g1 TG},\ref{g1 asypmt}
) does not support long-range order and excludes the existence of Bose-Einstein
condensation in one dimension even at zero temperature \cite{Schultz63}. The
behavior of the momentum distribution for $|k|\ll 1/\xi$ follows immediately from
(\ref{g1 asypmt})
\begin{eqnarray}
n(k) = C_{asympt} \left|\frac{2n_{1D}}{k}\right|^{1-\alpha}
\frac{\sqrt\pi \Gamma\left(\frac{1}{2}-\frac{\alpha}{2}\right)}{\Gamma\left(\frac{\alpha}{2}\right)}
\label{nk small}
\end{eqnarray}

Furthermore, the hydrodynamic theory allows one to calculate the static structure
factor in the long-wavelength regime $|k|\ll 1/\xi$. One finds the well-known
Feynman result \cite{Feynman54}
\begin{eqnarray}
S(k) = \frac{\hbar|k|}{2mc}
\label{Sk small}
\end{eqnarray}

Recently, the short range behavior of the one-, two-, and three-body correlation
functions has also been described. The value at $z=0$ of the pair correlation
function at arbitrary density can be obtained from the equation of state through the
Hellmann-Feynman theorem \cite{Gangardt03}:
\begin{eqnarray}
g_2(0)=-\frac{(n_{1D}|a_{1D}|)^2}{2} e',
\label{g2 small}
\end{eqnarray}
where the derivative of $e$ is taken with the respect to the gas parameter
$n_{1D}|a_{1D}|$.

This quantity vanishes in the TG regime and approaches unity in the GP regime. The
``excluded volume'' correction (\ref{exclvol}) allows one to specify its behavior
close to the TG region:
\begin{eqnarray}
g_2(0) = \frac{\pi^2n_{1D}^2|a_{1D}|^2}{3},\qquad n_{1D}|a_{1D}|\ll 1
\label{g2TG}
\end{eqnarray}

The $z=0$ value of the three-body correlation function was obtained in a
perturbative manner in the regions of strong and weak interactions
\cite{Gangardt03}. Similarly to $g_2(0)$, it quickly decays in the TG limit
\begin{eqnarray}
g_3(0) = \frac{(\pi n a_{1D})^6}{60}, \quad n_{1D}a_{1D}\ll 1,
\label{g3 TG}
\end{eqnarray}
and goes to a constant value in the GP regime:
\begin{eqnarray}
g_3 = 1-\frac{6\sqrt{2}}{\pi\sqrt{n a_{1D}}},\quad n_{1D}a_{1D}\gg 1.
\label{g3 GP}
\end{eqnarray}

Furthermore, recently the first few terms of the short-range series expansion of the
one-body correlation function have been calculated in \cite{Olshanii03}
\begin{eqnarray}
g_1(z) = 1 -\frac{1}{2}(e+e' n_{1D}|a_{1D}|)|n_{1D}z|^2 + \frac{e'}{6} |n_{1D}z|^3
\label{g1 small}
\end{eqnarray}

This expansion is applicable for distances $|n_{1D}z| \ll 1$ and arbitrary
densities.

\section{Quantum Monte Carlo Method\label{SMC}}

We use Diffusion Monte Carlo (DMC) technique in order to obtain the ground state
properties of the system.
A good choice of the trial wave-function is crucial for the efficiency of the
calculation. In order to prove that our trial wave-function is indeed very close to
the true ground state wave function we perform calculation of the variational energy
$E_{VMC}$ which provides an upper-bound to the ground state energy (see
Table~\ref{LLtable1}). We find that the variational energy is at maximum $2\%$
higher than the energy of the DMC calculation, which coincides with the exact
solution based on the use of the Bethe {\it ansatz} (see, also, Fig.~\ref{Fig
Energy}).

In a homogeneous system we use the Bijl-Jastrow construction (\ref{Jastrow}) of the
trial wave function. The construction of the the two-body term $f_2(z)$ is described
in Sec.~\ref{secWFphonon}. The $|z|<R$ part corresponds to the exact solution of the
two-body problem and provides a correct description of short-range correlations.
Long-range correlations arising from phonon excitations are instead accounted for by
the functional dependence of $f(z)$ for $z>Z$ \cite{Reatto67}.
The value of the matching point $R$ is a variational parameter which we optimize
using the variational Monte Carlo (VMC). The TG wave function (\ref{wf1Dtonks}) is
obtained as a special case of our trial wave function (\ref{wfphonon}) for $R = B =
L/2$ and $kL = \pi$.

\begin{table}
\begin{center}
\begin{tabular}{|l|l|l|}
\hline
$n_{1D}a_{1D}$&$E_{LL}/N$&$E_{VMC}/N$\\
\hline
$10^{-3} $&$1.6408~10^{-6} $&$1.64(1)~10^{-6}$\\
$0.03    $&$1.3949~10^{-3} $&$1.3956(3)~10^{-3}$\\
$0.3     $&$9.0595~10^{-2} $&$9.089(3)~10^{-2}$\\
$1       $&$0.5252         $&$0.535(3)     $\\
$30      $&$26.842         $&$27.121(3)    $\\
$10^3    $&$981.15         $&$981.72(6)    $\\
\hline
\end{tabular}
\caption{Energy per particle for different values of the dimensionless density
$n_{1D}a_{1D}$: exact result $E_{LL}$ obtained by solving Lieb-Liniger equations,
variational result $E_{VMC}$ obtained by optimization the trial wave function
\ref{wfphonon}. Variational energy gives the upper bound to the exact energy. DMC
calculation recovers the exact result $E_{LL}$.}
\label{LLtable1}
\end{center}
\end{table}

The level of accuracy of the trial wave function is particularly important for the
calculation of the of $g_1(z)$. Instead, the pair distribution function $g_2(z)$ is
calculated using the method of ``pure'' estimators, unbiased by the choice of the
trial wave function \cite{Casulleras95}. Due to non local property of the one-body
density matrix, the function $g_1(z)$ can instead by obtained only through the
extrapolation technique. For an operator $\hat A$, which does not commute with the
Hamiltonian, the output of the DMC method is a ``mixed'' estimator
$\langle\Psi_0|\hat A|\psi_T\rangle$. Combined together with the variational
estimator $\langle\psi_T|\hat A|\psi_T\rangle$ obtained from the VMC calculation it
can be used for extrapolation to the ``pure'' estimator by the rule
$\langle\Psi_0|\hat A|\Psi_0\rangle = 2\,\langle\Psi_0|\hat
A|\psi_T\rangle-\langle\psi_T|\hat A|\psi_T\rangle$. Of course, this procedure is
very accurate only if $\psi_T \simeq
\Psi_0$. We find that DMC and VMC give results for $g_1(z)$ which are very close and
we believe that the extrapolation technique is in this case exact.

\begin{figure}[ht!]
\begin{center}
\includegraphics[angle=-90,width=0.6\columnwidth]{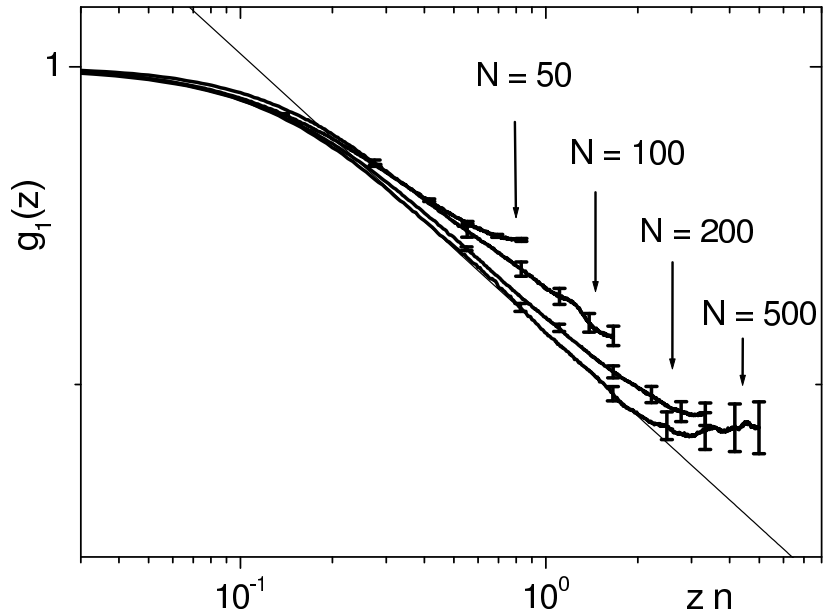}
\caption{Example of the finite size effects in the calculation
of the one-body density matrix at $n_{1D}a_{1D} = 30$.}
\label{Fig finitesize}
\end{center}
\end{figure}

We consider $N$ particles in a box of size $L$ with periodic boundary conditions. In
the construction of the trial wave function we have ensured that the two-body term
$f_2$ is uncorrelated at the boundaries $f_2(\pm L/2) = 1$. In order to estimate
properties of an infinite system we we increase number of particles and study
convergence in the quantities of interest. The dependence on the number of particles
(finite size effects) are more pronounced at the large density where the
correlations extend up to large distances. Out of the quantities we measured, the
one-body density matrix is the most sensitive to finite size corrections. As an
example in Fig.~\ref{Fig finitesize} we show $g_1(z)$ at density $n_{1D}a_{1D} = 30$
for 50, 100, 200 and 500 particles and make comparison it with the asymptotic
$z\to\infty$ behavior. We find largest finite size effects near the maximal distance
$L/2$ for which the one-body density matrix can be calculated. Also we see a
dependence of the slope on the number of particles. Already for 500 particles we
find the correct slope of the one-body density matrix. For the smaller densities,
where finite size effects are smaller, it is sufficient to have $N = 500$.

\section{Homogeneous system \label{SH}}

\begin{figure}[ht!]
\begin{center}
\includegraphics[width=0.6\columnwidth]{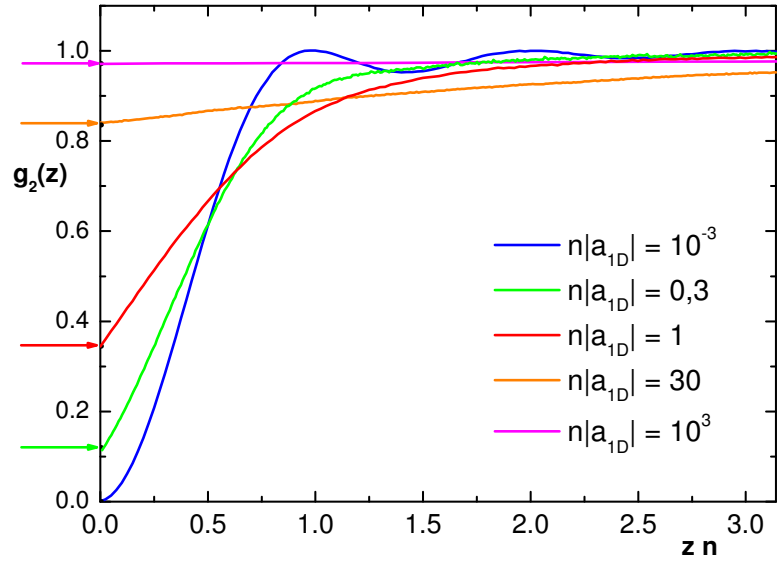}
\caption{Pair distribution function function for different values of the gas
parameter. In ascending order of the value at zero $n_{1D}|a_{1D}|
= 10^{-3}, 0.3, 1, 30, 10^3$. Arrows indicate the value of $g_2(0)$ as obtained
from Eq.~\ref{g2 small}.}
\label{Figg2}
\end{center}
\end{figure}

We calculated the pair distribution matrix for a number of densities ranging from
very small value of the gas parameter $n_{1D}a_{1D}\ll 1$ (TG regime) up large
densities $n_{1D}a_{1D}\gg 1$ (GP regime). The results are presented in the
Fig.~\ref{Figg2}. In the GP regime the correlations between particles are very weak
and $g_2(z)$ arrives very quickly at the bulk value. Decreasing the $|a_{1D}|$ (thus
making the coupling constant $g_{1D}$ larger) we enhance beyond-mean field effects
and enforce correlations. For the smallest considered value of the gas parameter
$n_{1D}a_{1D} = 10^{-3}$ we see oscillations in the pair distribution function,
which, in this sense becomes more similar to the one of a liquid, rather than of a
gas. At the same density we compare the pair distribution function with the one of
the TG gas and find no visible difference.

On the same Figure we plot predictions for the value of the pair distribution
function at zero and find perfect agreement with the analytical prediction. In the
TG regime particles never meet each other and consequently $g_2(0) = 0$. Making the
interaction between the particles weaker we find finite probability of two particles
coming close to each other according to the Eq.~\ref{g2 small}. As we go further
into the direction of the GP the interaction potential becomes more transparent and
we approach the ideal gas limit $g_2(0) = 1$.

\begin{figure}[ht!]
\begin{center}
\includegraphics[width=0.6\columnwidth]{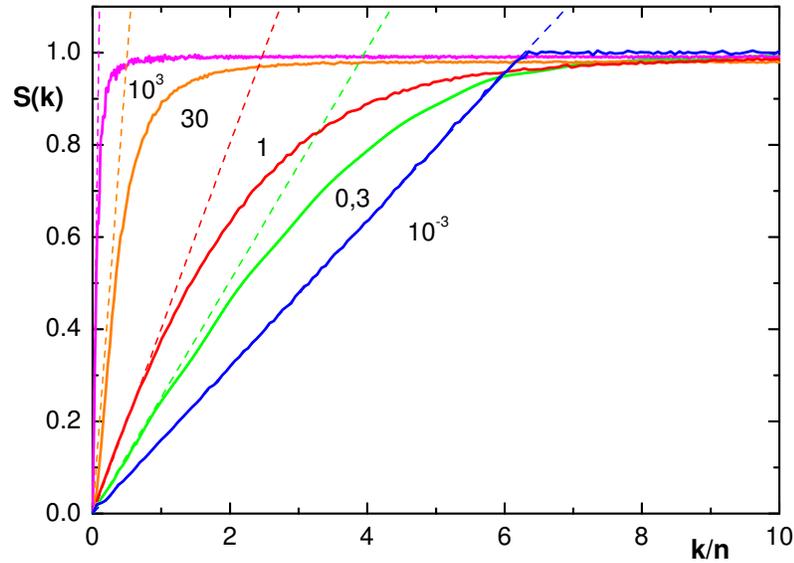}
\caption{Static structure factor for the same values of $n_{1D}|a_{1D}|$
as in Fig.~\ref{Figg2}
(solid lines). The dashed lines are the
corresponding long-wavelength asymptotics from Eq.~\ref{Sk small}.}
\label{Fig Sk}
\end{center}
\end{figure}

In the Fig.~\ref{Fig Sk} we present the static structure factor obtained from
$g_2(z)$ according to relation (\ref{Fourier Sk}). At the smallest density
$n_{1D}|a_{1D}| = 10^{-3}$ our points lie exactly on the top of the $S(k)$ of the TG
gas (Eq.~\ref{Sk tonks}). For all densities the small-momenta part of the structure
factor comes from generation of a phonon. We compare DMC results with the Feynman
prediction (Eq.~\ref{Sk small}). We see that in the strongly correlated regime
phononic description works well even to values of the momenta of the order of
inverse density $n_{1D}^{-1}$, although in the MF regime the healing length becomes
significantly larger than the mean interparticle distance leading to earlier
deviations.

\begin{figure}[ht!]
\begin{center}
\includegraphics[width=0.6\columnwidth]{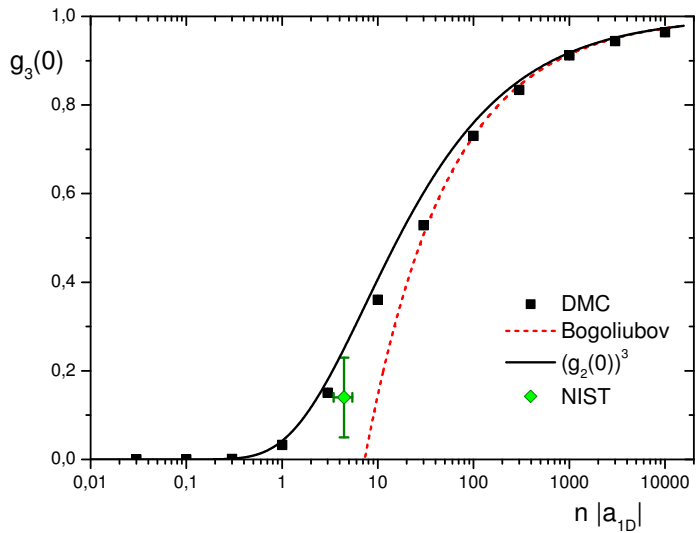}
\caption{Value at zero of the three-body correlation function
$g_3(0)$ (squares), Bogoliubov limit Eq.~\ref{g3 GP} (dashed line), $g_2^3(0)$
Eq.~\ref{g2 small} (solid line), experimental result of \cite{Tolra04} (diamond).}
\label{Figg3}
\end{center}
\end{figure}

\begin{figure}[ht!]
\begin{center}
\includegraphics[width=0.6\columnwidth]{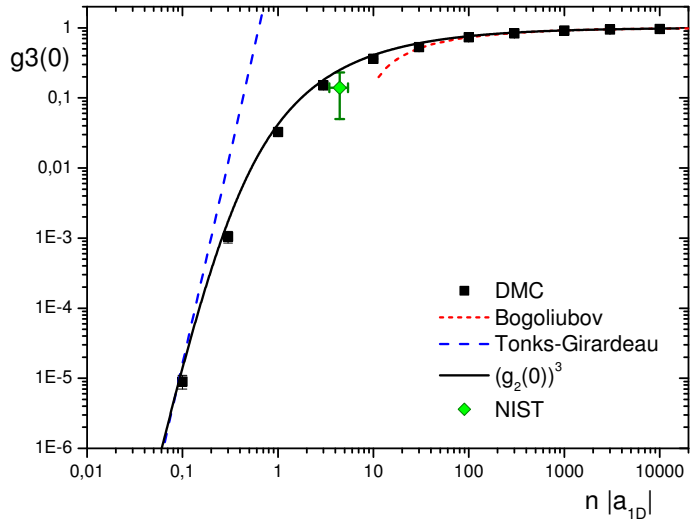}
\caption{Value at zero of the three-body correlation function,
log-log scale, $g_3(0)$ (squares), TG limit
Eq.~\ref{g3 TG} (dashed line), Bogoliubov limit Eq.~\ref{g3 GP} (short-dashed line),
$g_2^3(0)$ Eq.~\ref{g2 small} (solid line), experimental result of \cite{Tolra04}
(diamond).}
\label{Figg3log}
\end{center}
\end{figure}

We calculate the value at zero of the three-body correlation function (\ref{g3})
over a large range of densities. At large density $n_{1D}|a_{1D}| = 10^4$ the
probability of three-body collisions is high. Making the density smaller we find
decrease in the value of $g_3(0)$. Close to the MF limit the result of the
Bogoliubov theory (Eq.~\ref{g3 GP}) provides fairly good description of $g_3(0)$
(see. Fig.~\ref{Figg3}). Further decrease in the density leads to fast decay of
three-body collision rate and it becomes vanishing at values of the gas parameter
smaller than one. In order to resolve the law of the decay we plot same data on the
log-log scale (Fig.~\ref{Figg3log}) and show that the decay goes with the forth
power of the gas parameter in agreement with Eq.~\ref{g3 TG}. Numerical estimation
of $g_3(0)$ at smaller densities becomes very difficult due to very small value of
the measured quantity itself. It is interesting to note that $g_2^3(0)$ follows
closely to $g_3(0)$.

We compare the result of an experiment done in NIST \cite{Tolra04} with the
theoretical prediction of the Lieb-Liniger model, see
Figs.~\ref{Figg3},\ref{Figg3log}.
In this experiment the three body recombination rate was measured and the
value of $g_3(0)$ was extracted. We find an agreement between experiment and theory.
The result of DMC calculation is slightly closer to the experimental data point than
the estimation $g_2^3(0)$, although the error bars of the experimental measurement
cover both values.

We calculated the spatial dependence of the one-body density matrix for different
densities. At small distances we compare the function with the short range expansion
(\ref{g1 small}) and find an agreement for $zn_{1D}\ll 1$ (see Fig.~\ref{Fig
g1small}). For distances larger than the healing length we expect the hydrodynamic
theory to provide a correct description. Indeed, we see that the long-range decay
has a power-law form in agreement with the prediction Eq.~\ref{g1 asypmt} (see
Fig.~\ref{Fig g1}). We fix the coefficient of proportionally in Eq.~\ref{g1 asypmt}
by fitting the data. By doing it we conclude complete description of the one-body
density matrix starting from small distances up to large ones. The deviations on
Fig.~\ref{Fig g1} from power law-decay are at largest distances ($z\approx L/2$) are
due to finite size effects.

We derive a highly accurate expression for the coefficient $C_{asympt}$ from a
hydrodynamic approach considering weak interactions and low density fluctuations. In
terms of Euler's constant $\gamma=0.577$ we have (\ref{Casympt}):
%
\begin{eqnarray}
C_{asympt} = \left(\frac{e^{1-\gamma}}{8\pi\alpha}\right)^\alpha(1+\alpha)
\label{g1 MF}
\end{eqnarray}

Although this constant is formally derived in the limit $\alpha\ll 1$ ({\it i.e.} limit of
weak interaction $n_{1D}a_{1D}\gg 1$) it provides a very good description in the
whole range of densities. Indeed, the coefficient defined by fitting $g_{1}$ as
shown in Fig.~\ref{Fig g1} is always in agreement with prediction (\ref{g1 MF})
within the uncertainty errorbars we get from our DMC calculation. Further, in the
most strongly interacting TG regime we compare (\ref{g1 MF}) with the exact result
provided by formula (\ref{g1 TG}) and find only $0.3 \%$ difference.
A different expression was obtained by Popov \cite{Popov80} (and later recovered in
\cite{Castin02}) $C^{Popov}_{asympt} =
\left(\frac{e^{2-\gamma}}{8\pi\alpha}\right)^\alpha$. Both expressions coincide for
small values of $\alpha$, but Popov's coefficient lead up to larger $10\%$ maximal
error, as it was pointed out in \cite{Cazalilla04}. The comparison of different
coefficients is presented in Table~\ref{LLtable2}.

\begin{table}
\begin{center}
\begin{tabular}{|l|l|l|l|}
\hline
$n_{1D}a_{1D}$ & $C^{DMC}_{asympt}$ &  $C^{Popov}_{asympt}$ & $C_{asympt}$\\
\hline
1000  & 1.02  & 1.0226 & 1.0226 \\
30    & 1.06  & 1.0588 & 1.0579 \\
1     & 0.951 & 0.9646 & 0.9480 \\
0.3   & 0.760 & 0.8145 & 0.7814 \\
0.001 & 0.530 & 0.5746 & 0.5227 \\
\hline
\end{tabular}
\caption{Coefficient of the long-range decay of the one-body density matrix
defined as in (\ref{g1 asypmt}). First column is the one-dimensional
gas parameter, second column is the fitting coefficient extracted from
Eq.~\ref{Fig g1}, third column is Popov's prediction, fourth
column is formula~\ref{g1 MF}. Density $n_{1D}a_{1D} = 0.001$ is deeply
in the TG regime and here one can apply Eq.~\ref{g1 TG} leading to
$C^{TG}_{asympt} = 0.5214$}
\label{LLtable2}
\end{center}
\end{table}

\begin{figure}[ht!]
\begin{center}
\includegraphics[width=0.6\columnwidth]{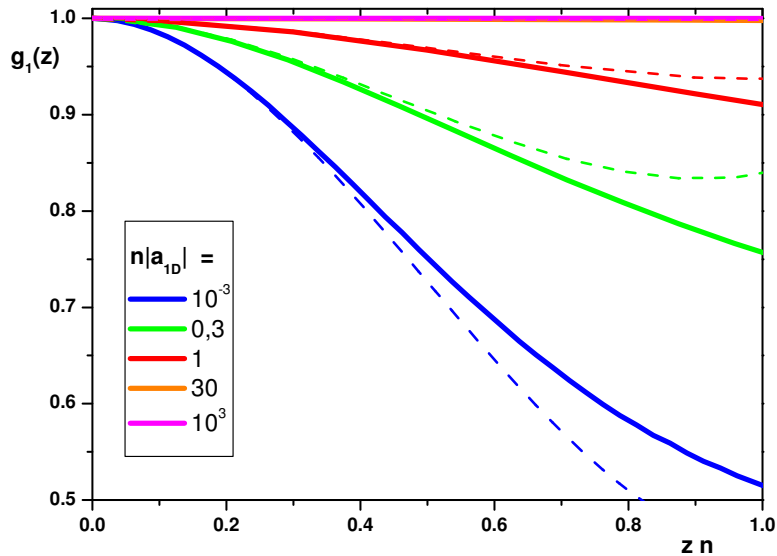}
\caption{Short range behavior of the one-body density matrix for different values
of the gas parameter $n_{1D}|a_{1D}| = 10^{-3}, 0.3, 1, 30, 10^3$ (the lowest curve
corresponds to $n_{1D}|a_{1D}| = 10^{-3}$, the uppermost to $n_{1D}|a_{1D}| =
10^{3}$, $g_1(z)$ (solid lines), series expansion at zero (eq.~\ref{g1 small})
(dashed lines).}
\label{Fig g1small}
\end{center}
\end{figure}

\begin{figure}[ht!]
\begin{center}
\includegraphics[width=0.6\columnwidth]{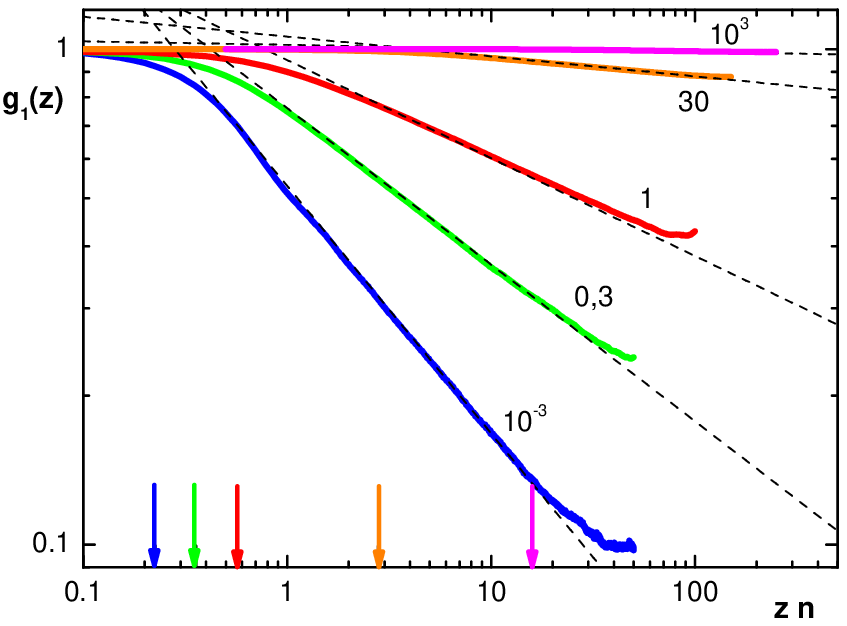}
\caption{Large range behavior of the one-body density matrix (solid lines),
fits to the long-wavelength asymptotics from eq.~\ref{g1 asypmt} (dashed lines).
Values of the density are same as in Fig.~\ref{Fig g1}. The arrows indicate the
value of $\xi n$: the leftmost corresponds to $n_{1D}|a_{1D}| = 10^{-3}$, the
rightmost to $n_{1D}|a_{1D}| = 10^3$}
\label{Fig g1}
\end{center}
\end{figure}

\begin{figure}[ht!]
\begin{center}
\includegraphics[width=0.6\columnwidth]{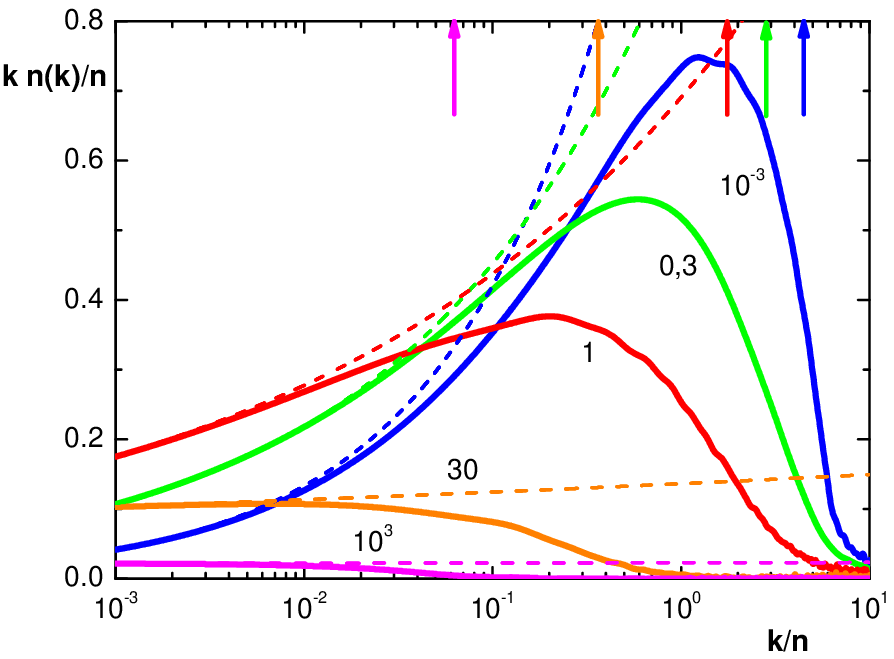}
\caption{Momentum distribution for the same values of $n_{1D}|a_{1D}|$ as
in Fig.~\ref{Fig g1} (solid lines). The dashed lines correspond to the infrared
behavior of Eq.~\ref{nk small}. The arrows indicate the value of $1/\xi n_{1D}$: the
rightmost corresponds to $n_{1D}|a_{1D}| = 10^{-3}$, the leftmost to $n_{1D}|a_{1D}|
= 10^3$.}
\label{Fig nk}
\end{center}
\end{figure}

We obtain the momentum distribution from the Fourier transform (see
Eq.~\ref{nk}) of the one-body density matrix at short distances and the fit
power-law decay at large distances. In an infinite homogeneous system the momentum
distribution has an infrared divergence (Eq.~\ref{nk small}). In order to present
the momentum distribution in the most efficient way we plot in Fig.~\ref{Fig nk} a
combination $k n(k)$, where this divergence is absent. We notice that the infrared
asymptotic behavior is recovered for values of $k$ considerably smaller than the
inverse healing length $1/\xi$. At large $k$ the numerical noise of our results is
too large to extract evidences of $1/k^4$ behavior predicted in \cite{Olshanii03}.

\section{Trapped system \label{ST}}

Now let us consider effects of the external trap. We consider the trapping potential
to be a harmonic oscillator. The effective one-dimensional Hamiltonian is then given
by
\begin{eqnarray}
\hat H^{trap}_{LL} = -\frac{\hbar^2}{2m}\sum\limits_{i=1}^N\frac{\partial^2}{\partial z^2_i}
+\frac{m\omega_z^2}{2}\sum\limits_{i=1}^N z_i^2
+g_{1D}\sum\limits_{i<j}\delta(z_i-z_j),
\end{eqnarray}
where the effective coupling constant depends both on the value of the 3D s-wave
scattering length and the oscillator length of the transverse confinement $a_\perp =
\sqrt{\hbar/m\omega_\perp}$ through relation $g_{1D} = 2\hbar^2a/ma^2_\perp$
(\ref{g1DMF}). The relevant parameters are: the ratio $a_{3D}/a_\perp$, the anisotropy
parameter $\lambda = \omega_z/\omega_\perp$ and the number of particles $N$.

In the construction of the trial wave function used in our DMC calculation we
introduce one-body Jastrow term $f_1(z_i)$ in addition to the two-body terms
$f_2(z_{ij})$ already contained in homogeneous trial wave function (\ref{wfphonon}).
Taking into account the harmonic nature of the external potential we choose the
one-body term in the Gaussian form $f_1(z) = \exp(-\alpha_z z^2)$ with $\alpha_z$
being the variational parameter. The correlations at distances much larger than the
longitudinal oscillator length $a_z = \sqrt{\hbar/m\omega_z}$ are dominated by the
oscillator confinement and two-body correlations become irrelevant.

We consider the following configurations: $a_{3D}/a_\perp = 0.2$, $\lambda = 10^{-3}$ and
number of particles $N = 5, 20, 100$. In Sec.~\ref{sec3D1D} we have proven
that in these conditions the ground-state energy and structure of the cloud is
correctly described by the Lieb-Liniger equation of state in local density
approximation.

\begin{figure}[ht!]
\begin{center}
\includegraphics[width=0.6\columnwidth]{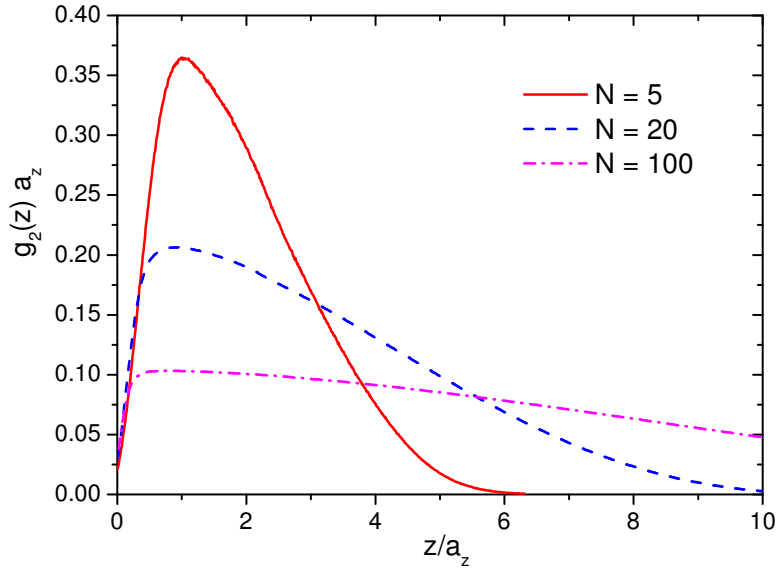}
\caption{Pair distribution of a trapped system for
5, 20, 100 particles and $a_{3D}/a_\perp = 0.2, \lambda = 10^{-3}$.}
\label{Fig trap PD}
\end{center}
\end{figure}

In Fig.~\ref{Fig trap PD} we plot the pair distribution function (\ref{DMCg2trap}) for
5, 20 and 100 particles. The short-range dependence is dominated by two-body
interactions. We do not find oscillations which means that strong shell structure is
absent. At large distances the external trapping suppresses density.

We refer to general definition of the static structure factor in terms of the
momentum distribution $n_k$ (\ref{Skdef}):
\begin{eqnarray}
S(k) = \frac{1}{N}(\langle n_{-k} n_k\rangle - |\langle n_k\rangle|^2)
\end{eqnarray}

\begin{figure}[ht!]
\begin{center}
\includegraphics[angle=-90,width=0.6\columnwidth]{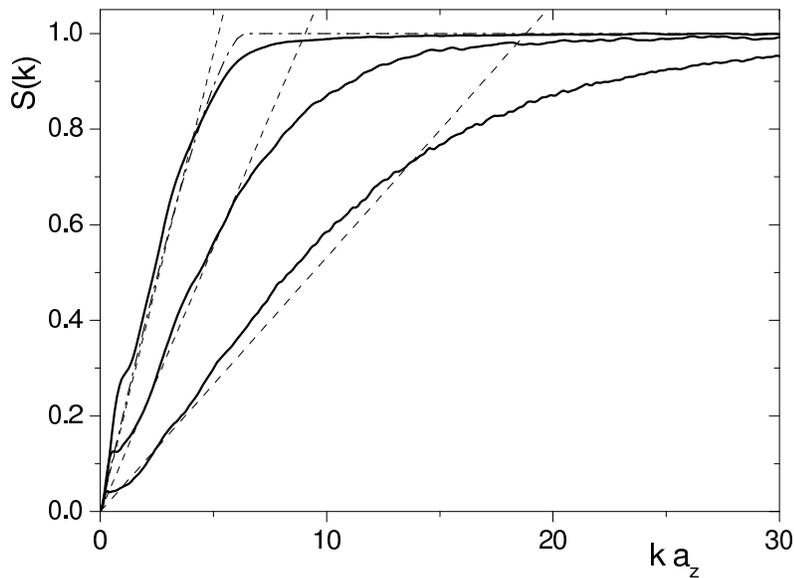}
\caption{Static structure factor of a trapped system for
$a_{3D}/a_\perp = 0.2, \lambda = 10^{-3}$ and 5, 20, 100 particles (solid lines from up
to down). The dashed lines show linear behavior residual of the phononic part of
$S(k)$ in a homogeneous system and given by formula (\ref{Sk small}). We use the
density in the center of the trap to estimate the sound velocity. The dash-dotted
line for $5$ particles is obtained within the local density approximation for the
TG-equation of state and is given by Eq.~\ref{SkTGLDA}.}
\label{Fig trap Sk}
\end{center}
\end{figure}

On the contrary uniform case, the last term is no longer vanishing for $k\ne 0$in a
trap. In Fig.~\ref{Fig trap Sk} we present the static structure factor obtained for
the same set of parameters. We are interested in evidences of the linear behavior
characteristic for the phonon propagation. We discover that Feynman formula
(\ref{Sk small}) with the speed of sound taken at the center of the trap provides
relatively good description also for the trapped systems. Of course, the very low
momenta part is different due to the finite size effects.

For the smallest number of particles considered ($N=5$), the density is always small
$n_{1D}a_{1D}<0.18$ and we can derive an explicit expression for the $S(k)$
exploiting knowledge of the static structure factor in the limit of small density
(\ref{Sk tonks}) as explained in Sec.~\ref{secSkLDA}. The resulting expression is
given by formula (\ref{SkTGLDA}). an be calculated.
We find that thr linear behavior at small $k$ matches the asymptotic constant in a
smoother way than it happens in a homogeneous system (see TG static structure factor
in Fig.~\ref{Fig Sk}).

\begin{figure}[ht!]
\begin{center}
\includegraphics[angle=-90,width=0.6\columnwidth]{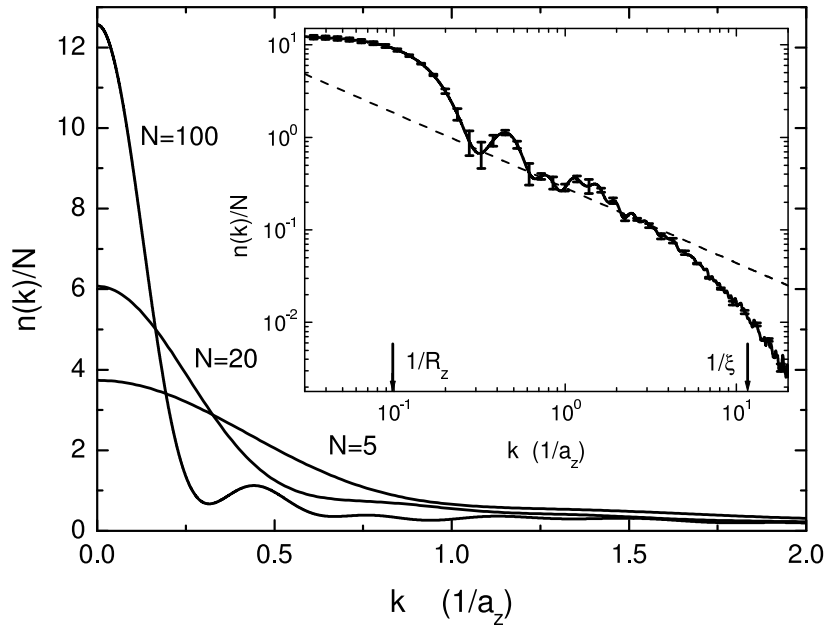}
\caption{Momentum distribution of a trapped system. Inset: momentum distribution
for $N=100$ (solid line) on a log-log scale. The dashed line is a fit to
$1/k^{1-\alpha}$ with $\alpha=0.19$. The momentum distribution is in units of
$a_z=\sqrt{\hbar/(m\omega_z)}$.}
\label{Fig trap Nk}
\end{center}
\end{figure}

In Fig.~\ref{Fig trap Nk} we show the results for the momentum distribution $n(k)$.
On the contrary to homogeneous case, $n(k)$ in a finite system always remain finite
due to natural limitations on the minimal possible value of wave vector $k_{min}
\simeq 1/R_z$, where $R_z$ is the size of the cloud in the axial direction. We are
looking for traces of the divergent behavior at small momenta similar to (\ref{nk
small}). In the case of $N=5$ and $N=20$ the rounding off of $n(k)$ at $k\sim
k_{min}$ washes out completely the divergent behavior. For the largest system with
$N=100$ we find some evidence of the infrared behavior (see inset in Fig.~\ref{Fig
trap Nk}) in the region of wave vectors $1/R_z < k < 1/\xi$. The healing length is
estimated by the density in the center of the trap $n_0|a_{1D}| \simeq 1.1$. We also
plot a power law function with the exponent $\alpha \simeq 0.19$ which corresponds
to the value in a homogeneous system with same density and the coefficient of
proportionality obtained by best fit. In order to see a cleaner signature of the
infrared behavior one should consider much larger systems.

\section{Conclusions\label{SC}}

This paper presents a thorough study of correlation properties of a one-dimensional
gas of bosons at zero temperature. In a homogeneous system the behavior is fixed by
the product of linear density $n_{1D}$ and one-dimensional scattering length
$a_{1D}$. In the strongly interacting regime $n_{1D}a_{1D}\ll 1$ the bosonic system
behaves effectively as a system of non interacting fermions. In this limit the
energy, pair distribution function $g_2(z)$, static structure factor $S_k$ are known
explicitly and are same as the ones of the corresponding fermionic system. For
arbitrary value of the gas parameter no complete description was known so far.
Switching on an external harmonic potential leads to modification in properties as
new length, the oscillator length $a_z$ is introduced.

Quasi one-dimensional systems have been already realized in a number of experiments
with elongated traps. Many new experiments with condensates in a same geometry, in a
waveguide or on a chip are expected to appear. The characteristic parameter
$n_{1D}a_{1D}$ can be varied by changing number of atoms in the condensate, trapping
frequencies or by adjusting the scattering length using the Feshbach resonance.
Momentum distribution is accessible from ballistic expansion and static factor can
be measured by the Bragg scattering.

We find for the first time full description of the correlation functions in a wide
range of the characteristic parameter $n_{1D}a_{1D}$ starting from Tonks-Girardeau
regime and up to Gross-Pitaevskii regime. We benchmark our Diffusion Monte Carlo
calculations by recovering the ground state energy known from solution of the
Lieb-Liniger integral equations. We completely recover all properties of the
Tonks-Girardeau gas and known asymptotic behavior of the momentum distribution and
correlation functions. We obtain the one-body density matrix $g_1(z)$ and pair
distribution function $g_2(z)$ for all densities. In particular we have the
description of the most nontrivial regime $n_{1D}a_{1D}\approx 1$ which is relevant
for current experiments.

We study the dependence of the value at zero of the three-body correlation function
$g_3(0)$ on the density $n_{1D}a_{1D}$.
This function is of a great interest as it governs the three-body recombination
rate, which leads to loss of the atoms out of the condensate. The data of an
experimental measurement of $g_3(0)$ is available \cite{Tolra04} and is compared
with the predictions of the Lieb-Liniger theory. An agreement between theory and
experiment is found.

By the means of Fourier transform we extract the momentum distribution $n(k)$
and static structure factor $S(k)$. Low momentum part is described by phonon
hydrodynamic theory which is expected to be applicable at distances $|z|$ larger
than the healing length $\xi$. We judge that $n(k)$ shows phononic power-law
divergence for values of $k$ considerably smaller than $1/\xi$.

Finally we discuss how the presence of a harmonic trapping modifies the correlation
functions. We plot the pair distribution function in typical experimental
configurations. We discuss possibility of finding in $n(k)$ traces of divergent
behavior, which is characteristic for a one-dimensional infinite system, in a
trapped system.

\chapter{Beyond Tonks-Girardeau: super-Tonks gas\label{secST}\label{secHRstability}}
\section{Introduction}

The study of quasi-1D Bose gases in the quantum-degenerate regime has become a very
active area of research. The role of correlations and of quantum fluctuations is
greatly enhanced by the reduced dimensionality and 1D quantum gases constitute well
suited systems to study beyond mean-field effects \cite{Petrov00}. Among these,
particularly intriguing is the fermionization of a 1D Bose gas in the strongly
repulsive Tonks-Girardeau (TG) regime, where the system behaves as if it consisted
of noninteracting spinless fermions \cite{Girardeau60}. The Bose-Fermi mapping of
the TG gas is a peculiar aspect of the universal low-energy properties which are
exhibited by bosonic and fermionic gapless 1D quantum systems and are described by
the Luttinger liquid model \cite{Voit95}. The concept of Luttinger liquid plays a
central role in condensed matter physics and the prospect of a clean testing for its
physical implications using ultracold gases confined in highly elongated traps is
fascinating \cite{Monien98,Recati03}.

Bosonic gases in 1D configurations have been realized experimentally. Complete
freezing of the transverse degrees of freedom and fully 1D kinematics has been
reached for systems prepared in a deep 2D optical lattice \cite{Moritz03,Tolra04}.
The strongly interacting regime has been achieved by adding a longitudinal periodic
potential and the transition from a 1D superfluid to a Mott insulator has been
observed \cite{Stoferle04}. A different technique to increase the strength of the
interactions, which is largely employed in both bosonic and fermionic 3D systems
\cite{Inouye98,O'Hara02} but has not yet been applied to 1D configurations, consists
in the use of a Feshbach resonance. With this method one can tune the effective 1D
coupling constant $g_{1D}$ to essentially any value, including $\pm\infty$, by
exploiting a confinement induced resonance \cite{Olshanii98,Bergeman03}. For large
and positive values of $g_{1D}$ the system is a TG gas of point-like impenetrable
bosons. On the contrary, if $g_{1D}$ is large and negative, we will show that a new
gas-like regime is entered (super-Tonks) where the hard-core repulsion between
particles becomes of finite range and correlations are stronger than in the TG
regime. In this Chapter we investigate using Variational Monte Carlo techniques
(Sec.~\ref{secVMC}) the equation of state and the correlation functions of a
homogeneous 1D Bose gas in the super-Tonks regime. We find that the
particle-particle correlations decay faster than in the TG gas and that the static
structure factor exhibits a pronounced peak. The momentum distribution and the
structure factor of the gas are directly accessible in experiments by using,
respectively, time-of-flight techniques and two-photon Bragg spectroscopy
\cite{Stoferle04}. The study of collective modes also provides a useful experimental
technique to investigate the role of interactions and beyond mean-field effects
\cite{Moritz03}. Within a local density approximation (LDA) for systems in harmonic
traps we calculate the frequency of the lowest compressional mode as a function of
the interaction strength in the crossover from the TG gas to the super-Tonks regime.

\section{The model and method}

We consider a 1D system of $N$ spinless bosons described by the following
contact-interaction Hamiltonian
\begin{equation}
H=-\frac{\hbar^2}{2m}\sum_{i=1}^{N}\frac{\partial^2}{\partial z_i^2}+g_{1D}\sum_{i<j}\delta(z_{ij}) \;,
\label{Hamiltonian}
\end{equation}
where $m$ is the mass of the particles, $z_{ij}=z_i-z_j$ denotes the interparticle
distance between particle $i$ and $j$ and $g_{1D}$ is the coupling constant which we
take large and negative. The study of the scattering problem of two particles in
tight waveguides yields the a relation of the effective 1D coupling constant
$g_{1D}$ in terms of the 3D $s$-wave scattering length $a_{3D}$ \cite{Olshanii98}.
The relation is given by the formula (\ref{g1Dolshanii}), where
$a_\perp=\sqrt{\hbar/m\omega_\perp}$ is the characteristic length of the transverse
harmonic confinement producing the waveguide. The confinement induced resonance is
located at the critical value $a_{3D}^c$ and corresponds to the abrupt change of
$g_{1D}$ from large positive values ($a_{3D}\lesssim a_{3D}^c$) to large negative
values ($a_{3D}\gtrsim a_{3D}^c$). The renormalization (\ref{g1Dolshanii}) of the
effective 1D coupling constant has been recently confirmed in a many-body
calculation of Bose gases in highly elongated harmonic traps using quantum Monte
Carlo techniques \cite{Astrakharchik04a,Astrakharchik04c}.

For positive $g_{1D}$, the Hamiltonian (\ref{Hamiltonian}) corresponds to the
Lieb-Liniger (LL) model (\ref{LL}). The ground state and excited states of the LL
Hamiltonian have been studied in detail \cite{Lieb63,Lieb63b} and, in particular,
the TG regime corresponds to the limit $g_{1D}=+\infty$. The ground state of the
Hamiltonian (\ref{Hamiltonian}) with $g_{1D}<0$ has been investigated by McGuire
\cite{McGuire64} and one finds a soliton-like state with energy
$E/N=-mg_{1D}^2(N^2-1)/24\hbar^2$. The lowest-lying gas-like state of the
Hamiltonian (\ref{Hamiltonian}) with $g_{1D}<0$ corresponds to a highly-excited
state that is stable if the gas parameter $na_{1D}\ll 1$, where $n$ is the density
and $a_{1D}$ is the 1D effective scattering length defined in Eq.
(\ref{g1Dolshanii}). This state can be realized in tight waveguides by crossing
adiabatically the confinement induced resonance. The stability of the gas-like state
can be understood from a simple estimate of the energy per particle. For a contact
potential the interaction energy is given by (\ref{meanEcontact}) $E_{int}/N=g_{1D}
n g_2(0)/2$, where the
$g_2(0)=\langle\hat\Psi^\dagger(z)\hat\Psi^\dagger(z)\hat\Psi(z)\hat\Psi(z)\rangle/n^2$,
is the value at zero of the two-body correlation function (\ref{g2}) and
$\hat\Psi^\dagger$, $\hat\Psi$ are the creation and annihilation particle operators
(\ref{Psi}). In the limit $g_{1D}\to-\infty$ one can use for the correlation
function the result in the TG regime (\ref{g2TG})\cite{Gangardt03}, which does not
depend on the sign of $g_{1D}$. In the same limit the kinetic energy can be
estimated by (\ref{ETG}): $E_{kin}/N\simeq\pi^2\hbar^2n^2/(6m)$, corresponding to
the energy per particle of a TG gas. For the total energy $E=E_{kin}+E_{int}$ one
finds the result (\ref{Eexclud})
$E/N\simeq\pi^2\hbar^2n^2/(6m)-\pi^2\hbar^2n^3a_{1D}/(3m)$\label{HRdiscussion},
holding for $na_{1D}\ll 1$. For $na_{1D}<0.25$ this equation of state yields a
positive compressibility $mc^2=n\partial\mu/\partial n$, where $\mu=dE/dN$ is the
chemical potential and $c$ is the speed of sound, corresponding to a stable gas-like
phase. We will show that a more precise estimate gives that the gas-like state is
stable against cluster formation for $na_{1D}\lesssim 0.35$.

The analysis of the gas-like equation of state is carried out using the VMC
technique. The trial wave function employed in the calculation is of the form
(\ref{wfST}). For $g_{1D}<0$ ($a_{1D}>0$) the wave function $f(z)$ changes sign at a
nodal point which, for $\Rm\gg a_{1D}$, is located at $|z|=a_{1D}$. In the
attractive case the wave function has a node which means that the variational
calculation can be easily done, while without additional modifications the DMC can
not be done. The variational energy is calculated through the expectation value of
the Hamiltonian (\ref{Hamiltonian}) on the trial wave function (\ref{Eloc}). In the
calculations we have used $N=100$ particles with periodic boundary conditions and
because of the negligible dependence of the variational energy on the parameter
$\Rm$ we have used in all simulations the value $\Rm=L/2$, where $L$ is the size of
the simulation box. Calculations carried out with larger values of $N$ have shown
negligible finite size effects.

\section{Energy}

\begin{figure}[ht!]
\begin{center}
\includegraphics*[width=0.65\textwidth]{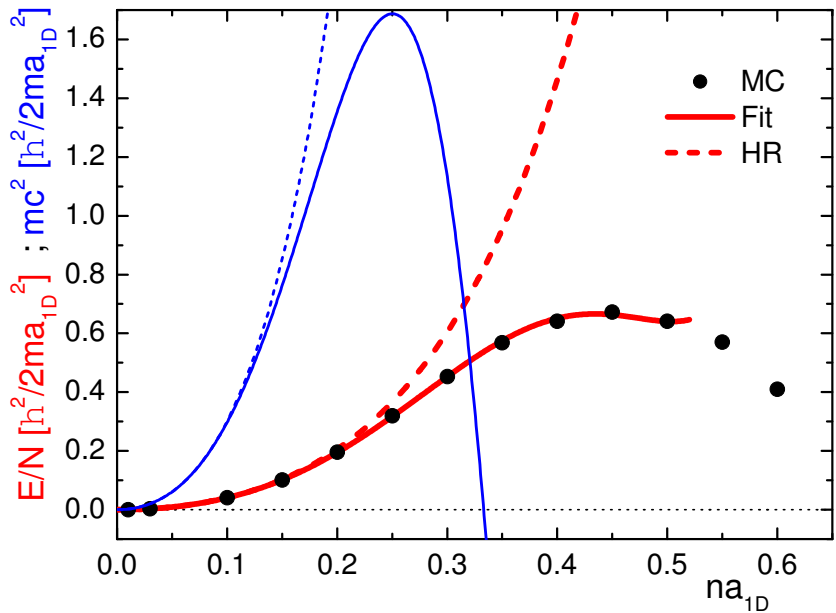}
\caption{Energy per particle and compressibility as a function of the gas parameter $na_{1D}$. Solid symbols and
thick solid line: VMC results and polynomial best fit; thick dashed line: HR equation of state [Eq. (\ref{EHR})].
Thin solid and dashed line: compressibility from the best fit to the variational equation of state and from the HR
equation of state respectively.}
\label{figHR1}
\end{center}
\end{figure}

The results for the variational energy as a function of the gas parameter $na_{1D}$
are shown in Fig.~\ref{figHR1} with solid symbols. For small values of the gas
parameter our variational results agree very well with the equation of state of a
gas of hard-rods (HR) of size $a_{1D}$ (thick dashed line). The HR energy per
particle can be calculated exactly from the energy of a TG gas by accounting for the
excluded volume (\ref{EHR}) \cite{Girardeau60}.

For larger values of $na_{1D}$, the variational energy increases with the gas
parameter more slowly than in the HR case and deviations are clearly visible. By
fitting a polynomial function to our variational results we obtain the best fit
shown in Fig.~\ref{figHR1} as a thick solid line. The compressibility obtained from
the best fit is shown in Fig.~\ref{figHR1} as a thin solid line and compared with
$mc^2$ of a HR gas (thin dashed line). As a function of the gas parameter the
compressibility shows a maximum and then drops abruptly to zero. The vanishing of
the compressibility implies that the system is mechanically unstable against cluster
formation. Our variational estimate yields $na_{1D}\simeq 0.35$ for the critical
value of the density where the instability appears. This value coincides with the
critical density for collapse calculated in the center of the trap for harmonically
confined systems \cite{Astrakharchik04c,Astrakharchik04a}. It is worth noticing that
the VMC estimate of the energy of the system can be extended beyond the instability
point, as shown in Fig.~\ref{figHR1}. This is possible since the finite size of the
simulation box hinders the long-range density fluctuations that would break the
homogeneity of the gas. This feature is analogous to the one observed in the quantum
Monte-Carlo characterization of the spinodal point in liquid $^4$He
\cite{Boronat94}.

As shown in Fig.~\ref{figHR1}, the HR model describes accurately the equation of
state for small values of the gas parameter. A similar accuracy is therefore
expected for the correlation functions of the system. The correlation functions of a
HR gas of size $a_{1D}$ can be calculated from the exact wave function
\cite{Nagamiya40} (\ref{wf1DHR}). We calculate the static structure factor $S(k)$
(\ref{Sk}) and the one-body density matrix $g_1(z)$ (\ref{OBDM})

\section{One-body density matrix and static structure factor}

\begin{figure}[ht!]
\begin{center}
\includegraphics*[width=0.4\textwidth,angle=-90]{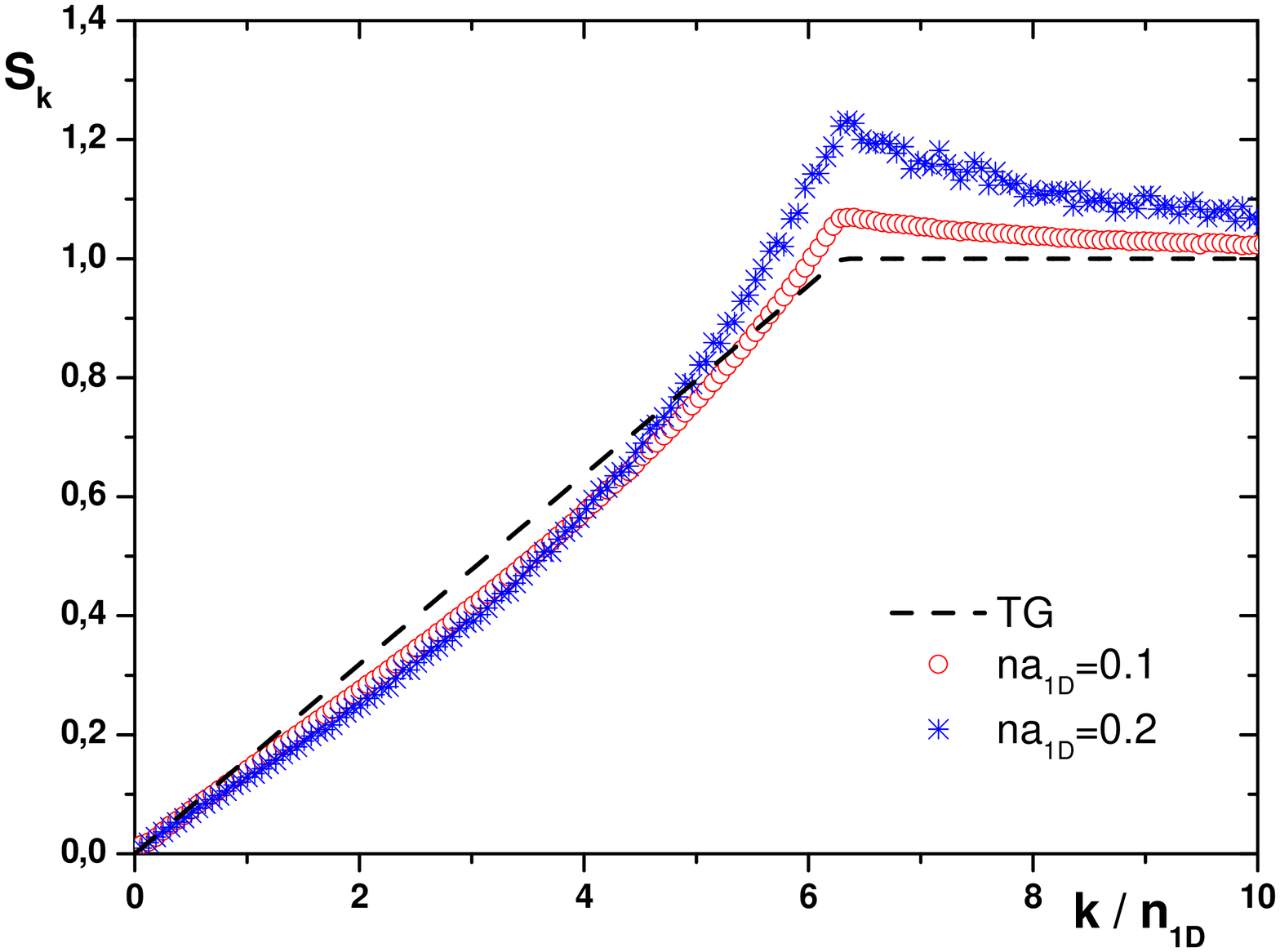}
\caption{Static structure factor $S(k)$ for a gas of HR at different values of the gas parameter $na_{1D}$
(symbols) and for a TG gas (dashed line).}
\label{figHR2}
\end{center}
\end{figure}

\begin{figure}[ht!]
\begin{center}
\includegraphics*[width=0.59\textwidth]{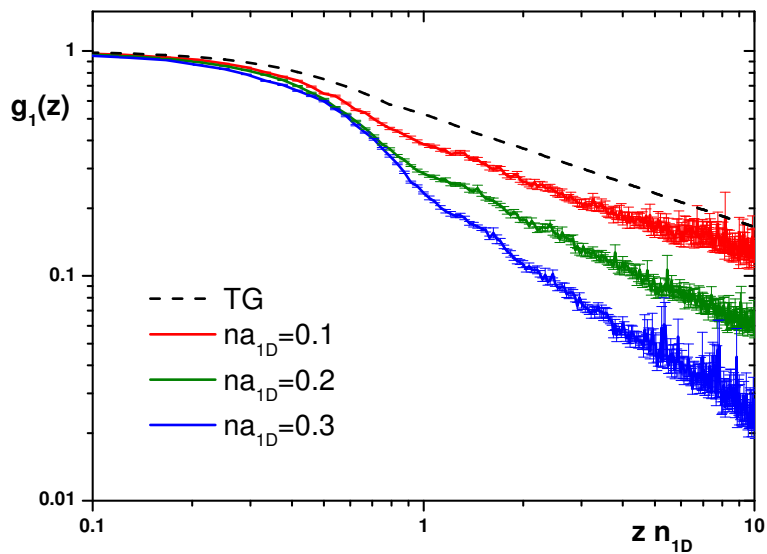}
\caption{One-body density matrix $g_1(z)$ for a gas of HR at different values of the gas parameter $na_{1D}$
(solid lines) and for a TG gas (dashed line). Higher values of density correspond
to a faster decay of $g_1(z)$.}
\label{figHR3}
\end{center}
\end{figure}

Contrarily to the TG case, it is not possible to obtain analytical expressions for
$g_1(z)$ and $S(k)$ in the HR problem. We have calculated them using configurations
generated by a Monte Carlo simulation according to the exact probability
distribution function $|\psi_{HR}|^2$. The results for the static structure factor
are shown in Fig.~\ref{figHR2}. Compared to $S(k)$ in the TG regime, a clear peak is
visible for values of $k$ of the order of twice the Fermi wave vector $k_F=\pi n$
and the peak is more pronounced as $na_{1D}$ increases. The change of slope for
small values of $k$ reflects the increase of the speed of sound $c$ with $na_{1D}$.
The long-range behavior of $g_1(z)$ can be obtained from the hydrodynamic theory of
low-energy excitations \cite{Reatto67,Schwartz77,Haldane81}. For $|z|\gg\xi$, where
$\xi=\hbar/(\sqrt{2}mc)$ is the healing length of the system, one finds the
following power-law decay (\ref{Casympt}):
\begin{equation}
g_1(z)\propto1/|z|^\alpha,
\label{OBDM1}
\end{equation}
where the exponent $\alpha$ is given by $\alpha=mc/(2\pi\hbar n)$. For a TG gas
$mc=\pi\hbar n$, and thus $\alpha_{TG}=1/2$. For a HR gas one finds
$\alpha=\alpha_{TG}/(1-na_{1D})^2$ and thus $\alpha>\alpha_{TG}$. This behavior at
long range is clearly shown in Fig.~\ref{figHR3} where we compare $g_1(z)$ of a gas
of HR with $na_{1D}=0.1$, 0.2 and 0.3 to the result of a TG gas \cite{Jimbo80}. The
long-range power-law decay of $g_1(z)$ is reflected in the infrared divergence of
the momentum distribution $n(k)\propto 1/|k|^{1-\alpha}$ holding for $|k|\ll 1/\xi$.
A gas of HR exhibits a weaker infrared divergence compared to a TG gas. The
correlation functions of a HR gas at $na_{1D}=0.1$, 0.2 should accurately describe
the physical situation of a Bose gas with large and negative $g_{1D}$. For
$na_{1D}=0.3$ we expect already some deviations from the HR model, as it is evident
from the equation of state in Fig.~\ref{figHR1}, which should broaden the peak in
$S(k)$ and decrease the slope of the power-law decay in $g_1(z)$ at large distances.
The analysis of correlation functions clearly shows that the super-Tonks regime
corresponds to a Luttinger liquid where short range correlations are significantly
stronger than in the TG gas.

\section{Collective modes}

\begin{figure}[ht!]
\begin{center}
\includegraphics*[width=0.6\textwidth]{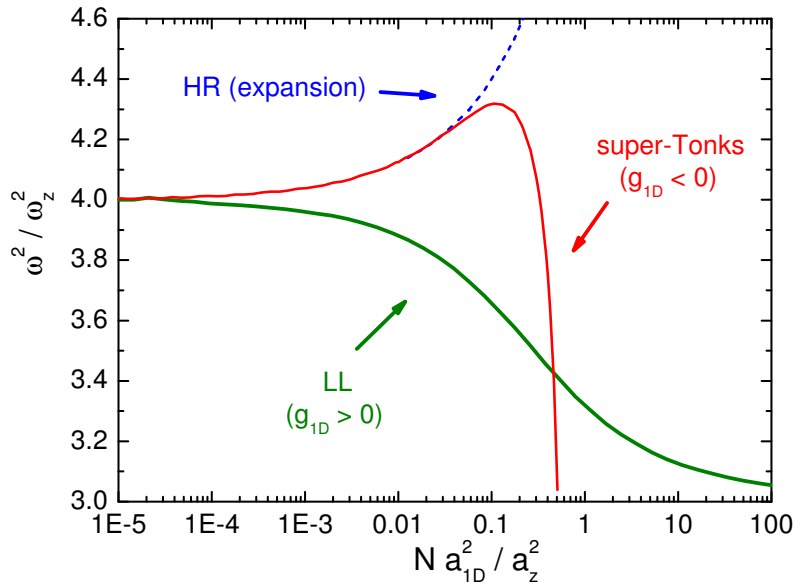}
\caption{Square of the lowest breathing mode frequency, $\omega^2$,
as a function of the coupling strength $Na_{1D}^2/a_z^2$ for the LL Hamiltonian
($g_{1D}>0$) and in the super-Tonks regime ($g_{1D}<0$). The dashed line is obtained
from the HR expansion (see Table~\ref{tableFrequencies}).}
\label{figHR4}
\end{center}
\end{figure}

Another possible experimental signature of the super-Tonks regime can be provided by
the study of collective modes. To this aim, we calculate the frequency of the lowest
compressional mode of a system of $N$ particles in a harmonic potential
$V_{ext}=\sum_{i=1}^N m\omega_z^2z_i^2/2$. We make use of LDA (Sec.~\ref{secLDA})
which allows us to calculate the chemical potential of the inhomogeneous system
$\tilde{\mu}$ and the density profile $n(z)$ from the local equilibrium equation
$\tilde{\mu}=\mu[n(z)]+m\omega_z^2z^2/2$, and the normalization condition
$N=\int_{-R}^R n(z) dz$, where $R=\sqrt{2\tilde{\mu}/(m\omega_z^2)}$ is the size of
cloud. For densities $n$ smaller than the critical density for cluster formation,
$\mu[n]$ is the equation of state of the homogeneous system derived from the fit to
the VMC energies (Fig.~\ref{figHR1}). From the knowledge of the density profile
$n(z)$ one can obtain the mean square radius of the cloud $\langle
z^2\rangle=\int_{R}^R n(z)z^2 dz /N$ and thus, making use of the result
\cite{Menotti02}
\begin{equation}
\omega^2=-2\frac{\langle z^2\rangle}{d\langle z^2\rangle/d\omega_z^2} \;,
\label{cmode}
\end{equation}
one can calculate the frequency $\omega$ of the lowest breathing mode. Within LDA,
the result will depend only on the dimensionless parameter $Na_{1D}^2/a_z^2$, where
$a_z=\sqrt{\hbar/m\omega_z}$ is the harmonic oscillator length. For $g_{1D}>0$, {\it
i.e.} in the case of the LL Hamiltonian, the frequency of the lowest compressional
mode increases from $\omega=\sqrt{3}\omega_z$ in the weak-coupling mean-field regime
($Na_{1D}^2/a_z^2\gg 1$) to $\omega=2\omega_z$ in the strong-coupling TG regime
($Na_{1D}^2/a_z^2\ll 1$). The results for $\omega$ in the super-Tonks regime are
shown in Fig.~\ref{figHR4} as a function of the coupling strength. In the regime
$Na_{1D}^2/a_z^2\ll 1$, where the HR model is appropriate, we can calculate
analytically the first correction to the frequency of a TG gas (refer to
Table~\ref{tableFrequencies}).
One finds the result $\omega=2\omega_z [1+(16\sqrt{2}/15\pi^2)(N
a_{1D}^2/a_z^2)^{1/2}+...]$. Fig.~\ref{figHR4} shows that this expansion accurately
describes the frequency of the breathing mode when $Na_{1D}^2/a_z^2\ll 1$, for
larger values of the coupling strength the frequency reaches a maximum and drops to
zero at $Na_{1D}^2/a_z^2\simeq 0.6$. The observation of a breathing mode with a
frequency larger than $2\omega_z$ would be a clear signature of the super-Tonks
regime.

\section{Conclusions}

In conclusion we have pointed out the existence of a strongly correlated regime in
quasi-1D Bose gases beyond the Tonks-Giradeau regime. This regime can be entered by
exploiting a confinement induced resonance of the effective 1D scattering amplitude.
We calculate the equation of state of the gas in the super-Tonks regime using VMC
and we estimate the critical density for the onset of instability against cluster
formation. The static structure factor and one-body density matrix are calculated
exactly within the hard-rod model, which provides the correct description of the
system for small values of the gas parameter. For harmonically trapped systems we
provide explicit predictions for the frequency of the lowest compressional mode.

\chapter{Motion of a heavy impurity through a Bose-Einstein condensate\label{secdelta}}
\section{Introduction}

\newcommand{\gi}{{g_{i}}}

One of the most important peculiarities of Landau theory superfluidity is the
existence of a finite critical velocity. If a body moves in a superfluid at $T=0$
with velocity $V$ less then $v_c$, the motion is dissipationless. At $V>v_c$ a drag
force arises because of the possibility of emission of elementary excitations.
However, both theoretical and experimental investigation in superfluid $^4$He are
difficult. The critical velocity in $^4$He is related to creation of rotons, for
which one has no simple theoretical description. Further, an important role is
played by complicated processes involving vortex rings production.

The situation in low-density weakly-interacting Bose-Einstein condensed (BEC) gases
is simpler. The Landau critical velocity in this case is due to Cherenkov emission
of phonons which can be described by mean-field theory. Due to the presence in the
theory of an intrinsic length parameter, the correlation length $\xi$, the friction
force for a small body does not depend on its structure. Vortex rings in the BEC
cannot have radius less then $\xi$ (see, {\it e.g.}, \cite{Jones82}) and it is reasonable
to believe that probability of their creation by a small body is small. Thus
quantitative investigation of critical velocities in BEC are very interesting and
can be used to probe the superfluidity of a quantum gas.

Recently existence of the critical velocity in a Bose-Einstein Condensed gas was
confirmed in a few experiments. At MIT a trapped condensate was stirred by a blue
detuned laser beam \cite{Raman99} and the energy of dissipation was measured.
The critical velocity was found to be smaller than the speed of sound due to
emission of vortices. The diameter of the laser spot in this experiment was of a
macroscopic size and was large compared to the healing length. An improved technique
allowed measurement of the drag force acting on the condensate in a subsequent
experiment \cite{Onofrio00}.

The analytical study of flow of the condensate over an impurity is highly nontrivial
due to the intrinsic nonlinearity of the problem arising from the interaction of the
particles in the condensate. In one dimension the dissipation could occur at
velocities smaller than predicted by Landau's approach due to emission of solitons
\cite{Hakim97}. The dependence of the critical velocity on the type of the potential
was studied both by using a perturbative approach and numerical integration in
\cite{Leboeuf01,Pavloff02}. The effective two dimensional problem was considered
in \cite{Kagan00}. In this work generation of excitations in the oscillating
condensate in a time dependent parabolic trap in the presence of a static impurity
was studied analytically. A three-dimensional flow of a condensate around an
obstacle was calculated numerically by integration of the Gross-Pitaevskii (GP)
equation and emission of vortices was observed \cite{Frisch92,Winiecki99}.

There are different definitions of the superfluidity. It is possible to make
following experiment. Move a small body through the system. According to Landau if
there is no normal part (we consider zero temperature or small enough) no
dissipation will happen if the speed is smaller than the speed of sound. Our goal is
to calculate effect of such a probe, a small impurity moving through a condensate
which is described by the Gross-Pitaevskii equation.

We want to find an answer to a question which is rather complicated. From one side
we know from usual considerations in the mean field regime that the system is
superfluid. From the other side we know that in the Tonks-Girardeau regime the
system is mapped on the fermions, which are, definitely, not superfluid. Indeed
what we find is that the situation is somewhere between.

\section{Three-dimensional system \label{secDelta3D}}

Let us consider an impurity moving through a three-dimensional condensate at $T=0$.
One of the possible realizations of this model could be scattering of heavy neutral
molecules by the condensate.

\subsection{Perturbed solution\label{total energy}}

We start from the three-dimensional energy functional (\ref{H1st}) of a homogeneous
weakly-interacting Bose gas in the presence of a $\delta $-function perturbation (an
impurity) moving with a constant velocity ${\bf V}$
\begin{equation}
E=\int\left(\frac{\hbar^2}{2m}|\nabla\psi|^2+(\mu-\gi\delta({\bf r}-{\bf V}t))|\psi |^2+\frac{g}2|\psi|^4\right) d^3r,
\end{equation}
where $\psi$ is the condensate wave function, $\mu$ is the chemical potential, $m$
mass of a particle in the condensate, $g=4\pi\hbar^2a/m$ and $\gi=2\pi\hbar^2b/m$
are particle-particle and particle-impurity coupling constants, with $a$ and $b$
being the respective scattering lengths\footnote{Considering a homogenious condensate, we assume that its
size is large enough, particularly that the gas is the Thomas-Fermi
conditions. Then results for a homogenious gas will be approximately valid for a gas in a trap.}. We will assume that the
interaction with impurity is small and we will use perturbation theory. By splitting
the wave function into a sum of the unperturbed solution and a small correction
$\psi ({\bf r},t)=\phi_0+\delta\psi ({\bf r},t)$ and linearizing the time-dependent GP equation
with respect to $\delta\psi$, we obtain an equation describing the time
evolution of $\delta\psi$
\begin{eqnarray}
i\hbar \frac{\partial}{\partial t}\delta\psi=
\left(-\frac{\hbar^2}{2m}\triangle -\mu +2g|\phi_0|^2\right) \delta\psi +
g|\phi_0|^2\,\delta\psi^*+\gi\,\delta ({\bf r}-{\bf V}t)\phi_0
\label{GPE delta Psi}
\end{eqnarray}

In a homogeneous system $\phi_0$ is a constant fixed by the particle density
$\phi_0=\sqrt{n}$ and $\mu =gn=mc^2$.

The perturbation follows the moving impurity, {\it i.e.} $\delta\psi$ is a function of
$({\bf r}-{\bf V}t)$, so the coordinate derivative is related to
the time derivative
\begin{eqnarray}
\partial\delta\psi({\bf r}-{\bf V}t)/\partial t=-{\bf V}\vec{\nabla}\delta\psi ({\bf r}-{\bf V}t)
\label{rule}
\end{eqnarray}

We shall work in the frame moving with the impurity ${\bf r}^{\prime}={\bf r}-{\bf
V}t$ and the subscript over ${\bf r}$ will be dropped.

Eq. (\ref{GPE delta Psi}) for a perturbation in a homogeneous system can be
conveniently solved in momentum space. In order to do this we introduce the
Fourier transform of the wave function $\delta\psi_{{\bf k}}=\int e^{-i
{\bf k\cdot}{\bf r}}\delta\psi ({\bf r})\,d^3r$. Eq. (\ref{GPE delta Psi}
) becomes
\begin{equation}
\left\{
\begin{array}{ccc}
\left(-\hbar{\bf k\cdot V}+\frac{\hbar^2k^2}{2m}+\mu\right)\delta\psi_{{\bf k}}+\mu\,(\delta\psi_{{\bf -k}})^*+\gi\,\phi_0&=&0\\
\mu\,\delta\psi_{{\bf k}}+\left(\hbar{\bf k\cdot V}+\frac{\hbar^2k^2}{2m}+\mu\right)(\delta\psi_{{\bf -k}})^*+\gi\,\phi_0&=&0\\
\end{array}
\right.
\label{GPE delta Psi fourier}
\end{equation}
Here the second equation is obtained by doing the substitution ${\bf k}\rightarrow
-{\bf k}$ and and complex conjugation. We also use property of the Fourier
transformation $\delta (\psi^*)_{{\bf k}} = \delta (\psi_{-{\bf k}})^*$. The system
of linear equations (\ref{GPE delta Psi fourier}) can be easily solved
\begin{equation}
\delta\psi_{{\bf k}}=\gi\phi_0\frac{\hbar {\bf k\cdot V}+
\frac{\hbar^2k^2}{2m}}{(\hbar {\bf k\cdot V})^2-\frac{\hbar^2k^2}{2m}
\left(\frac{\hbar^2k^2}{2m}+2\mu\right)}
\label{dPsi}
\end{equation}

\subsection{Total energy}

For a fixed value of the chemical potential $\mu$ the energy $E' = E-\mu N$ reaches
minimum on the ground state function $\phi_0$. From this it follows that $E'$ does
not have terms linear in $\delta\psi_{{\bf k}}$ and $\delta\psi_{{\bf k}}^*$, and
\begin{eqnarray}
E'= E^{(0)}+E^{(2)}+\gi(\phi_0^*\delta\psi(0)+\phi_0\delta\psi^*(0))
\label{Eprime}
\end{eqnarray}
Here $E^{(0)}=Ngn/2+\gi\phi_0^2$ is the energy of the system in absence of the
perturbation plus the mean-field shift in the energy due to the impurity. The next
term comes from the linear expansion of the energy $\int |\psi({\bf r})|^2 \gi
\delta({\bf r}){\bf dr}$. The term $E^{(2)}$ being quadratic in
$\delta\psi_{{\bf k}}$ and $\delta\psi_{{\bf k}}^*$ satisfies the Euler identity:
\begin{eqnarray}
2 E^{(2)} =
\int \left[
\delta \psi({\bf r}) \frac{\delta E^{(2)}}{\delta(\delta \psi({\bf r}))}
+\delta \psi^*({\bf r}) \frac{\delta E^{(2)}}{\delta(\delta
\psi^*({\bf r}))}
\right]{\bf dr}
\end{eqnarray}
which using the variational equation
\begin{eqnarray}
i \hbar \frac{\partial\delta(\delta \psi)}{\partial t} =
\frac{\delta E^{(2)}}{\delta(\delta\psi^*)} + \gi\phi_0\delta({\bf r})
\end{eqnarray}
can be rewritten as
\begin{eqnarray}
E^{(2)}=
\displaystyle\frac{i\hbar}{2} \int \left[
\delta \psi^*({\bf r}) \frac{\partial\delta \psi({\bf r})}{\partial t}
-\frac{\partial\delta \psi^*({\bf r})}{\partial t}\delta \psi({\bf r})
\right]{\bf dr}-\frac{\gi}{2}(\phi_0^*\delta\psi(0)+\phi_0\delta\psi^*(0))
\label{deltaE2b}
\end{eqnarray}

To start with, let us Fourier transform the first term. Exchanging time derivatives
with gradients by the rule (\ref{rule}) one obtains
\begin{eqnarray}
E^{(2)} = \int\hbar{\bf k\cdot V}|\delta\psi_{{\bf k}}|^2\frac{d^3k}{(2\pi)^3}+\frac{\gi\phi_0}2
(\delta\psi +\delta\psi^*)_{{\bf r}=0}
\label{E Psi}
\end{eqnarray}

In the energy calculation we assume that the velocity $V$ is small and will make an
expansion in powers of $V$ up to quadratic terms. It means that in the calculation
of $|\delta \psi_{{\bf k}}|^2$ the term $(\hbar{\bf kV})^2$ in the denominator of
(\ref{dPsi}) can be neglected and $|\delta \psi_{{\bf k}}|^2$ is written as
\begin{eqnarray}
|\delta \psi_{{\bf k}}|^2 = \frac{\gi ^2 |\phi_0|^2
\left[\left(\frac{\hbar^2 k^2}{2m}\right)^2 +
2\frac{\hbar^2 k^2}{2m} \hbar{\bf k V} \right]}
{\left[\frac{\hbar^2 k^2}{2m}
\left( \frac{\hbar^2 k^2}{2m}+2 \mu\right)\right]^2}
+ {\cal O}(V^4)
\label{dPsi2}
\end{eqnarray}

The energy does not have terms linear in ${\bf V}$, because all terms independent of
${\bf V}$ in (\ref{dPsi2}) are even in ${\bf k}$, so once multiplied by ${\bf k}$
and integrated over momentum space they provide zero contribution to the energy. The
only term that is left is the following
\begin{eqnarray}
E^{(2)} =
2\gi ^2|\phi_0|^2
\int \frac{(\hbar{\bf kV})^2}
{\frac{\hbar^2 k^2}{2m}
\left( \frac{\hbar^2 k^2}{2m}+2 \mu\right)^2}
\frac{\bf dk}{(2\pi)^3}
\label{E2}
\end{eqnarray}

For the calculation of $\delta\psi(0)$ in (\ref{Eprime}) and (\ref{deltaE2b}) one
should consider $\delta\psi_{{\bf k}}$ taking into account that $\hbar{\bf
kV}\ll\mu$ and then integrate it over the momentum space
\begin{eqnarray}
\begin{array}{rcl}
\delta \psi_{k}&=&
\displaystyle
 -\frac{\gi \left(\hbar{\bf k V} + \frac{\hbar^2 k^2}{2m}\right)\phi_0}
{\frac{\hbar^2 k^2}{2m} \left(\frac{\hbar^2 k^2}{2m}+2\mu\right)}
\left[
1-\frac{(\hbar{\bf k V})^2}{\frac{\hbar^2 k^2}{2m} \left(\frac{\hbar^2 k^2}{2m}+2\mu\right)}
\right]^{-1} \approx\\
&\approx&\displaystyle
\frac{(\hbar{\bf kV})^2}{\frac{\hbar^2 k^2}{2m}\left(\frac{\hbar^2 k^2}{2m}+2\mu\right)^2}
\gi \phi_0-\left\{
\frac{\hbar{\bf kV} \left[\frac{\hbar^2 k^2}{2m}\left(\frac{\hbar^2 k^2}{2m}+2\mu\right)\right]
+(\hbar{\bf kV})^3}
{\left[\frac{\hbar^2 k^2}{2m}\left(\frac{\hbar^2 k^2}{2m}+2\mu\right)\right]^2}
+\frac{1}{\frac{\hbar^2 k^2}{2m}+2\mu}
\right\}
\gi \phi_0
\end{array}
\label{expansion of deltaPsi}
\end{eqnarray}

The integral of the second term over momentum space is equal to zero.

The third term suffers from large-$k$ divergency and one should renormalize the
scattering amplitude. It is sufficient to express the coupling constant in the
$2^{nd}$ term of eq. (\ref{E Psi}) in terms of the scattering amplitude $b$ using
the second order Born approximation:
\begin{equation}
\gi=\frac{2\pi \hbar^2b}{m}\left(1+\frac{2\pi \hbar^2b}{m}\int
\left(\frac{\hbar^2k^2}{2m}\right)^{-1}\frac{d^3k}{(2\pi)^3}\right)
\end{equation}
Now the third term in (\ref{E Psi}) is converging and can be calculated
\begin{eqnarray}
\gi^2n\int\frac{2\mu}{\frac{\hbar^2 k^2}{2m}\left(\frac{\hbar^2 k^2}{2m}+2\mu\right)}\frac{d^3k}{(2\pi)^3}
=8\pi\sqrt{\pi} (na^3)^{3/2}\left(\frac{b}{a}\right)^2\frac{\hbar^2}{ma^2}
\end{eqnarray}

The energy shift quadratic in velocity $V$ is defined by the following integral
\begin{eqnarray}
\begin{array}{rcl}
\delta E&=&
\displaystyle
E^{(2)} \gi (\phi_0^*\delta\psi(0)+\phi_0\delta\psi^*(0))
= \gi^2n \int \frac{(\hbar{\bf kV})^2}
{\frac{\hbar^2 k^2}{2m}
\left(\frac{\hbar^2 k^2}{2m}+2 \mu\right)^2}
\frac{\bf dk}{(2\pi)^3}
\end{array}
\label{int}
\end{eqnarray}
Here we make use of relation $\phi_0 = \sqrt{n}$. In a three-dimensional case the
term $(\bf{kV})^2 d{\bf k}$ in the integral (\ref{int}) can be replaced by
$1/3~k^2 V^2 4\pi k^2 dk$ due to the equivalence of different directions.
\begin{eqnarray}
\delta E =
\frac{\gi^2n \hbar^2 V^2}{6\pi^2\left(\frac{\hbar^2}{2m}\right)^{5/2}}
\int\limits_0^\infty \frac{\frac{\hbar^2k^2}{2m}
\,d\left(\frac{\hbar k}{\sqrt{2m}}\right)}
{\left(\frac{\hbar^2k^2}{2m}+2 gn\right)^2}
\end{eqnarray}

This integral can be easily calculated if one recall the integral identity $\int
\frac{x^2dx}{(x^2+a^2)^2} = -\frac{x}{2(x^2+a^2)^2} +\frac{1}{2a}
\arctg\frac{x}{a}$. Finally, by collecting everything together
and considering $N_{imp}$ impurities with a concentration given by $\chi=N_{imp}/N$
we obtain the energy per particle
\begin{eqnarray}
\frac{E}{N}=\left\{2\pi na^3\left(1+\chi \frac{b}{a}\right)
+8\pi^{3/2}(na^3)^{3/2}\chi \left(\frac{b}{a}\right)^2\right\}
\frac{\hbar^2}{ma^2}
+\frac{2\sqrt{\pi}}3(na^3)^{1/2}\chi\left(\frac{b}{a}\right)^2\frac{mV^2}{2}
\label{dE}
\end{eqnarray}

If we set $V=0$ we recover Bogoliubov's corrections to the energy in the presence of
quenched impurities \cite{Huang92,Astrakharchik02a}. Note that even if the ``mean-field''
energy obtained from the GP equation in the absence of impurities ($\chi =0$) leaves
out terms of the order of $(na^3)^{3/2}$, the equations we obtain in the presence of
impurities in a perturbative manner still correctly describe the effect of the
disorder up to the terms of the order of $(na^3)^{3/2}$.

\subsection{Effective mass and normal fraction\label{effective mass}}

If $V\neq 0$ a quadratic term in the impurity contribution to the energy is present.
It can be denoted as $\chi m^* V^2/2$ with
\begin{equation}
m^*=\frac{2\sqrt{\pi}}3(na^3)^{1/2}\left(\frac{b}{a}\right)^2m
\label{mstar}
\end{equation}
being the induced mass, i. e. the mass of particles dragged by an impurity
\cite{Astrakharchik01}. Applicability of the perturbation theory demands $m^*$ to be small
compared to $m.$ This gives the condition $(na^3)^{1/2}\left(\frac{b}{a}\right)^2\ll
1$. At zero temperature the interaction between particles does not lead to depletion
of the superfluid density and the suppression of the superfluidity comes only from
the interaction of particles with impurities. Thus $\left(\ref{mstar}\right)$
defines the normal density
\begin{eqnarray}
\frac{\rho_n}{\rho} = \frac{m^*}{m} \chi =
\frac{2\sqrt{\pi}}{3} (na^3)^{1/2} \chi\left(\frac{b}{a}\right)^2
\label{SD}
\end{eqnarray}

This result is in agreement with the one obtained by the means of Bogoliubov
transformation starting from the Hamiltonian written in the second-quantized form in
the presence of disorder\cite{Huang92,Astrakharchik02a}. The normal density of a superfluid
is an observable quantity. It was evaluated in liquid $^4$He by measuring of the
moment of inertia of a rotating liquid or by measuring of the second sound velocity.
Both methods can be, in principle, developed for BEC gases.

\subsection{Drag force and energy dissipation \label{energy dissipation}}

The force with which the impurity acts on the system is
\begin{equation}
{\bf F}=-\int |\psi \left( {\bf r}\right) |^2\,\vec{\nabla}(\gi\delta
\left({\bf r}\right) )\,d^3r=\gi(\vec{\nabla}|\psi({\bf r})|^2)_{{\bf r}=0}
\end{equation}

Expanding the wave function into the sum of $\phi_0$ and $\delta\psi$ and neglecting
terms of order $\delta\psi^2$ we obtain
\begin{eqnarray}
{\bf F}=\gi\phi_0\int i{\bf k}\left[ \delta\psi_{{\bf k}}+\left(
\delta\psi_{{\bf -k}}\right)^*\right] \frac{d^3k}{(2\pi )^3}
=\int \frac{2\left(\gi\phi_0\right)^2i{\bf k}\left(\hbar
^2k^2/2m\right)}{(\hbar {\bf k\cdot V}+i0)^2-\frac{\hbar^2k^2}{2m}
\left(\frac{\hbar^2k^2}{2m}+2\mu\right)}\frac{d^3k}{\left( 2\pi\right)^3},
\label{force}
\end{eqnarray}
where we added an infinitesimal positive imaginary part $+i0$ to the frequency ${\bf
k\cdot V}$according to the usual Landau causality rule. The drag force is obviously
directed along to the velocity ${\bf V}$. The integration (\ref{force}) can be done
by using the formula $\frac{1}{x+i0}={\cal P}\frac{1}{x}-i\pi \delta (x)$. Due to
the integration between symmetric limits, only the imaginary part contributes to the
integral and the final value is real.
\begin{eqnarray}
{\bf F}=2(\gi\phi_0)^2 \int
\frac{i{\bf k}\frac{\hbar^2 k^2}{2m}(-i\pi)}{2\hbar{\bf k V}}
\mbox{\small $
\left[
\delta\!\left(\hbar{\bf k V}\!\!-\!\!\sqrt{\frac{\hbar^2 k^2}{2m} \left(\frac{\hbar^2
k^2}{2m}\!+\!2\mu\right)}\right)
\!+\!\delta\!\left(\hbar{\bf k V}\!\!+\!\!\sqrt{\frac{\hbar^2 k^2}{2m} \left(\frac{\hbar^2
k^2}{2m}\!+\!2\mu\right)}\right)
\right]$}
\,\frac{d^3k}{(2\pi)^3}
\label{y}
\end{eqnarray}

It is convenient to do the integration in spherical coordinates with $\vartheta$
being angle between $\bf k$ and $\bf V$. There is no dependence on the angle $\phi$
and it can be immediately integrated out
\begin{eqnarray}
\int f({\bf k})\,d^3k =
\int\limits_{0}^{\infty}\!\!\int\limits_0^\pi
f(k, \vartheta)\,2\pi k^2 \sin \vartheta \; dk d\vartheta
= \int\limits_{0}^{\infty}\int\limits_{-1}^1
f(k, \cos\vartheta)\,2\pi k^2 dk d(\cos\vartheta)
\label{3DintegrRule}
\end{eqnarray}

The $\delta$-function can be further developed
\begin{eqnarray}
\delta\left(\hbar{\bf k V}\pm\sqrt{\frac{\hbar^2 k^2}{2m} \left(\frac{\hbar^2
k^2}{2m}+2\mu\right)}\right) =
\frac{1}{\hbar kV}\delta\left(\cos\vartheta\pm
\frac{1}{\hbar kV}\sqrt{\frac{\hbar^2 k^2}{2m} \left(\frac{\hbar^2
k^2}{2m}+2\mu\right)}\right)
\end{eqnarray}

The poles in the integration over $\cos\vartheta$ appear if the square root in the
denominator is smaller than one, which leads to the restriction on the values of
momentum which contribute
\begin{equation}
|k| \leq k_{\max} = 2m (V^2-c^2)^{1/2}/\hbar
\label{kmax}
\end{equation}

Thus the energy dissipation takes place only if the impurity moves with a speed
larger than the speed of sound.

Let us calculate the projection $F_V$ of the force ${\bf F}$ to the direction $\bf
V$of the movement of the perturbation. It means that we have to multiply formula
(\ref{y}) on ${\bf V}/V$
\begin{eqnarray}
\nonumber
F_V =
(\gi\phi_0)^2~2\!\!\!\!\!\!\!\!\int\limits_0^{2m(V^2-c^2)}
\frac{\frac{\hbar^2 k^2}{2m}\pi}{\hbar V}
\frac{1}{\hbar kV} \frac{2\pi k^2dk}{(2\pi)^3}
=\frac{(\gi\phi_0)^2m}{2\pi \hbar^4 V^2}
\frac{1}{2}\left(2m(V^2-c^2)\right)^2
\end{eqnarray}

Now we can use that square of the unperturbed wave function gives the density
$\phi_0^2 = n$ and the coupling constant can be expressed as
\begin{eqnarray}
\gi = \frac{2\pi\hbar^2 b}{m},
\end{eqnarray}
which can be obtained from the formula (\ref{g3D}) recalling that the reduced mass
is $\mu = m$ for the scattering on a quenched impurity.

Finally, we obtain following expression for the projection of the force
\begin{equation}
F_{V}=4\pi nb^2mV^2(1-c^2/V^2)^2
\label{F3D}
\end{equation}

The energy dissipation, $\dot{E}=-F_{V}V$, can be evaluated by measuring the heating
of the gas.

For large $V$ the force is proportional to $V^2$. The energy dissipation per unit
time can then be presented as $\dot{E}=-\gamma E$ with the damping rate $\gamma \sim
nb^2V.$

Note in conclusion that our perturbative calculations can not describe processes
involving dissipation of energy due to creation of quantized vortex rings. Such a
creation is possible at $V<c$ but has a small probability for low velocity and for a
weak point-like impurity.

\section{Low dimensional systems \label{low dimensional systems}}

In this type of experiment the role of the impurity can also be played by a laser
beam with small enough size and intensity. The Fourier components of the perturbed
wave function $\delta\psi_{{\bf k}}$ are given by the formula (\ref{dPsi}), which is
derived in an arbitrary number of dimensions. The only difference is in the
substitution of $d^3k/(2\pi)^3$ with $d^Dk/(2\pi)^D$ in the integrals:
\begin{eqnarray}
F_V =
\frac{2i\gi^2\phi_0^2}{V}
\int \frac{{\bf k V} \frac{\hbar^2 k^2}{2m}}
{(\hbar{\bf k V})^2-\frac{\hbar^2 k^2}{2m} \left(\frac{\hbar^2 k^2}{2m}+2mc^2\right)}
\,\frac{\bf dk}{(2\pi)^D}
\label{FVlowD}
\end{eqnarray}

In the expression for the energy (\ref{expansion of deltaPsi}), the term quadratic
in velocity is of a great interest, as the coefficient in front of $V^2/2$ has
physical meaning of an effective mass. We develop further this term:
\begin{eqnarray}
\Delta E = \gi^2\phi_0^2\int\frac{(\hbar{\bf kV})^2}{\frac{\hbar^2 k^2}{2m}
\left(\frac{\hbar^2 k^2}{2m}+2 mc^2\right)^2}\frac{d^Dk}{(2\pi)^D}
= \frac{\gi^2\phi_0^2(2m)^3V^2}{D\hbar^4}
\int \frac{1}{(k^2+(2mc/\hbar)^2)^2}\frac{d^Dk}{(2\pi)^D},
\label{dED}
\end{eqnarray}
where we used symmetry properties
$\int f(k) ({\bf kV})^2 d^Dk = \frac{1}{D}\int f(k) (kV)^2 d^Dk$.

\subsection{Two-dimensional system \label{2D}}

There are different possible geometries of the experiment. One can create a
two-dimensional perturbation in the three-dimensional condensate. Such a
two-dimensional impurity can be created, analogously to the MIT experiment
\cite{Raman99,Onofrio00}, by a thin laser beam. Such a beam creates a
cylindrical hole in the condensate, which is stirred by moving the position of the
laser beam. Another possibility is to fix the position of the laser beam along the
long axis of an elongated condensate, so that the dissipation can be studied by
shaking the trap and exciting the breathing modes. The problem is to create a beam
with a diameter which is small with compared to the correlation length. The theory
can be easily generalized for beams of finite diameter. The intensity of the beam
can be tuned to satisfy the condition of a weak perturbation.

The more interesting possibility is the investigation of true two-dimensional
condensates, which can be created in plane optical traps, produced by a standing
light wave. If the light intensity is large enough, tunneling between planes is
small and the condensates behave as independent two dimensional systems. The
impurity can again be created by a laser beam perpendicular to the condensate plane.
Another possibility is to use impurity atoms, which can be drive by a laser beam,
with a frequency close to the atomic resonance of the impurity.

We expand the two dimensional differential $d^2k$ by its representation in the
polar coordinates
$d^2k=kdkd\vartheta=-\frac{k}{\sqrt{1-\cos^2\vartheta}}dk\,d\!\cos\vartheta$.
The 3D integrate rule (\ref{3DintegrRule}) should be substituted by
\begin{eqnarray}
\int_{-\infty}^\infty dk_x\int_{-\infty}^\infty dk_y
f(k_x,k_y) = 2\int_0^\infty dk\int_{-1}^1 d\cos\vartheta
\left(\frac{kf(k,\cos\vartheta)}{\sqrt{1-\cos^2\vartheta}}\right)
\label{2DintegrRule}
\end{eqnarray}

\subsubsection{Dragg force}

Now the projection of the force ${\bf F}$ onto the direction of motion ${\bf V}$
is given by the integral (\ref{FVlowD})
\begin{eqnarray}
\label{F2Da}
F^{2D}_V = -\frac{4i\gi^2\phi_0^2}{V}
\int\limits_0^\infty\int\limits_{-1}^1\frac{kV\cos\vartheta\frac{\hbar^2 k^2}{2m}}
{(\hbar kV\cos\vartheta )^2-\frac{\hbar^2 k^2}{2m} \left(\frac{\hbar^2 k^2}{2m}+2mc^2\right)}
\frac{k}{\sqrt{1-\cos^2\vartheta}}
\frac{dk\,d\!\cos\vartheta}{(2\pi)^2} = \qquad\\
=\!-\!\frac{i\gi^2\phi_0^2}{4\pi^2 mV^2}
\int\limits_0^\infty\!\!\!\int\limits_{-1}^1\!\!
\left(\frac{1}{\cos\vartheta\!+\!\sqrt{\left(\frac{\hbar^2k^2}{2m}+2mc^2\right)/2mV^2}}
+\frac{1}{\cos\vartheta - \sqrt{\left(\frac{\hbar^2k^2}{2m}+2mc^2\right)/2mV^2}}\right)\!\!
\frac{k^2\,dk\,d\!\cos\vartheta}{\sqrt{1-\cos^2\vartheta}}
\nonumber
\end{eqnarray}

In the following we will introduce a two-dimensional density $n_{2D} = N/L^2$. The
square of the unperturbed homogeneous solution equals to it $\phi_0^2 = n_{2D}$. The
integral (\ref{F2Da}) is different from zero only if integrand has poles, which
means that the velocity $V$ must be larger than the speed of sound $c$. Only momenta
smaller than $k_{max}$ (see eq.(\ref{kmax})) contribute to the integral
\begin{eqnarray}
F^{2D}_V = -\frac{i\gi^2n_{2D}}{4\pi^2mV^2} \int\limits_0^{k_{max}}\frac{2\pi i k^2\,dk}
{\sqrt{1-\left(\frac{\hbar^2k^2}{2m}+2mc^2\right)/2mV^2}}
= \frac{\gi^2n_{2D}}{2\pi mV^2} \int\limits_0^{k_{max}}
\frac{2mV k^2\,dk} {\hbar\sqrt{\frac{4m^2(V^2-c^2)}{\hbar^2}-k^2}}
\end{eqnarray}

We recall simple integral equality
$\int_0^\varkappa\frac{k^2\,dk}{\sqrt{\varkappa^2-k^2}}=\frac{\pi}{4}\varkappa^2$
and finally have
\begin{eqnarray}
F^{2D}_V = \frac{\gi^2 n_{2D}}{\hbar^3V}(V^2-c^2).
\label{FV2}
\end{eqnarray}

In a quasi two-dimensional system, {\it i.e.} when the gas is confined in the
$z$-direction by the harmonic potential $m\omega_z^2z^2/2$, the two-dimensional
coupling constant equals (see \ref{g2D})
\begin{eqnarray}
\gi^{2D}=\sqrt{2\pi}\frac{\hbar^2b}{ma_z},
\end{eqnarray}
where $a_z=\sqrt{\hbar /m\omega_z}$ is the oscillator length and $b$ is the
three-dimensional scattering length. We consider here only the mean-field $2D$
situation. See \cite{Pitaevskii03},\S17 and \cite{Petrov04b} for a more detail
discussion.

Notice again that our calculations do not take into account creation of vortex pairs
which is possible at $V<c$.

\subsubsection{Effective mass}

The energy depending on the velocity contribution is given by an integral
(\ref{dED}), which can be easily calculated
\begin{eqnarray}
\Delta E^{2D}(V)
=\frac{\gi^2n_{2D}(2m)^3V^2}{2\hbar^4}
\int_0^\infty \frac{1}{(k^2+(2mc/\hbar)^2)^2}
\frac{k dk}{(2\pi)^2}
= \frac{\gi^2n_{2D}m}{4\pi^2\hbar^2c^2}
\frac{V^2}{2}
\end{eqnarray}

From this result we infer the effective mass
\begin{equation}
m^*=\gi^2n_{2D} m/(4\pi^2\hbar^2c^2)
\label{mstar2}
\end{equation}

\subsection{One-dimensional system. Mean-field theory}
\subsubsection{Dragg force}

In one dimension the integration is straightforward. From (\ref{FVlowD}) we find
\begin{equation}
F^{1D} =
-\frac{imn \gi^2}{\pi\hbar^2}
\int\limits_{-\infty}^\infty
\left(
\frac{1}{k+2m\sqrt{V^2-c^2}/\hbar}
+\frac{1}{k-2m\sqrt{V^2-c^2}/\hbar}
\right)\,dk
\end{equation}

The integration over $k$ gives $2\pi i$ if $V>c$ and zero otherwise. So, the force is
\begin{eqnarray}
F^{1D} = \frac{2\gi^2n_{1D}m}{\hbar^2},
\end{eqnarray}
where $n_{1D} = N/L$ is the linear density. In a quasi one dimensional system ({\it
i.e.} a very elongated trap or a waveguide) there are no excitations in the radial
harmonic confinement and the coupling constant is obtained from (\ref{g1D}) keeping
in mind that the reduced mass equals to the mass of an incident particle $\mu=m$ for
the scattering on a heavy impurity
\begin{eqnarray}
\gi=-\frac{\hbar^2}{mb_{1D}}
\label{g1Dimp}
\end{eqnarray}

For the non-resonance scattering $b_{1D}=-a_{\perp}^2/b$, where
$a_{\perp}=\sqrt{\hbar/m\omega_{\perp}}$. The expression of the force
in terms of the scattering length reads as
\begin{eqnarray}
F^{1D}=2n_{1D}\hbar^2/mb_{1D}^2
\label{F1D}
\end{eqnarray}

An interesting peculiarity is that the result does not depend on the velocity $V$
(where, of course, the velocity must be larger than the speed of sound). This
phenomenon comes from particular properties of a $\delta$-potential, namely that the
Fourier transform of this potential is a constant. Numerical solutions by
Pavloff\cite{Pavloff02} for finite-range potentials in $1D$ show no friction for
$V<c$, maximal friction for $V\ge c$ and smaller friction for $V\gg c$, although the
constant result (\ref{F1D}) was found for the $\delta$-potential.

In a 1D system energy dissipation is possible at $V<c$ due to creation of the ``gray
solitons'' first considered in \cite{Tsuzuki71}. Non-linear calculations \cite{Hakim97}
show that the critical velocity for this process decreases with increasing coupling
constant $\gi$.

This theory can be checked in an experiment in a three-dimensional condensate. The
impurity can be presented by a moving light sheet.

\subsubsection{Effective mass}

The energy term (\ref{dED}), which depends on the velocity ${\bf V}$ can be
trivially calculated by using of the integral equality
$\int \frac{dx}{(x^2+a^2)^2} =\frac{1}{2a^3}\arctg\frac{x}{a}+\frac{x}{2a^2(x^2+a^2)}$.
The result of the integration is $\Delta E^{1D}(V) = \frac{\gi^2 nV^2}{4\hbar c^3}$
and the effective mass is given by
\begin{eqnarray}
m^*=\gi^2n_{1D}/2\hbar c^3.
\end{eqnarray}

It can be expressed in terms of the particle-particle $a$ and
particle-impurity $b$ scatering lengths
$m^* = \frac{1}{\sqrt{32n_{1D}a}}\left(\frac{a}{b}\right)^2m$

\subsubsection{Density profile}

The wave function of the perturbation, $\delta\psi_k$, was obtained in the momentum
representation and is given by expression (\ref{dPsi}). The spatial dependence,
$\delta\psi(x)$, is related to $\delta\psi_k$ by means of the Fourier
transformation. We will find the density profile $n(x)= |\psi(x)|^2$. Within the
same level of accuracy, as in the calculations above, $n(x)$ is given by
\begin{eqnarray}
n(x) \approx \phi_0^2+\phi_0(\delta\psi(x)+\delta\psi^*(x))
\end{eqnarray}

In terms of Fourier components one has
$\delta\psi(x) = \int e^{ikx}\delta\psi_k\frac{dk}{2\pi}$
\begin{eqnarray}
n(x)
= n_0 +\phi_0\int e^{ikx}(\delta\psi_k+(\delta\psi_{-k})^*)\frac{dk}{2\pi}
\end{eqnarray}
where we used property of the Fourier transform $(\delta\psi^*)_k =
(\delta\psi_{-k})^*$. Together with (\ref{dPsi}) and (\ref{g1Dimp}) we obtain
a simple expression
\begin{eqnarray}
n(x)
= n_0\left(1 + \frac{4}{b_{1D}}\int\limits_{-\infty}^\infty
\frac{e^{ikx}}{k^2+\frac{4m^2(c^2-V^2)}{\hbar^2}}\frac{dk}{2\pi}
\right)
\label{n}
\end{eqnarray}

There are two cases to be considered separately:
\begin{itemize}
\item[1)] The impurity moves with velocity smaller than the speed of sound.
We introduce the notation $\varkappa=2m\sqrt{c^2-V^2}/\hbar>0$ and note that the
integral has form of the inverse Fourier transform of the Yukawa potential:
\begin{eqnarray}
\int\limits_{-\infty}^\infty \frac{e^{ikx}}{k^2+\varkappa^2}\frac{dk}{2\pi}
=\frac{\exp\{-\varkappa|x|\}}{2\varkappa}
\end{eqnarray}

Thus, the density perturbation has a form of a bump and decays exponentially fast:
\begin{eqnarray}
n(x)
= n_0 \left(1 + \frac{2e^{-\varkappa\,|x|}}{\varkappa b_{1D}} \right)
\end{eqnarray}

For a repulsive interaction with the impurity the scattering length is negative
$b_{1D}<0$ and the density is suppressed by the presence of the impurity. Instead an
attractive interaction $b_{1D}>0$ leads to an increase in the density.

\item[2)] The impurity moves with velocity larger than the speed of sound.
In this case we introduce $\varkappa$ in the following way
$\varkappa=2m\sqrt{V^2-c^2}/\hbar>0$. There are poles appearing in the function in
the integral. We use Landau casuality rule $k\to k+i 0$ in order to modify the
integration contour.

In this case for $x>0$ the pole is absent and the integral vanishes. This means that
there is no perturbation in front of the impurity (impurity moves to the right).

Instead for $x<0$ the pole is present and the integral is different from zero.
\begin{eqnarray}
\int\limits_{-\infty}^\infty \frac{e^{ikx}}{k^2-\varkappa^2}\frac{dk}{2\pi}
= \frac{\sin \varkappa x}{\varkappa}
\end{eqnarray}
so the density profile behind the perturbation is oscillating and corresponds to
the wake generated by the moving impurity
\begin{eqnarray}
n(x)=
\left\{
\begin{array}{ll}
n_0 \left(1 + \frac{4}{\varkappa\,b_{1D}}\sin\varkappa x\right)& x<0\\
n_0 & x>0 \\
\end{array}
\right.
\end{eqnarray}
\end{itemize}

The condition of the applicability of the perturbation theory demands the
perturbation $|n(x)-n_0|$ be small compared to the unperturbed solution $n_0$. This
condition is satisfied if the velocity of the impurity $V$ is not to close to the
speed of sound $c$.

\subsection{One-dimensional system. Bethe-ansatz theory}

We saw in the previous subsection that for a weakly interacting impurity the drag
force appears only when the impurity velocity $V$ is larger than the Landau critical
velocity, which is equal to the velocity of sound $c$. The situation is, however,
different in the Bethe-ansatz Lieb-Liniger theory of a 1D Bose gas \cite{Lieb63}.
According to this theory excitations in the system actually have a fermionic nature.
Even a low frequency perturbation can create a particle-hole pair with a total
momentum near $2p_{F}\equiv 2\hbar k_{F}=\hbar 2\pi n_{1D}$. To calculate the drag
force for this case we will use the dynamic form factor of the system $\sigma
\left(\omega, k\right)$ (we follow notation of \cite{Lifshitz80}, \S87). The dissipated
energy at $T=0$ can be calculated as
\begin{equation}
\dot{E}=-\int\limits_{-\infty}^{\infty}\frac{dk}{2\pi}\int\limits_0^\infty
\frac{d\omega}{\pi}\omega\frac{n_{1D}}{2\hbar}\sigma
\left(\omega,k\right) \left| U\left(\omega, k\right) \right|^2,
\label{Q}
\end{equation}
where $U(\omega, k)=2\pi \gi\delta (\omega -kV)$ is the Fourier transform of the
impurity potential{\bf}$U(t,z)=\gi\delta (z-Vt)$. One has
$\left|U\left(\omega,k\right)\right|^2=2\pi\gi^2t\delta(\omega-kV)$, where $t$ is
''time of observation''. Thus the energy dissipation per unit of time is
\begin{equation}
\dot{E}=-F_{V}V=-\frac{\gi^2n_{1D}V}{\hbar}\int\limits_0^{\infty}%
\frac{dk}{2\pi}k\sigma\left(kV,k\right),
\label{deltaF}
\end{equation}
where $F_{V}$ is the drag force. We will try to estimate the velocity dependence of
$F_{V}$.

For low frequency dissipation the important values of $k$ are near $2k_{F}$.
According to \cite{Neto94}
\begin{equation}
\sigma \left(\omega, 2k_{F}\right) \sim \omega^{\left(\eta -2\right)
},\omega \rightarrow 0,
\label{omega}
\end{equation}
where $\eta =\frac{2\hbar k_{F}}{mc}=\frac{2\pi \hbar n_{1D}}{mc}\geq 2$ is the
characteristic parameter of a 1D Bose gas. In the mean-field limit when
$n_{1D}\rightarrow\infty$ the parameter $\eta\rightarrow \infty$. In the opposite
case of a small density bosons behave as impenetrable particles (Tonks-Girardeau
limit \cite{Girardeau60}) and the dynamic form-factor coincides with the one of an
ideal Fermi gas. In this limit $\eta = 2$.

In the general case one can calculate $\sigma\left(\omega,k\right)$ at small
$\omega$ and $k\approx 2k_{F}$ generalizing the method of Haldane \cite{Haldane81} for
the case of time-dependent correlation functions. Calculations give
\begin{equation}
\sigma \left(\omega, k\right) =\frac{n_{1D}c}{\omega^2}\left(\frac{\hbar
\omega}{mc^2}\right)^{\eta}f\left(\frac{c\Delta k}{\omega}\right),
\omega>0,k>0,
\label{So}
\end{equation}
where $k=2k_{F}+\Delta k$ and the function $f(x)$ is
\begin{equation}
f\left(x\right) = A\left(\eta\right)\left(1-x^2\right)^{\eta/2-1}
\end{equation}
in the interval $\left| x\right| <1$ and is equal to zero at $|x|\geq 1$ (see also
\cite{Korepin93}). The constant $A\left(\eta \right)$ can be calculated in two
limiting cases: $A\left(\eta =2\right) =\pi/4$
(see \cite{Pitaevskii03} \S17.3) and $A\left(\eta\right)
\approx 4\pi^2/\left[\left(8C\right)^{\eta}\Gamma^2\left(\frac{\eta}2\right)\right]$,
where $C=1.78$... is the Euler 's constant (see Eq.~\ref{Skw}), for $\eta \gg 1$.

Substituting $(\ref{So})$ into $(\ref{deltaF})$ we finally find velocity dependence
of the drag force:
\begin{equation}
F_{V}=\frac{\Gamma \left(\frac{\eta}2\right)}{2\sqrt{\pi}\Gamma \left(
\frac{\eta +1}2\right)}A\left(\eta \right) \frac{\gi^2n_{1D}^2}{%
\hbar V}\left(\eta \frac{V}{c}\right)^{\eta}
\label{FV}
\end{equation}
Equation $\left(\ref{FV}\right)$ is valid for the condition $V\ll c$.

Thus in the Tonks-Girardeau strong-interaction limit $F_{V}\sim V$ and Bose gas
behaves, from the point of view of friction, as a normal system, where the drag
force is proportional to the velocity. On the contrary, in the mean-field limit the
force is very small and the behavior of the system is analogous to a 3D superfluid.
However, even in this limit the presence of the small force makes a great
difference. Let us imagine that our system is twisted into a ring, and that the
impurity rotates around the ring with a small angular velocity. If the system is
superfluid in the usual sense of the word, the superfluid part must stay at rest.
Presence of the drag force means that equilibrium will be reached only when the gas
as a whole rotates with the same angular velocity. From this point of view the
superfluid part of the 1D Bose gas is equal to zero even at $T=0$. Notice that in an
earlier paper \cite{Sonin71} the author concluded that $\rho_{s}=\rho$ at $T=0$ for
arbitrary $\eta $. We believe that this difference results from different
definitions of $\rho_{s}$ and reflects the non-standard nature of the system.

Equation $\left(\ref{FV}\right)$ is equivalent to a result which was obtained by a
different method in \cite{Buchler01}, with a model consisting of an impurity considered
as a Josephson junction. Notice that the process of dissipation, which in the
language of fermionic excitations can be described as creation of a particle-hole
pair, corresponds in the mean-field limit to creation of a phonon and a small-energy
soliton. It seems that such a process cannot be described in the mean-field approach
in the linear approximation.

Experimental confirmation of these quite non-trivial predictions demands a true
one-dimensional condensate, where non mean-field effects can be sufficiently large.
Such condensates have been investigated for the first time in experiments
\cite{Schreck01,Gorlitz01}. In experiments \cite{Gorlitz01,Stoferle04} condensates have been created in the
form of elongated independent ''needles'' in optical traps, consisting of two
perpendicular standing laser waves. The role of an impurity in this case must be
played by a light sheet, perpendicular to the axis of condensates and moving along
them.

Notice also, that application of the additional light waves in this experiments of
this type allows one to create a harmonic perturbation of the form
\begin{equation}
U(t,z)=U_0\cos \left(\omega t-kz\right),\qquad k=2k_{F}+\Delta k
\end{equation}
with small $\omega$ and $\Delta k$. Such potential with was used in
\cite{Gorlitz01,Stoferle04} for experiments with 1D condensate in a periodic
lattice. However, for a small amplitude $U_0$, measurement of the dissipation energy
$Q$ gives, according to $\left(\ref{Q}\right)$, the dynamic form-factor
$S\left(\omega, k\right)$ directly.

\section{Conclusions \label{conclusions}}

We have studied motion of an impurity through the condensate at zero temperature by
considering the perturbation of a stationary solution of the GP equation. We
calculated the induced mass which contributes to the mass of normal component. We
find that the motion at small velocities is dissipationless in one-, two-, and
three- dimensional systems, although movement with velocities larger than the speed
of sound leads to a non-zero drag force due to Cherenkov radiation of phonons. The
expressions for the drag force are calculated. We used results for the dynamic form
factor of exact Lieb-Liniger theory to investigate the velocity dependence of the
drag force in a 1D system. The form factor was calculated with the help of the
Haldane method of calculations of correlation functions. The drag force exists at an
arbitrarily small velocity of motion, but is very small in the mean-field limit. The
dynamic form-factor can be also directly measured by applying a harmonic
time-dependent perturbation on one-dimensional condensates \cite{Gorlitz01,Stoferle04}.

\chapter{Interacting fermions in highly elongated harmonic traps\label{sec1Dfermions}}
\section{Introduction}

The study of cold quasi one dimensional atomic quantum gases presents a very active
area of research. So far, most of the experimental
\cite{Gorlitz01,Schreck01,Greiner01,Tolra04,Moritz03} and theoretical
\cite{Olshanii98,Petrov00,Dunjko01,Menotti02,Girardeau01} investigations have been
devoted to quasi-one dimensional Bose gases and, in particular, to the
strongly-interacting Tonks-Girardeau gas, which can be mapped to a gas of
non-interacting fermions~\cite{Girardeau60,Olshanii98,Reichel03}. Quasi-1D
two-component atomic Fermi gases have not been realized experimentally yet; however,
their realization in highly-elongated, needle-shaped traps is within reach of
present-day techniques.
The behavior of quasi one dimensional two-component Fermi gases can, if the
confinement is chosen properly, be characterized to a very good approximation by an
effective 1D coupling constant, $g_{1D}$, which encapsulates the interspecies
atom-atom interaction strength. This coupling constant can be tuned to essentially
any value, including zero and $\pm\infty$, by varying the 3d $s$-wave scattering
length $a_{3d}$ through application of an external magnetic field in the proximity
of a Feshbach resonance.

The role of interactions in quasi one dimensional atomic Fermi gases has been
studied mainly in connection with Luttinger liquid theory
\cite{Xianlong02,Gleisberg04,Recati03,Recati03b}. Recati {\it et
al.}~\cite{Recati03,Recati03b} investigate the properties of a two-component Fermi
gas with {\em{repulsive}} interspecies interactions confined in highly-elongated
harmonic traps. In the limit of weak and strong coupling these authors relate the
parameters of the Luttinger Hamiltonian, which describe the low-energy properties of
the gas, to the microscopic parameters of the system. The prospect of realizing
Luttinger liquids with cold fermionic atoms is fascinating since it would allow
detailed investigations of strongly correlated many-body systems, which play a
central role in condensed matter physics\footnote{See, {\it e.g.}, \cite{Voit95}},
to be conducted.

In homogeneous 1D Fermi gases with attractive interactions, sound waves propagate
with a well defined velocity, while spin waves exhibit a gap \cite{Krivnov75}.
Furthermore, in the strong-coupling regime, the ground state is comprised of bosonic
molecules (consisting of two fermions with different spin), whose spatial size is
much smaller than the average intermolecular distance \cite{Krivnov75}.
Consequently, BCS-type equations have been discussed for effectively attractive 1D
interactions \cite{Casas91}. The quasi one dimensional molecular Bose gas discussed here (see
also Ref.~\cite{Tokatly04,Fuchs04}) has similarities with the formation of a
molecular Bose-Einstein condensate (BEC) from a 3d Fermi sea close to a magnetic
atom-atom Feshbach resonance \cite{Greiner03, Zwierlein03}.

This Chapter investigates the properties of inhomogeneous quasi one dimensional
two-component Fermi gases under harmonic confinement with {\it attractive} and {\it
repulsive} interspecies interactions. Our study is based on the exact equation of
state of a homogeneous 1D system of fermions with zero-range attractive
\cite{Gaudin67,Krivnov75} and repulsive \cite{Yang67} interactions treated within
the local density approximation (Sec.~\ref{secLDA}). We calculate the energy per
particle, the size of the cloud, and the frequency of the lowest compressional mode
as a function of the effective 1D coupling constant, including infinitely strong
attractive and repulsive interactions. Our predictions for the size of the cloud and
for the breathing mode frequency have immediate implications for experimental
studies. It has been shown recently for quasi one dimensional Bose
gases~\cite{Moritz03} that precise measurements of collective mode frequencies can
provide evidence for beyond mean-field effects. For attractive interactions we
discuss the cross-over from the weak- to the strong-coupling regime and point out
the possibility of forming a mechanically stable molecular Tonks-Girardeau gas.


\section{Model}

Consider a two-component atomic Fermi gas confined in a highly-elongated trap. The
fermionic atoms are assumed to belong to the same atomic species, that is, to have
the same mass $m$, but to be trapped in different hyperfine states $\sigma$, where
$\sigma$ represents a generalized spin or angular momentum, $\sigma=\uparrow$ or
$\downarrow$. The trapping potential is assumed to be harmonic and axially
symmetric,
\begin{equation}
V_{trap}=\sum_{i=1}^N\frac{1}{2}m\left( \omega_\rho^2\rho_i^2 + \omega_z^2 z_i^2 \right)
\label{trap}
\end{equation}
Here, $\rho_i=\sqrt{x_i^2+y_i^2}$ and $z_i$ denote, respectively, the radial and
longitudinal coordinate of the $i$th atom; $\omega_{\rho}$ and $\omega_z$ denote,
respectively, the angular frequency in the radial and longitudinal direction; and
$N$ denotes the total number of atoms. We require the anisotropy parameter
$\lambda$, $\lambda=\omega_z/\omega_\rho$, to be so small that the transverse motion
is ``frozen'' to zero point oscillations. At zero temperature this implies that the
Fermi energy associated with the longitudinal motion of the atoms in the absence of
interactions, $\epsilon_F=N\hbar\omega_z/2$, is much smaller than the separation
between the levels in the transverse direction, $\epsilon_F\ll\hbar\omega_\rho$.
This condition is fulfilled if $\lambda\ll 1/N$. The outlined scenario can be
realized experimentally with present-day technology using optical traps.

\begin{figure}
\begin{center}
\includegraphics*[width=7cm]{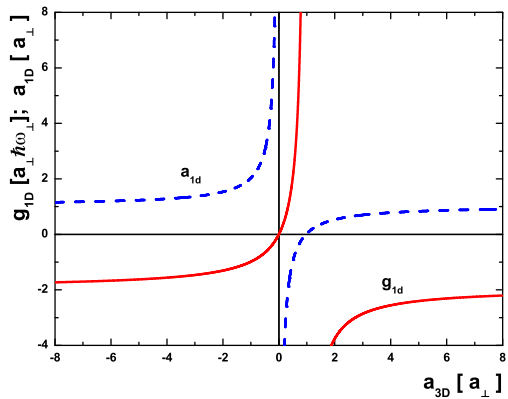}
\caption{Effective 1D coupling constant $g_{1D}$ [solid line,
Eq.~\ref{g1Dolshanii}], together with effective 1D scattering length $a_{1D}$
[dashed line, Eq.~\protect\ref{1DFa1D}] as a function of $a_{3d}$.}
\label{fig1Df1}
\end{center}
\end{figure}

If the Fermi gas is kinematically in 1D, it can be described by an effective 1D
Hamiltonian with contact interactions,
\begin{equation}
H=N\hbar\omega_\rho + H_{1D}^0 + \sum_{i=1}^N\frac{1}{2}m\omega_z^2z_i^2,
\label{hamiltonian1}
\end{equation}
where
\begin{equation}
H_{1D}^0=-\frac{\hbar^2}{2m}\sum_{i=1}^N\frac{\partial^2}{\partial z_i^2} + g_{1D}\sum_{i=1}^{N_\uparrow}
\sum_{j=1}^{N_\downarrow}\delta(z_i-z_j)
\label{hamiltonian2}
\end{equation}
and $N=N_{\uparrow}+N_{\downarrow}$. This effective Hamiltonian accounts for the
interspecies atom-atom interactions, which are parameterized by the 3d $s$-wave
scattering length $a_{3d}$, through the effective 1D coupling constant
$g_{1D}$~\cite{Olshanii98,Bergeman03},
\begin{equation}
g_{1D}=\frac{2\hbar^2a_{3d}}{m a_\rho^2} \frac{1}{1-A a_{3d}/a_\rho},
\label{g1Dolshanii}
\end{equation}
but neglects the typically much weaker $p$-wave interactions. In
Eq.~\ref{g1Dolshanii}, $a_\rho=\sqrt{\hbar/m\omega_\rho}$ is the characteristic
oscillator length in the transverse direction and $A=|\zeta(1/2)|/\sqrt{2}\simeq
1.0326$. Alternatively, $g_{1D}$ can be expressed through the effective 1D
scattering length $a_{1D}$, $g_{1D}=-2\hbar^2/(ma_{1D})$, where
\begin{equation}
a_{1D}=-a_\rho\left(\frac{a_\rho}{a_{3d}}-A\right)
\label{1DFa1D}
\end{equation}
Figure~\ref{fig1Df1} shows $g_{1D}$ and $a_{1D}$ as a

function of the 3d $s$-wave scattering length $a_{3d}$, which can be varied
continuously by application of an external field. The effective 1D interaction is
repulsive, $g_{1D}>0$, for $0<a_{3d}<a_{3d}^c$ ($a_{3d}^c=0.9684 a_\rho$), and
attractive, $g_{1D}<0$, for $a_{3d}>a_{3d}^c$ and for $a_{3d}<0$.
By varying $a_{3d}$, it is possible to go adiabatically from the weakly-interacting
regime ($g_{1D}\sim 0$) to the strongly-interacting repulsive regime ($g_{1D}\to
+\infty$ or $a_{3d}\lesssim a_{3d}^c$), as well as from the weakly-interacting
regime to the strongly-interacting attractive regime ($g_{1D}\to -\infty$ or
$a_{3d}\gtrsim a_{3d}^c$)\footnote{Note that Eqs.~\ref{g1D}-\ref{1DFa1D} are valid
only if the condition $|a_{1D}|\gg a_\rho^3 n_{1D}^2$ is
satisfied~\cite{Olshanii98}.}

For two fermions with different spin the Hamiltonian $H_{1D}^0$,
Eq.~\ref{hamiltonian2}, supports one bound state with binding energy
$\epsilon_{bound}=-\hbar^2/(ma_{1D}^2)$ and spatial extent $\sim a_{1D}$ for
$g_{1D}<0$, and no bound state for $g_{1D}>0$, that is, the molecular state becomes
exceedingly weakly-bound and spatially-delocalized as $g_{1D}\to 0^-$
\cite{Olshanii98,Bergeman03}. In the following we investigate the properties of a
gas with $N$ fermions, $N_{\uparrow}= N_{\downarrow}$, for both effectively
{\em{attractive and repulsive}} 1D interactions {\em{with and without}} longitudinal
confinement.

\section{Homogeneous system}

Consider the Hamiltonian $H_{1D}^0$, Eq.~\ref{hamiltonian2}, which describes a
homogeneous 1D two-component Fermi gas. The ground state energy $E_{hom}$ of
$H_{1D}^0$ has been calculated exactly using Bethe's ansatz for
attractive \cite{Gaudin67} and repulsive \cite{Yang67} interactions, and can be
expressed in terms of the linear number density $n_{1D}=N/L$, where $L$ is the size
of the system,
\begin{equation}
\frac{E_{hom}}{N}=\frac{\hbar^2n_{1D}^2}{2m}e(\gamma)
\label{homenergy1}
\end{equation}
The dimensionless parameter $\gamma$ is proportional to the coupling constant
$g_{1D}$, $\gamma=m g_{1D}/(\hbar^2n_{1D})$, while its absolute value is inversely
proportional to the 1D gas parameter $n_{1D}|a_{1D}|$, $|\gamma|=2/n_{1D}|a_{1D}|$.
The function $e(\gamma)$ is obtained by solving a set of integral
equations\footnote{Details on the numerical solution of the integral equations and a
table for the function $e(\gamma)$ can be downloaded from
\url{http://www.science.unitn.it/~astra/1Dfermions/}.}, which is similar to that
derived by Lieb and Liniger \cite{Lieb63} for 1D bosons with repulsive contact
interactions.
To obtain the energy per particle, Eq.~\ref{homenergy1}, we solve these integral
equations for $\gamma<0$~\cite{Gaudin67} and for $\gamma>0$~\cite{Yang67}.

\begin{figure}
\begin{center}
\includegraphics*[width=7cm]{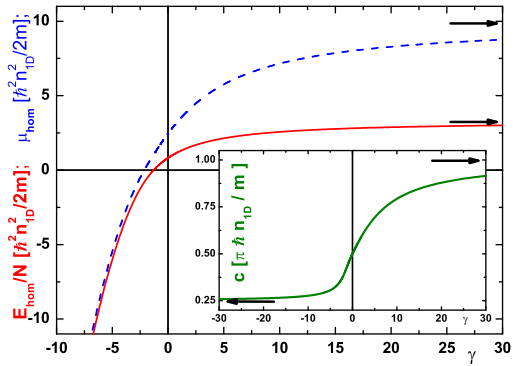}
\caption{$E_{hom}/N$ (solid line), $\mu_{hom}$ (dashed line)
and $c$ (inset) for a homogeneous two-component 1D Fermi gas as a function of
$\gamma$ (horizontal arrows indicate the asymptotic values of $E_{hom}/N$,
$\mu_{hom}$ and $c$, respectively).}
\label{fig1Df2}
\end{center}
\end{figure}

Figure~\ref{fig1Df2} shows
the energy per particle, $E_{hom}/N$ (solid line), the chemical potential
$\mu_{hom}$, $\mu_{hom}=dE_{hom}/dN$ (dashed line), and the velocity of sound $c$
(inset), which is obtained from the inverse compressibility
$mc^2=n_{1D}\partial\mu_{hom}/\partial n_{1D}$, as a function of the interaction
strength $\gamma$. In the weak coupling limit, $|\gamma|\ll 1$, $\mu_{hom}$ is given
by
\begin{equation}
\mu_{hom}=\frac{\pi^2}{4} \; \frac{\hbar^2n_{1D}^2}{2m}+
\gamma \; \frac{\hbar^2 n_{1D}^2}{2m} +
\cdots\;,
\label{limit1}
\end{equation}
where the first term on the right hand side is the energy of an ideal two-component
atomic Fermi gas, and the se\-cond term is the mean-field energy, which accounts for
interactions. The chemical potential increases with increasing $\gamma$, and reaches
an asymptotic value for $\gamma \rightarrow \infty$ (indicated by a horizontal arrow
in Fig.~\ref{fig1Df2}),
\begin{equation}
\mu_{hom}= \pi^2 \; \frac{\hbar^2n_{1D}^2}{2m}-
\frac{16 \pi^2 \ln(2)}{3 \gamma} \; \frac{\hbar^2n_{1D}^2}{2m} + \cdots\;.
\label{limit2}
\end{equation}
The first term on the right hand side coincides with the chemical potential of a
one-component ideal 1D Fermi gas with $N$ atoms, the second term has been calculated
in~\cite{Recati03}. Interestingly, for $\gamma\gg 1$, the strong atom-atom repulsion
between atoms with different spin plays the role of an effective Pauli
principle~\cite{Recati03,Recati03b}.

For attractive interactions and large enough $|\gamma|$ the energy per particle is
negative (see Fig.~\ref{fig1Df2}), reflecting the existence of a molecular Bose gas,
which consists of $N/2$ diatomic molecules with binding energy $\epsilon_{bound}$.
Each molecule is comprised of two atoms with different spin. In the limit $\gamma
\rightarrow -\infty$, the chemical potential becomes
\begin{equation}
\mu_{hom}=-\frac{\hbar^2}{2ma_{1D}^2}+
\frac{\pi^2}{16} \; \frac{\hbar^2n_{1D}^2}{2m}-
\frac{\pi^2}{12 \gamma} \;
\frac{\hbar^2n_{1D}^2}{2m} + \cdots
\label{limit3}
\end{equation}
The first term is simply $\epsilon_{bound}/2$, one half of the binding energy of the
1D molecule, while the second term is equal to half of the chemical potential of a
bosonic Tonks-Girardeau gas with density $n_{1D}/2$, consisting of $N/2$ molecules
with mass $2m$\footnote{The coefficient and sign of the third term on the right hand
side of Eq.~\ref{limit3} differ from Eq.~A7 in Ref.~\cite{Casas91}.}.
Importantly, the compressibility remains positive for $\gamma \rightarrow -\infty$
[a horizontal arrow in the inset of Fig.~\ref{fig1Df2} indicates the asymptotic
value of $c$, $c=\pi\hbar n_{1D}/(4m)$], which implies that two-component 1D Fermi
gases are mechanically stable even in the strongly-attractive regime. In contrast,
the ground state of 1D Bose gases with $g_{1D}<0$ has negative
compressibility~\cite{McGuire64} and is hence mechanically unstable.

\section{Trapped system}

Using the solutions for the homogeneous two-component 1D Fermi gas,
\begin{figure}
\begin{center}
\includegraphics*[width=7cm]{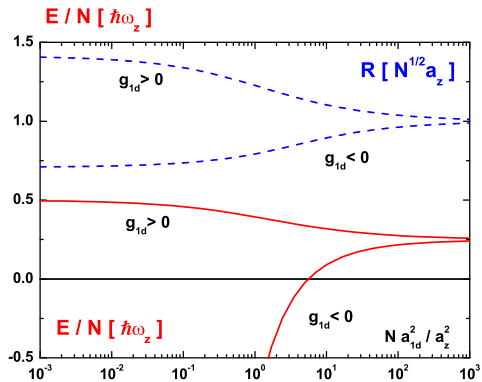}
\caption{Energy per particle, $E/N-\hbar\omega_\rho$ (solid lines),
and size of the cloud, $R$ (dashed lines), for an inhomogeneous two component 1D
Fermi gas as a function of $Na_{1D}^2/a_z^2$ for repulsive ($g_{1D}>0$) and
attractive ($g_{1D}<0$) interactions.}
\label{fig1Df3}
\end{center}
\end{figure}
we now describe the inhomogeneous gas, Eq.~\ref{hamiltonian1}, within the
LDA~\cite{Dunjko01,Menotti02,Recati03}. This approximation is applicable if the size
$R$ of the cloud is much larger than the harmonic oscillator length $a_z$ in the
longitudinal direction, $a_z=\sqrt{\hbar/m\omega_z}$, implying $\epsilon_F\gg
\hbar\omega_z$ and $N\gg 1$. The chemical potential $\mu$ of the inhomogeneous
system can be determined from the local equilibrium condition,
\begin{equation}
\mu=\mu_{hom}[n_{1D}(z)]+\frac{1}{2}m\omega_z^2z^2,
\label{lda}
\end{equation}
and the normalization condition $N=\int_{-R}^R n_{1D}(z)dz$, where $z$ is measured
from the center of the trap, $R=\sqrt{2\mu^{\prime}/(m\omega_z^2)}$, and
$\mu^{\prime}=\mu$ for $g_{1D}>0$ and $\mu^{\prime}=\mu+|\epsilon_{bound}|/2$ for
$g_{1D}<0$. The normalization condition can be reexpressed in terms of the
dimensionless chemical potential $\tilde{\mu}$ and the dimensionless density
$\tilde{n}_{1D}$ [$\tilde{\mu}=\mu^{\prime}/(\hbar^2/2ma_{1D}^2)$ and
$\tilde{n}_{1D}=|a_{1D}|n_{1D}$],
\begin{equation}
N\frac{a_{1D}^2}{a_z^2}=
\int_0^{\tilde{\mu}} \frac{\tilde{n}_{1D}(\tilde{\mu}-x)}{\sqrt{x}} dx\;.
\label{normalization2}
\end{equation}
This expression emphasizes that the coupling strength is determined by
$Na_{1D}^2/a_z^2$; $Na_{1D}^2/a_z^2\gg 1$ corresponds to the weak coupling and
$Na_{1D}^2/a_z^2\ll 1$ to the strong coupling regime, irrespective of whether the
interactions are attractive or repulsive~\cite{Menotti02}.

Figure~\ref{fig1Df3} shows the energy per particle $E/N$ and the size $R$ of the
cloud as a function of the coupling strength $Na_{1D}^2/a_z^2$ for positive and
negative $g_{1D}$ calculated within the LDA for an inhomogeneous two-component 1D
Fermi gas. Compared to the non-interacting gas, for which $R=\sqrt{N}a_z$, $R$
increases for repulsive interactions and decreases for attractive interactions. For
$Na_{1D}^2/a_z^2 \ll 1$, $R$ reaches the asymptotic value $\sqrt{2N}a_z$ for the
strongly repulsive regime, $g_{1D} \rightarrow +\infty$, and the value
$\sqrt{N/2}a_z$ for the strongly attractive regime, $g_{1D} \rightarrow -\infty$.
The shrinking of the cloud for attractive interactions reflects the formation of
tightly bound molecules. In the limit $g_{1D} \rightarrow -\infty$, the energy per
particle approaches $\epsilon_{bound}/2+N \hbar \omega_z/8 + \hbar \omega_{\rho}$,
indicating the formation of a molecular bosonic Tonks-Girardeau gas, consisting of
$N/2$ molecules. The size of the cloud shrinks from $R=\sqrt{2N}a_z$ in the strongly
repulsive regime ($Na_{1D}^2/a_z^2\ll 1$ and $g_{1D}>0$) to $R=\sqrt{N/2}a_z$ in the
strongly attractive regime ($Na_{1D}^2/a_z^2\ll 1$ and $g_{1D}<0$). We also notice
that, similarly to the homogeneous case, for large attractive interactions the
energy per particle approaches the molecular binding energy $\epsilon_{bound}$.

Using a sum rule approach, the frequency $\omega$ of the lowest compressional
(breathing) mode of harmonically trapped 1D gases can be calculated from the
mean-square size of the cloud $\langle z^2 \rangle$~\cite{Menotti02},
\begin{equation}
\omega^2=-2\frac{\langle z^2\rangle }{d\langle z^2\rangle/d\omega_z^2}
\label{collmode}
\end{equation}
In the weak and strong coupling regime ($Na_{1D}^2/a_z^2 \gg 1$ and $\ll 1$,
respectively), $\langle z^2 \rangle$ has the same dependence on $\omega_z$ as the
ideal 1D Fermi gas. Consequently, $\omega$ is in these limits given by $2\omega_z$,
irrespective of whether the interaction is repulsive or attractive.
%
Solid lines in Fig.~\ref{fig1Df3} show $\omega^2$, determined numerically from
Eq.~\ref{collmode}, as a function of the interaction strength $N a_{1D}^2/a_z^2$.
A non-trivial behavior of $\omega^2$ as a function of $Na_{1D}^2/a_z^2$ is visible.
To gain further insight, we calculate the first correction $\delta\omega$ to the
breathing mode frequency $\omega$
\begin{figure}
\begin{center}
\includegraphics*[width=0.7\textwidth]{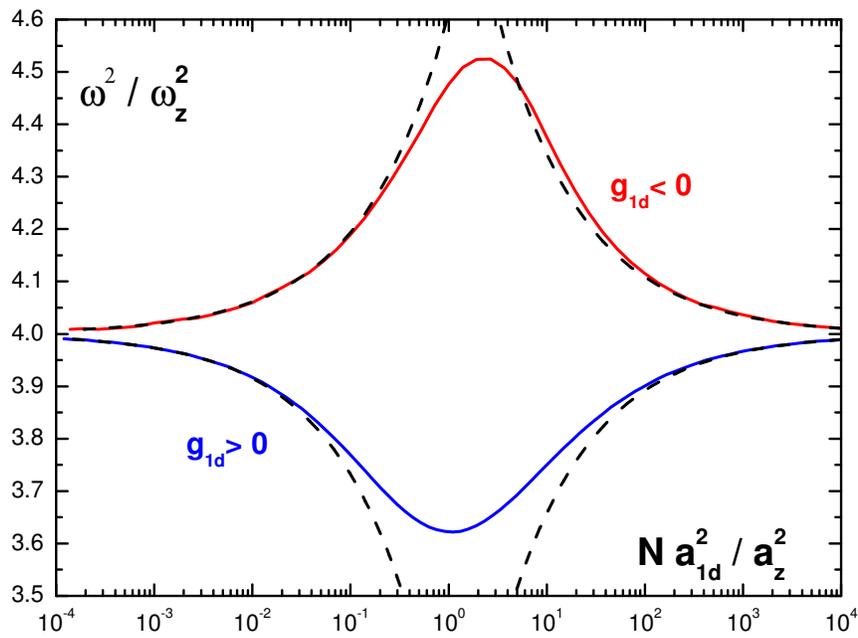}
\caption{Square of the lowest breathing mode frequency,
$\omega^2$, as a function of the coupling strength $Na_{1D}^2/a_z^2$ for an
inhomogeneous two-component 1D Fermi gas with repulsive ($g_{1D}>0$) and attractive
($g_{1D}<0$) interactions determined numerically from Eq.~\ref{collmode} (solid
lines). Dashed lines show analytic expansions.}
\label{fig1Df4}
\end{center}
\end{figure}
[$\omega=2\omega_z(1+\delta\omega/\omega_z+\cdots)$] analytically for weak repulsive
and attractive interactions, as well as for strong repulsive and attractive
interactions. For the weak coupling regime, we find $\delta\omega/\omega_z=\pm
(4/3\pi^2)/(Na_{1D}^2/a_z^2)^{1/2}$, where the minus sign applies to repulsive
interactions and the plus sign to attractive interactions. For the strong coupling
regime, we find
$\delta\omega/\omega_z=-[16\sqrt{2}\ln(2)/15\pi^2](Na_{1D}^2/a_z^2)^{1/2}$ for
repulsive interactions and
$\delta\omega/\omega_z=(8\sqrt{2}/15\pi^2)(Na_{1D}^2/a_z^2)^{1/2}$ for attractive
interactions (see Table~1.1).
Dashed lines in Fig.~\ref{fig1Df4} show the resulting analytic expansions for
$\omega^2$, which describe the lowest breathing mode frequency quite well over a
fairly large range of interaction strengths but break down for $Na_{1D}^2/a_z^2\sim
1$.

\section{Conclusions}

In conclusion, we have investigated the cross-over from weak to strong coupling of
quasi one dimensional harmonically trapped two-component Fermi gases with both
repulsive and attractive effective interactions. The frequency of the lowest
breathing mode, which can provide an experimental signature of the cross-over, is
calculated. We predict the existence of a stable molecular Tonks-Girardeau gas in
the strongly attractive regime.
%
%

\chapter{BEC-BCS crossover\label{secBECBCS}}
\section{Introduction}

Recent experiments on two-component ultracold atomic Fermi gases near a Feshbach
resonance have opened the possibility of investigating the crossover from a
Bose-Einstein condensate (BEC) to a Bardeen-Cooper-Schrieffer (BCS) superfluid. In
these systems the strength of the interaction can be varied over a very wide range
by magnetically tuning the two-body scattering amplitude. For positive values of the
$s$-wave scattering length $a$, atoms with different spins are observed to pair into
bound molecules which, at low enough temperature, form a Bose
condensate~\cite{Jochim03,Greiner03,Zwierlein03}. The molecular BEC state is
adiabatically converted into an ultracold Fermi gas with $a<0$ and $k_F|a|\ll
1$~\cite{Bartenstein04,Bourdel04}, where standard BCS theory is expected to apply.
In the crossover region the value of $|a|$ can be orders of magnitude larger than
the inverse Fermi wave vector $k_F^{-1}$ and one enters a new strongly-correlated
regime known as unitary limit~\cite{O'Hara02,Bartenstein04b,Bourdel04}. In dilute
systems, for which the effective range of the interaction $R_0$ is much smaller than
the mean interparticle distance, $k_FR_0\ll 1$, the unitary regime is believed to be
universal~\cite{Heiselberg01,Bruun04,Palo04,Diener04}. In this regime, the only
relevant energy scale should be given by the energy of the noninteracting Fermi gas,
\begin{equation}
\epsilon_{FG}=\frac{3}{10}\frac{\hbar^2k_F^2}{m} \;.
\label{fermienergy}
\end{equation}

The unitary regime presents a challenge for many-body theoretical approaches because
there is not any obvious small parameter to construct a well-posed theory. The first
theoretical studies of the BEC-BCS crossover at zero temperature are based on the
mean-field BCS equations~\cite{Leggett80,Nozieres85,Engelbrecht97}. More
sophisticated approaches take into account the effects of
fluctuations~\cite{Pieri00,Pieri04}, or include explicitly the bosonic molecular
field~\cite{Holland01,Ohashi03}. These theories provide a correct description in the
deep BCS regime, but are only qualitatively correct in the unitary limit and in the
BEC region. In particular, in the BEC regime the dimer-dimer scattering length has
been calculated exactly from the solution of the four-body problem, yielding
$a_m=0.6 a$~\cite{Petrov04}. Available results for the equation of state in this
regime do not describe correctly the repulsive molecule-molecule
interactions~\cite{Holland04}.

Quantum Monte Carlo techniques are the best suited tools for treating
strongly-correlated systems. These methods have already been applied to ultracold
degenerate Fermi gases in a recent work by Carlson {\it et al.}~\cite{Carlson03}. In
this study the energy per particle of a dilute Fermi gas in the unitary limit is
calculated with the fixed-node Green's function Monte Carlo method (FN-GFMC) giving
the result $E/N=\xi\epsilon_{FG}$ with $\xi=0.44(1)$. In a subsequent
work~\cite{Chang04}, the same authors have extended the FN-GFMC calculations to
investigate the equation of state in the BCS and BEC regimes. Their results in the
BEC limit are compatible with a repulsive molecular gas, but the equation of state
has not been extracted with enough precision.

In the present Chapter, we report results for the equation of state of a Fermi gas
in the BEC-BCS crossover region using the fixed-node diffusion Monte Carlo method
(FN-DMC). The interaction strength is varied over a very broad range from $-6\le
-1/k_Fa\le 6$, including the unitary limit and the deep BEC and BCS regimes. In the
unitary and in the BCS limit we find agreement, respectively, with the results of
Ref.~\cite{Carlson03} and with the known perturbation expansion of a weakly
attractive Fermi gas~\cite{Huang57,Lee57}. In the BEC regime, we find a gas of
molecules whose repulsive interactions are well described by the dimer-dimer
scattering length $a_m=0.6 a$. Results for the pair correlation functions of
parallel and antiparallel spins are reported in the various regimes. In the BEC
regime we find agreement with the pair correlation function of composite bosons
calculated using the Bogoliubov approximation.

\section{Model}

The homogeneous two-component Fermi gas is described by the Hamiltonian
\begin{equation}
\hat H=-\frac{\hbar^2}{2m}\left( \sum_{i=1}^{N_\uparrow}\nabla^2_i + \sum_{i^\prime=1}^{N_\downarrow}\nabla^2_{i^\prime}\right)
+\sum_{i,i^\prime}V(r_{ii^\prime}) \;,
\label{hamiltonian}
\end{equation}
where $m$ denotes the mass of the particles, $i,j,...$ and $i^\prime,j^\prime,...$
label, respectively, spin-up and spin-down particles and
$N_\uparrow=N_\downarrow=N/2$, $N$ being the total number of atoms. We model the
interspecies interatomic interactions using an attractive square-well potential:
$V(r)=-V_0$ for $r<R_0$, and $V(r)=0$ otherwise. In order to ensure that the mean
interparticle distance is much larger than the range of the potential we use
$nR_0^3=10^{-6}$, where $n=k_F^3/(3\pi^2)$ is the gas number density. By varying the
depth $V_0$ of the potential one can change the value of the $s$-wave scattering
length, which for this potential is given by $a=R_0[1-\tan(K_0R_0)/(K_0R_0)]$, where
$K_0^2=mV_0/\hbar^2$. We vary $K_0$ is the range: $0<K_0<\pi/R_0$. For
$K_0R_0<\pi/2$ the potential does not support a two-body bound state and $a<0$. For
$K_0R_0>\pi/2$, instead, the scattering length is positive, $a>0$, and a molecular
state appears whose binding energy $\epsilon_b$ is determined by the trascendental
equation $\sqrt{|\epsilon_b|m/\hbar^2}R_0\tan(\bar{K}R_0)/(\bar{K}R_0)=-1$, where
$\bar{K}^2=K_0^2-|\epsilon_b|m/\hbar^2$. The value $K_0=\pi/(2R_0)$ corresponds to
the unitary limit where $|a|=\infty$ and $\epsilon_b=0$.

In the present study we resort to the Fixed Node Monte Carlo technique described in
Sec.~\ref{fndmc}. We make use of the following trial wave functions. A BCS wave
function
\begin{equation}
\psi_{BCS}({\bf R})={\cal A} \left( \phi(r_{11^\prime})\phi(r_{22^\prime})...\phi(r_{N_\uparrow N_\downarrow})\right) \;,
\label{psiBCS}
\end{equation}
and a Jastrow-Slater (JS) wave function
\begin{equation}
\psi_{JS}({\bf R})=\prod_{i,i^\prime}\varphi(r_{ii^\prime}) \left[ {\cal A}\prod_{i,\alpha}
e^{i{\bf k}_\alpha\cdot{\bf r}_i} \right] \left[ {\cal A}\prod_{i^\prime,\alpha}
e^{i{\bf k}_\alpha\cdot{\bf r}_{i^\prime}} \right] \;,
\label{psiJS}
\end{equation}
where ${\cal A}$ is the antisymmetrizer operator ensuring the correct antisymmetric
properties under particle exchange. In the JS wave function, Eq. (\ref{psiJS}), the
plane wave orbitals have wave vectors ${\bf k}_\alpha=2\pi/L(\ell_{\alpha x}
\hat{x}+\ell_{\alpha y}\hat{y}+\ell_{\alpha z}\hat{z})$, where $L$ is the size of the periodic cubic box fixed by $nL^3=N$,
and $\ell$ are integer numbers. The correlation functions $\phi(r)$ and $\varphi(r)$
in Eqs. (\ref{psiBCS})-(\ref{psiJS}) are constructed from solutions of the two-body
Schr\"odinger equation with the square-well potential $V(r)$. In particular, in the
region $a>0$ we take for the function $\phi(r)$ the bound-state solution
$\phi_{bs}(r)$ with energy $\epsilon_b$ and in the region $a<0$ the unbound-state
solution corresponding to zero scattering energy:
$\phi_{us}(r)=(R_0-a)\sin(K_0r)/[r\sin(K_0R_0)]$ for $r<R_0$ and
$\phi_{us}(r)=1-a/r$ for $r>R_0$. In the unitary limit, $|a|\to\infty$,
$\phi_{bs}(r)=\phi_{us}(r)$.

\begin{figure}[ht!]
\begin{center}
\includegraphics*[width=0.6\textwidth]{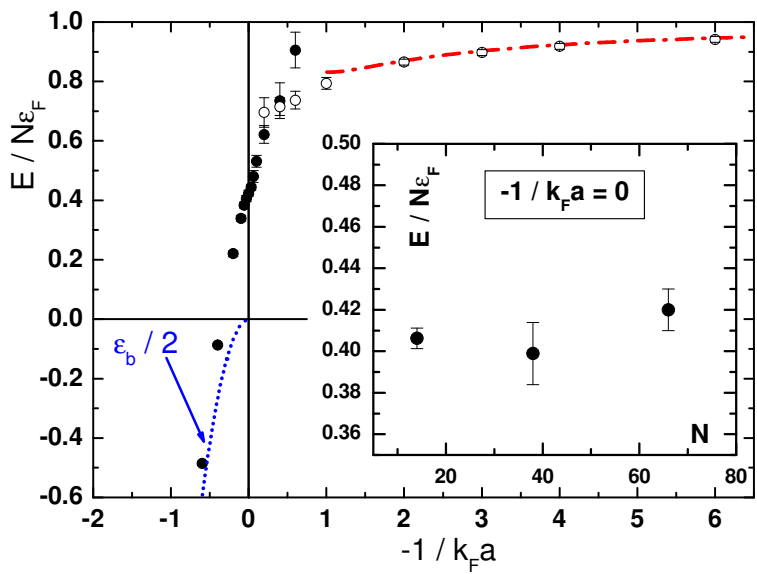}
\caption{Energy per particle in the BEC-BCS crossover. Solid symbols refer to
results obtained with the trial wave function $\psi_{BCS}$, open symbols refer to
the ones obtained with $\psi_{JS}$. The red dot-dashed line is the expansion
(\ref{BCSexp}) holding in the BCS region and the blue dotted line corresponds to the
binding energy $\epsilon_b/2$. Inset: finite size effects in the unitary limit
$-1/k_Fa=0$.}
\label{figBECBCS1}
\end{center}
\end{figure}

The JS wave function $\psi_{JS}$, Eq. (\ref{psiJS}), is used only in the region of
negative scattering length, $a<0$, with a Jastrow factor $\varphi(r)=\phi_{us}(r)$
for $r<\bar{R}$. In order to reduce possible size effects due to the long range tail
of $\phi_{us}(r)$, we have used $\varphi(r)=C_1+C_2\exp(-\alpha r)$ for $r>\bar{R}$,
with $\bar{R}<L/2$ a matching point. The coefficients $C_1$ and $C_2$ are fixed by
the continuity condition for $\varphi(r)$ and its first derivative at $r=\bar{R}$,
whereas the parameter $\alpha>0$ is chosen in such a way that $\varphi(r)$ goes
rapidly to a constant. Residual size effects have been finally determined carrying
out calculations with an increasing number of particles $N=14$, 38, and 66. In the
inset of Fig.~\ref{figBECBCS1} we show the dependence of the energy per particle
$E/N$ on $N$ in the unitary limit. Similar studies carried out in the BEC and BCS
regime show that the value $N=66$ is optimal since finite-size corrections in the
energy are below the reported statistical error in the whole BEC-BCS crossover. We
have also checked that effects due to the finite range $R_0$ of the potential are
negligible.

\section{Results}

The FN-DMC energies for $N=66$ atoms and the potential $V(r)$ with $nR_0^3=10^{-6}$
are shown in Fig.~\ref{figBECBCS1} and in Table~\ref{tab1} as a function of the
interaction parameter $-1/k_Fa$. The numerical simulations are carried out both with
the BCS wave function, Eq. (\ref{psiBCS}), and with the JS wave function, Eq.
(\ref{psiJS}). For $-1/k_Fa>0.4$ we find that $\psi_{JS}$ gives lower energies,
whereas for smaller values of $-1/k_Fa$, including the unitary limit and the BEC
region, the function $\psi_{BCS}$ is preferable. This behavior reflects the level of
accuracy of the variational {\it ansatz} for the nodal structure of the trial wave
function. We believe that in the intersection region, $-1/k_Fa\sim 0.4$, both wave
functions $\psi_{BCS}$ and $\psi_{JS}$ give a poorer description of the exact nodal
structure of the state, resulting in a less accurate estimate of the energy. In the
BCS region, $-1/k_Fa>1$, our results for $E/N$ are in agreement with the
perturbation expansion of a weakly attractive Fermi gas\footnote{Note that for
$k_F|a|\ll 1$ the nonanalytic correction to the ground-state energy due to the
superfluid gap is exponentially small.}~\cite{Huang57,Lee57}
\begin{equation}
\frac{E}{N\epsilon_{FG}}= 1+\frac{10}{9\pi}k_Fa+\frac{4(11-2\log2)}{21\pi^2}(k_Fa)^2+... \;.
\label{BCSexp}
\end{equation}

\begin{table}[ht!]
\label{tab1}
\center{
\begin{tabular}{|c|c|c|c|}
\hline
$-1/k_Fa$ & $E/N$ & $\epsilon_b/2$ & $E/N-\epsilon_b/2$ \\
\hline
\hline
-6   & -73.170(2)  & -73.1804 & 0.010(2)  \\
-4   & -30.336(2)  & -30.3486 & 0.013(2)  \\
-2   & -7.071(2)   &  -7.1018 & 0.031(2)  \\
-1   & -1.649(3)   &  -1.7196 & 0.071(3)  \\
-0.4 & -0.087(6)   &  -0.2700 & 0.183(6)  \\
-0.2 &  0.223(1)   &  -0.0671 & 0.29(1)   \\
 0   &  0.42(1)    &   0      & 0.42(1)   \\
 0.2 &  0.62(3)    &   0      & 0.62(3)   \\
 0.4 &  0.72(3)    &   0      & 0.72(3)   \\
 1   &  0.79(2)    &   0      & 0.79(2)   \\
 2   &  0.87(1)    &   0      & 0.87(1)   \\
 4   &  0.92(1)    &   0      & 0.92(1)   \\
 6   &  0.94(1)    &   0      & 0.94(1)   \\
\hline
\end{tabular}}
\caption{Energy per particle and binding energy in the BEC-BCS crossover
(energies are in units of $\epsilon_{FG}$).}
\end{table}

\begin{figure}[ht!]
\begin{center}
\includegraphics*[width=0.6\textwidth]{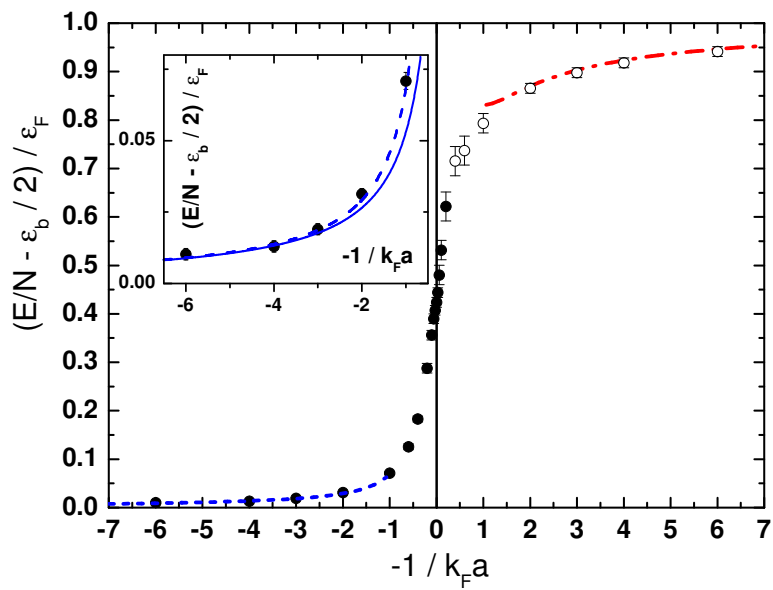}
\caption{Energy per particle in the BEC-BCS crossover with the binding energy
subtracted from $E/N$. Solid symbols: results with $\psi_{BCS}$, open symbols:
results with $\psi_{JS}$. The red dot-dashed line is as in Fig.~\ref{figBECBCS1} and
the blue dashed line corresponds to the expansion (\ref{BECexp}) holding in the BEC
regime. Inset: enlarged view of the BEC regime $-1/k_Fa\le-1$. The solid blue line
corresponds to the mean-field energy [first term in the expansion (\ref{BECexp})],
the dashed blue line includes the beyond mean-field correction Eq.~\ref{BECexp}).}
\label{figBECBCS2}
\end{center}
\end{figure}

\begin{figure}[ht!]
\begin{center}
\includegraphics*[width=0.6\textwidth]{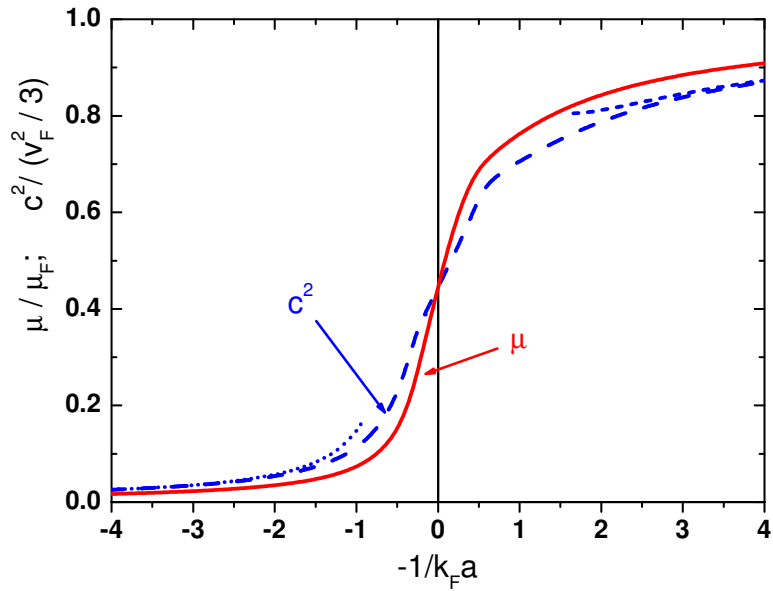}
\caption{Chemical potential $\mu$ (red solid line) and square of the speed of sound
$c^2$ (blue long dashed line) in the BEC-BCS crossover calculated from a best fit to
the equation of state. The blue short-dashed line and the blue dotted line
correspond to $c^2$ calculated respectively from the expansion (\ref{BCSexp}) and
(\ref{BECexp}).}
\label{figBECBCS3}
\end{center}
\end{figure}

In the unitary limit we find $E/N=\xi\epsilon_{FG}$, with $\xi=0.42(1)$. This result
is compatible with the findings of Refs.~\cite{Carlson03,Chang04} obtained using a
different trial wave function which includes both Jastrow and BCS correlations. The
value of the parameter $\beta=\xi-1$ has been measured in experiments with trapped
Fermi gases~\cite{O'Hara02,Bartenstein04b,Bourdel04}, but the precision is too low
to make stringent comparisons with theoretical predictions. In the region of
positive scattering length $E/N$ decreases by decreasing $k_Fa$. At approximately
$-1/k_Fa\simeq-0.3$, the energy becomes negative, and by further decreasing $k_Fa$
it rapidly approaches the binding energy per particle $\epsilon_b/2$ indicating the
formation of bound molecules~\cite{Chang04}. The results with the binding energy
subtracted from $E/N$ are shown in Fig.~\ref{figBECBCS2}. In the BEC region,
$-1/k_Fa<-1$, we find that the FN-DMC energies agree with the equation of state of a
repulsive gas of molecules
\begin{equation}
\frac{E/N-\epsilon_b/2}{\epsilon_{FG}}=\frac{5}{18\pi}k_Fa_m\left[1+\frac{128}{15\sqrt{6\pi^3}}(k_Fa_m)^{3/2}+...\right] \;,
\label{BECexp}
\end{equation}
where the first term corresponds to the mean-field energy of a gas of molecules of
mass $2m$ and density $n/2$ interacting with the positive molecule-molecule
scattering length $a_m$, and the second term corresponds to the first beyond
mean-field correction~\cite{Lee57b}. If for $a_m$ we use the value calculated by
Petrov {\it et al.}~\cite{Petrov04} $a_m=0.6a$, we obtain the curves shown in
Fig.~\ref{figBECBCS2}. If, instead, we use $a_m$ as a fitting parameter to our
FN-DMC results in the region $-1/k_Fa\le-1$, we obtain the value $a_m/a=0.62(1)$.
From a best fit to the equation of state we calculate the chemical potential
$\mu=dE/dN$ and the inverse compressibility $mc^2=n\partial\mu/\partial n$, where
$c$ is the speed of sound. The results in units of the Fermi energy
$\mu_F=\hbar^2k_F^2/2m$ and of the Fermi velocity $v_F=\hbar k_F/m$ are shown in
Fig.~\ref{figBECBCS3}. A detailed knowledge of the equation of state of the
homogeneous system is important for the determination of the frequencies of
collective modes in trapped systems~\cite{Stringari04}, which have been recently
measured in the BEC-BCS crossover regime~\cite{Kinast04b,Kinast04,Bartenstein04}.

\begin{figure}[ht!]
\begin{center}
\includegraphics*[width=0.6\textwidth]{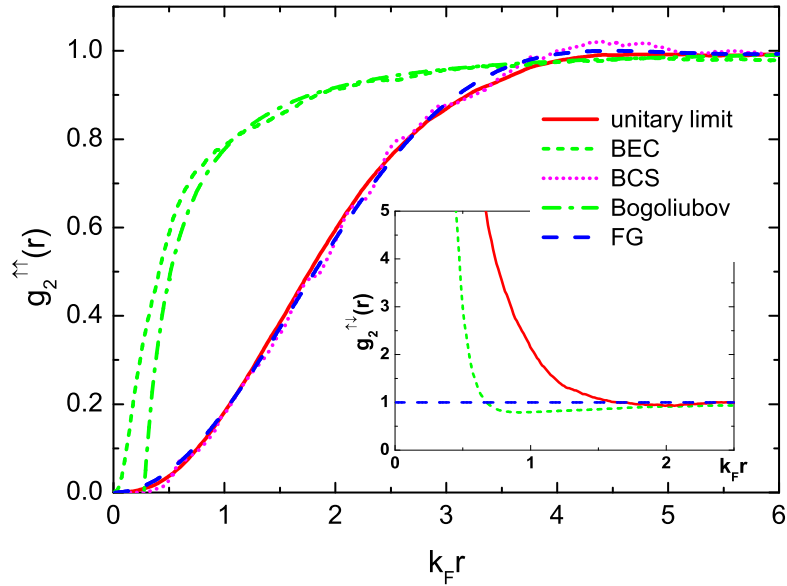}
\caption{Pair correlation function of parallel, $g_2^{\uparrow\uparrow}(r)$,
and (inset) of antiparallel spins, $g_2^{\uparrow\downarrow}(r)$, for $-1/k_Fa=0$
(unitary limit), $-1/k_Fa=-4$ (BEC regime), $-1/k_Fa=4$ (BCS regime) and for a
noninteracting Fermi gas (FG). The dot-dashed line corresponds to the pair
correlation function of a Bose gas with $a_m=0.6a$ and $-1/k_Fa=-4$ calculated using
the Bogoliubov approximation.}
\label{figBECBCS4}
\end{center}
\end{figure}

In Fig.~\ref{figBECBCS4} we show the results for the pair correlation function of
parallel, $g_2^{\uparrow\uparrow}(r)$, and antiparallel spins,
$g_2^{\uparrow\downarrow}(r)$. For parallel spins, $g_2^{\uparrow\uparrow}(r)$ must
vanish at short distances due to the Pauli principle. In the BCS regime the effect
of pairing is negligible and $g_2^{\uparrow\uparrow}(r)$ coincides with the
prediction of a noninteracting Fermi gas
$g_2^{\uparrow\uparrow}(r)=1-9/(k_Fr)^4[\sin(k_Fr)/k_Fr-\cos(k_Fr)]^2$. This result
continues to hold in the case $-1/k_Fa=0$, where it is consistent with the picture
of a gas in the unitary regime as a noninteracting Fermi gas with effective mass
$m^\star=m/\xi$. In the BEC regime the static structure factor $S(k)$ of composite
bosons can be estimated using the Bogoliubov result:
$S(k)=\hbar^2k^2/[2M\omega(k)]$, where
$\omega(k)=(\hbar^4k^4/4M^2+gn_m\hbar^2k^2/M)^{1/2}$ is the Bogoliubov dispersion
relation for particles with mass $M=2m$, density $n_m=n/2$ and coupling constant
$g=4\pi\hbar^2a_m/M$. The pair distribution function $g_2(r)$ of composite bosons,
obtained through $g_2(r)=1+2/N\sum_{\bf k}[S(k)-1]e^{-i{\bf k}\cdot{\bf r}}$ using
the value $a_m=0.6 a$, is shown in Fig.~\ref{figBECBCS4} for $-1/k_Fa=-4$ and
compared with the FN-DMC result. For large distances $r\gg a_m$, where Bogoliubov
approximation is expected to hold, we find a remarkable agreement. This result is
consistent with the equation of state in the BEC regime and shows that structural
properties of the ground state of composite bosons are described correctly in our
approach. For antiparallel spins, $g_2^{\uparrow\downarrow}(r)$ exhibits a large
peak at short distances due to the attractive interaction. In the BEC regime the
short range behavior is well described by the exponential decay
$g_2^{\uparrow\downarrow}(r)\propto\exp(-2r\sqrt{|\epsilon_b|m}/\hbar)/r^2$ fixed by
the molecular wave function $\phi_{bs}(r)$. In the unitary regime correlations extend
over a considerably larger range compared to the tightly bound BEC regime. In the
BCS regime the range of $g_2^{\uparrow\downarrow}(r)$ is much larger than $k_F^{-1}$
and is determined by the coherence length $\xi_0=\hbar^2 k_F/(m\Delta)$, where
$\Delta$ is the gap parameter. In this regime the wave function we use does not
account for pairing and is inadequate to investigate the behavior of
$g_2^{\uparrow\downarrow}(r)$.

\section{Conclusions}

In conclusion, we have carried out a detailed study of the equation of state of a
Fermi gas in the BEC-BCS crossover using FN-DMC techniques. In the BCS regime and in
the unitary limit our results are in agreement with known perturbation expansions
and with previous FN-GFMC calculations~\cite{Carlson03,Chang04}, respectively. In
the BEC regime, we recover the equation of state of a gas of composite bosons with
repulsive effective interactions which are well described by the molecule-molecule
scattering length $a_m=0.6a$ recently calculated in Ref.~\cite{Petrov04}.


\chapter*{Conclusions\label{secConclusions}}
\addcontentsline{toc}{chapter}{Conclusions}
This Dissertation presents results of a thorough study of ultracold bosonic and
fermionic gases in three-dimensional and quasi-one-dimensional systems. Although the
analyses are carried out within various theoretical frameworks (Gross-Pitaevskii,
Bethe {\it ansatz}, local density approximation, etc.) the main tool of the study is
the Quantum Monte Carlo method in different modifications (variational MC, diffusion
MC, fixed-node MC). We benchmark our Monte Carlo calculations by recovering known
analytical results (perturbative theories in dilute limits, exactly solvable models,
etc.) and extend calculations to regimes, where the results are so far unknown. In
particular we calculate the equation of state and correlation functions for gases in
various geometries and with various interatomic interactions.

The main novel results can be summarized as follows.

We present exact Quantum Monte Carlo results of the ground-state energy and
structure of a Bose gas confined in highly anisotropic harmonic traps. Starting from
a 3D Hamiltonian, where interparticle interactions are modeled by a hard-sphere or a
soft-sphere potentials, we show that the system exhibits striking features due to
particle correlations. By reducing the anisotropy parameter $\lambda$, while the
number of particles $N$ and the ratio $a/a_\perp$ of scattering to transverse
oscillator length are kept fixed, the system crosses from a regime where
Gross-Pitaevskii mean-field theory applies to a regime which is well described by
the 1D Lieb-Liniger equation of state in local density approximation. In the
cross-over region both theories fail and one must resort to exact methods to account
properly for both finite size effects and residual 3D effects. For very small values
of $\lambda$ we find clear evidence, both in the energy per particle and in the
longitudinal size of the cloud, of the fermionization of the system in the
Tonks-Girardeau regime.

We use different methods for studying 
the resonant scattering of a Bose gas in a highly elongated trap, when the system
enters a quasi one dimensional regime. We make a fully three dimensional calculation
of the lowest-lying gas-like state of the many body system using a microscopic
Fixed-Node Monte Carlo method. In order to prove the presence of the confined
induced resonance predicted by Olshanii in a many-body system we make a full
microscopic one-dimensional calculation for contact interactions with renormalized
coupling constant $g_{1D}$. The resulting energies are in excellent agreement. This
agreement proves that a properly chosen many-body 1D Hamiltonian describes well 3D
Bose gases in the quasi-one dimensional regime. We consider the Lieb-Liniger and the
hard-rod equation of state of a 1D system treated within the local density
approximation, which is expected to be correct for large number of particles. Our
detailed microscopic studies suggest that these LDA treatments provide a good
description of quasi-1D Bose gases. In particular, we suggest a simple treatment of
1D systems with negative $g_{1D}$ using the hard-rod equation of state. We address
the question of stability of an inhomogeneous gas in this regime utilizing a
variational many-body framework. We find that the lowest-lying gas-like state is
stable for negative coupling constants, up to a minimum critical value of
$|g_{1D}|$. Our numerical results suggest that the stability condition can be
expressed as $n_{1D} a_{1D}
\simeq0.35$.

Properties of the Lieb-Liniger gas are investigated in details. We calculate for the
first time the behaviour of correlation functions in a wide range of the
characteristic parameter $na_{1D}$ covering Gross-Pitaevskii and Tonks-Girardeau
regimes.
We obtain the one-body density matrix $g_1(z)$ and pair distribution function
$g_2(z)$ for all densities. In particular we investigate the nontrivial regime
$na_{1D}\approx 1$ which is relevant for current experiments. We study the
dependence of the value at zero of the three-body correlation function $g_3(0)$ on
the gas parameter $na_{1D}$ and compare it with experimental results obtained at
NIST\cite{Tolra04}. We find agreement between theory and experiment.
We extract the momentum distribution $n(k)$ and static structure factor $S(k)$ for
all densities. We discuss how the presence of a harmonic trapping modifies the
properties of the system. Using the Haldane approach for one-dimensional liquids we
calculate the asymptotic behaviour of the one-body density matrix, density-density
correlation function, dynamic form factor. In particular a direct comparison with
the DMC calculation shows that the accuracy of the obtained coefficient of the
one-body density matrix decay is better than $0.3\%$ in the whole range of
densities.

We propose a novel technique of creating a metastable gas-like state of attractive
bosons by crossing a confinement induced resonance.
Such a gas has correlations even stronger than in the Tonks-Girardeau regime where
the coupling constant is very large $g_{1D}\to\infty$. We calculate the equation of
state in this ``super-Tonks'' regime using the Variational Monte Carlo method and
estimate the critical density for the onset of instability against cluster
formation. The static structure factor and one-body density matrix are calculated
exactly within the hard-rod model, which provides the correct description of the
system for small values of the gas parameter. For harmonically trapped systems we
provide explicit predictions for the frequency of the lowest compressional mode.

We have studied the motion of an impurity through the condensate at zero temperature
by solving the Gross-Pitaevskii equation in a perturbative way. We calculated the
energy of a slow impurity. We find that the $V=0$ energy agrees with Bogoliubov
theory, the velocity contribution can be written as $m^*V^2/2$, where the effective
mass $m^*$ contributes to the mass of the normal component. We find that the motion
at small velocities is dissipationless in one-, two-, and three- dimensional
systems, although motion with velocities larger than the speed of sound leads to a
non-zero drag force due to Cherenkov radiation of phonons. The expressions for the
drag force are calculated. We used results for the dynamic form factor of the exact
Lieb-Liniger theory to investigate the velocity dependence of the drag force in a 1D
system. The form factor is calculated with the help of the Haldane
method\cite{Haldane81}.
The drag force exists for arbitrarily small velocity of motion, but is very small in
the mean-field limit.

We considered a quasi-one-dimensional system of two component Fermi gas with contact
potential between fermions of different spins. We have investigated the cross-over
from weak to strong coupling of harmonically trapped gases with both repulsive and
attractive effective interactions. The frequency of the lowest breathing mode, which
can provide an experimental signature of the cross-over, is calculated. We predict
the existence of a stable molecular Tonks-Girardeau gas in the strongly attractive
regime.
We obtain description of trapped one- and three- dimensional gas in the local
density approximation for a perturbative equation of state. Obtained predictions for
the frequencies of the lowest breathing mode are compared with numerical solutions.

We have carried out a detailed study of the equation of state of a Fermi gas in the
BEC-BCS crossover using Fixed Node Monte Carlo techniques. In the BCS regime and in
the unitary limit our results are in agreement with known perturbation expansions
and with previous Fixed Node Green Function MC
calculations~\cite{Carlson03,Chang04}. In the BEC regime, in our many body
calculation we recover the equation of state of a gas of composite bosons with
repulsive effective interactions which are well described by the molecule-molecule
scattering length $a_m=0.6a$ recently calculated in Ref.~\cite{Petrov04}.

The results obtained in this dissertation are relevant for present and future
experiments. We make direct comparison the three-body loss rate of a 1D Bose system
measured in experiments\cite{Tolra04} finding good agreement. The equation of state
obtained here for the BEC-BCS crossover in a two component Fermi gase can be used to
determine frequencies of collective modes, which have been recently measured in
experiments\cite{Bartenstein04,Kinast04,Kinast04b}. It is important to note that the
methods of obtaining quasi-one-dimensional cigar-shaped systems have been developed
considerably in the last years and it is expected that many more experiments on
low-dimensional systems will appear soon. Another important point is that the
strength of interactions can be tuned in a controlled way through the application of
an external magnetic field in the proximity of a Feshbach resonance. Strengths of
interaction in quasi-one-dimensional systems can be controlled by means of
confinement induced resonance. This allows to hope that many new properties of
low-dimensional quantum systems will be measured soon and compared to theoretical
predictions.

\addcontentsline{toc}{chapter}{Referencies}
\newcommand{\etalchar}[1]{$^{#1}$}


\addcontentsline{toc}{chapter}{Appendix:}
\begin{appendix}

\chapter{Bethe {\it ansatz} solutions}
In this Section the integral equations of exactly solvable bosonic and fermionic
one-dimensional models are presented. The equations were derived by using Bethe {\it
ansatz} method (see, for example, book \cite{Gaudin83}).

\section{Lieb-Liniger equations}

A gas of repulsive bosons interacting via $\delta$-potential in one-dimensional
system is described by the Hamiltonian (\ref{LL}) with the relation between the
coupling constant $g_{1D}$ and the scattering length $a_{1D}$ given by (\ref{g1D}).
It was shown by Lieb and Liniger \cite{Lieb63} that the ground state energy can be
found from the solution of the integral equation (see for example,
\cite{Lieb63,Dunjko01}):
\begin{eqnarray}
\rho(k) =
\frac{1}{2\pi}
+\int\limits_{-1}^1\frac{2\lambda\rho(\varkappa)}{\lambda^2+(k-\varkappa)^2}\,\frac{d\varkappa}{2\pi}
\label{LLintegraleqs}
\end{eqnarray}

Normalization of the function $\rho(k)$ is related to the density $n|a_{1D}|$:
\begin{eqnarray}
\gamma = \frac{2}{n|a_{1D}|} = \frac{\lambda}{\int\limits_{-1}^1 \rho(k)\,dk,}
\label{LLna1D}
\end{eqnarray}
here we use parameter $\gamma$ which is often introduced for solving the Bethe
equations and is inversely proportional to the density.

The energy per particle $E/N = n^2 e(n|a_{1D}|) \hbar^2/2m$ is obtained from integral
\begin{eqnarray}
e(n|a_{1D}|) = 
\frac{\gamma^3}{\lambda^3}\int\limits_{-1}^1 k^2\rho(k)\,dk
\label{LLe}
\end{eqnarray}

The procedure of solving the integral equations can be following:
\begin{enumerate}
\item Fix some value of $\lambda$
\item Obtain $\rho(k)$ from (\ref{LLintegraleqs})
\item Obtain density $n|a_{1D}|$ from (\ref{LLna1D})
\item Obtain energy $e(n|a_{1D}|)$ from (\ref{LLe})
\end{enumerate}

\section{Attractive Fermi gas}

The Hamiltonain of a two-component fermi gas reads as follows:
\begin{eqnarray}
\hat H  =
\frac{\hbar^2n^2}{m}
\left[-\sum\limits_{i, \sigma}\frac{1}{2}\frac{\partial^2}{\partial z^2_{i,\sigma}}
+\frac{g_{1D}m}{\hbar^2n}\sum\limits_{i<j}^N\delta(z_{i,\uparrow}-z_{j,\downarrow})
\right]
\end{eqnarray}

In following we will express all energies in units of $\hbar^2 n^2/2m$ and all
distances in units of $|a_{1D}|$. Let us introduce notation
$\gamma = -\frac{g_{1D}m}{\hbar^2n}>0$.

The integral equation for the equation of state is
(see \cite{Krivnov75} with notation $\gamma=u/2$, $\rho=\sigma/2$):
\begin{eqnarray}
\rho(k)
= \frac{2}{\pi}
-\int\limits_{-K}^K\frac{2\gamma\,\rho(\varkappa)}{\gamma^2+(k-\varkappa)^2}
\,\frac{d\varkappa}{2\pi}
\end{eqnarray}

The normalization condition is written as
\begin{eqnarray}
na_{1D} = \int\limits_{-K}^K \rho(k)\,dk
\end{eqnarray}

Once the density $\rho(k)$ is known, the energy can be written as
\begin{eqnarray}
e(\gamma) = \frac{1}{na_{1D}}\int\limits_{-K}^K k^2\rho(k)\,dk
-\frac{\gamma^2}{4}
\end{eqnarray}

\section{Repulsive Fermi gas}

This Hamiltonain can be solved for an arbitrary number of particles spin up
$N_{\uparrow}$ and spins down $N_\downarrow$. The corresponding integral equations
are ($\gamma = \frac{g_{1D}m}{\hbar^2n}>0$)\cite{Yang67}:
\begin{eqnarray}
\left\{
{\begin{array}{ccc}
\sigma(k) &=&
-\int\limits_{-B}^B\frac{2\gamma\sigma(\varkappa)}{\gamma^2+(k-\varkappa)^2}
\frac{d\varkappa}{2\pi}
+\int\limits_{-Q}^Q\frac{4\gamma\rho(y)}{\gamma^2+4(k-\varkappa)^2}
\frac{d\varkappa}{2\pi}\\
\rho(k) &=& 1+\int\limits_{-B}^B\frac{4\gamma\sigma(\varkappa)}{\gamma^2+4(k-\varkappa)^2}
\frac{d\varkappa}{2\pi}
\end{array}}
\right.
\label{Yang}
\end{eqnarray}

The limit $B\to\infty$ correspond to $N_\downarrow = N_\uparrow$. In this limit one
can simplify further the system of integral equations by introducing a Fourier
transformation (see also discrete lattice model \cite{Coll74}):
\begin{eqnarray}
\sigma(x) = \int\limits_{-\infty}^\infty e^{-ikx}\sigma(k)\frac{dk}{2\pi}\\
\sigma(k) = \int\limits_{-\infty}^\infty e^{ikx}\sigma(x)\,dx
\label{Fourier2}
\end{eqnarray}

By multiplying first equation from (\ref{Yang}) by $e^{-ikx}/2\pi$
and integrating over $k$ one obtains expression for the
$\sigma(x)$\footnote{
It is convenient to use following equality
$\int\limits_{-\infty}^\infty\frac{e^{\pm ikx}}{c^2+a^2(\varkappa-k)^2}
\frac{dk}{2\pi}=\frac{1}{2ac}e^{\pm i\varkappa x -\frac{c}{a}|x|}$
}

\begin{eqnarray}
\sigma(x) = \frac{1}{2\ch \frac{\gamma|x|}{2}} \int\limits_{-Q}^Q e^{-ikx}\rho(k)\frac{dk}{2\pi}
\label{sigma_k}
\end{eqnarray}

Setting $x=0$ one immediately sees that number of spin-down
particles is half of the total number of particles
$\int\sigma(k)\,dk = \frac{1}{2}\int\rho(k)\,dk$.

Inserting (\ref{Fourier2}) into second equation from (\ref{Yang}),
taking into account formula (\ref{sigma_k}) and carrying out two
integrations one obtains the integral equation involving only $\rho(x)$
\begin{eqnarray}
\rho(k) = \frac{1}{2\pi} + \int\limits_{-Q}^QK(k-\varkappa)\rho(\varkappa)\,\frac{d\varkappa}{2\pi},
\label{Repulsive fermions}
\end{eqnarray}
where the kernel is

\begin{eqnarray}
K(\xi) = 
2\int\limits_0^\infty \frac{\cos\xi x}{1+e^{\gamma x}}\,dx
\end{eqnarray}

Ones this equation is solved the density and energy are given by
\begin{eqnarray}
na_{1D} = \int\limits_{-Q}^Q\rho(k)\,dk,\\
e(\gamma) = \frac{1}{na_{1D}}\int\limits_{-Q}^Qk^2\rho(k)\,dk
\end{eqnarray}

In the strongly interacting limit $\gamma\to\infty$ and the kernel can be simplified
\begin{eqnarray}
K(\xi) = 2\int\limits_0^\infty \frac{\cos\xi x}{1+e^{\gamma x}}\,dx
=2\sum\limits_{n=1}^\infty (-1)^{n+1}\frac{n\gamma}{(n\gamma)^2+\xi^2}
\approx\frac{2}{\gamma}\sum\limits_{n=1}^\infty \frac{(-1)^{n+1}}{n}
=\frac{2\ln 2}{\gamma}
\label{K1}
\end{eqnarray}

The energy per particle in units of $\left[\frac{\hbar^2}{2ma^2}\right]$ is given by
\begin{eqnarray}
E = \frac{\pi^2 n^2}{3} - \frac{2\ln(2)\pi^3 n}{3}
\end{eqnarray}
which equals to the energy of gas of $N$ free fermions of the same spin.

It is possible to express the kernel in terms of $\beta$-function (see
Gradstein-Ryzhik). Taking into account the series representation $\beta(z) =
\sum\limits_{k=0}^\infty\frac{(-1)^k}{z+k}$ one obtain following result from the
(exact) sum (\ref{K1})
\begin{eqnarray}
K(\xi) = -\frac{1}{\gamma}\left(\beta\left(\frac{i\xi}{\gamma}\right)
+\beta\left(-\frac{i\xi}{\gamma}\right)\right).
\end{eqnarray}

The $\beta$-function is defined using the digamma function $\beta(z) =
\frac{1}{2}\left(\Psi(\frac{x+1}{2})-\Psi(\frac{x}{2})\right)$. The digamma function
is defined as logarithmic derivative of the Gamma function $\Psi(z) =
\frac{\partial}{\partial z} \ln \Gamma(z)$.

The kernel can be expanded at small and large values of the argument:
\begin{eqnarray}
K(\xi) = \frac{2}{\gamma}\ln 2 -\frac{3}{2\gamma^3} \xi^2 + {\cal O}(\xi^4),\\
K(\xi) = \frac{\gamma}{2}\xi^{-2} +\frac{2\gamma^3}{4}\xi^{-4} + {\cal O}(\xi^{-6})
\end{eqnarray}

%
%

\section{Numerical solution}

In the most general form the integral equations we have to solve is written as
\begin{eqnarray}
f(x)+\int\limits_{-Y}^Y K(x-y)f(y)\,dy = g(x),
\end{eqnarray}
where $f(x)$ is so far unknown solution, $K(x)$ is the kernel, $Y$ defines the
integration limit, $g(x)$ defines the normalization (in LL case it is constant). The
function $f(x)$ enters twice: once inside the integral and second time outside,
this can be remedied inserting the $\delta$-function:
\begin{eqnarray}
\int\limits_{-Y}^Y(\delta(x-y)+K(x-y))f(y)\,dy = g(x),
\label{BetheI}
\end{eqnarray}

Now we do discretization with spacing $\Delta x$. The equation (\ref{BetheI}) now
can be expressed in the matrix form:
\begin{eqnarray}
(I+K\Lambda) \vec f \Delta x = \vec g,
\end{eqnarray}
here $I$ stands for a unity matrix and the diagonal matrix $\Lambda$ is defined by
the integration method. Now the vector $f$ is obtained by multiplication of the
inverse matrix on $\vec g$:
\begin{eqnarray}
\vec f = \frac{1}{\Delta x} (I+K\Lambda)^{-1} \vec g,
\end{eqnarray}

For a uniform grid very good precision is achieved using the Simpson method. The
matrix $\Lambda$ in this case is defined as
$\Lambda = \mbox{diag}\{\frac13,\frac13,\frac43,\frac23,\frac43,...,
\frac43,\frac23,\frac43,\frac13\}$. The residual term of the integration is very
small and can be estimated as $I_{err} = \max f^{(4)}(x)\frac{(\Delta x)^5}{2880}$
and the error in the energy (which is defined by integrating the solution $f(x)$
with the weight proportional to $x^2$) is proportional to the spacing $\Delta x$ to
the forth power.

\newpage
\section{Expansions}

Energy expansion, unit of energy $\hbar^2/2ma_{1D}^2$:

\begin{eqnarray}
\nonumber
\begin{array}{|c|c|c|c|c|}
\hline
 term    & 0th & 1st & 2nd & 3rd\\
\hline
\hline
$Attractive gas: strong interaction$&\displaystyle
-1&\displaystyle
\frac{\pi^2(na_{1D})^2}{48}&\displaystyle
\frac{\pi^2(na_{1D})^3}{96}&\displaystyle
\frac{\pi^2(na_{1D})^4}{256}\\
\hline
$Attractive gas: weak interaction$&\displaystyle
\frac{\pi^2(na_{1D})^2}{12}&\displaystyle
-n|a_{1D}|&\displaystyle
-\frac{\ln^2(na_{1D}/2)}{\pi^2}&\\
\hline
$Repulsive gas: strong interaction$&\displaystyle
\frac{\pi^2(na_{1D})^2}{3}&\displaystyle
-\frac{2\ln(2) \pi^2(na_{1D})^3}{3}&&\\
\hline
$Repulsive gas: weak interaction$&\displaystyle
\frac{\pi^2(na_{1D})^2}{12}&\displaystyle
n|a_{1D}|&\displaystyle
&\\
\hline
\end{array}
\end{eqnarray}

Expansion of the chemical potential, unit of energy $\hbar^2/2ma_{1D}^2$:

\begin{eqnarray}
\nonumber
\begin{array}{|c|c|c|c|c|}
\hline
 term    & 0th & 1st & 2nd & 3rd\\
\hline
\hline
$Attr. gas: strong interaction$&\displaystyle
-1&\displaystyle
\frac{\pi^2(na_{1D})^2}{16}&\displaystyle
\frac{\pi^2(na_{1D})^3}{24}&\displaystyle
\frac{5\pi^2(na_{1D})^4}{256}\\
\hline
$Attr. gas: weak interaction$&\displaystyle
\frac{\pi^2(na_{1D})^2}{4}&\displaystyle
-2|na_{1D}|&
\displaystyle-\frac{\ln^2(\frac{na_{1D}}{2})+2\ln(\frac{na_{1D}}{2})}{\pi^2}&\\
\hline
$Rep. gas: strong interaction$&\displaystyle
\pi^2(na_{1D})^2&\displaystyle
-\frac{8\pi^2\ln(2) (na_{1D})^3}{3}&&\\
\hline
$Rep. gas: weak interaction$&\displaystyle
\frac{\pi^2(na_{1D})^2}{4}&\displaystyle
2|na_{1D}|&\displaystyle
&\\
\hline
$LL gas: strong interaction$&\displaystyle
\pi^2(na_{1D})^2&\displaystyle
-\frac{8\pi^2}{3} (na_{1D})^3&\displaystyle
&\\
\hline
$HR gas (small density)$&\displaystyle
\pi^2(na_{1D})^2&\displaystyle
\frac{8\pi^2}{3} (na_{1D})^3&\displaystyle
&\\
\hline
\end{array}
\end{eqnarray}

The frequency of the oscillations
$\frac{\omega^2}{\omega_z^2} =4(1+\triangle\omega)$

\begin{eqnarray}
\nonumber
\begin{array}{|l|c|}
\hline
$limit$&\displaystyle
\triangle\omega\\
\hline
\hline
$Attractive gas: strong interaction$&\displaystyle
\frac{16\sqrt{2}}{15\pi^2} \frac{\sqrt{N}a_{1D}}{a_{z}}\\
\hline
$Attractive gas: weak interaction$&\displaystyle
\frac{8}{3\pi^2}/\frac{\sqrt{N}a_{1D}}{a_{z}}\\
\hline
$Repulsive gas: strong interaction$&\displaystyle
-\frac{32\sqrt{2}\ln 2}{15\pi^2} \frac{\sqrt{N}a_{1D}}{a_{z}}\\
\hline
$Repulsive gas: weak interaction$&\displaystyle
-\frac{8}{3\pi^2}/\frac{\sqrt{N}a_{1D}}{a_{z}}\\
\hline
$Lieb-Liniger gas: strong interaction$&\displaystyle
-\frac{32\sqrt{2}}{15\pi^2} \frac{\sqrt{N}a_{1D}}{a_{z}}\\
\hline
$Gas of Hard-Rods$&\displaystyle
\frac{32\sqrt{2}}{15\pi^2} \frac{\sqrt{N}a_{1D}}{a_{z}}\\
\hline
\end{array}
\end{eqnarray}

Speed of sound in units of $m/\pi\hbar n$

\begin{eqnarray}
\nonumber
\begin{array}{|c|c|c|c|c|}
\hline
 term    & 0th & 1st & 2nd & 3rd\\
\hline
\hline
$Attractive gas: strong interaction$&\displaystyle
\frac{1}{4} &\displaystyle
\frac{na_{1D}}{8}&\displaystyle
\frac{3(na_{1D})^2}{64} &\displaystyle
-\frac{3(na_{1D})^3}{128}\\
\hline
$Attractive gas: weak interaction$&\displaystyle
\frac{1}{2} &\displaystyle
-\frac{1}{\pi^2na_{1D}}&\displaystyle
\frac{\ln (2/na_{1D})-2}{\pi^4(na_{1D})^2}&\displaystyle
\\
\hline
$Repulsive gas: strong interaction$&\displaystyle
1&\displaystyle
-2\ln 2\, na_{1D}&&
\\
\hline
$Repulsive gas: weak interaction$&\displaystyle
\frac{1}{2}&\displaystyle
\frac{1}{\pi^2na_{1D}}&\displaystyle
&\\
\hline
\end{array}
\end{eqnarray}

Energy and chemical potential (LDA) in units of $N\hbar\omega_z$

\begin{eqnarray}
\nonumber
\begin{array}{|l|c|c|}
\hline
$limit$&\displaystyle \mu&\displaystyle E/N\\
\hline
\hline
$Repulsive gas: strong interaction$&\displaystyle
1-\frac{32\sqrt 2\ln 2}{9\pi^2}\frac{\sqrt{N}a_{1D}}{a_z}&\displaystyle
\frac{1}{2}\left(1-\frac{128\sqrt 2\ln 2}{45\pi^2}\frac{\sqrt{N}a_{1D}}{a_z}\right)\\
\hline
$Attractive gas: strong interaction$&\displaystyle
\frac{1}{4}\left(1+\frac{16\sqrt{2}}{9\pi^2}\frac{\sqrt N a_{1D}}{a_z}\right)&\displaystyle
\frac{1}{8}\left(1+\frac{64\sqrt{2}}{45\pi^2}\frac{\sqrt N a_{1D}}{a_z}\right)\\
\hline
$Repulsive: weak interaction$&\displaystyle
\frac{1}{2}\left(1+\frac{8}{\pi^2}/\frac{\sqrt{N}a_{1D}}{a_z}\right)&\displaystyle
\frac{1}{4}\left(1+\frac{32}{3\pi^2}/\frac{\sqrt{N}a_{1D}}{a_z}\right)\\
\hline
$Attractive gas: weak interaction$&\displaystyle
\frac{1}{2}\left(1-\frac{8}{\pi^2}/\frac{\sqrt{N}a_{1D}}{a_z}\right)&\displaystyle
\frac{1}{4}\left(1-\frac{32}{3\pi^2}/\frac{\sqrt{N}a_{1D}}{a_z}\right)\\
\hline
\end{array}
\end{eqnarray}

Size of the condensate and mean $z^2$ in units of $a_z^2$

\begin{eqnarray}
\nonumber
\begin{array}{|l|c|c|}
\hline
$limit$&\displaystyle R^2& z^2\\
\hline
\hline
$Repulsive gas: strong interaction$&\displaystyle
2N\left(1-\frac{32\sqrt 2\ln 2}{9\pi^2}\frac{\sqrt{N}a_{1D}}{a_z}\right)&\displaystyle
\frac{N}{2}\left(1-\frac{64\sqrt 2\ln 2}{15\pi^2}\frac{\sqrt{N}a_{1D}}{a_z}\right)\\
\hline
$Attractive gas: strong interaction$&\displaystyle
\frac{N}{2}\left(1+\frac{16\sqrt{2}}{9\pi^2}\frac{\sqrt N a_{1D}}{a_z}\right)&\displaystyle
\frac{N}{8}\left(1+\frac{32\sqrt{2}}{15\pi^2}\frac{\sqrt N a_{1D}}{a_z}\right)\\
\hline
$Repulsive: weak interaction$&\displaystyle
N\left(1+\frac{8}{\pi^2}/\frac{\sqrt{N}a_{1D}}{a_z}\right)&\displaystyle
\frac{N}{4}\left(1+\frac{16}{3\pi^2}/\frac{\sqrt{N}a_{1D}}{a_z}\right)\\
\hline
$Attractive gas: weak interaction$&\displaystyle
N\left(1-\frac{8}{\pi^2}/\frac{\sqrt{N}a_{1D}}{a_z}\right)&\displaystyle
\frac{N}{4}\left(1-\frac{16}{3\pi^2}/\frac{\sqrt{N}a_{1D}}{a_z}\right)\\
\hline
\end{array}
\end{eqnarray}

\chapter{Obtaining the momentum distribution from $g_1(r)$}

The asymptotic behaviour of the one body density matrix of the Lieb gas is
\begin{equation}
\rho(x) =  \frac{C}{x^\alpha}, \qquad x \gg 1
\label{OBDM_as}
\end{equation}

In order to calculate the momentum distribution one has to calculate the Fourier
transform of it
\begin{equation}
n(k) = 2 \int_0^\infty \cos kx \rho(x) dx
\end{equation}

This integral can be calculated numerically up to some cut-off distance $L$. Let us
suppose, that at distances larger than $L$ the asymptotic behavior is valid
(\ref{OBDM_as}). Than one can calculate the ``tail'' integral
analytically\footnote{Compare with \cite{Gradstein80} $\int_L^\infty x^{\mu-1} \cos x
dx = \frac{1}{2}[e^{-i\mu\pi/2}\Gamma(\mu, iL)+e^{i\mu\pi/2}\Gamma(\mu, -iL)]$}
by a substitution $t = e^{i\pi/2}kx$
\begin{eqnarray}
\int\limits_L^\infty \frac{\cos kx\,dx}{x^\alpha}
= Re \int\limits_{ikL}^\infty
\frac{e^{-i\frac{\pi}{2}(1-\alpha)}}{k^{1-\alpha}}
e^{-t}t^{-\alpha}\,dt
= Re \frac{e^{-i\frac{\pi}{2}(1-\alpha)}\Gamma(1-\alpha, ikL)}{k^{1-\alpha}}
\end{eqnarray}

Here the incomplete Gamma function is defined as
%
\begin{eqnarray}
\Gamma(\alpha, L) = \int\limits_L^\infty e^{-t} t^{\alpha-1}dt
\end{eqnarray}


If the if set $L=0$ then the integral can be simplified\footnote{See
\cite{Gradstein80} 3.761.7 $\int_0^\infty x^{\mu-1} \cos(ax) dx =
\frac{\Gamma(\mu)}{a^\mu} \cos \frac{\mu\pi}{2}, a>0, 0<Re \mu<1$}
%
\begin{equation}
\int_0^\infty \frac{\cos kx}{x^{\alpha}} dx =
\frac{\Gamma(1-\alpha)}{k^{1-\alpha}} \cos \frac{\pi(1-\alpha)}{2}, k>0, 0<\alpha<1
\end{equation}

Let us derive an expansion of the incomplete Gamma function in terms of $1/(kL)$.
For us it is convenient to use following definition of the function
\begin{eqnarray}
f(k,L,\alpha) = \int\limits_L^\infty \frac{\cos\,kx}{x^\alpha}\,dx
\label{fkk}
\end{eqnarray}

Integrating it by parts two times we obtain\footnote{In dimensionless units $y=kL$
formulae (\ref{fkk}, \ref{kff2}) look like $f(k,L,\alpha) =
k^{\alpha-1}\int\limits_{kL}^\infty \frac{\cos\,y}{y^\alpha}\,dy$ and $f(k,L,\alpha)
= \frac{1}{kL^\alpha}
\left(-\sin\,kL+\frac{\alpha}{kL}\cos\,kL\right)
-\alpha(\alpha+1) k^{\alpha-1}
\int\limits_{kL}^\infty \frac{\cos\,y}{y^{\alpha+2}}\,dy$}
\begin{eqnarray}
f(k,L,\alpha) =
\frac{1}{kL^\alpha}
\left(-\sin\,kL+\frac{\alpha}{kL}\cos\,kL\right)
-\frac{\alpha(\alpha+1)}{k^2}
\int\limits_L^\infty \frac{\cos\,kx}{x^{\alpha+2}}\,dx
\label{kff2}
\end{eqnarray}

Here the last term has the same form as (\ref{fkk}). And can be expanded in a
similar way. Continuation of this expansion leads to formula
\begin{eqnarray}
f(k,K,\alpha) =
\sum\limits_{n=0}^\infty
\frac{(-1)^n}{kL^\alpha}
\left(
\cos\,kL \frac{\alpha(\alpha+1)...(\alpha+2n)}{(kL)^{2n+1}}
-\sin\,kL\frac{\alpha(\alpha+1)...(\alpha+2n-1)}{(kL)^{2n}}
\right)
\end{eqnarray}

Another way to present it is
\begin{eqnarray}
f(k,K,\alpha) =
\frac{1}{kL^\alpha}
\sum\limits_{n=0}^\infty
\frac{\mathop{\rm Im} i^ne^{-ikL}}{(kL)^n}
\frac{(\alpha+n-1)!}{(\alpha-1)!}
\label{expansion}
\end{eqnarray}

Let us write explicitly
\begin{eqnarray}
\mathop{\rm Im} i^ne^{-ikL} = (-1)^{\mathop{\rm mod}(n+1,2)+1} f_n(kL)
\end{eqnarray}
where $\mathop{\rm mod}$ operation is integer division and the function $f_n$ is
defined as
\begin{eqnarray}
f_n(x) =
\left\{
\begin{array}{lr}
\sin x,&n=0,2,4,...\\
\cos x,&n=1,3,5,...
  \end{array}
\right.
\end{eqnarray}

Looking at the structure of the expansion (\ref{expansion}) one finds out that the
sum converges only is $kL>1$. The factorial dependence of the numerator on the order
of the term $n$ leads to divergence of the entire sum. Let us find order of the term
$n_{cr}$ when the summation procedure should be stopped. The condition is
\begin{eqnarray}
\frac{\partial}{\partial n}
\frac{(\alpha+1-n)!}{(kL)^n} = 0
\end{eqnarray}

In order to proceed further we will take use of the Stirling formula
\begin{eqnarray}
n! = \sqrt{2\pi n} n^n e^{-n}
\end{eqnarray}

Simple calculation gives
\begin{eqnarray}
\ln(\alpha+n_{cr}-1)+\frac{1}{2(\alpha+n_{cr}-1)} = \ln kL -1
\end{eqnarray}

Now we assume that $n$ is much larger than one, so we can neglect the second term.
Finally we obtain
\begin{eqnarray}
n_{cr} = kL/e
\end{eqnarray}

Another approach for the calculation of the integral is by modifying the integration
contour. First of all, let us expand the cosine into sum of complex exponents
\begin{eqnarray}
\int_L^\infty \frac{\cos x\,dx}{x^\alpha} =
\frac{1}{2}\int_L^\infty \frac{e^{ix}\,dx}{x^\alpha} +
\frac{1}{2}\int_L^\infty \frac{e^{-ix}\,dx}{x^\alpha}
\label{cont}
\end{eqnarray}

Let us calculate the first integral in this sum.
\begin{eqnarray}
I_1 = \int\limits_L^\infty \frac{e^{ix}\,dx}{x^\alpha} =
\int\limits_0^\infty \frac{e^{-y} e^{iL}\,idy}{(iy+L)^\alpha},
\end{eqnarray}
where we introduced notation $x=iy+L$, which is a complex variable $x =
\sqrt{y^2+L^2} e^{i\arctg \frac{y}{L}}$, so
\begin{eqnarray}
I_1 =
\int\limits_0^\infty e^{-y}(y^2+L^2)^{-\frac{\alpha}{2}}
e^{i(L-\alpha\arctg\frac{y}{L})}\,idy=\\
= \int\limits_0^\infty e^{-y}(y^2+L^2)^{-\frac{\alpha}{2}}
\left[
\sin\left(\alpha\arctg\frac{y}{L}-L\right)+
i\cos\left(\alpha\arctg\frac{y}{L}-L\right)
\right]
\,dy
\label{I1}
\end{eqnarray}

The second integral in (\ref{cont}) can be calculated by means of the substitution
$x = -iy + L = \sqrt{y^2+L^2}e^{-i\arctg\frac{y}{L}}$
\begin{eqnarray}
I_2 = \int\limits_L^\infty \frac{e^{-ix}\,dx}{x^\alpha} =
\int\limits_0^\infty \frac{e^{-y} e^{-iL}\,(-i)dy}{(iy-L)^\alpha}=
\int\limits_0^\infty e^{-y}(y^2+L^2)^{-\frac{\alpha}{2}}
e^{i(-L+\alpha\arctg\frac{y}{L})}\,(-i)dy
= \\=
\int\limits_0^\infty e^{-y}(y^2+L^2)^{-\frac{\alpha}{2}}
\left[
\sin\left(\alpha\arctg\frac{y}{L}-L\right)
-\cos\left(\alpha\arctg\frac{y}{L}-L\right)
\right]\,dy
\label{I2}
\end{eqnarray}

The imaginary parts of the integrals (\ref{I1}, \ref{I2}) cancel each other and the
result is real
\begin{eqnarray}
\int_L^\infty \frac{\cos kx\,dx}{x^\alpha}
= k^{\alpha-1}\int\limits_0^\infty e^{-y}(y^2+(kL)^2)^{-\frac{\alpha}{2}}
\sin\left(\alpha\arctg\frac{y}{kL}-kL\right)\,dy
\end{eqnarray}


\end{appendix}

\chapter*{Acknowledgements}
\addcontentsline{toc}{chapter}{Acknowledgements}
I would like to gratefully acknowledge the enthusiastic supervision of Dr. Stefano
Giorgini and Prof. Lev P. Pitaevskii. They taught me so many new things and were
always ready to answer any of my countless questions.

I owe special thanks to
Prof. Sandro Stringari without whom this Dissertation would not exist. I want to
thank him for giving me the opportunity to work in his group, for being a great
mentor, for giving me invaluable help in solving the organizational problems.


I am grateful to the many people I have met in the
Trento BEC group, in the University of Trento and who have assisted me so much in
the course of this work.

\end{document}